%% file: shear_bands_lattice.tex
\documentclass[10pt,a4paper]{article}

\usepackage[T1]{fontenc}
\usepackage[utf8]{inputenc}
\usepackage{amssymb,amsmath,amsthm,amsfonts}    
\usepackage{stmaryrd}                           
\usepackage{bm}                                 
\usepackage{mathtools}                          
\usepackage{breqn}                              
\usepackage[textfont=footnotesize,labelfont={footnotesize,bf}]{caption}
\usepackage{subcaption}
\usepackage{booktabs}
\usepackage[inline]{enumitem}
\usepackage{graphicx}
\usepackage[dvipsnames]{xcolor}
\usepackage[top=22mm,right=22mm,left=22mm,bottom=22mm]{geometry}
\usepackage[colorlinks=true,linkcolor=blue,urlcolor=blue]{hyperref}
\usepackage[affil-it,auth-sc]{authblk}
\usepackage[colorinlistoftodos,textwidth=25mm,textsize=footnotesize]{todonotes}
\usepackage{lipsum}                             
\usepackage[
    backend=bibtex,
    style=numeric,
    citestyle=numeric-comp,
    maxbibnames=99,
    minbibnames=99, 
    maxcitenames=2,
    mincitenames=1,
    giveninits=true,
    sorting=none,
    sortlocale=auto,
    natbib=true,
    url=false,
    isbn=false,
    doi=true,
    eprint=true
]{biblatex}
\input{definitions}		                                
\graphicspath{{./}{./figures/}}                         

\title{Incremental constitutive tensors and strain localization for prestressed elastic lattices: Part I - quasi-static response}

\author[1]{G. Bordiga}
\author[2]{L. Cabras}
\author[1]{A. Piccolroaz}
\author[1]{D. Bigoni\footnote{Corresponding author: e-mail: \href{mailto:bigoni@ing.unitn.it}{bigoni@ing.unitn.it}; phone: +39\,0461\,282507.}}

\affil[1]{DICAM, University of Trento, Trento, Italy}
\affil[2]{DICATAM, University of Brescia, Brescia, Italy}

\date{}

\begin{document}

\maketitle

\begin{abstract}
\noindent
A lattice of elastic rods organized in a parallelepiped geometry can be axially loaded up to an arbitrary amount without distortion and then be subject to incremental displacements.
Using quasi-static homogenization theory, this lattice can be made equivalent to a prestressed elastic solid subject to incremental deformation, in such a way to obtain extremely localized mechanical responses. 
These responses can be analyzed with reference to a mechanical model which can, \textit{in principle}, be realized, so that features such as for instance shear bands inclination, or emergence of a single shear band, or competition between micro (occurring in the lattice but not in the equivalent solid) and macro (present in both the lattice and the equivalent continuum) instabilities become all designable features. 
The analysis of localizations is performed using a Green's function-based perturbative approach to highlight the correspondence between micromechanics of the composite and homogenized response of the equivalent solid. 
The presented results, limited to quasi-static behaviour, provide a new understanding of strain localization in a continuum and open new possibilities for the realization and experimentation of materials exhibiting these extreme mechanical behaviours.
Dynamic homogenization and vibrational localization are deferred to Part II of this study.
\end{abstract}

\paragraph{Keywords} 
Homogenization \textperiodcentered\ 
Ellipticity loss \textperiodcentered\
Shear bands \textperiodcentered\
Lattice buckling

\section{Introduction}
\label{sec:introduction}
Shear banding and strain localizations, usually found to emerge before failure of materials, are typically accompanied by large plastic deformation, damage, and possibly fracture.
Mechanical features of shear bands strongly depend on the tested material, so that for instance shear bands are normally inclined (to the direction of tensile stress) less in rocks than in metals. 
As a consequence, from the modelling point of view, the analysis of these material instabilities is complicated by the fact that (complex and often phenomenological)  elastoplastic constitutive laws are to be used for a material which has to be brought through and beyond several bifurcation thresholds  (corresponding for instance to surface instability or  cavitation), before encountering shear band formation, the latter typically complicated by the simultaneous emergence of elastic unloading zones adjacent to zones of intense plastic loading. 
From the experimental point of view, samples have to be brought to failure, so that experiments cannot be repeated on the same sample and the material forming the latter cannot be easily changed to analyzed different instability manifestations, for instance in such a way to alter the shear band inclination. 

Imagine now a material in which shear banding and other instabilities may occur well inside the elastic range and far from failure. 
A material that can be designed to produce shear bands with a desired inclination, or in which shear bands are the first instability occurring at increasing stress, or in which the anisotropy (not imperfections) allows the formation of only one shear band. 
Imagine that this material would be characterized by rigorously determined elastic constitutive laws (thus avoiding complications such as the double branch of the incremental constitutive laws of plasticity) and would be, at least in principle, a material realizable (for instance via 3D printing technology) and testable in laboratory conditions. 
This material would be ideal not only to \textit{theoretically} analyze instabilities, but also to \textit{practically} realize the `architected materials' which are preconized to yield extreme mechanical properties such as foldability, channelled response, and surface effects~\cite{overvelde_2017,kochmann_2017,rafsanjani_2019}.
The crucial step towards the definition of a class of these materials was made by Triantafyllidis~\cite{triantafyllidis_1985,geymonat_1993,triantafyllidis_1993,triantafyllidis_1998,nestorovic_2004,santisidavila_2016} and Ponte~Casta\~neda~\cite{pontecastaneda_1989,pontecastaneda_1991,pontecastaneda_1996,pontecastaneda_1997,pontecastaneda_2002,pontecastaneda_2002a,lopez-pamies_2006,lopez-pamies_2006a,avazmohammadi_2016}, who laid down a general framework for the homogenization of elastic composites and for the analysis of bifurcation and strain localization in these materials.
In particular, 
\begin{enumerate*}[label=(\roman*)]
    \item they showed how to realize an elastic material displaying a prestress-sensitive incremental response, exactly how it is \textit{postulated} for nonlinear elastic solids subject to incremental deformation, and
    \item provided a new understanding of strain localization phenomena, showing that a global bifurcation of a lattice structure corresponds to a loss of ellipticity of the equivalent continuum, while the latter is unaffected by a local bifurcation occurring in the composite. 
\end{enumerate*}

The aim of the present article is to extend the mentioned findings to lattices of elastic rods of arbitrary geometry and subject to nonlinear axial deformation of the elements, so to explore shear band formation and localization by applying a perturbative approach \cite{bigoni_2002a}, both to the lattice and to the equivalent continuum. 
In particular, a lattice of elastic rods organized in a parallelepiped network is an example of a composite which may be arbitrarily preloaded without introducing grid distortion, so that homogenization allows to obtain rigorous results, showing how a prestressed composite material can react to incremental displacements as an equivalent elastic continuum. 
A quasi-static approach to homogenization, based on a strain energy equivalence between the lattice and the continuum, is developed to analyze a generic lattice\footnote{For the geometries investigated in \cite{triantafyllidis_1993} our homogenization approach provides exactly the same results. Moreover, the energy equivalence provides the same results that will be derived in Part II of this article using a Floquet-Bloch dynamic approach.}, so that it becomes possible to obtain the infinite-body Green's function for the homogenized solid and compare the response of this solid to an applied concentrated force with the behaviour of the lattice at various levels of preload.
This comparison reveals, first of all, the excellent quality of the homogenization approach (so that the incremental displacement fields found in the lattice and in the homogenized material are practically coincident) and highlights the features of shear banding, so that this instability is on the one hand given a clear interpretation in terms of structural global instability of the lattice and on the other sharply discriminated from local instabilities in the composite, which remain undetected in the continuum. 
Examples of such instabilities, `invisible' in the equivalent material, are provided, which exhibit an `explosive' character, so that extend from a punctual  perturbation to the whole lattice. 
%
\begin{figure}[htb!]
    \centering
    \begin{subfigure}{0.32\textwidth}
        \centering
        \phantomcaption{\label{fig:straws_0}}
        \includegraphics[width=0.98\linewidth]{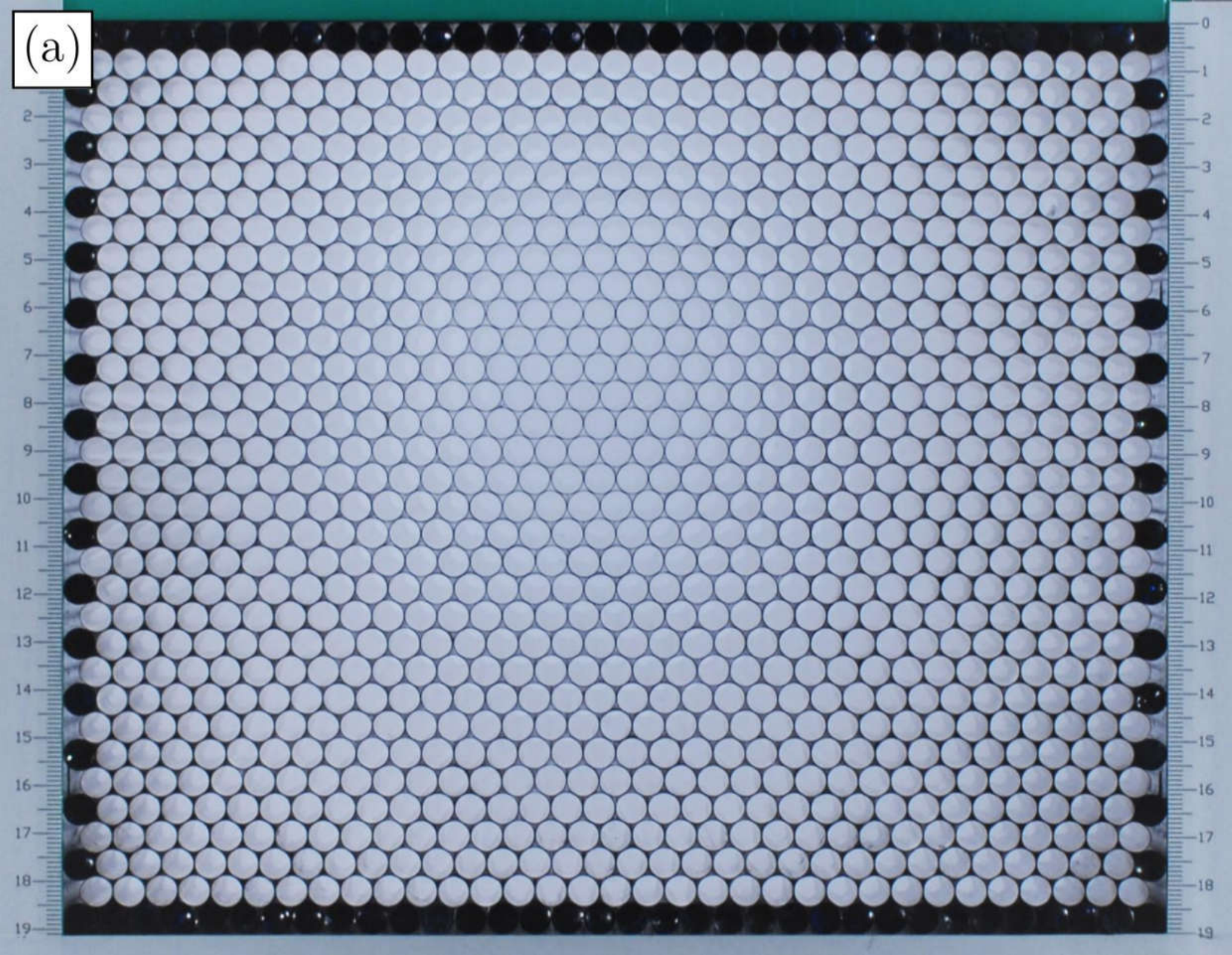}
    \end{subfigure}
    \begin{subfigure}{0.32\textwidth}
        \centering
        \phantomcaption{\label{fig:straws_1}}
        \includegraphics[width=0.98\linewidth]{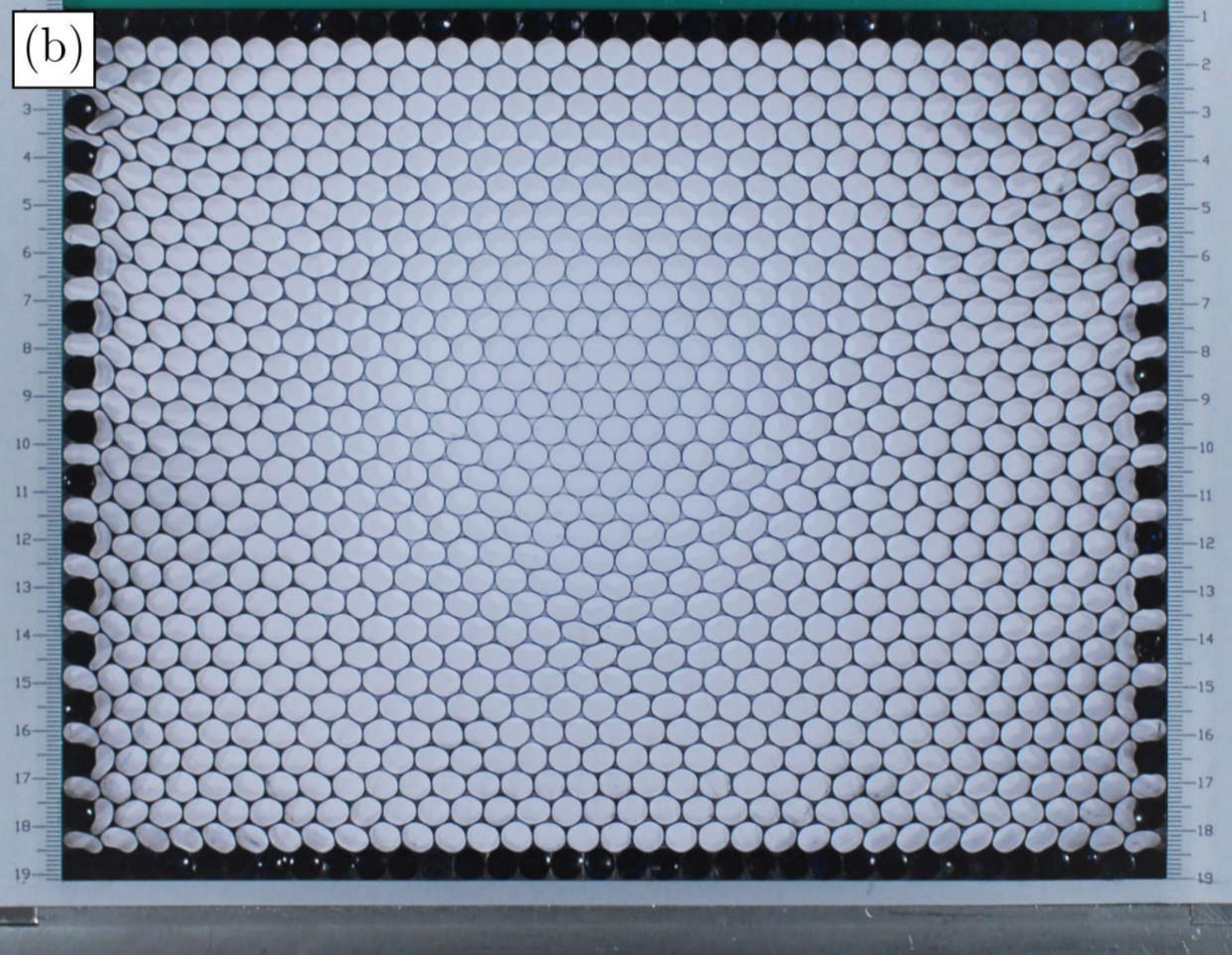}
    \end{subfigure}
    \begin{subfigure}{0.32\textwidth}
        \centering
        \phantomcaption{\label{fig:straws_micro_buckling}}
        \includegraphics[width=0.98\linewidth]{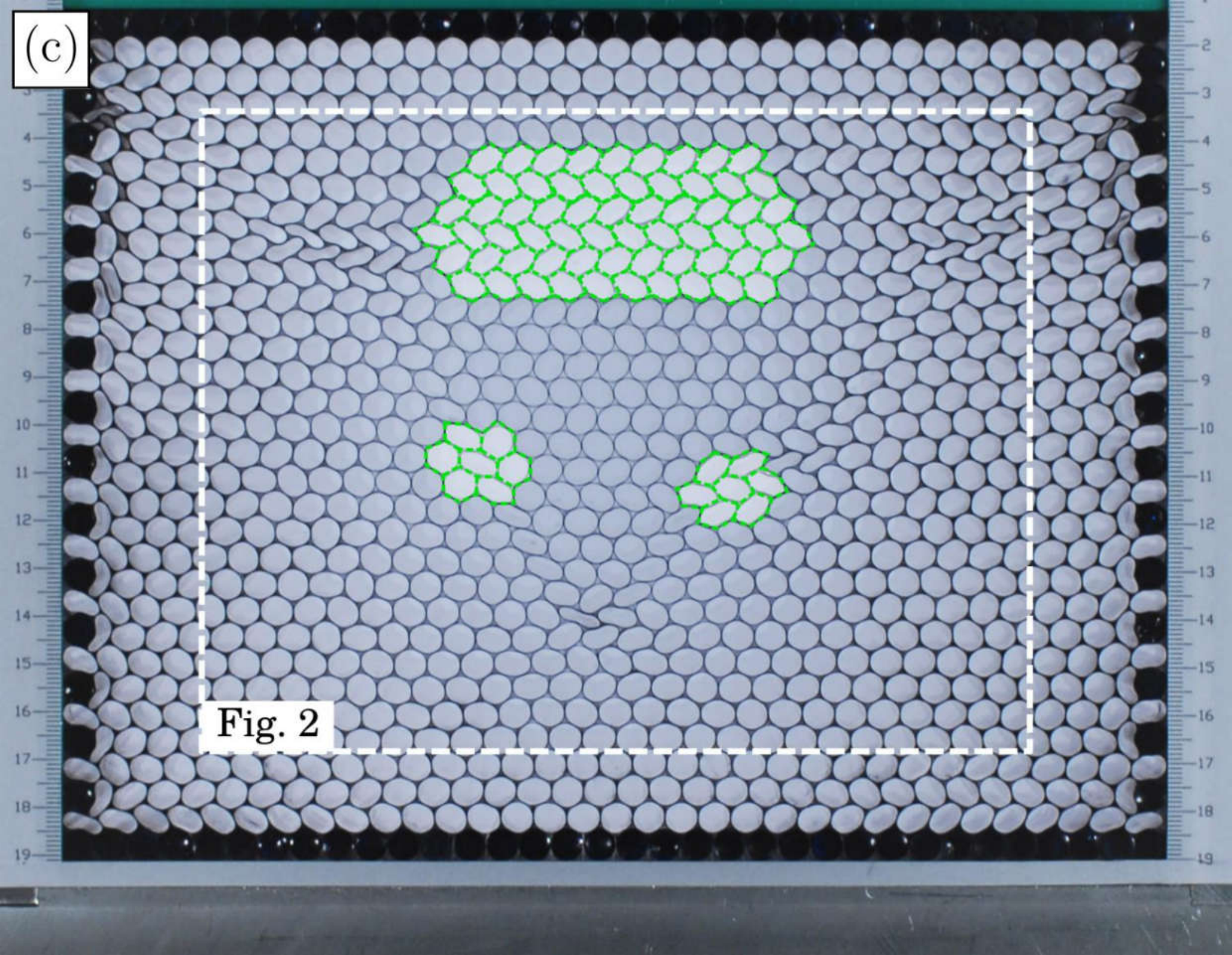}
    \end{subfigure}\\ \vspace{2mm}
    \begin{subfigure}{0.32\textwidth}
        \centering
        \phantomcaption{\label{fig:straws_band_formation}}
        \includegraphics[width=0.98\linewidth]{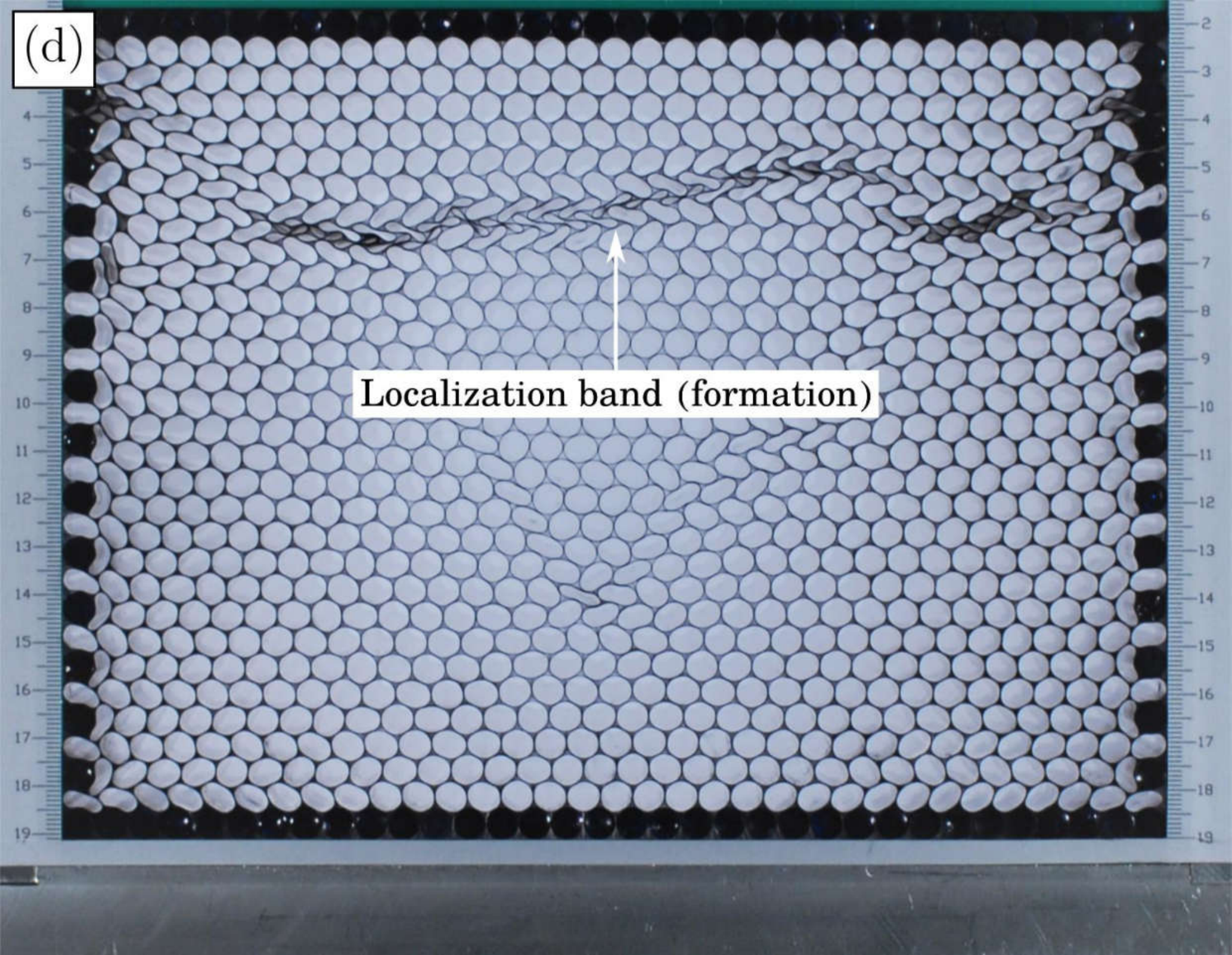}
    \end{subfigure}
    \begin{subfigure}{0.32\textwidth}
        \centering
        \phantomcaption{\label{fig:straws_4}}
        \includegraphics[width=0.98\linewidth]{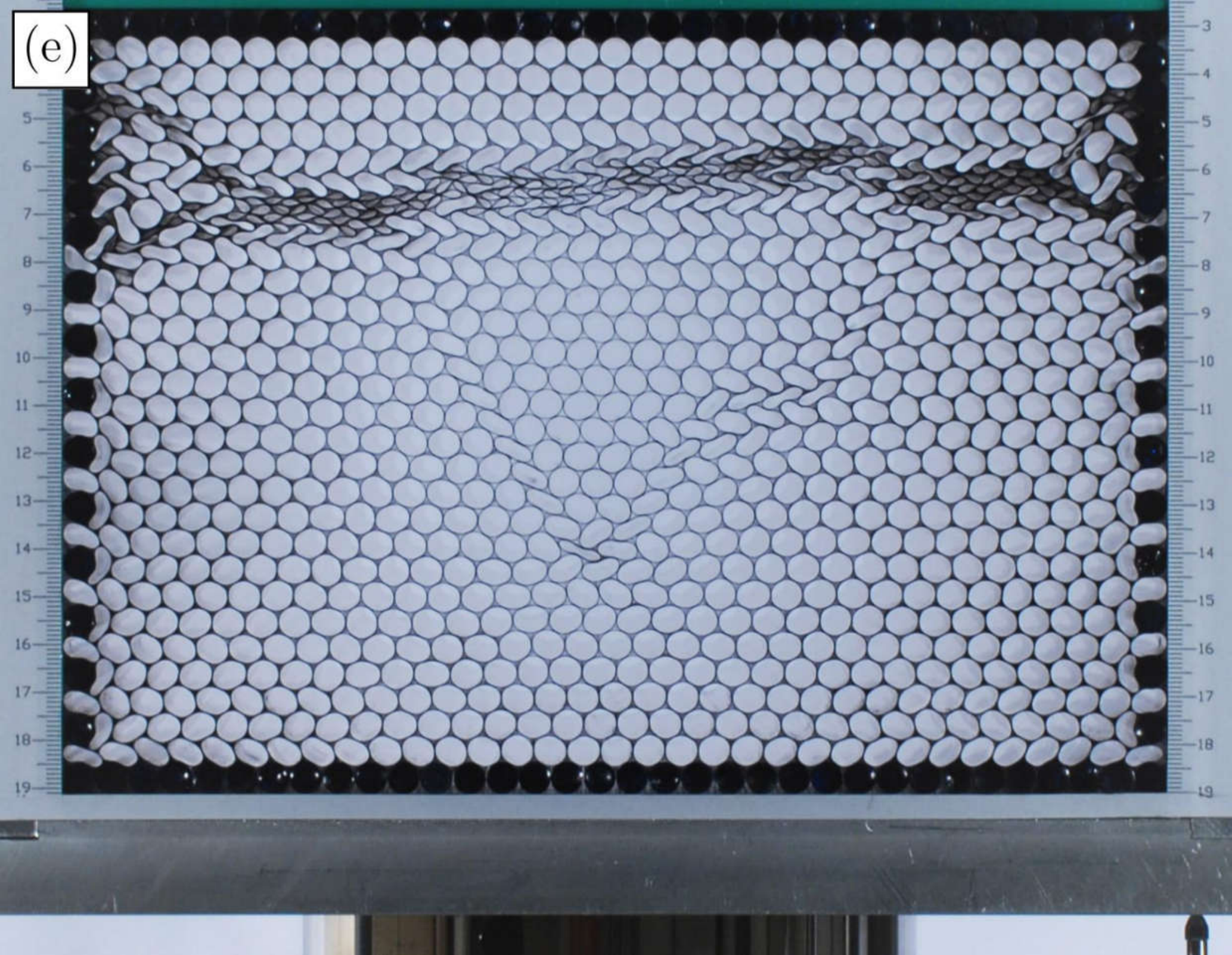}
    \end{subfigure}
    \begin{subfigure}{0.32\textwidth}
        \centering
        \phantomcaption{\label{fig:straws_band_accumulation}}
        \includegraphics[width=0.98\linewidth]{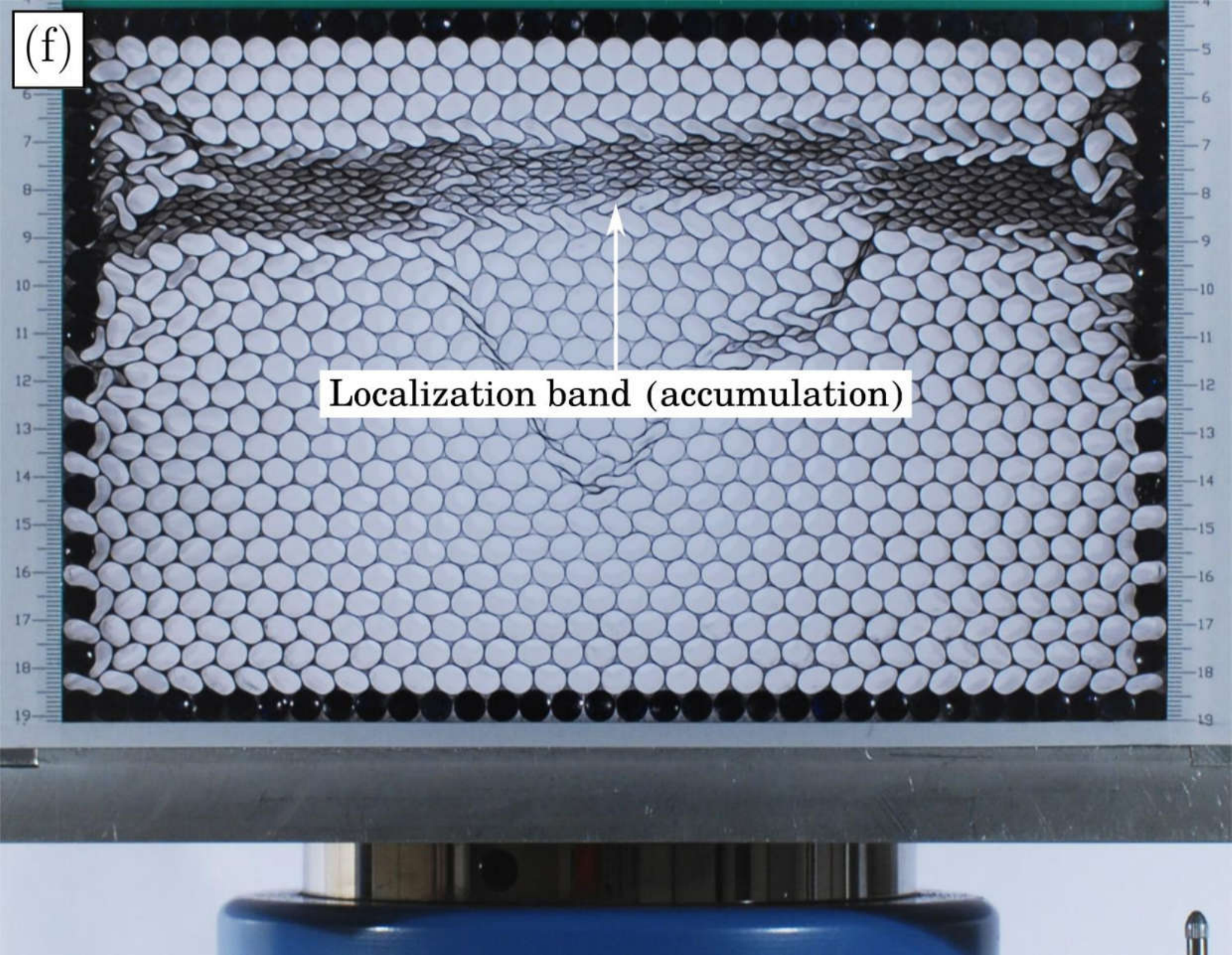}
    \end{subfigure}
    \caption{\label{fig:straws_localization}
        Emergence of a periodic micro-bifurcation (ovalization of the straws' cross sections, part~\subref{fig:straws_micro_buckling}), subsequent strain localization (collapse of the straws' cross sections, part~\subref{fig:straws_band_formation}), and final strain accumulation (parts~\subref{fig:straws_4} and~\subref{fig:straws_band_accumulation}) during uniaxial deformation of an initially (parts~\subref{fig:straws_0} and~\subref{fig:straws_1}) hexagonal  packing of drinking straws.
    }
\end{figure}
%
\begin{figure}[htb!]
    \centering
    \begin{minipage}[c]{0.274\textwidth}
        \begin{subfigure}{\textwidth}
            \centering
            \caption{\label{fig:geometry_cluster_hexagonal}}
            \includegraphics[width=0.98\linewidth]{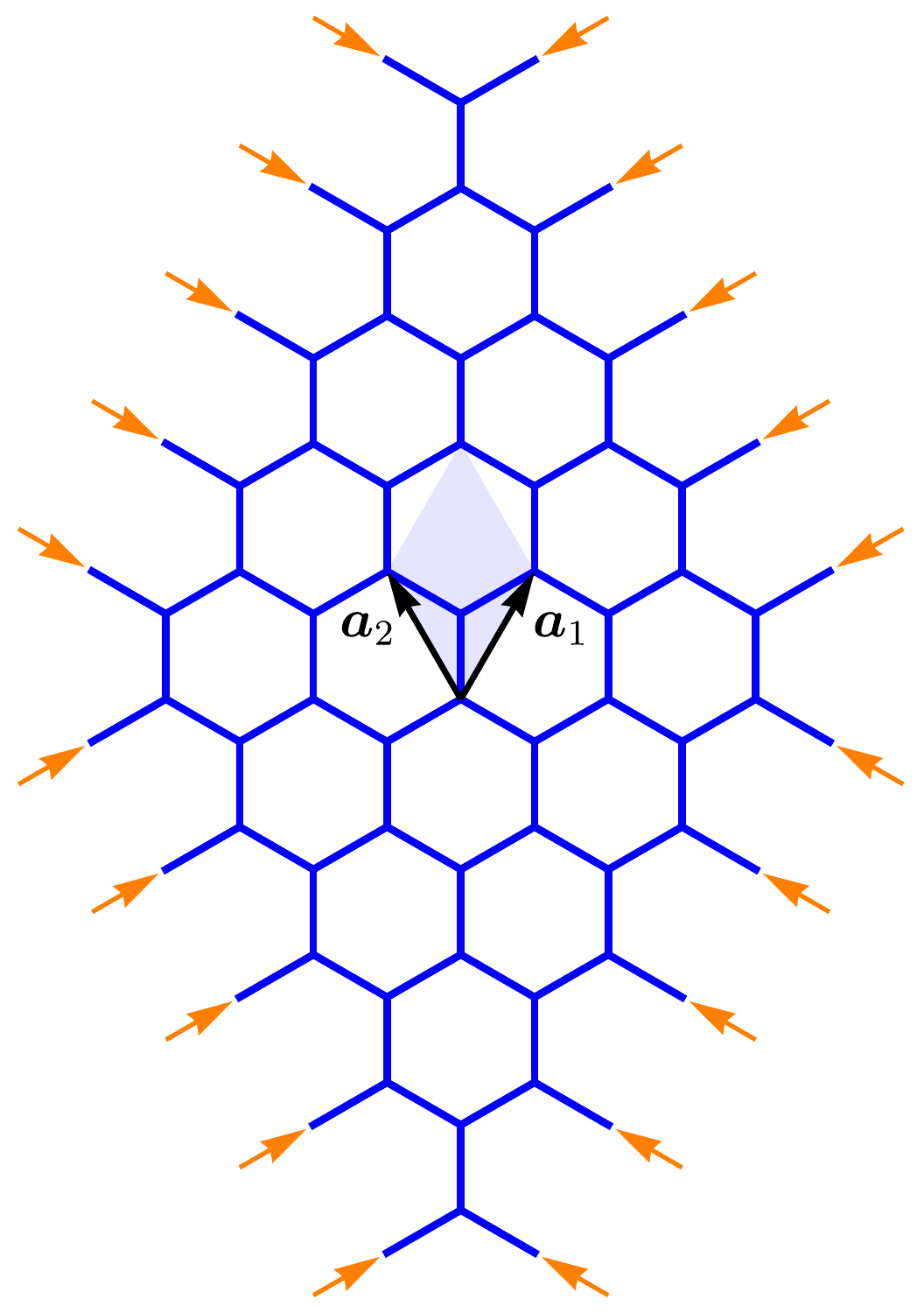}
        \end{subfigure}
    \end{minipage}%
    \begin{minipage}[c]{0.139\textwidth}
        \begin{subfigure}{\textwidth}
            \centering
            \caption{\label{fig:buckling_mode_1_cell_hexagonal}}
            \includegraphics[width=0.98\linewidth]{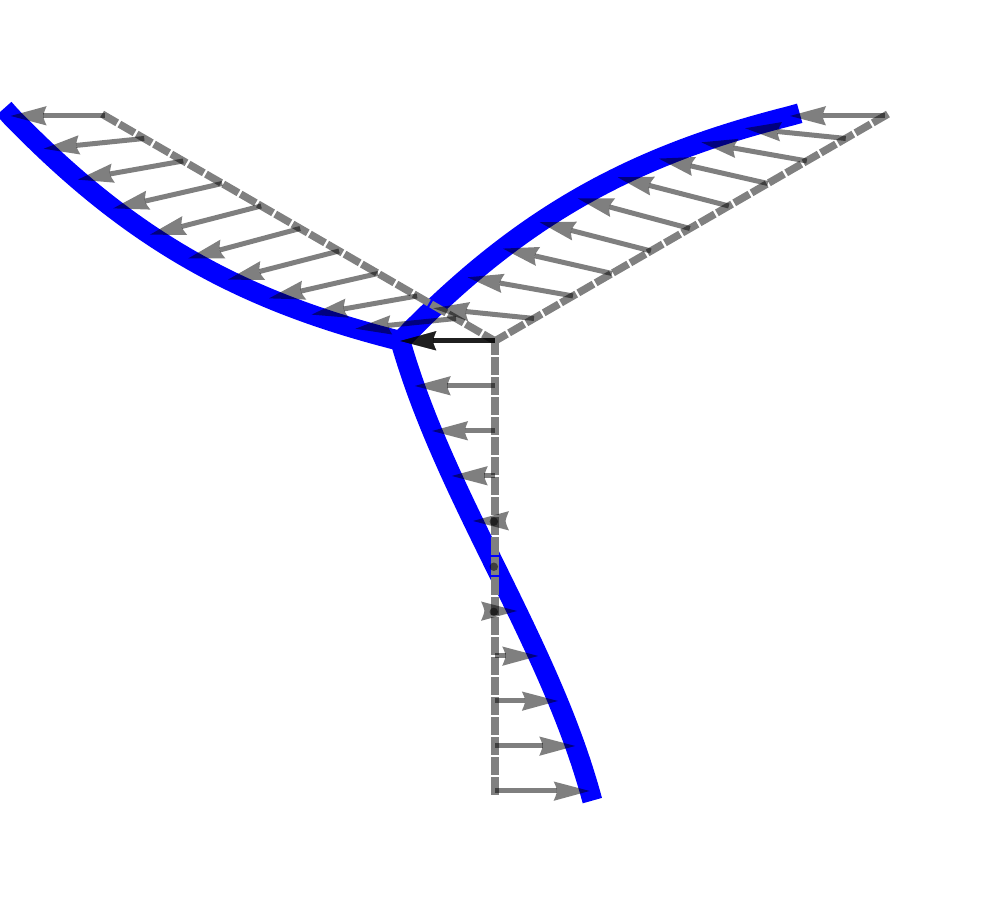}
        \end{subfigure}
        \begin{subfigure}{\textwidth}
            \centering
            \includegraphics[width=0.98\linewidth]{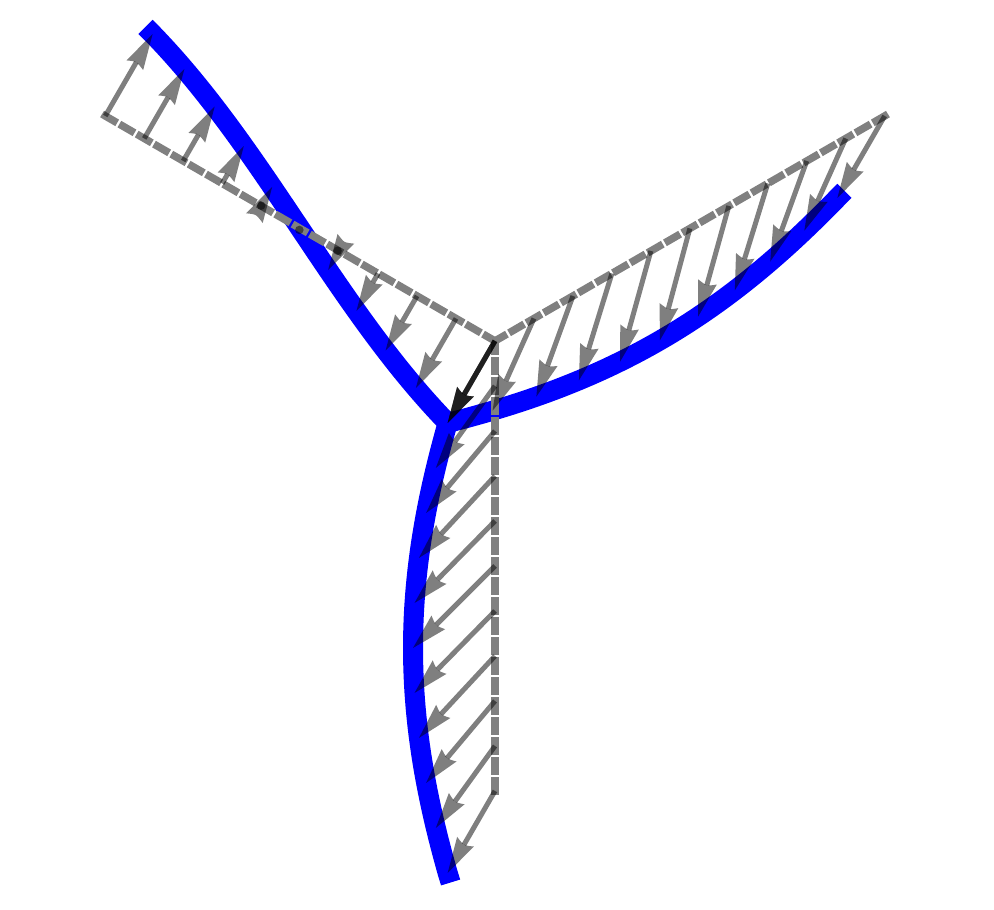}
        \end{subfigure}
        \begin{subfigure}{\textwidth}
            \centering
            \includegraphics[width=0.98\linewidth]{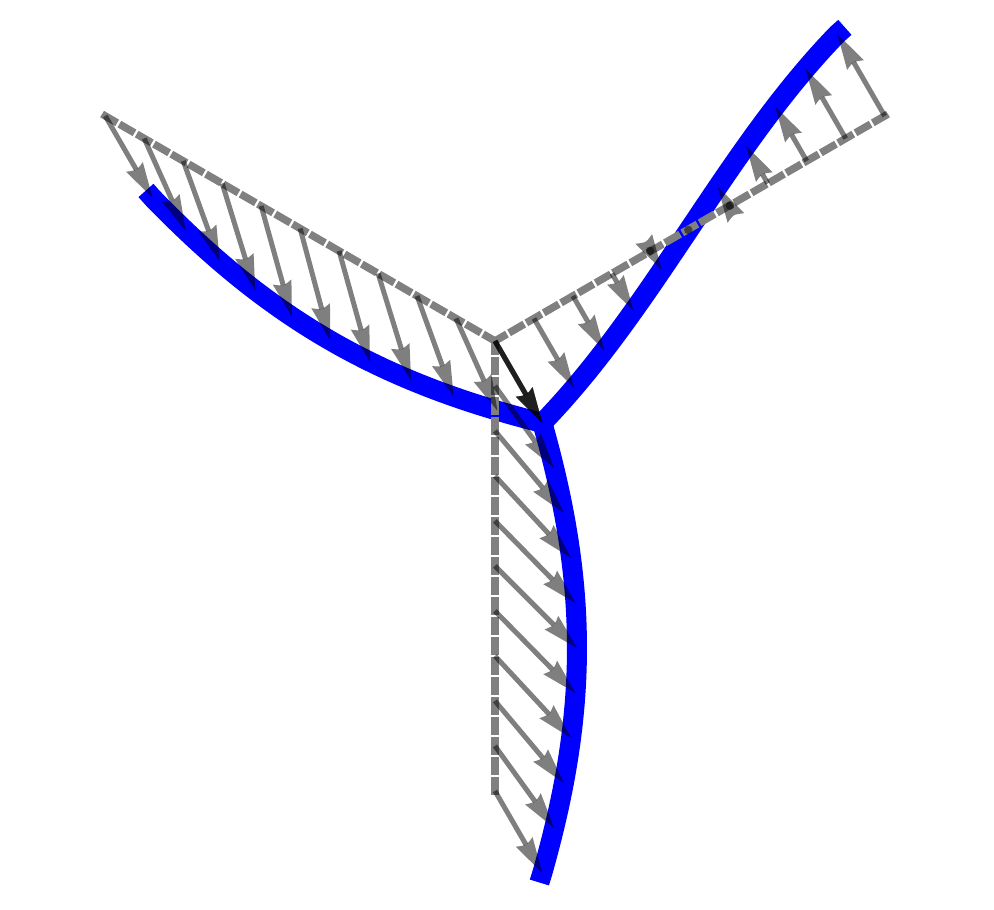}
        \end{subfigure}
    \end{minipage}%
    \begin{minipage}[c]{0.586\textwidth}
        \begin{subfigure}{\textwidth}
            \centering
            \caption{\label{fig:straws_micro_buckling_detailed}}
            \includegraphics[width=0.98\linewidth]{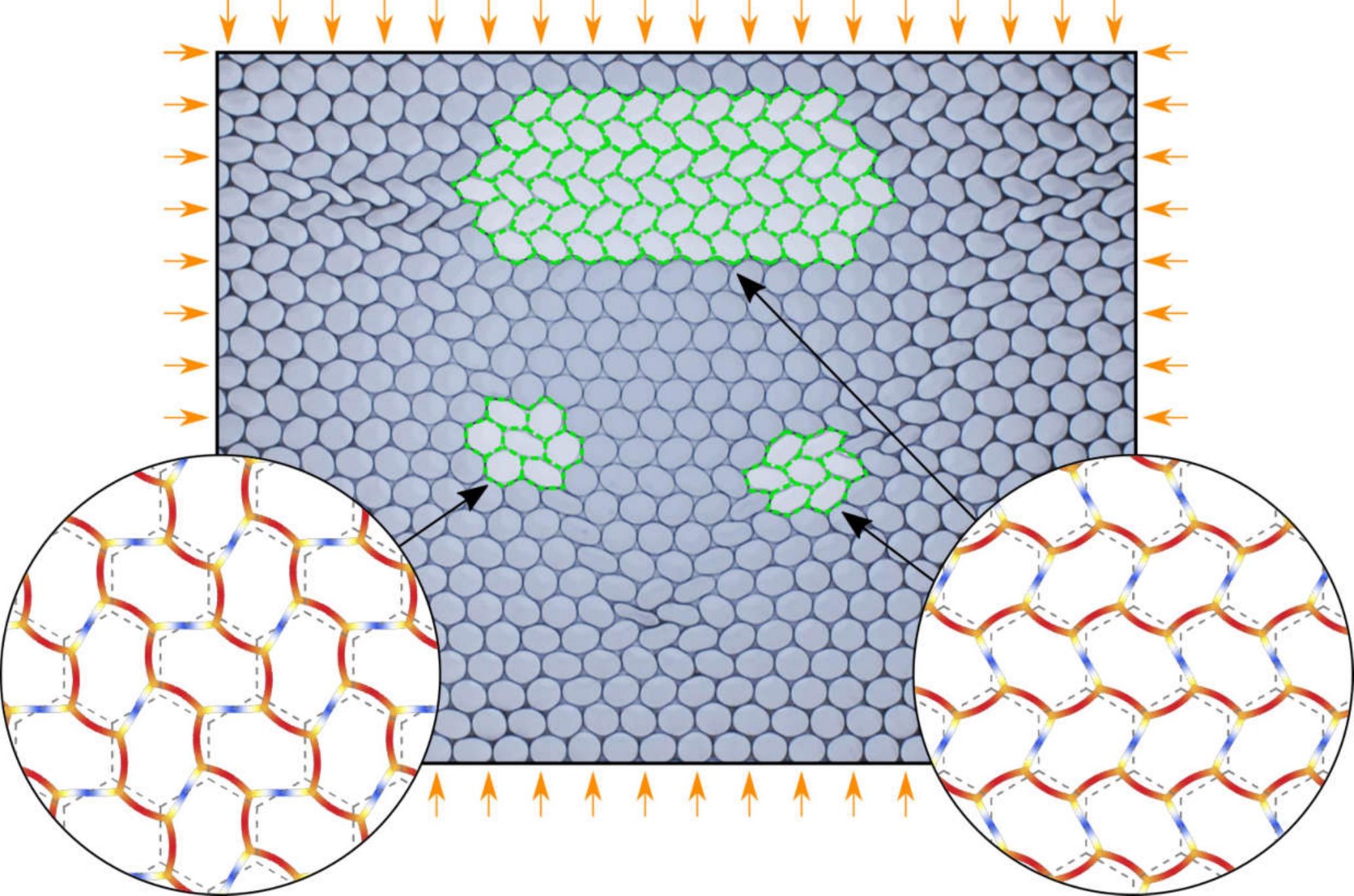}
        \end{subfigure}
    \end{minipage}%
    \caption{\label{fig:straws_buckling detailed}
        The micro-bifurcation mode emerging during the uniaxial deformation of the package of drinking straws shown in Fig.~\ref{fig:straws_localization} is  modelled (with the tools provided in this article) as the micro-buckling of an honeycomb lattice of elastic rods, isotropically loaded with compressive forces.
        The equilibrium of the honeycomb structure~(\subref{fig:geometry_cluster_hexagonal}) bifurcates displaying three critical modes~(\subref{fig:buckling_mode_1_cell_hexagonal}), which induces a periodic ovalization pattern, explaining the regular and diffuse buckled zones in the array of drinking straws~(\subref{fig:straws_micro_buckling_detailed}).
    }
\end{figure}

An example of local instability, undetected in the homogenized material, but revealed through the analysis of the microstructure, is provided in Fig.~\ref{fig:straws_localization}, where photos of experiments (performed at the Instabilities Lab of the University of Trento) are shown in which a package of drinking straws, initially in a regular hexagonal disposition, is subject to an overall uniaxial strain. 
The unloaded configuration (Fig.~\ref{fig:straws_0}) 
is not particularly different from the configuration subject to a light loading (Fig.~\ref{fig:straws_1}). 
An increase of the loading yields a micro-bifurcation in terms of a periodic ovalization of the straws' cross sections (Fig.~\ref{fig:straws_micro_buckling}), while at higher load strain localization occurs (in terms of collapse of the cross sections, Fig.~\ref{fig:straws_band_formation}), with subsequent strain band accumulation (Figs.~\ref{fig:straws_4} and \ref{fig:straws_band_accumulation}). 
The periodic ovalization is perfectly captured by a bifurcation analysis of the hexagonal rods' grid (Fig.~\ref{fig:straws_buckling detailed}) subject to isotropic compression and displaying a periodic bifurcation mode which is compared with a detail of the photo shown in Fig.~\ref{fig:straws_micro_buckling}.\footnote{
	The bifurcation occurs at an axial load in the grid (that was analytically calculated to be $-\arccos^2{(-1/3)} EJ/l^2\approx -3.6EJ/l^2$) smaller than the load corresponding to loss of ellipticity in the equivalent material (which was calculated through the homogenization scheme developed in this article to be $\approx -7.014 EJ/l^2$). 
}

Homogenization is shown to provide a tool to select the geometry and loading of a lattice in a way to produce an equivalent solid with arbitrary incremental anisotropy, so that the shear band inclination, or the emergence of a singular shear band can be designed.
The results that will be presented also demonstrate how lattice models of heterogeneous materials can be highly effective to obtain analytical expressions for homogenized properties, thus allowing an efficient analysis of the influence of the microstructural parameters.
This is a clear advantage over continuum formulations of composites, where analytical results can only be obtained for simple geometries and loading configurations (as for instance in the case of laminated solids~\cite{nestorovic_2004,santisidavila_2016,bacigalupo_2013}).
Several new features are found, including a `super-sensitivity' of the localization direction to the preload state and the conditions in which a perfect correspondence between the lattice and the continuum occurs (so that the discrete system and the equivalent solid share all the same bifurcation modes). 
The microscopic features found for the strain localization are shown to share remarkable similarities with the localized failure patterns observed in honeycombs (as Fig.~\ref{fig:straws_localization} demonstrates), foams and wood~\cite{papka_1994,papka_1998,papka_1999,jang_2010}, while the highly localized deformation bands emerging at macroscopic loss of ellipticity are reminiscent of the failure modes observed in balsa wood~\cite{dasilva_2007}.

This article is organized as follows.
The derivation of the incremental equilibrium is presented in Section 2 for a lattice of elastic rods organized in an arbitrary periodic geometry, while the homogenization is developed in Section 3, providing the incremental constitutive tensor of the effective Cauchy continuum. 
The stability of lattice structure and its relation with the strong ellipticity of the equivalent solid is given in Section 4, while examples and comparisons with the perturbative approach are presented in Sections 5 and 6, where the analysis is specialized to a grid of elastic rods arbitrarily inclined and equipped with diagonal springs. 

Results presented in this article are restricted to quasi-static behaviour, while the important case of dynamic homogenization (with the Floquet-Bloch technique) and dynamic shear banding is deferred to Part II of this study.

\section{Incremental response of lattices of axially preloaded elastic rods}
\label{sec:prestressed_lattice}
\begin{figure}[htb!]
    \centering
    \includegraphics[width=0.98\linewidth]{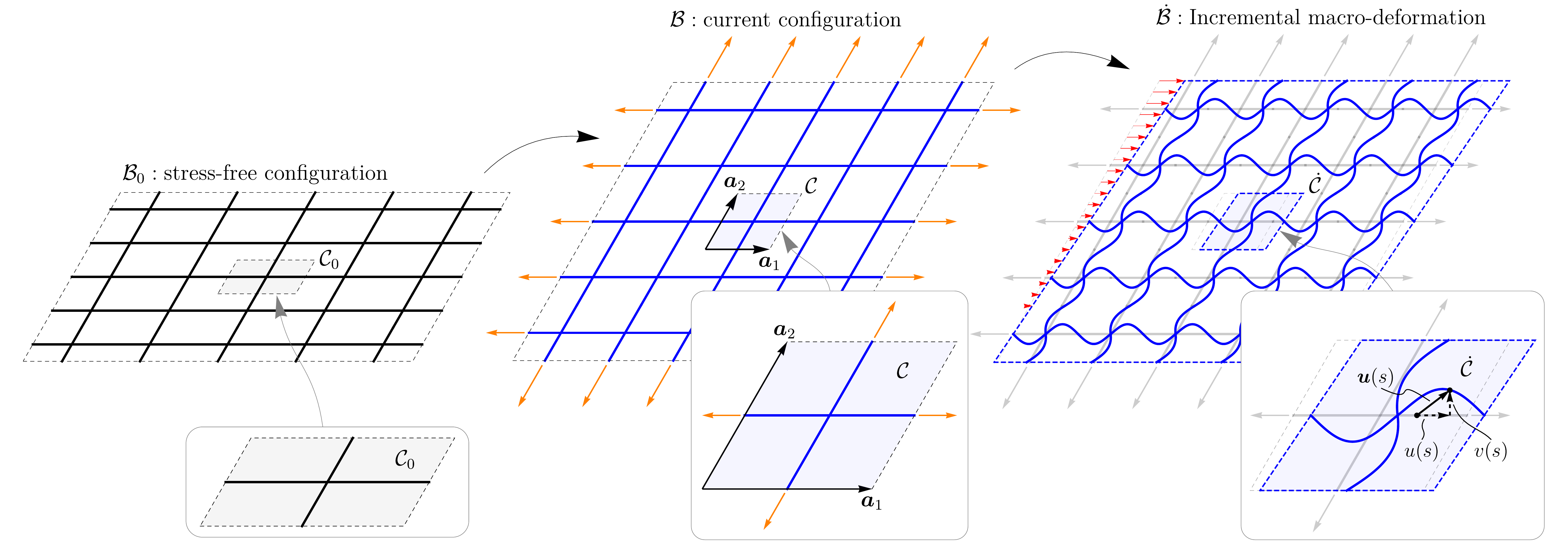}
    \caption{\label{fig:lattice_configurations}
    A periodic two-dimensional lattice of (axially and flexurally deformable) elastic rods is considered prestressed from the stress-free configuration $\mB_0$ (left) by means of a purely axial loading state.
    The prestressed configuration $\mB$ (center) can be represented as the tessellation of a single unit cell along the vectors of the direct basis $\{\ba_1,\ba_2\}$.
    Upon the current prestressed configuration, the incremental response (right) is defined by the incremental displacement field of each rod $\bu(s)$, here decomposed in an axial and transverse component, $u(s)$ and $v(s)$.
    }
\end{figure}

A two-dimensional periodic lattice of elastic rods, deformable in the plane both axially and flexurally, is considered, in which all structural members are axially prestressed from an unloaded reference configuration $\mB_0$. 
The prestress is assumed to be produced by dead loading acting at infinity, while body forces in the lattice are excluded for simplicity. 
It is assumed that the preload not only satisfies equilibrium, but also preserves periodicity and leaves the structure free of flexure. 
The prestressed configuration $\mB$ is periodic along two linearly independent vectors $\{\ba_1,\ba_2\}$, defining the direct basis of the lattice, so that the structure can be constructed from a single unit cell $\mC$, assumed to be composed of $N_b$ nonlinear elastic rods with Euler-Bernoulli incremental kinematics, as sketched in Fig.~\ref{fig:lattice_configurations}.

By considering in-plane flexural and axial incremental deformations, the incremental displacement field of the $k$--th rod in a given unit cell is defined by the vector field (Fig.~\ref{fig:lattice_configurations}) 
\begin{equation}
\label{eq:displacement_beams}
    \bu_k(s_k)=\trans{\{ u_k(s_k), v_k(s_k)\}}, \qquad \forall k \in \{1,...,N_b\} \,,
\end{equation}
where $s_k$ is the coordinate along the $k$--th rod, $u_k(s_k)$ and $v_k(s_k)$ are the axial and transverse incremental displacements.
The incremental rotation of the rod's cross-section $\theta_k(s_k)$ is assumed to satisfy the unshearability condition of the Euler-Bernoulli kinematics, namely, $\theta_k(s_k)=v_k'(s_k)$.

\subsection{Analytic solution for the prestressed elastic rod}
\label{sec:solution_beam}
The incremental equilibrium equations for an elastic rod obeying Euler-Bernoulli kinematics and prestressed with an axial load $P$ (assumed positive in tension) and pre-stretched by $\lambda_0>0$, are the following 
\begin{subequations}
\label{eq:governing_beam_EB}
    \begin{gather}
    \label{eq:governing_beam_EB_u}
    A(\lambda_0)\, u''(s) = 0 \,, \\
    \label{eq:governing_beam_EB_v}
    B(\lambda_0) \, v''''(s) - P(\lambda_0)\,v''(s) = 0 \,, 
    \end{gather}
\end{subequations}
where $A(\lambda_0)$ and $B(\lambda_0)$ are the \textit{current} axial and bending stiffnesses, respectively, and $s\in(0,l)$ with $l$ being the \textit{current} length of the rod. 
It is worth noting that the current axial and bending stiffnesses are, in general, function of the current pre-stretch $\lambda_0$, which in turn depends on the axial load $P$.
In fact, Eqs.~\eqref{eq:governing_beam_EB} govern the incremental equilibrium of an axially pre-stretched rod, and their analytic derivation, accompanied with the evaluation of $A(\lambda_0)$ and $B(\lambda_0)$ from given strain-energy functions, is deferred to  Appendix~\ref{sec:linearized_elastica}.
In the following, the parameters $A(\lambda_0)$ and $B(\lambda_0)$ will simply be denoted as $A$ and $B$, and treated as independent quantities for generality. 
The specific example in which the rods composing the lattice are made up of a Mooney-Rivlin elastic incompressible material is explicitly reported in Appendix~\ref{sec:linearized_elastica}. 

Eqs.~\eqref{eq:governing_beam_EB} is a system of linear ODEs for the functions $u(s)$ and $v(s)$.
As the system is fully decoupled, the solution is easily obtained in the form
\begin{equation}
\label{eq:u_v_sol}
    u(s) = C_{1}^u + C_{2}^u\,s \,, \qquad v(s) = C_{1}^v\, e^{-\beta\,s} + C_{2}^v\, e^{\beta\,s} + C_{3}^v\,s + C_{4}^v \,,
\end{equation}
where $\{C_{1}^u,C_{2}^u,C_{1}^v,...,C_{4}^v\}$ are 6 arbitrary complex constants and $\beta=\sqrt{P/B}$.

\subsection{Exact shape functions and stiffness matrix}
\label{sec:exact_shape_functions}
For a rod of length $l$, the following nomenclature can be introduced
\begin{equation}
\label{eq:beam_bcs}
    u(0) = u_1 \,, \quad v(0) = v_1 \,, \quad \theta(0) = \theta_1 \,, \quad u(l) = u_2 \,, \quad v(l) = v_2 \,, \quad \theta(l) = \theta_2 \,,
\end{equation}
so that the vector $\bq=\trans{\{u_1,v_1,\theta_1,u_2,v_2,\theta_2\}}$ now collects the degrees of freedom of the rod expressed in terms of end displacements.
Solving the conditions~\eqref{eq:beam_bcs} for the constants $\trans{\{C_{1}^u,C_{2}^u,C_{1}^v,...,C_{4}^v\}}$ allows the solution~\eqref{eq:u_v_sol} to be rewritten as
\begin{equation}
    \label{eq:static_sf}
    \bu(s) = \bN(s;P)\, \bq \,,
\end{equation}
which is now a linear function of the nodal displacements $\bq$.
The $2{\times}6$ matrix $\bN(s;P)$ acts as a matrix of prestress-dependent `shape functions' and therefore the representation~\eqref{eq:static_sf} can also be considered as the definition of a `finite element' endowed with shape functions built from the exact solution.
Moreover, these shape functions reduce to the solution holding true in the absence of prestress, because in the limit 
\begin{equation*}
    \lim_{P\to0}\bN(s;P) =
        \begin{bmatrix}
            1-\frac{s}{l} & 0 & 0 & \frac{s}{l} & 0 & 0 \\[1mm]
            0 & \frac{(l-s)^2 (l+2 s)}{l^3} & \frac{(l-s)^2 s}{l^2} & 0 & \frac{(3 l-2 s) s^2}{l^3} & \frac{s^2 (s-l)}{l^2}
        \end{bmatrix} \,,
\end{equation*}
the usual shape functions (linear and Hermitian for axial and flexural displacements, respectively) are retrieved.

By employing Eq.~\eqref{eq:static_sf}, the incremental stiffness matrix of a prestressed rod can be computed, so that 
for the $k$-th rod the elastic strain energy is given by
\begin{equation}
\label{eq:strain_energy_beam}
    \mE_k = \frac{1}{2} \int_0^{l_k} \left(A_k\,u'_k(s_k)^2 + B_k\,v''_k(s_k)^2\right) \,ds_k 
        = \frac{1}{2} \,\trans{\bq_{k}} \left(\int_0^{l_k} \trans{\bB_{k}(s_k;P_k)}\bE_k\,\bB_{k}(s_k;P_k)\,ds_k \right) \bq_{k} \,,
\end{equation}
where $\bE_{k}$ is a matrix collecting the stiffness terms, while $\bB_{k}(s_k;P)$ is the strain-displacement matrix, defined as follows
\begin{equation*}
    \bE_{k}=\begin{bmatrix}
        A_k 	& 0 	 \\
        0  		            & B_k \\
    \end{bmatrix} \,,
    \qquad
        \bB_{k}(s_k;P_k) = \begin{bmatrix}
        \deriv{}{s_k} 	& 0 \\
        0  	 	        & \deriv{^2}{s_k^2} \\
    \end{bmatrix} \bN_{k}(s_k;P_k) \,.
\end{equation*}

The `geometric' contribution of the axial prestress is now included in the potential energy (details are provided in Appendix~\ref{sec:linearized_elastica}), 
\begin{equation}
\label{eq:prestress_energy_beam}
    \mV_k^g = \frac{1}{2} P_k \int_0^{l_k} v'_k(s_k)^2 \,ds_k 
            = \frac{1}{2} \,\trans{\bq_{k}} \left(P_k \int_0^{l_k} \trans{\bb_{k}(s_k;P_k)} \bb_{k}(s_k;P_k) \,ds_k \right) \bq_{k} \,,
\end{equation}
where $\bb_{k}(s_k;P_k)=\begin{bmatrix}0 & \deriv{}{s_k}\end{bmatrix}\bN_{k}(s_k;P_k)$ is a vector collecting the derivatives of the shape functions describing the transverse displacement $v$.
A combination of Eqs.~\eqref{eq:strain_energy_beam} and~\eqref{eq:prestress_energy_beam}, yields the potential energy for the $k$-th rod in the form
\begin{equation}
    \label{eq:potential_energy_beam}
    \mV_k = \mE_k + \mV_k^g \,.
\end{equation}
Note that, as the equilibrium equations for the rods have been linearized around an axially pre-loaded configuration, the potential~\eqref{eq:potential_energy_beam} represents the incremental potential energy with respect to the current configuration.
See the Appendix~\ref{sec:linearized_elastica} for details on the derivation of Eqs.~\eqref{eq:strain_energy_beam}--\eqref{eq:potential_energy_beam}.

From Eqs.~\eqref{eq:strain_energy_beam},~\eqref{eq:prestress_energy_beam} and~\eqref{eq:potential_energy_beam} the prestress-dependent stiffness matrix is defined as 
\begin{equation*}
\label{eq:K_beam}
    \bK_{k}(P_k) =
    \int_0^{l_k} \trans{\bB_{k}(s_k;P_k)}\bE_{k}\,\bB_{k}(s_k;P_k)\,ds_k +
    P_k \int_0^{l_k} \trans{\bb_{k}(s_k;P_k)} \bb_{k}(s_k;P_k) \,ds_k , 
\end{equation*}
so that
\begin{equation*}
    \bK_{k}=
    \begin{bmatrix}
        \frac{A_k}{l_k} & 0 & 0 & -\frac{A_k}{l_k} & 0 & 0 \\[2mm]
        0 & \frac{12 B_k}{l_k^3}\varphi_1(p_k) & \frac{6 B_k}{l_k^2}\varphi_2(p_k) & 0 & -\frac{12 B_k}{l_k^3}\varphi_1(p_k) & \frac{6 B_k}{l_k^2}\varphi_2(p_k) \\[2mm]
        0 & \frac{6 B_k}{l_k^2}\varphi_2(p_k) & \frac{4 B_k}{l_k}\varphi_3(p_k) & 0 & -\frac{6 B_k}{l_k^2}\varphi_2(p_k) & \frac{2 B_k}{l_k}\varphi_4(p_k) \\[2mm]
        -\frac{A_k}{l_k} & 0 & 0 & \frac{A_k}{l_k} & 0 & 0 \\[2mm]
        0 & -\frac{12 B_k}{l_k^3}\varphi_1(p_k) & -\frac{6 B_k}{l_k^2}\varphi_2(p_k) & 0 & \frac{12 B_k}{l_k^3}\varphi_1(p_k) & -\frac{6 B_k}{l_k^2}\varphi_2(p_k) \\[2mm]
        0 & \frac{6 B_k}{l_k^2}\varphi_2(p_k) & \frac{2 B_k}{l_k}\varphi_4(p_k) & 0 & -\frac{6 B_k}{l_k^2}\varphi_2(p_k) & \frac{4 B_k}{l_k}\varphi_3(p_k) 
    \end{bmatrix} 
    ,
\end{equation*}
where the $\varphi_j$ are functions of the non-dimensional measure of prestress $p_k=P_k l_k^2/B_k$ given by
\begin{align*}
    \varphi_1(p_k) &= \frac{p_k^{3/2}}{12 \left(\sqrt{p_k}-2 \tanh \left(\sqrt{p_k}/2\right)\right)} \,, &
    \varphi_2(p_k) &= \frac{p_k}{6 \sqrt{p_k} \coth \left(\sqrt{p_k}/2\right)-12} \,, \\
    \varphi_3(p_k) &= \frac{p_k \cosh \left(\sqrt{p_k}\right)-\sqrt{p_k} \sinh \left(\sqrt{p_k}\right)}{4 \sqrt{p_k} \sinh \left(\sqrt{p_k}\right)-8 \cosh \left(\sqrt{p_k}\right)+8} \,, &
    \varphi_4(p_k) &= \frac{\sqrt{p_k} \left(\sinh \left(\sqrt{p_k}\right)-\sqrt{p_k}\right)}{\left(4 \sqrt{p_k} \coth \left(\sqrt{p_k}/2\right)-8\right)\sinh^2\left(\sqrt{p_k}/2\right)} \,.
\end{align*}

Note that the stiffness matrix representative of the lattice reduces, in the limit of vanishing prestress (or unitary pre-stretch $\lambda_{0k}=1$), to the usual stiffness matrix of an Euler-Bernoulli beam with Hermitian shape functions, so that 
\begin{equation*}
	\lim_{p\to0} \varphi_j(p) = 1, \quad \forall j\in\{1,...,4\}.
\end{equation*}

\subsection{Incremental equilibrium of the lattice of elastic rods}
\label{sec:equilibrium_lattice}
The current configuration of the lattice unit cell is subject, on the boundary, to the nominal internal \textit{incremental} actions and \textit{incremental} displacements transmitted by the rest of the lattice. 
Therefore, the incremental potential energy of the cell can be evaluated through a direct summation of all contributions from the rods, equation \eqref{eq:potential_energy_beam}, over the set of structural elements present inside the cell, plus the incremental work done on the unit cell boundary by the internal actions $\bef$, 
\begin{equation}
\label{eq:potential_energy_unit_cell}
    \mV(\bq) = \sum_{k=1}^{N_b} \mV_k(\bC_k \bq) - \bef\scalp\bq \,,
\end{equation}
where $\bq$ is the vector collecting the degrees of freedom of the unit cell, $\bC_k$ is the connectivity matrix of the $k$-th rod, such that $\bq_k = \bC_k \bq$, and $\bef$ is the vector collecting the \textit{generalized} (incremental and nominal) internal actions (including bending moments) at the nodes of the unit cell.

The (referential) incremental equilibrium equations (in the absence of body forces) are therefore obtained from the stationarity of the potential energy~\eqref{eq:potential_energy_unit_cell} yielding
\begin{equation}
\label{eq:equilibrium_unit_cell}
    \bK(\bP)\,\bq = \bef \,,
\end{equation}
where $\bK(\bP) = \partial/\partial \bq\, \sum_{k=1}^{N_b} \mV_k(\bC_k \bq)$ is the symmetric (as derived from a potential) stiffness matrix of the unit cell, function of the vector $\bP=\{P_1,...,P_{N_b}\}$ collecting the axial prestress of the rods.
The dimension of the system~\eqref{eq:equilibrium_unit_cell} is $3N_j$ where $N_j$ is the number of nodes in the unit cell.

\section{Incremental response of the effective Cauchy continuum}
\label{sec:homogenization}
As the preloaded configuration of the lattice is assumed to be spatially periodic, the homogenized incremental response of an equivalent prestressed elastic solid can be defined by computing the average strain-energy density, associated to an incremental displacement field (defined for the $j$-th node by the displacement and rotation components, respectively, $\bq_u^{(j)} = \trans{\{u^{(j)},v^{(j)}\}}$ and $\bq_\theta^{(j)} = \{\theta^{(j)}\}$) which obeys the Cauchy-Born hypothesis~\cite{born_1955,willis_2002,hutchinson_2006}. 
The latter, for a single unit cell, prescribes that the displacement of the lattice nodes be decomposed into the sum of an affine incremental deformation (ruled by a second-order tensor $\bL$) and a periodic field (defined by a displacement $\Tilde{\bq}_u^{(j)}$ and a rotational $\Tilde{\bq}_\theta^{(j)}$ component) as
\begin{equation}
\label{eq:cauchy_born_hypothesis}
    \bq_u^{(j)} = \Tilde{\bq}_u^{(j)} + \bL\,\bx_j \,, \qquad \bq_\theta^{(j)} = \Tilde{\bq}_\theta^{(j)} \,, \qquad \forall j\in\{1,...,N_j\}
\end{equation}
where $\bx_j$ is the position of the $j$-th node.
\begin{figure}[htb!]
    \centering
    \includegraphics[height=40mm]{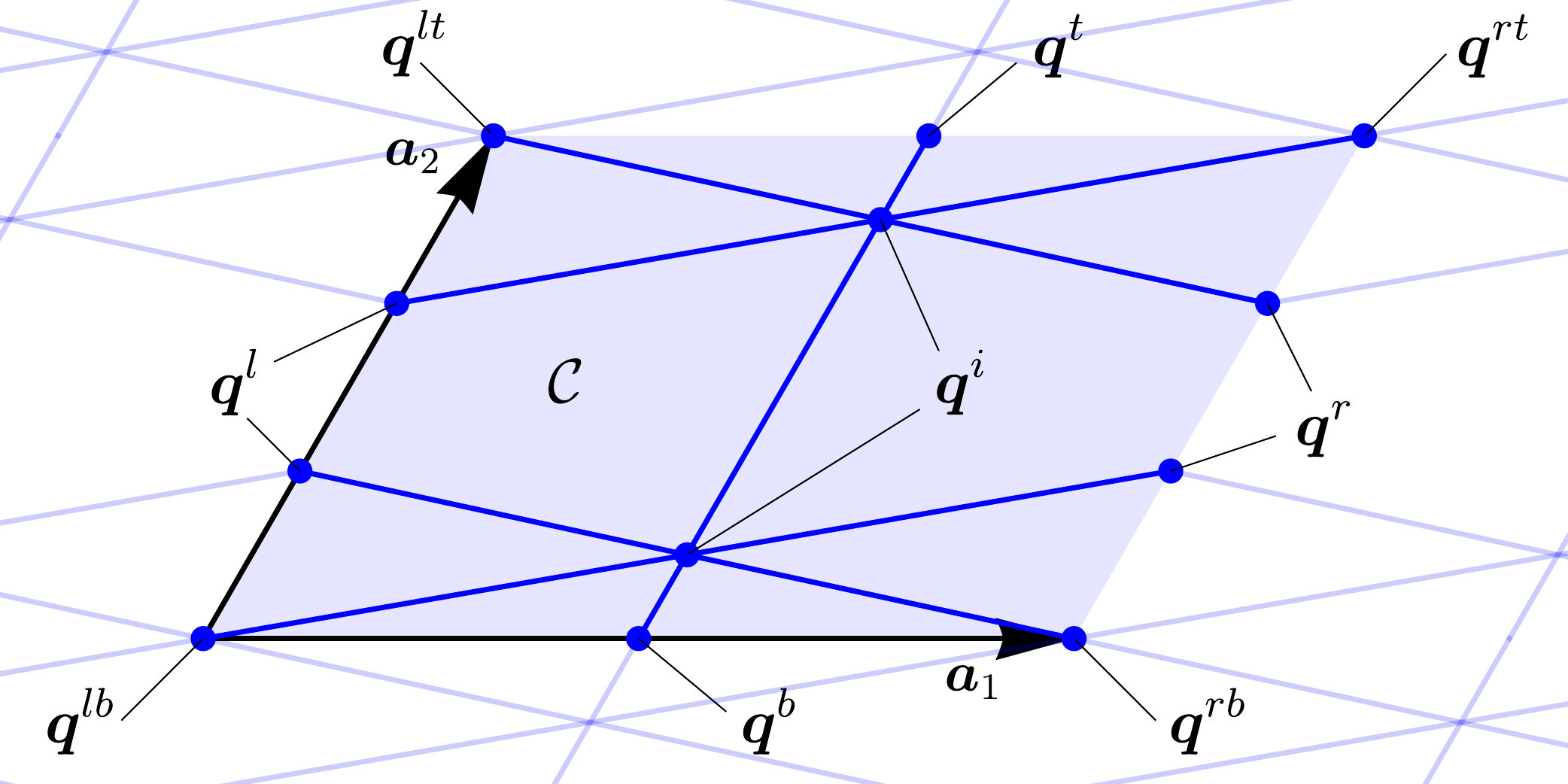}
    \caption{\label{fig:FB_joints}
    The vector $\bq$ collects the degrees of freedom of the unit cell.
    In order to impose the periodic boundary conditions required by the Cauchy-Born hypothesis~\eqref{eq:cauchy_born_hypothesis} (or the quasi-periodic boundary conditions required by the Floquet-Bloch hypothesis~\eqref{eq:FB_conditions} that will  be used in the bifurcation analysis), $\bq$ is partitioned between sets of inner nodes $\bq^i$, boundary nodes located at corners $\{\bq^{lb}, \bq^{lt}, \bq^{rb}, \bq^{rt}\}$, and boundary nodes on the edges $\{\bq^l, \bq^b, \bq^r, \bq^t\}$.
    The corresponding vector of incremental internal nominal actions $\bef$ is partitioned in the same way.
    }
\end{figure}

The periodic term $\Tilde{\bq}$ satisfies $\Tilde{\bq}^{(p)}=\Tilde{\bq}^{(q)}$ for all $\{p,q\}$ such that $\bx_q-\bx_p = n_j\ba_j$ (with $n_j\in\{0,1\}$), and can be expressed in terms of its independent components through a partition of the degrees of freedom, made in accordance with the location of the nodes present inside the unit cell (see Fig.~\ref{fig:FB_joints}), as
\begin{subequations}
\label{eq:periodic_conditions}
\begin{equation}
    \label{eq:periodic_conditions_long}
    \Tilde{\bq} =
    \begin{Bmatrix}
        \Tilde{\bq}^i \\
        \Tilde{\bq}^l \\
        \Tilde{\bq}^b \\
        \Tilde{\bq}^{lb} \\
        \Tilde{\bq}^r \\
        \Tilde{\bq}^t \\
        \Tilde{\bq}^{rb} \\
        \Tilde{\bq}^{lt} \\
        \Tilde{\bq}^{rt}
    \end{Bmatrix}
    =
    \begin{bmatrix}
        \bI & \bzero & \bzero & \bzero \\
        \bzero & \bI & \bzero & \bzero \\
        \bzero & \bzero & \bI & \bzero \\
        \bzero & \bzero & \bzero & \bI \\
        \bzero & \bI & \bzero & \bzero \\
        \bzero & \bzero & \bI & \bzero \\
        \bzero & \bzero & \bzero & \bI \\
        \bzero & \bzero & \bzero & \bI \\
        \bzero & \bzero & \bzero & \bI
    \end{bmatrix}
    \begin{Bmatrix}
        \Tilde{\bq}^i \\
        \Tilde{\bq}^l \\
        \Tilde{\bq}^b \\
        \Tilde{\bq}^{lb}
    \end{Bmatrix} \,,
\end{equation}
which may succinctly be rewritten as 
\begin{equation}
    \label{eq:periodic_conditions_short}
    \Tilde{\bq} = \bZ_0\,\Tilde{\bq}^* \,,
\end{equation}
\end{subequations}
where $\bZ_0$ and $\Tilde{\bq}^*$ are defined according to Eq.~\eqref{eq:periodic_conditions_long}.
The vector $\Tilde{\bq}$ in Eq.~\eqref{eq:periodic_conditions_long} has been partitioned to highlight the inner and boundary nodes according to the notation introduced in Fig.~\ref{fig:FB_joints}.
The same partitioning is also used for the vectors $\bq$ and $\bef$.

In order to enforce the Cauchy-Born conditions into the equations of incremental equilibrium~\eqref{eq:equilibrium_unit_cell}, it is convenient to rewrite Eq.~\eqref{eq:cauchy_born_hypothesis} as
\begin{equation}
\label{eq:cauchy_born_hypothesis_vector}
    \bq(\Tilde{\bq}^*,\bL) = \bZ_0\,\Tilde{\bq}^* + \hat{\bq}(\bL) ,
\end{equation}
where the affine part of the deformation $\hat{\bq}(\bL)$ is a vector-valued function linear in $\bL$ and such that 
\begin{equation*}
    \hat{\bq}(\bL)_u^{(j)}=\bL\,\bx_j \,, \qquad \hat{\bq}(\bL)_\theta^{(j)}=\bzero  \,, \qquad \forall j\in\{1,...,N_j\},
\end{equation*}
where the same notation introduced with Eq. \eqref{eq:cauchy_born_hypothesis} has been used do that the subscript $u$ (subscript $\theta$) denotes displacement (rotations) components.

Note that, since the lattice is subject to a non-vanishing prestress state, the macroscopic incremental deformation gradient defined by $\bL$ must be an arbitrary second-order tensor and \textit{not constrained to be symmetric} (as it happens in the absence of prestress~\cite{pontecastaneda_1997,willis_2002,hutchinson_2006}).
As explained in the next section, this unsymmetry is essential for the correct evaluation of the incremental fourth-order tensor defining the effective continuum, `macroscopically equivalent' to the lattice.

\subsection{Incremental constitutive tensor for the equivalent continuum}
\label{sec:homogenization_constitutive_tensor}
Before introducing the homogenization technique, it is important to recall that, as shown in Section~\ref{sec:prestressed_lattice}, the equilibrium equations for the lattice are
\begin{enumerate*}[label=(\roman*)]
    \item obtained in the context of a linearized theory, and
    \item referred to a prestressed reference configuration, 
\end{enumerate*}
therefore, the unknown `equivalent' continuum has to be formulated in the context of the incremental theory of nonlinear elasticity by means of a relative Lagrangian description~\cite{bigoni_2012}.
As a consequence, the response of the effective material is defined by an \textit{incremental constitutive law} in the form
\begin{equation}
    \label{eq:incremental_constitutive_law}
    \dot{\bS} = \fC[\bL] \,,
\end{equation}
relating the increment of the first Piola-Kirchhoff stress $\dot{\bS}$ to the gradient of incremental displacement $\bL$, through the elasticity tensor $\fC$.
The most general form for the constitutive tensor $\fC$ is
\begin{equation}
    \label{eq:constitutive_operator}
    \fC = \fE + \bI \boxtimes \bT \qquad \mbox{in components} \qquad \fC_{ijkl} = \fE_{ijkl} + \delta_{ik} T_{jl} \,,
\end{equation}
where $\delta_{ik}$ is the Kronecker delta, $\bT$ is the Cauchy stress, defining the \textit{prestress}, and $\fE$ is a fourth-order elastic tensor, endowed with all usual (left and right minor and major) symmetries
\begin{equation}
    \label{eq:minor_symmetries_E}
    \fE_{ijkl} = \fE_{jikl} = \fE_{ijlk} = \fE_{klij} \,,
\end{equation}
so that $\fC$ lacks the minor symmetries but has the major symmetry. 
The symmetries of $\fC$ explain the reason why the full incremental deformation gradient $\bL$, and \textit{not} only its symmetric part, appears in the Cauchy-Born hypothesis~\eqref{eq:cauchy_born_hypothesis} of the lattice.
Moreover, Eq.~\eqref{eq:constitutive_operator} shows that $\bL$ can be restricted to be symmetric \textit{only in the absence of prestress}, $\bT=\bzero$. 

The incremental strain-energy density for the prestressed continuum is referred to the prestressed configuration and can be expressed in terms of a second-order expansion with respect to the incremental deformation gradient $\bL$ as follows
\begin{equation}
\label{eq:incremental_energy_continuum}
    \mW(\bL) = \underbrace{\bT\scalp\bL}_{\mW_1(\bL)} + \underbrace{\fC[\bL]\scalp\bL/2}_{\mW_2(\bL)} \,,
\end{equation}
where the first-order increment $\mW_1(\bL)$ accounts for the work expended by the current prestress state $\bT$ (due to the relative Lagrangean description the first Piola-Kirchhoff stress coincides with the Cauchy stress), while the second-order term $\mW_2(\bL)$ is the strain-energy density associated with the incremental first Piola-Kirchhoff stress given by Eq.~\eqref{eq:incremental_constitutive_law}.

It is also worth noting that taking the second gradient of the incremental energy density~\eqref{eq:incremental_energy_continuum} with respect to $\bL$ yields the constitutive fourth-order tensor $\fC$ relating the stress increment to the incremental displacement gradient, while the first gradient provides, when evaluated at $\bL=\bzero$, the prestress $\bT$.
The latter property will be used to dissect the effect of prestress in the homogenized response of the lattice.

\subsection{First and second-order matching of the incremental strain-energy density}
\label{sec:energy_matching}
The homogenization of the lattice response is based on the equivalence between the average incremental strain-energy associated to a \textit{macroscopic} incremental displacement gradient applied to the lattice and the incremental strain-energy density of the effective elastic material subject to the same deformation.
In the classical homogenization theory, this condition is known as \textit{macro-homogeneity} condition, or Hill-Mandel theorem,~\cite{hill_1972,sanchez-palencia_1987,pontecastaneda_1997,willis_2002}, which provides the link between the microscopic and macroscopic scale.

In the following, the macro-homogeneity condition is enforced to obtain the incremental energy density~\eqref{eq:incremental_energy_continuum} that matches the effective behaviour of the prestressed lattice at first- $\mW_1(\bL)$ and at second- $\mW_2(\bL)$ order.
Thus, the homogenization scheme is based on the following steps:
\begin{enumerate}[label=(\roman*)]
	\item An incremental deformation gradient $\bL$ is considered, so that the incremental energy density for the unknown equivalent continuum is defined by Eq.~\eqref{eq:incremental_energy_continuum};
	\item following the Cauchy-Born hypothesis,  Eq.~\eqref{eq:cauchy_born_hypothesis_vector}, the incremental displacement field for the lattice is prescribed by the given tensor $\bL$ and the periodic vector $\Tilde{\bq}^*$ necessary to enforce the  equilibrium of the lattice; 
	\item with the solution of the lattice in terms of $\bL$ (the periodic vector $\Tilde{\bq}^*$ becomes in solution a function of $\bL$) the incremental energy density is calculated for the lattice;
	\item the two incremental energy densities in the continuum and in the lattice are matched, so to obtain the parameters defining the equivalent solid.
\end{enumerate}

\paragraph{Determination of the periodic displacement field for the lattice.}
By substituting condition~\eqref{eq:cauchy_born_hypothesis_vector} into Eqs.~\eqref{eq:equilibrium_unit_cell} and pre-multiplying by $\trans{\bZ_0}$, the incremental equilibrium becomes
\begin{equation}
\label{eq:equilibrium_internal_strain_f}
    \trans{\bZ_0}\bK(\bP)\,\bZ_0\,\Tilde{\bq}^* + \trans{\bZ_0}\bK(\bP)\,\hat{\bq}(\bL) = \trans{\bZ_0}\bef \,,
\end{equation}
where the right-hand side can be written more explicitly using the partitioning introduced in Fig.~\ref{fig:FB_joints} as
\begin{equation*}
    \trans{\bZ_0}\bef = 
    \begin{Bmatrix}
        \bef^i \\
        \bef^l + \bef^r \\
        \bef^b + \bef^t \\
        \bef^{lb} + \bef^{rb} + \bef^{lt} + \bef^{rt}
    \end{Bmatrix} \,.
\end{equation*}
The fact that the only non-vanishing forces are assumed to be the internal actions transmitted at the unit cell boundary by the neighboring cells implies  $\bef^i=\bzero$.
Moreover, as the displacement field satisfying the Cauchy-Born hypothesis generates \textit{internal} forces in the infinite lattice that are \textit{periodic} along the direct basis $\{\ba_1,\ba_2\}$, any single unit cell is subject to \textit{external} boundary forces that are \textit{anti-periodic}.
Consequently $\bef^l=-\bef^r$, $\bef^b=-\bef^t$ and $\bef^{lb}=-\bef^{rb}-\bef^{lt}-\bef^{rt}$, so that the term $\trans{\bZ_0}\bef$ vanishes and Eq.~\eqref{eq:equilibrium_internal_strain_f} becomes 
\begin{equation}
\label{eq:equilibrium_internal_strain}
    \trans{\bZ_0}\bK(\bP)\,\bZ_0\,\Tilde{\bq}^* = -\trans{\bZ_0}\bK(\bP)\,\hat{\bq}(\bL) \,.
\end{equation}
The solution of the linear system~\eqref{eq:equilibrium_internal_strain} provides the incremental strain $\Tilde{\bq}^*$ internal to the lattice for every given $\bL$. 
As a consequence of the linearity of $\hat{\bq}(\bL)$, the solution $\Tilde{\bq}^*(\bL)$ is, in turn, linear in $\bL$.

A few considerations have to be made about the solvability of the system~\eqref{eq:equilibrium_internal_strain}.
In fact, it is easy to show that the matrix $\trans{\bZ_0}\bK(\bP)\,\bZ_0$ is always singular, regardless of the specific lattice structure under consideration.
This is proved by considering a vector $\Tilde{\bq}^* = \bt$ defining a pure rigid-body translation and observing that $\bK(\bP)\,\bZ_0\,\bt=\bzero$, which, in turn, implies that the dimension of the nullspace of $\trans{\bZ_0}\bK(\bP)\,\bZ_0$ is \textit{at least} 2, as two linearly independent rigid-body translations exist for a 2D lattice. 
Any other deformation mode, possibly contained in $\ker(\trans{\bZ_0}\bK(\bP)\,\bZ_0)$, is therefore a zero-energy mode~(called also with the pictoresque name `floppy mode' \cite{mao_2018,zhang_2018}).
These modes are excluded in the following analysis to ensure solvability of system~\eqref{eq:equilibrium_internal_strain}, so that $\ker(\trans{\bZ_0}\bK(\bP)\,\bZ_0)$ contains \textit{only} two
(in the present 2D formulation) rigid-body translations. 
This exclusion does not affect generality, as the analysis of floppy modes can always be recovered in the limit of vanishing stiffness of appropriate structural elements. 
Note also that sometimes floppy modes can be eliminated or introduced simply playing with the prestress state (which may induce stiffening or softening~\cite{pellegrino_1986,pellegrino_1990}).

Having excluded floppy modes and observing that the right-hand side of Eq.~\eqref{eq:equilibrium_internal_strain} is orthogonal to $\ker(\trans{\bZ_0}\bK(\bP)\,\bZ_0)$, 
\begin{equation*}
    \bt\scalp\trans{\bZ_0}\bK(\bP)\,\hat{\bq}(\bL) = 0, 
\end{equation*}
for all rigid-body translations $\bt$, the solution $\Tilde{\bq}^*(\bL)$ can be determined.

\paragraph{Match of the second-order incremental strain-energy density and determination of the incremental constitutive tensor.}
The solution of the linear system~\eqref{eq:equilibrium_internal_strain} allows the incremental displacement~\eqref{eq:cauchy_born_hypothesis_vector} to be expressed only in terms of the macroscopic displacement gradient $\bL$ as $\bq(\Tilde{\bq}^*(\bL),\bL)$.
Therefore, the second-order incremental strain-energy stored in a single unit cell of the lattice undergoing a macroscopic strain can be evaluated as follows
\begin{equation}
\label{eq:strain_energy_lattice}
    \mE(\bL) = \frac{1}{2} \, \bq(\Tilde{\bq}^*(\bL),\bL) \scalp \bK(\bP)\,\bq(\Tilde{\bq}^*(\bL),\bL) \,,
\end{equation}
which is a quadratic form in $\bL$, because $\bq(\Tilde{\bq}^*(\bL),\bL)$ is linear in $\bL$.
By equating the second-order strain-energy density of the continuum 
$\mW_2(\bL) = \fC[\bL] \scalp \bL/2$
to the average energy of the lattice~\eqref{eq:strain_energy_lattice}, the following equivalence condition is obtained
\begin{equation}
\label{eq:strain_energy_density_equivalence}
    \underbrace{ \frac{1}{2}\,\fC[\bL] \scalp \bL}_{\text{Continuum}} = \underbrace{\frac{1}{\lvert\mC\lvert}\,\mE(\bL)}_{\text{Lattice}} \,,
\end{equation}
where $\lvert\mC\lvert$ is the area of the unit cell.

Finally, the second gradient of~\eqref{eq:strain_energy_density_equivalence} with respect to $\bL$ yields the incremental constitutive tensor for the effective Cauchy material, equivalent to the lattice, in the form
\begin{equation}
\label{eq:constitutive_tensor_lattice}
    \fC = \frac{1}{\lvert\mC\lvert}\,\frac{\partial^2\,\mE(\bL)}{\partial\bL\,\partial\bL}
        = \frac{1}{2\lvert\mC\lvert}\,\frac{\partial^2}{\partial\bL\,\partial\bL} \Big[ \bq(\Tilde{\bq}^*(\bL),\bL) \scalp \bK(\bP)\,\bq(\Tilde{\bq}^*(\bL),\bL) \Big] \,,
\end{equation}
which becomes now an \textit{explicit function of the prestress state}, as well as of all the mechanical parameters defining the lattice.

\paragraph{Match of the first-order incremental strain-energy density and homogenization of the prestress state.}
\label{sec:homogenization_prestress}
So far, the incremental constitutive tensor $\fC$ of a continuum equivalent to a prestressed elastic lattice, Eq.~\eqref{eq:constitutive_tensor_lattice}, has been obtained through homogenization.
It is important now to `dissect' from  $\fC$ the effect of the prestress $\bT$ and, as a consequence, to obtain the tensor $\fE$.

It will be shown below that the current prestress state $\bT$ of the homogenized material can be directly linked to the prestress state $\bP=\{P_1,...,P_{N_b}\}$ of the lattice. 
In fact, by observing that~\eqref{eq:strain_energy_density_equivalence} represents the \textit{second-order} incremental strain energy, equal to $\mW_2(\bL)=\dot{\bS}(\bL)\scalp\bL/2$, an equivalence analogous to~\eqref{eq:strain_energy_density_equivalence} can be obtained considering the \textit{first-order} increment of the strain energy, $\mW_1(\bL)=\bT\scalp\bL$.
Thus, the first-order term can be identified as the \textit{average work done by the prestress state} $\bP$ \textit{during the lattice deformation} $\bq(\Tilde{\bq}^*(\bL),\bL)$ \textit{induced by} $\bL$ so that the following equivalence can be stated
\begin{equation}
\label{eq:work_prestress_equivalence}
    \underbrace{\bT\scalp\bL}_{\text{Continuum}} = \underbrace{\frac{1}{\lvert\mC\rvert}\,\bef_\bP\scalp\bq(\Tilde{\bq}^*(\bL),\bL)}_{\text{Lattice}} \,,
\end{equation}
where the vector $\bef_\bP$ collects the forces that emerge at the nodes of the unit cell and are in equilibrium with the axial preload of the elastic rods $\bP$ \textit{in the current configuration assumed as reference}. 
As a consequence, the forces $\bef_\bP$ are independent of $\bL$ and linear in $\bP$.

Equation \eqref{eq:work_prestress_equivalence} requires that the work done by axial loads $\bef_\bP$ for nodal displacements $\bq$ associated to a skew-symmetric velocity gradient $\bL = \bW$ be zero, namely
\begin{equation}
\label{eq:work_prestress_spin}
    \bef_\bP\scalp\bq(\Tilde{\bq}^*(\bW),\bW) = 0 \,.
\end{equation}
This statement is a direct consequence of the principle of virtual work for rigid body incremental motions, because $\bq(\Tilde{\bq}^*(\bW),\bW)$ represents an incremental rotation of the lattice and $\bef_\bP$ satisfies equilibrium. 
Hence, taking into account the property~\eqref{eq:work_prestress_spin}, the homogenized prestress $\bT$ can be obtained as the gradient of the equivalence condition~\eqref{eq:work_prestress_equivalence} with respect to the symmetric part of $\bL$, denoted as $\bD$, 
\begin{equation}
\label{eq:prestress_tensor_lattice}
    \bT = \frac{1}{\lvert\mC\rvert}\,\deriv{}{\bD} \Big[ \bef_\bP\scalp\bq(\Tilde{\bq}^*(\bD),\bD) \Big]\,.
\end{equation}

\section{Stability of prestressed lattices of elastic rods, strong ellipticity, and ellipticity of the effective continuum}
\label{sec:ellipticity_stability}

\paragraph{Lattice bifurcations} are governed by the value of the preload $\bP$ and they can exhibit deformation modes with different wavelength.
When the wavelength becomes infinite, a `global bifurcation' occurs. 
While in the homogenization procedure periodic conditions are used, the systematic investigation of bifurcations occurring in the lattice can be conducted by complementing the incremental equilibrium of the lattice~\eqref{eq:equilibrium_unit_cell} with Floquet-Bloch boundary conditions, which involve displacement fields of arbitrary wavelength~\cite{hutchinson_2006}.

Denoting the wave vector as $\bk\in\Reals^2$ and applying Bloch's theorem (see~\cite{phani_2006,bordiga_2019a} for details), Eq.~\eqref{eq:equilibrium_unit_cell} becomes
\begin{equation}
\label{eq:equilibrium_local_buckling}
\conjtrans{\bZ(\bk)} \bK(\bP)\, \bZ(\bk) \, \bq^* = \bzero \,,
\end{equation}
where 
symbol $\conjtrans{}$ denotes the complex conjugate transpose operation
and the matrix-valued function $\bZ(\bk)$ generalizes $\bZ_0$ in Eq.~\eqref{eq:periodic_conditions} as 
\begin{equation}
\label{eq:FB_conditions}
    \bq =
        \begin{Bmatrix}
            \bq^i \\
            \bq^l \\
            \bq^b \\
            \bq^{lb} \\
            \bq^r \\
            \bq^t \\
            \bq^{rb} \\
            \bq^{lt} \\
            \bq^{rt}
        \end{Bmatrix}
    =
        \begin{bmatrix}
            \bI & \bzero & \bzero & \bzero \\
            \bzero & \bI & \bzero & \bzero \\
            \bzero & \bzero & \bI & \bzero \\
            \bzero & \bzero & \bzero & \bI \\
            \bzero & z_1\bI & \bzero & \bzero \\
            \bzero & \bzero & z_2\bI & \bzero \\
            \bzero & \bzero & \bzero & z_1\bI \\
            \bzero & \bzero & \bzero & z_2\bI \\
            \bzero & \bzero & \bzero & z_1 z_2 \bI
        \end{bmatrix}
        \begin{Bmatrix}
            \bq^i \\
            \bq^l \\
            \bq^b \\
            \bq^{lb}
        \end{Bmatrix} \,,
    \quad
    \bq = \bZ(\bk)\, \bq^* \,,
\end{equation}
in which $z_j = e^{i\,\bk\scalp\ba_j}\quad\forall j\in\{1,2\}$.

Note that conditions~\eqref{eq:FB_conditions} represents the generalization of Eq.~\eqref{eq:periodic_conditions} to displacement fields of arbitrary wavelengths, in fact $\bZ(\bzero)=\bZ_0$, therefore they allow bifurcations of arbitrary wavelength to be detected and not only those occurring at the long-wavelength limit, $\norm{\bk}\to 0$.

For a given $\bk$, the associated preload state $\bP$ leading to a bifurcation can be obtained by searching for non-trivial solutions of the incremental equilibrium~\eqref{eq:equilibrium_local_buckling}.
Hence, by introducing the notation $\bK^*(\bP,\bk)=\conjtrans{\bZ(\bk)} \bK(\bP)\, \bZ(\bk)$, a bifurcation becomes possible when
\begin{equation}
\label{eq:det_local_buckling}
    \det \bK^*(\bP,\bk) = 0 \,.
\end{equation}
It is worth noting that the matrix $\bK^*(\bP,\bk)$ is Hermitian,  $\bK^*(\bP,\bk)=\conjtrans{\bK^*(\bP,\bk)}$, which implies that the determinant~\eqref{eq:det_local_buckling} is always real.
Moreover, the periodicity of $\bZ(\bk)$ implies that this determinant is periodic in the $\bk$-space with period $[0,2\pi]{\times}[0,2\pi]$ in the basis $\{\bb_1,\bb_2\}$ reciprocal to $\{\ba_1,\ba_2\}$, so that  $\bb_i\scalp\ba_j=\delta_{ij}$.

In order to construct the \textit{stability domain} of a lattice, the \textit{critical} (in other words first) bifurcation needs to be selected by solving Eq.~\eqref{eq:det_local_buckling} for the smallest preload spanning over all possible wavelengths.
Specifically, by introducing the unit vector $\hat{\bP}$, which singles out a direction in the preload space, the prestress state is defined as $\gamma \hat{\bP}$ for a radial loading, so that the critical bifurcation corresponds to the value $\gamma_{\text{B}}$ defined as 
\begin{equation}
\label{eq:local_buckling_multiplier_simpler}
    \gamma_{\text{B}} = \inf_{\gamma \geq 0} 
    \left\{ \gamma \, \Big\lvert \, \det\bK^*(\gamma\hat{\bP},\, \eta_1\bb_1+\eta_2\bb_2) = 0\,,\, 0 < \eta_1 < 2\pi\,,\, 0 < \eta_2 < 2\pi \right\} . 
\end{equation}
where the periodicity of $\bK^*(\bP,\bk)$ is used to conveniently restrict to one period the search for the infimum over the $\bk$-space. 
It is worth noting that for a vanishing wave vector, Eq.~\eqref{eq:det_local_buckling} is always satisfied regardless of the preload state, because the nullspace of $\bK^*(\bP,\bzero)$ always contains rigid-body translations.
These trivial solutions clearly need to be excluded.

\paragraph{Strong ellipticity} enforces uniqueness of the incremental problem of a homogeneous and homogeneously deformed material subject to prescribed incremental displacement on the whole boundary \cite{hill_1962} and corresponds to the positive definiteness of the acoustic tensor (associated to the incremental constitutive tensor $\fC$) defined with reference to every unit vectors $\bn$ and $\bg$ as 
\begin{equation}
	\bA^{(\fC)}(\bn)\,\bg = \fC[\bg\otimes\bn]\,\bn \,.
\end{equation}
When the prestress state is null and except in the case of an extreme material, where the stiffness of the rods becomes vanishing small~\cite{gourgiotis_2016}, the homogenized material response is strong elliptic, which in turn implies ellipticity.

\paragraph{Failure of ellipticity} corresponds to macro (or global) instabilities, where the bifurcation is characterized by a wavelength long when compared to the period of the lattice structure, which models a localization of deformation in the equivalent continuum. 
The homogenized material is elliptic (E) as long as the the acoustic tensor $\bA^{(\fC)}(\bn)$  is non-singular for every pair of unit vectors $\bn$ and $\bg$, namely, 
\begin{equation}
    \label{eq:failure_E}
     \bA^{(\fC)}(\bn)\,\bg \neq \bzero \,. 
\end{equation}
When the acoustic tensor becomes singular, a localization of deformation may occur corresponding to a dyad $\bg \otimes \bn$. The localization is called \lq shear band' in the special case $\bg\scalp\bn=0$, or \lq compaction band' or \lq splitting mode' when $\bg\scalp\bn=\pm 1$.

It is assumed that the elastic lattice under consideration is equivalent, at null prestress, to a strong elliptic elastic solid, characterized by a constitutive tensor which is function of the prestress $\bT$, in turn through the axial preload $\bP$ in the elastic rods, equation~\eqref{eq:prestress_tensor_lattice}, namely, $\bA^{(\fC)}(\bP, \bn)$. 
Therefore, using again the previously defined unit vector $\hat{\bP}$ and with reference to an infinite material (or to a material with prescribed displacement on the whole boundary) bifurcations are excluded as long as the response remains strong elliptic, while failure of this condition is simultaneous to failure of ellipticity, which occurs at the value $\gamma_{\text{E}}$ defined as 
\begin{equation}
\label{eq:failure_E_multiplier}
    \gamma_{\text{E}} = \min_{\gamma\geq0}\left\{ \gamma \, \Big\lvert \, \min_{\bn,\norm{\bn}=1} \left[\det\bA^{(\fC)}(\gamma\hat{\bP},\bn)\right]=0\right\}. 
\end{equation}

\paragraph{Relation between bifurcations in the lattice and in the effective continuum} is that failure of ellipticity of the latter corresponds to long-wavelength bifurcations of the former, $\norm{\bk}\to0$, while \textit{all} bifurcations are scanned through equation~\eqref{eq:local_buckling_multiplier_simpler}, a circumstance which implies $\gamma_{\text{B}}\leq\gamma_{\text{E}}$.
Moreover, whenever $\gamma_{\text{B}}<\gamma_{\text{E}}$ \textit{the bifurcation occurs at microscopic level and is not detectable in the homogenized material, which can still be strong elliptic}~\cite{geymonat_1993,triantafyllidis_1993,lopez-pamies_2006}.

\section{Derivation of the incremental constitutive operator, failure of ellipticity and micro-bifurcation for a specific elastic lattice}
\label{sec:grid}
\begin{figure}[htb!]
    \centering
    \begin{subfigure}{0.47\textwidth}
        \centering
        \caption{\label{fig:geometry_grid}}
        \includegraphics[height=50mm]{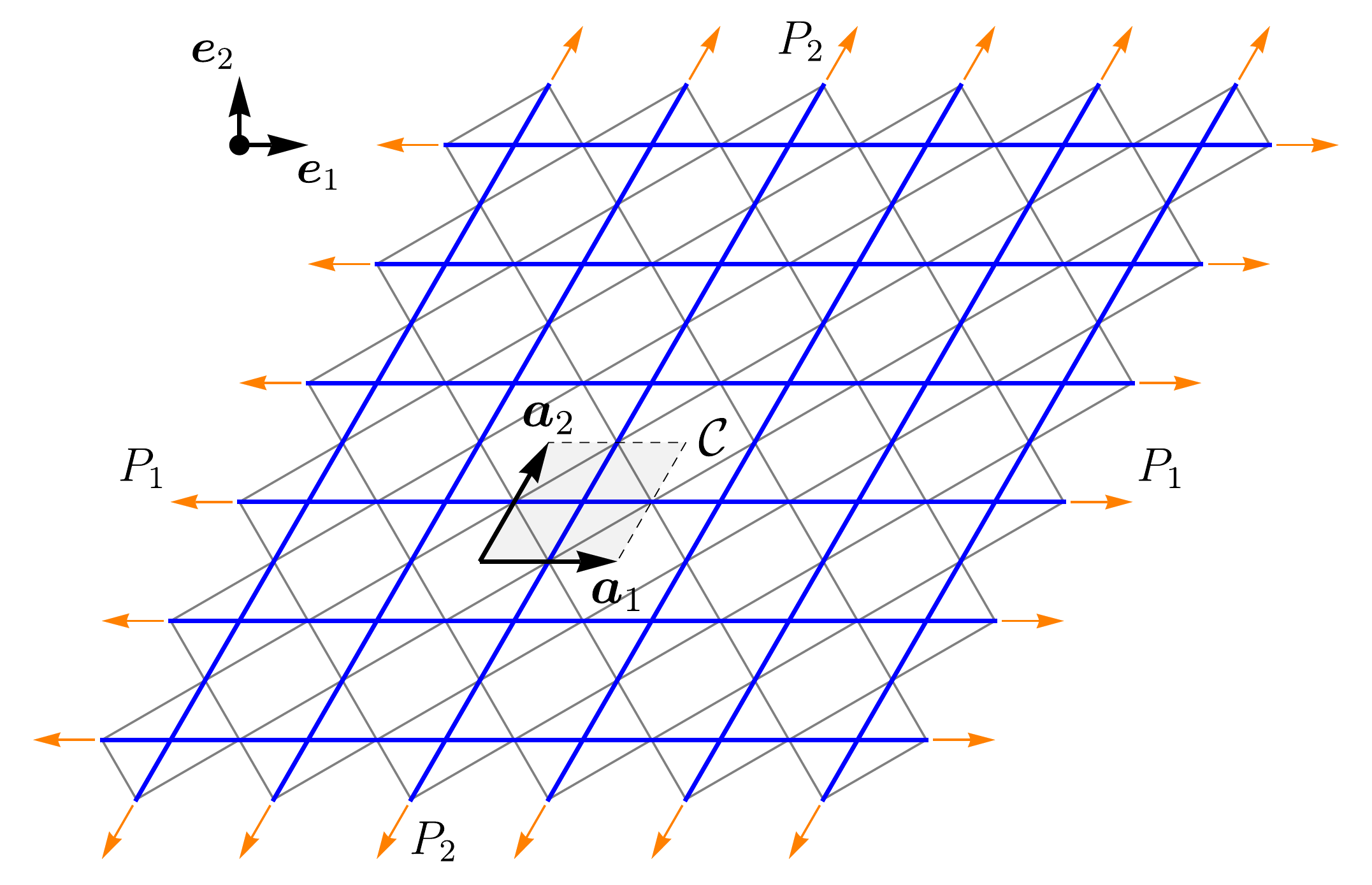}
    \end{subfigure}\hspace{3mm}%
    \begin{subfigure}{0.44\textwidth}
        \centering
        \caption{\label{fig:geometry_grid_unit_cell}}
        \includegraphics[height=50mm]{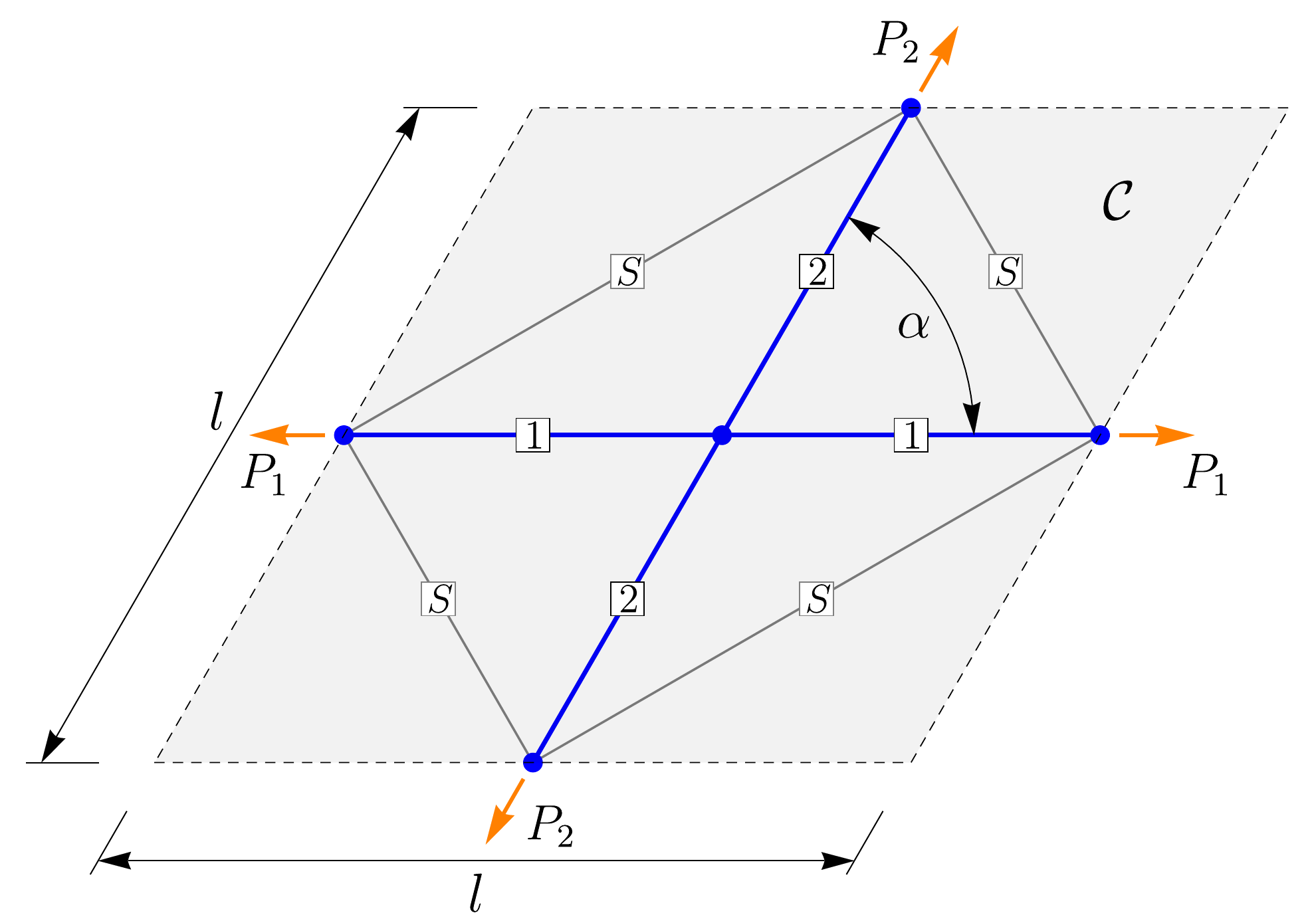}
    \end{subfigure}%
    \caption{\label{fig:geometry_grid_and_unit_cell}
    Current configuration of a rhombic lattice of preloaded elastic rods~(\subref{fig:geometry_grid}), with the associated unit cell $\mC$~(\subref{fig:geometry_grid_unit_cell}).
    The direct basis of the lattice is denoted by the pair of vectors $\{\ba_1,\ba_2\}$ (\subref{fig:geometry_grid}).
    Labels 1, 2, and $S$ denote the horizontal rods, the inclined rods, and the diagonal springs, respectively~(\subref{fig:geometry_grid_unit_cell}).
    The spring stiffness, the axial and flexural rigidity of the rods, the preloads $P_1$ and $P_2$, as well as the grid angle $\alpha$ can all be varied to investigate different incremental responses.
    }
\end{figure}

The geometry of the current, prestressed configuration of a preloaded lattice, selected to apply the previously developed formalism, is sketched in Fig.~\ref{fig:geometry_grid_and_unit_cell} and is composed of a rhombic grid (of side $l$) of elastic rods, inclined at an angle $\alpha$, and characterized by the following non-dimensional parameters and $A_2=A_1=A$, $\Lambda_1 = l \sqrt{A/B_1}$, $\Lambda_2 = l \sqrt{A/B_2}$, where the subscript $1$ and $2$ are relative to the horizontal and inclined rods, as depicted in Fig.~\ref{fig:geometry_grid_unit_cell}.
The direct basis of the periodic structure is denoted by the pair of vectors $\{\ba_1,\ba_2\}$ whose representation with respect to the basis $\{\be_1,\be_2\}$ (see Fig.~\ref{fig:geometry_grid}) is
\begin{equation*}
    \ba_1 = l\,\be_1 \,, \qquad \ba_2 = l\,(\be_1\cos\alpha+\be_2\sin\alpha) \,,
\end{equation*}
while the reciprocal basis $\{\bb_1,\bb_2\}$ is defined as $\ba_i\scalp\bb_j=\delta_{ij}$, so that
\begin{equation*}
    \bb_1 = (\be_1 - \be_2\cot\alpha)/l \,, \qquad \bb_2 = \be_2\csc\alpha/l \,,
\end{equation*}
and the wave vector can be written as $\bk=\eta_1\bb_1+\eta_2\bb_2$, with $\eta_{1,2}$ being dimensionless components.
The resulting `skewed' grid is also considered stiffened by a diagonal bracing realized by linear springs\footnote{These springs can be seen as added after the lattice has been deformed or as deformed together with the lattice. In the former case further assumptions need not be introduced, while in the latter, the effects of the preload on the springs has to be neglected in the interest of simplicity. The diagonal springs are used in this example to show that microscopic instabilities may occur before macroscopic.} connecting the midpoints of the horizontal and inclined rods, as sketched in Fig.~\ref{fig:geometry_grid_unit_cell}. 
The stiffness of the springs is assumed constant $k_{\text{S}}=\kappa A/l$, with $\kappa$ being a dimensionless measure of stiffness.

In the configuration shown in Fig. \ref{fig:geometry_grid_and_unit_cell}, the lattice is subject to a preload state defined by the axial forces $P_1$ and $P_2$, made dimensionless respectively as 
$p_1 = {P_1 l^2}/{B_1}$ and $p_2 = {P_2 l^2}/{B_2}$, so that a lattice is defined by the parameter set $\{\alpha,\Lambda_1,\Lambda_2,\kappa,p_1,p_2\}$.
Note also that the considered lattice structure includes as a particular case that of a rectangular grid, analyzed in~\cite{triantafyllidis_1993}.

\subsection{Incremental constitutive tensor of the equivalent continuum}
\label{sec:grid_C}
The homogenization technique outlined in Section~\ref{sec:homogenization} for prestressed lattices of arbitrary geometry can be directly applied to the grid of elastic rods shown in Fig.~\ref{fig:geometry_grid_and_unit_cell}.
The incremental constitutive tensor is computed via  Eq.~\eqref{eq:constitutive_tensor_lattice} and made dimensionless as follows
\begin{subequations}
\begin{equation}
\label{eq:grid_C}
    \fC =  \frac{A}{l} \, \Bar{\fC}(\underbrace{p_1,\,p_2}_{\text{prestress}},\underbrace{\Lambda_1,\,\phi,\,\kappa,\,\alpha}_{\text{microstructure}}) \,,
\end{equation}
where $\phi=B_2/B_1$ (note that $\Lambda_2=\Lambda_1/\sqrt{\phi}$), and the non-dimensional tensor-valued function $\Bar{\fC}(p_1,p_2,\Lambda_1,\phi,\kappa,\alpha)$ can be decomposed as
\begin{equation}
    \Bar{\fC}(p_1,p_2,\Lambda_1,\phi,\kappa,\alpha) = \Bar{\fC}^{\text{G}}(p_1,p_2,\Lambda_1,\phi,\alpha) + \Bar{\fC}^{\text{S}}(\kappa,\alpha) \,,
\end{equation}
with $\Bar{\fC}^{\text{G}}$ and $\Bar{\fC}^{\text{S}}$ being, respectively, the contribution of the rod's grid and the diagonal springs.
\end{subequations}
The full expression for the components of $\Bar{\fC}^{\text{G}}$ and $\Bar{\fC}^{\text{S}}$ with respect to the basis $\{\be_1,\be_2\}$ (sketched in Fig.~\ref{fig:geometry_grid}) 
is the following (components that have to be equal by symmetry are not reported) 
\begin{dgroup*}[style={\footnotesize},breakdepth={20}]
    \begin{dmath*}
        \Bar{\fC}^{\text{G}}_{1111} = \frac{1}{2 d \sin\alpha}
        \left(\sinh \left(\frac{\sqrt{p_2}}{2}\right) \left(\sqrt{p_1} p_2 \phi \cosh \left(\frac{\sqrt{p_1}}{2}\right) \left( \cos (4 \alpha) \left(\Lambda_1^2-p_2 \phi\right)+4 \Lambda_1^2 \cos (2 \alpha)+11\Lambda_1^2+p_2 \phi\right)
        -2 \sinh \left(\frac{\sqrt{p_1}}{2}\right) \left( \cos (4 \alpha) \left(\Lambda_1^2 \left(p_1+p_2 \phi\right)-p_2^2 \phi^2\right)
        +\Lambda_1^2 \left(p_1+p_2\phi\right) \left(4 \cos (2 \alpha) +11\right)
        +p_2^2 \phi^2\right)\right) + p_1 \sqrt{p_2} \sinh \left(\frac{\sqrt{p_1}}{2}\right) \cosh \left(\frac{\sqrt{p_2}}{2}\right) \left(\cos (4 \alpha) \left(\Lambda_1^2-p_2 \phi\right)+4 \Lambda_1^2 \cos (2 \alpha)+ 11\Lambda_1^2+p_2 \phi\right)\right)
        ,
    \end{dmath*}
    \begin{dmath*}
        \Bar{\fC}^{\text{G}}_{1122} = \frac{4 \sin\alpha \cos^2\alpha}{d}
        \left(\sinh \left(\frac{\sqrt{p_2}}{2}\right) \left(\sqrt{p_1} p_2   \phi \cosh \left(\frac{\sqrt{p_1}}{2}\right) \left(\Lambda_1^2-p_2 \phi\right)-2 \sinh \left(\frac{\sqrt{p_1}}{2}\right) \left(\Lambda_1^2 \left(p_1+p_2   \phi\right)-p_2^2 \phi^2\right)\right)+p_1 \sqrt{p_2} \sinh \left(\frac{\sqrt{p_1}}{2}\right) \cosh \left(\frac{\sqrt{p_2}}{2}\right) \left(\Lambda_1^2-p_2 \phi\right)\right)
        ,
    \end{dmath*}
    \begin{dmath*}
        \Bar{\fC}^{\text{G}}_{1112} = 
         \frac{-2 \cos\alpha}{d}
         \left( \sinh \left(\frac{\sqrt{p_2}}{2}\right) \left(2 \sinh \left(\frac{\sqrt{p_1}}{2}\right) \left(\cos (2 \alpha) \left(\Lambda_1^2 \left(p_1+p_2 \phi\right)-p_2^2 \phi^2\right)+\Lambda_1^2 \left(p_1+p_2 \phi\right)+p_2^2 \phi^2\right)-\sqrt{p_1} p_2 \phi \cosh \left(\frac{\sqrt{p_1}}{2}\right) \left(\cos (2 \alpha) \left(\Lambda_1^2-p_2 \phi\right)+\Lambda_1^2+p_2 \phi\right)\right)+p_1 \sqrt{p_2} \sinh \left(\frac{\sqrt{p_1}}{2}\right) \cosh \left(\frac{\sqrt{p_2}}{2}\right) \left(\cos (2 \alpha) \left(p_2 \phi-\Lambda_1^2\right)-\Lambda_1^2-p_2 \phi\right)\right)
        ,
    \end{dmath*}
    \begin{dmath*}
        \Bar{\fC}^{\text{G}}_{1121} = \frac{4 \cos\alpha}{d}
        \left(  \sinh \left(\frac{\sqrt{p_2}}{2}\right) \left(2 \sinh \left(\frac{\sqrt{p_1}}{2}\right) \left(p_1 p_2 \phi-\cos^2\alpha \left(\Lambda_1^2 \left(p_1+p_2   \phi\right)-p_2^2   \phi^2\right)\right)+\sqrt{p_1} p_2   \phi \cos^2\alpha \cosh \left(\frac{\sqrt{p_1}}{2}\right) \left(\Lambda_1^2-p_2 \phi\right)\right)+p_1 \sqrt{p_2}   \cos^2\alpha \sinh \left(\frac{\sqrt{p_1}}{2}\right) \cosh \left(\frac{\sqrt{p_2}}{2}\right) \left(\Lambda_1^2-p_2 \phi\right)\right)
        ,
    \end{dmath*}
    \begin{dmath*}
        \Bar{\fC}^{\text{G}}_{2222} = 
        \frac{2 \sin\alpha }{d} \left(  \sinh \left(\frac{\sqrt{p_2}}{2}\right) \left(\sqrt{p_1} p_2  \phi \cosh \left(\frac{\sqrt{p_1}}{2}\right) \left(\cos (2 \alpha) \left(p_2 \phi-\Lambda_1^2\right)+\Lambda_1^2+p_2 \phi\right)-2 \sinh \left(\frac{\sqrt{p_1}}{2}\right) \left(-\cos (2 \alpha) \left(\Lambda_1^2 \left(p_1+p_2   \phi\right)-p_2^2   \phi^2\right)+\Lambda_1^2 \left(p_1+p_2   \phi\right)+p_2^2   \phi^2\right)\right)+p_1 \sqrt{p_2}   \sinh \left(\frac{\sqrt{p_1}}{2}\right) \cosh \left(\frac{\sqrt{p_2}}{2}\right) \left(\cos (2 \alpha) \left(p_2 \phi-\Lambda_1^2\right)+\Lambda_1^2+p_2 \phi\right)\right)
        ,
    \end{dmath*}
    \begin{dmath*}
        \Bar{\fC}^{\text{G}}_{2212} = 
        \frac{4   \sin^2\alpha \cos\alpha }{d}
        \left(\sinh \left(\frac{\sqrt{p_2}}{2}\right) \left(\sqrt{p_1} p_2   \phi \cosh \left(\frac{\sqrt{p_1}}{2}\right) \left(\Lambda_1^2-p_2 \phi\right)-2 \sinh \left(\frac{\sqrt{p_1}}{2}\right) \left(\Lambda_1^2 \left(p_1+p_2   \phi\right)-p_2^2   \phi^2\right)\right)+p_1 \sqrt{p_2} \sinh \left(\frac{\sqrt{p_1}}{2}\right) \cosh \left(\frac{\sqrt{p_2}}{2}\right) \left(\Lambda_1^2-p_2 \phi\right)\right)
        ,
    \end{dmath*}
    \begin{dmath*}
        \Bar{\fC}^{\text{G}}_{2221} = 
        \frac{2   \cos\alpha }{d}
        \left(\sinh \left(\frac{\sqrt{p_2}}{2}\right) \left(\sqrt{p_1} p_2   \phi \cosh \left(\frac{\sqrt{p_1}}{2}\right) \left(\cos (2 \alpha) \left(p_2 \phi-\Lambda_1^2\right)+\Lambda_1^2+p_2 \phi\right)-2 \sinh \left(\frac{\sqrt{p_1}}{2}\right) \left(-\cos (2 \alpha) \left(\Lambda_1^2 \left(p_1+p_2   \phi\right)-p_2^2   \phi^2\right)+\Lambda_1^2 \left(p_1+p_2   \phi\right)+p_2 \phi \left(2 p_1+p_2   \phi\right)\right)\right)+p_1 \sqrt{p_2} \sinh \left(\frac{\sqrt{p_1}}{2}\right) \cosh \left(\frac{\sqrt{p_2}}{2}\right) \left(\cos (2 \alpha) \left(p_2 \phi-\Lambda_1^2\right)+\Lambda_1^2+p_2 \phi\right)\right)
        ,
    \end{dmath*}
    \begin{dmath*}
        \Bar{\fC}^{\text{G}}_{1212} = \frac{-2  \sin\alpha }{d}
        \left(\sinh \left(\frac{\sqrt{p_2}}{2}\right) \left(2 \sinh \left(\frac{\sqrt{p_1}}{2}\right) \left(\cos (2 \alpha) \left(\Lambda_1^2 \left(p_1+p_2   \phi\right)-p_2^2   \phi^2\right)+\Lambda_1^2 \left(p_1+p_2   \phi\right)+p_2^2   \phi^2\right)-\sqrt{p_1} p_2   \phi \cosh \left(\frac{\sqrt{p_1}}{2}\right) \left(\cos (2 \alpha) \left(\Lambda_1^2-p_2 \phi\right)+\Lambda_1^2+p_2 \phi\right)\right)+p_1 \sqrt{p_2} \sinh \left(\frac{\sqrt{p_1}}{2}\right) \cosh \left(\frac{\sqrt{p_2}}{2}\right) \left(\cos (2 \alpha) \left(p_2 \phi-\Lambda_1^2\right)-\Lambda_1^2-p_2 \phi\right)\right)
        ,
    \end{dmath*}
    \begin{dmath*}
        \Bar{\fC}^{\text{G}}_{1221} = \frac{-4   \sin\alpha }{d}
        \left(\sinh \left(\frac{\sqrt{p_2}}{2}\right) \left(2 \sinh \left(\frac{\sqrt{p_1}}{2}\right) \left(\cos^2\alpha \left(\Lambda_1^2 \left(p_1+p_2 \phi\right)-p_2^2 \phi^2\right)-p_1 p_2 \phi\right)+\sqrt{p_1} p_2 \phi \cos^2\alpha \cosh \left(\frac{\sqrt{p_1}}{2}\right) \left(p_2 \phi-\Lambda_1^2\right)\right)+p_1 \sqrt{p_2} \cos^2\alpha \sinh \left(\frac{\sqrt{p_1}}{2}\right) \cosh \left(\frac{\sqrt{p_2}}{2}\right) \left(p_2 \phi-\Lambda_1^2\right)\right)
        ,
    \end{dmath*}
    \begin{dmath*}
        \Bar{\fC}^{\text{G}}_{2121} = \frac{p_1 \sqrt{p_2} }{d}
        \sinh \left(\frac{\sqrt{p_1}}{2}\right) \cosh \left(\frac{\sqrt{p_2}}{2}\right) \left( \sin\alpha \left(\Lambda_1^2-5 p_2 \phi\right)+ \sin (3 \alpha) \left(\Lambda_1^2-p_2 \phi\right)+4 \csc (\alpha) \left(p_1+p_2   \phi\right)\right)-2 \sin\alpha \sinh \left(\frac{\sqrt{p_2}}{2}\right) \left(2 \sinh \left(\frac{\sqrt{p_1}}{2}\right) \left( \cos (2 \alpha) \left(\Lambda_1^2 \left(p_1+p_2 \phi\right)-p_2^2 \phi^2\right)+2 \csc^2(\alpha) \left(p_1+p_2   \phi\right)^2+\Lambda_1^2 \left(p_1+p_2 \phi\right)-p_2 \phi \left(4 p_1+3 p_2   \phi\right)\right)-\sqrt{p_1} p_2 \phi \cosh \left(\frac{\sqrt{p_1}}{2}\right) \left(\cos (2 \alpha) \left(\Lambda_1^2-p_2 \phi\right)+2 \csc ^2(\alpha) \left(p_1+p_2 \phi\right)+ \left(\Lambda_1^2-3 p_2 \phi\right)\right)\right)
        ,
    \end{dmath*}
\end{dgroup*}
where
\begin{dmath*}[style={\footnotesize}]
    d =
    e^{-\frac{1}{2} \left(\sqrt{p_1}+\sqrt{p_2}\right)}
    \Lambda_1^2 \left(\left(e^{\sqrt{p_1}} \left(\sqrt{p_1}-2\right)+\sqrt{p_1}+2\right) \left(e^{\sqrt{p_2}}-1\right) p_2   \phi-2 \left(e^{\sqrt{p_1}}-1\right) p_1 \left(e^{\sqrt{p_2}}-1\right)+\left(e^{\sqrt{p_1}}-1\right) p_1 \left(e^{\sqrt{p_2}}+1\right) \sqrt{p_2}\right)
    .
\end{dmath*}
The constitutive tensor ruling the effect of diagonal springs can be written as 
\begin{align*}
    \Bar{\fC}^{\text{S}}_{1111} &= \kappa \frac{5+3 \cos (2 \alpha)}{4\sin\alpha}  \,,\\
    \Bar{\fC}^{\text{S}}_{1112} &= \Bar{\fC}^{\text{S}}_{1121} = \Bar{\fC}^{\text{S}}_{1211} = \Bar{\fC}^{\text{S}}_{1121} = \Bar{\fC}^{\text{S}}_{2111} = \kappa  \cos \alpha  \,,\\
    \Bar{\fC}^{\text{S}}_{1122} &= \Bar{\fC}^{\text{S}}_{2211} = \Bar{\fC}^{\text{S}}_{1212} = \Bar{\fC}^{\text{S}}_{1221} = \Bar{\fC}^{\text{S}}_{2112} = \Bar{\fC}^{\text{S}}_{2121} = \Bar{\fC}^{\text{S}}_{2222} = \frac{1}{2} \kappa  \sin \alpha  \,, \\
    \Bar{\fC}^{\text{S}}_{1222} &= \Bar{\fC}^{\text{S}}_{2122} = \Bar{\fC}^{\text{S}}_{2212} = \Bar{\fC}^{\text{S}}_{2221} = 0 \,.
\end{align*}

\subsection{Prestress tensor of the equivalent continuum}
\label{sec:grid_T}
The prestress tensor $\bT$, equivalent in the continuum to the preload forces $\bP$ in the elastic lattice, can be either calculated using equation ~\eqref{eq:prestress_tensor_lattice} or, directly, by computing the average normal and tangential tractions along the faces with unit normal $\be_1$ and $\be_2$. With reference to Fig.~\ref{fig:geometry_grid_unit_cell} 
the following expression is obtained
\begin{equation}
\label{eq:grid_T_guessed}
    \bT = \left( \frac{P_1}{l \sin\alpha} + \frac{P_2 \cos^2\alpha}{l \sin \alpha} \right) \be_1\otimes\be_1 + \frac{P_2 \cos\alpha}{l} (\be_1\otimes\be_2 + \be_2\otimes\be_1) + \frac{P_2 \sin\alpha}{l} \be_2\otimes\be_2 \,. 
\end{equation}

\subsection{Loss of ellipticity vs micro-bifurcation}
\label{sec:grid_ellipticity_local_buckling}
With reference to the lattice sketched in Fig.~\ref{fig:geometry_grid_unit_cell}, 
the value of the prestress state, which is critical for bifurcation of the grid is determined by employing conditions~\eqref{eq:failure_E_multiplier} and~\eqref{eq:local_buckling_multiplier_simpler}, and computing numerically the prestress multipliers $\gamma_{\text{E}}$ and $\gamma_{\text{B}}$. 
Results are presented as \textit{uniqueness domains} in the non-dimensional prestress space $\{p_1,p_2\}$ by fixing the set of geometrical and mechanical parameters $\{\alpha,\Lambda_1,\Lambda_2,\kappa\}$.
The boundary of the stability domain identifies the `critical', namely, the first bifurcation of the incremental equilibrium of the lattice.

The dependence on $\alpha,\,\Lambda_1,\,\Lambda_2,\,\kappa$ has been analyzed by considering two grid configurations that will be referred to as the \textit{orthotropic grid}, with equal slenderness $\Lambda_1=\Lambda_2=10$, and the \textit{anisotropic grid}, characterized by different slendernesses, $\Lambda_1=7$ and $\Lambda_2=15$.
For each lattice, the influence of the rods' inclination is explored by setting $\alpha=\pi/2, \pi/3, \pi/4, \pi/6$, while the stiffness of the springs is investigated in the range $\kappa \in [0,1]$. 
In this way, the influence of the diagonal bracing on the critical bifurcation mode is analyzed.

\begin{figure}[!htb]
    \centering
    \begin{subfigure}[t]{0.32\textwidth}
        \centering
        \caption{\label{fig:macro_bifurcation_stab}Uniqueness domain $(\kappa=0)$}
        \includegraphics[width=\linewidth]{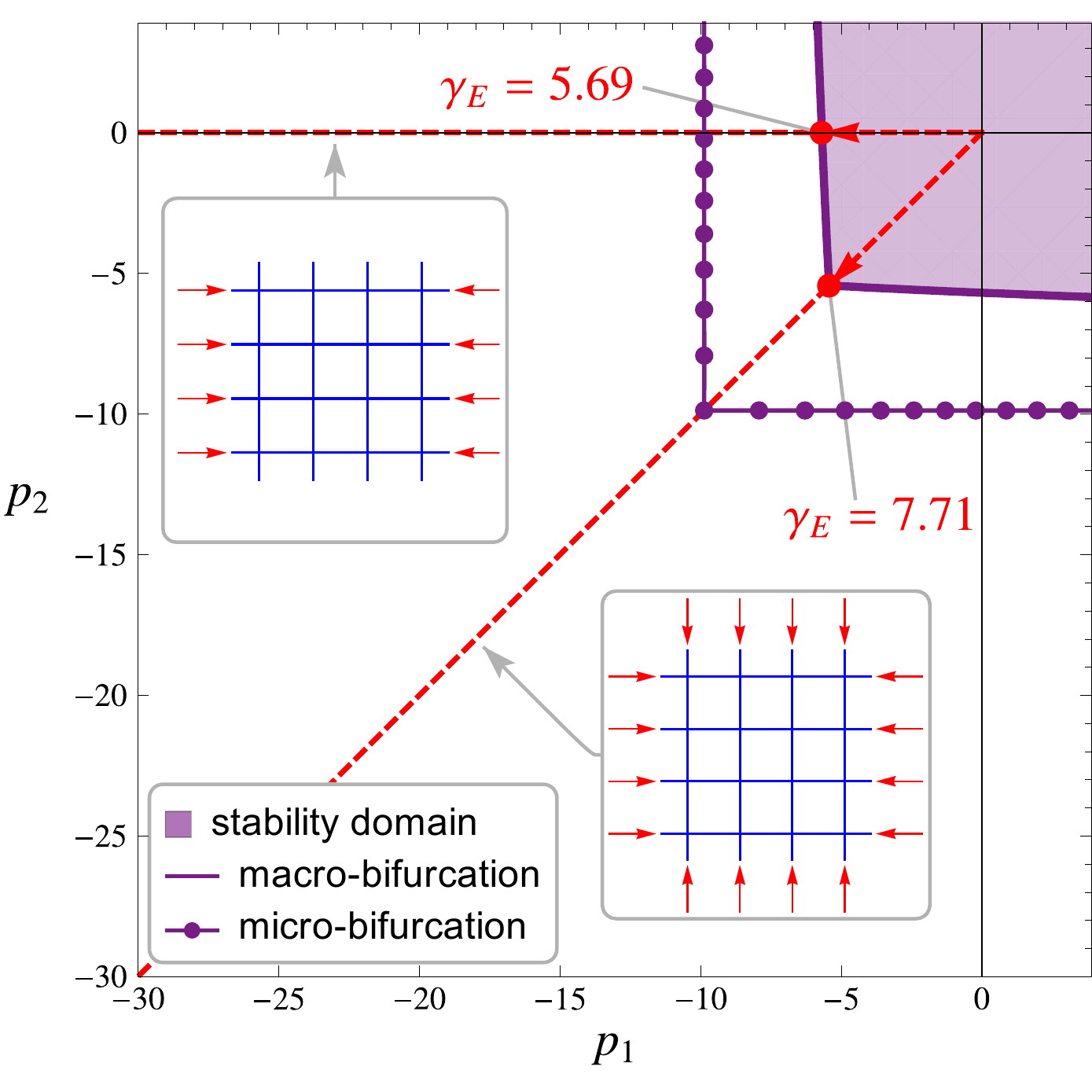}
    \end{subfigure}
    \hspace*{1mm}
    \begin{subfigure}[t]{0.31\textwidth}
        \centering
        \caption{\label{fig:macro_bifurcation_surf1}Equibiaxial compression}
        \includegraphics[width=0.95\linewidth]{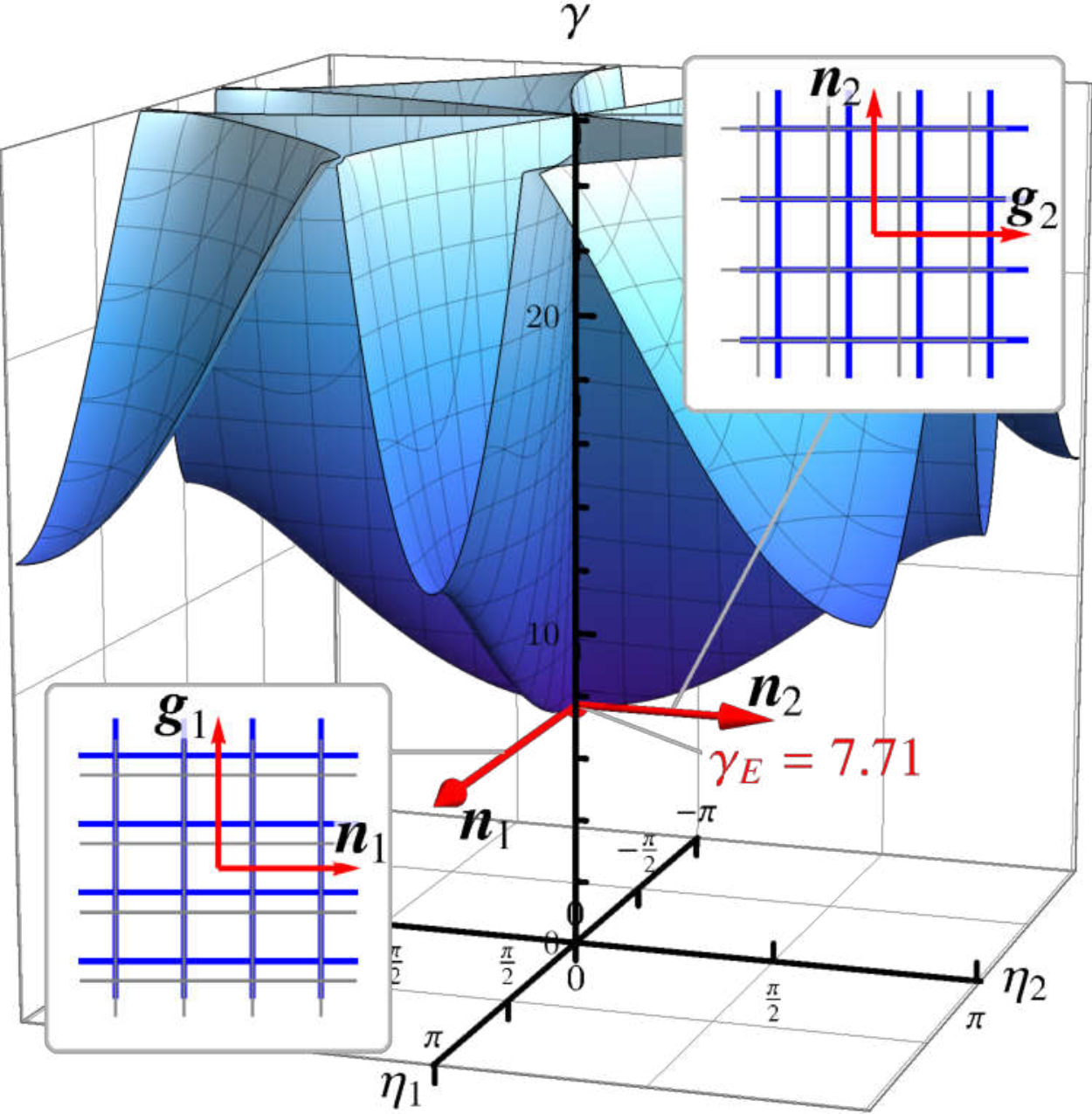}
    \end{subfigure}
    \begin{subfigure}[t]{0.31\textwidth}
        \centering
        \caption{\label{fig:macro_bifurcation_surf2}Uniaxial compression}
        \includegraphics[width=0.95\linewidth]{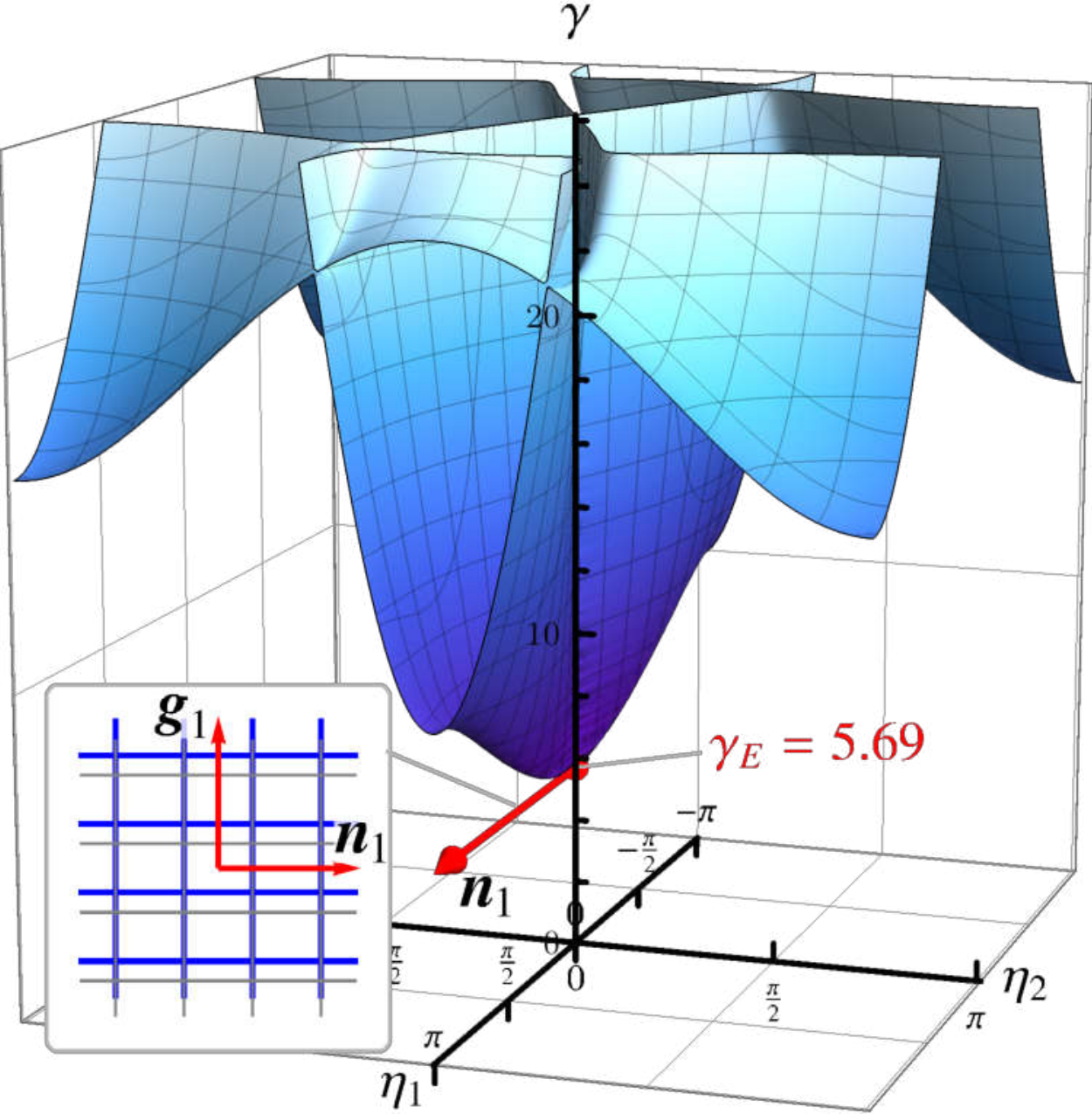}
    \end{subfigure}\\ \vspace{2mm}
    \begin{subfigure}[t]{0.32\textwidth}
        \centering
        \caption{\label{fig:macro_micro_bifurcation_stab}Uniqueness domain $(\kappa=0.2)$}
        \includegraphics[width=\linewidth]{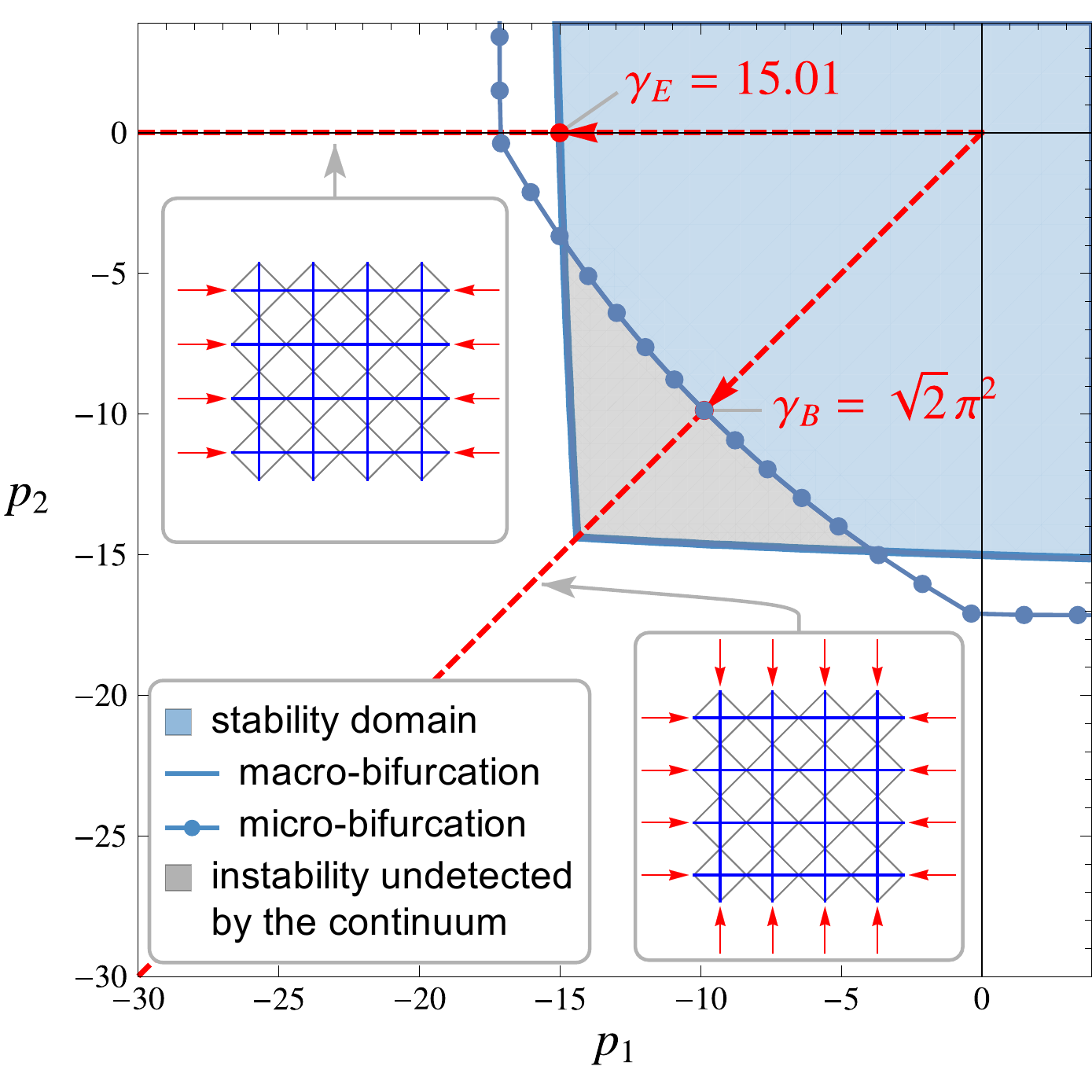}
    \end{subfigure}
    \hspace*{1mm}
    \begin{subfigure}[t]{0.31\textwidth}
        \centering
        \caption{\label{fig:macro_micro_bifurcation_surf1}Equibiaxial compression}
        \includegraphics[width=0.95\linewidth]{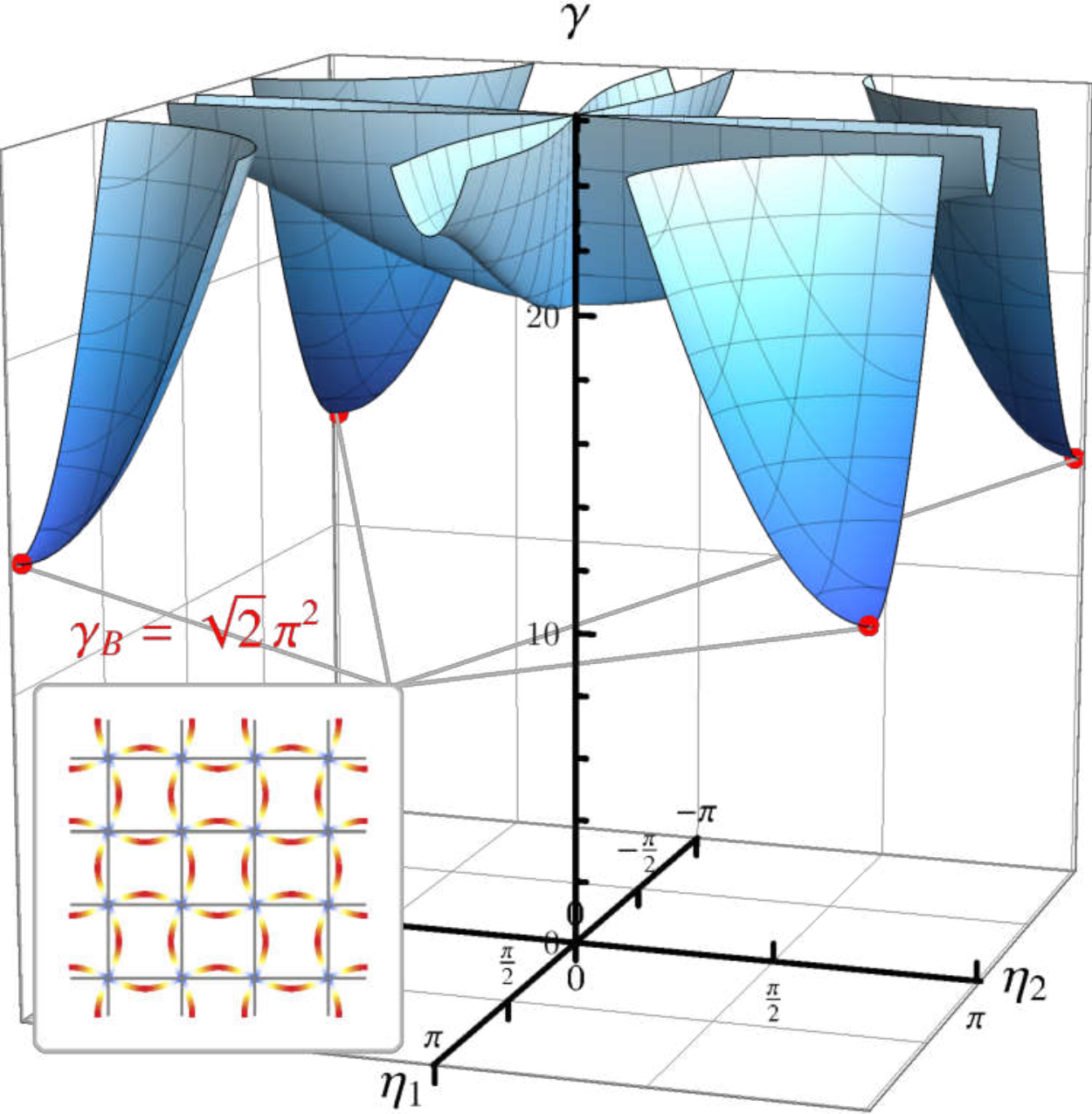}
    \end{subfigure}
    \begin{subfigure}[t]{0.31\textwidth}
        \centering
        \caption{\label{fig:macro_micro_bifurcation_surf2}Uniaxial compression}
        \includegraphics[width=0.95\linewidth]{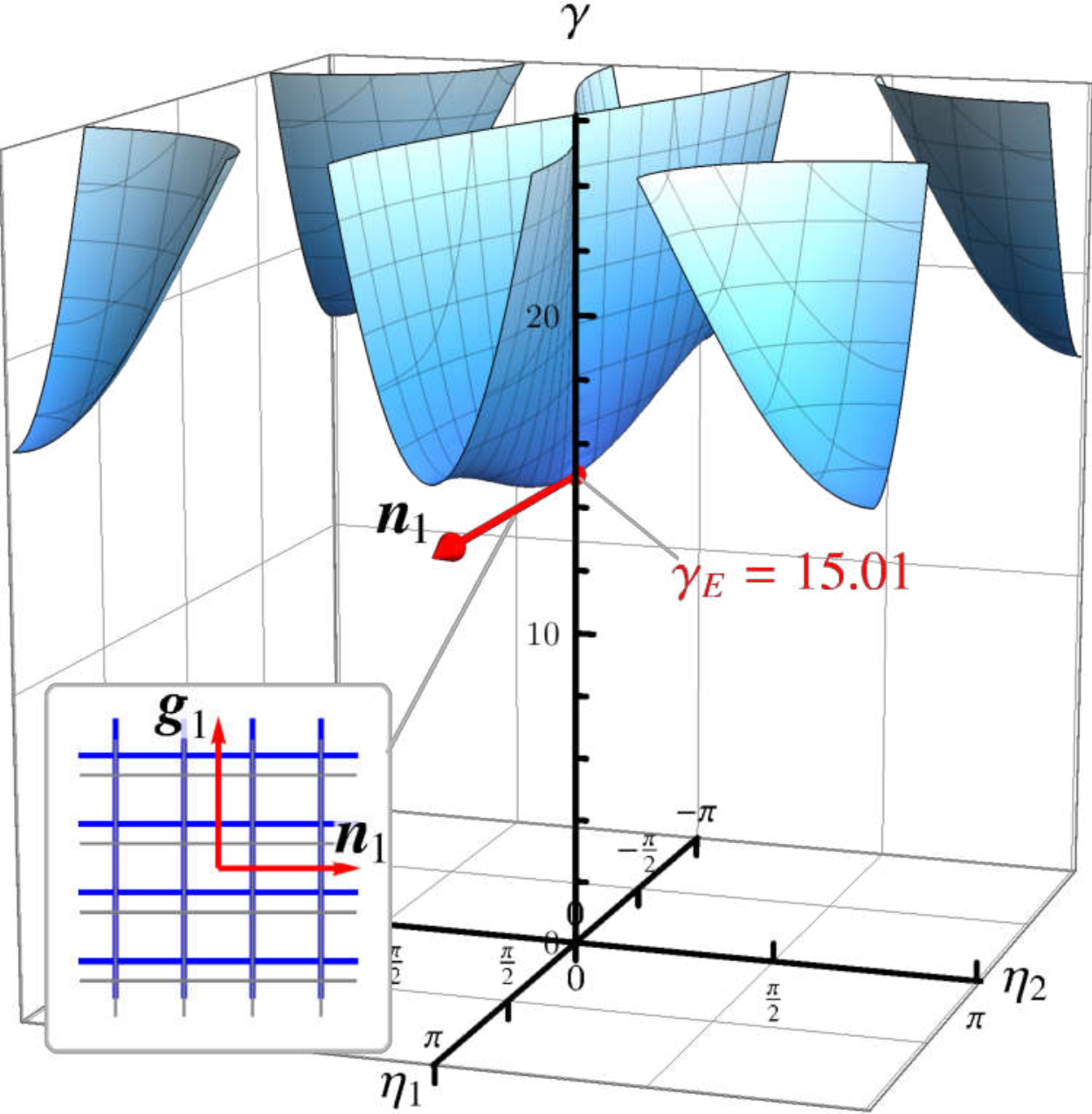}
    \end{subfigure}
    \caption{\label{fig:macro_micro_bifurcation}
    (\subref{fig:macro_bifurcation_stab}) and (\subref{fig:macro_micro_bifurcation_stab}):
    Uniqueness/stability domains in the loading space $\{p_1,p_2\}$ for a square grid (with equal slendernesses of the rods $\Lambda_1 = \Lambda_2 = 10$), when diagonal springs are absent (upper part) and present (with spring stiffness $\kappa=0.2$,  lower part).
    A continuous (a dotted) contour represents the occurrence of macro (of micro) bifurcations, so that the shaded regions correspond to strong ellipticity and uniqueness for the equivalent continuum. 
    In the absence of diagonal springs, macro-instabilities, corresponding to ellipticity loss, prevail and always occur before micro bifurcations, while when the diagonal springs are considered, the situation is more complex so that one or the other instability may be critical. 
    (\subref{fig:macro_bifurcation_surf1},~\subref{fig:macro_bifurcation_surf2}) and ~(\subref{fig:macro_micro_bifurcation_surf1},~\subref{fig:macro_micro_bifurcation_surf2}): with reference to two specific radial loading paths of equibiaxial and uniaxial compression, shown as red dashed lines in (\subref{fig:macro_bifurcation_stab}) and (\subref{fig:macro_micro_bifurcation_stab}), the bifurcation surfaces evidence the solutions for failure of ellipticity in terms of critical dyads $\bn \otimes \bg$. 
    }
\end{figure}

To investigate both macroscopic (infinite wavelength) and microscopic (finite wavelength) bifurcations, results for the orthotropic grid with $\alpha=\pi/2$ are reported in Fig.~\ref{fig:macro_micro_bifurcation}, where critical bifurcation loads $p_1$ and $p_2$ are reported for the cases in which diagonal springs are absent ($\kappa=0$, Fig.~\ref{fig:macro_bifurcation_stab},~\subref{fig:macro_bifurcation_surf1},~\subref{fig:macro_bifurcation_surf2}) and for a spring stiffness $\kappa=0.2$ (Fig.~\ref{fig:macro_micro_bifurcation_stab},~\subref{fig:macro_micro_bifurcation_surf1},~\subref{fig:macro_micro_bifurcation_surf2}).

The uniqueness domains (Fig.~\ref{fig:macro_bifurcation_stab} and~\ref{fig:macro_micro_bifurcation_stab}) have been computed by solving equation~\eqref{eq:local_buckling_multiplier_simpler} for radial loading paths in the non-dimensional load space $\{p_1,p_2\}$.
To clarify the results of this computation, two critical boundaries are reported, one with a continuous line and the other with a continuous-dotted line, referring to bifurcations of long (infinite) and `shortest possible' wavelength, respectively.
The former occurs when the infimum of~\eqref{eq:local_buckling_multiplier_simpler} is attained at $\bk=\bzero$, while the latter refers to the infimum computed on the boundary of the reciprocal unit cell, $\eta_{1,2}=\pm\pi$\footnote{Note that all the possible wavelengths have been considered in the computation of the stability domain (as expressed by Eq.~\eqref{eq:local_buckling_multiplier_simpler}), but in Fig.~\ref{fig:macro_bifurcation_stab} and~\ref{fig:macro_micro_bifurcation_stab} the critical wave vectors $\bk$ have been found to either be at the origin ($\bk=\bzero$) or on the boundary of the reciprocal unit cell (shortest wavelengths).}.
The location of the infimum can be visualized, by fixing the loading direction as $\bp=\gamma\,\hat{\bp}$, and then by numerically computing the bifurcation surface defined as   $\det\bK^*(\gamma\hat{\bp},\,\eta_1\bb_1+\eta_2\bb_2) = 0$ in the space $\{\eta_1,\eta_2,\gamma\}$.
Two radial paths are considered in Fig.~\ref{fig:macro_bifurcation_stab} and~\ref{fig:macro_micro_bifurcation_stab}, namely, equibiaxial $\hat{\bp}=\{-1/\sqrt{2},-1/\sqrt{2}\}$ and uniaxial $\hat{\bp}=\{-1,0\}$ compression (red dashed lines), and the corresponding bifurcation surfaces are reported in Fig.~\ref{fig:macro_bifurcation_surf1},~\subref{fig:macro_bifurcation_surf2} and Fig.~\ref{fig:macro_micro_bifurcation_surf1},~\subref{fig:macro_micro_bifurcation_surf2}, respectively.
%

In the absence of diagonal springs, Fig.~\ref{fig:macro_bifurcation_stab} reports the uniqueness domain, corresponding to strong ellipticity in the solid equivalent to the lattice,  showing that 
(for every loading direction $\hat{\bp}$) 
a macro-bifurcation, in other words an ellipticity loss (referred to the dyad $\bn\otimes\bg$), is always reached before micro-bifurcation. The latter represents 
a structural instability for the lattice that cannot be detected in the equivalent continuum. 

For the two radial loading paths shown in Fig.~\ref{fig:macro_bifurcation_stab}, the bifurcation surfaces  Figs.~\ref{fig:macro_bifurcation_surf1},\subref{fig:macro_bifurcation_surf2}, show that the minimum values of the load multiplier $\gamma$ are attained at $\{\eta_1,\eta_2\} = \{0,0\}$, which corresponds to a macro-bifurcation for the lattice (associated to an infinite wavelength mode), so that the critical prestress multipliers $\gamma_{\text{E}}=7.71$ and $\gamma_{\text{E}}=5.69$ lie on the border of ellipticity loss.
The two bifurcations correspond respectively to two orthogonal modes and a single mode.

The presence of diagonal springs complicates the situation as reported in Fig.~\ref{fig:macro_micro_bifurcation_stab}. In this case the uniqueness/stability domains show that micro-bifurcations may sometimes occur within the region of strong ellipticity, which is for instance the case of equibiaxial compression (radial path inclined at $45^\circ$) and not the case of uniaxial compression (horizontal radial path). 
In fact, when the diagonal springs are present, for equibiaxial compression a critical micro-bifurcation occurs, so that Fig.~\ref{fig:macro_micro_bifurcation_surf1} shows that the minimum value of the load multiplier, $\gamma_{\text{B}}=\sqrt{2} \pi^2$, is attained at four points, $\{\eta_1,\eta_2\} = \{\pm \pi,\pm \pi\}$, all associated to a bifurcation mode with a finite wavelength, as shown in the inset.
For uniaxial compression, Fig.~\ref{fig:macro_micro_bifurcation_surf2}, a macro-bifurcation of the grid, in other words a loss of ellipticity, occurs at $\gamma_{\text{E}} = 15.01$ and the tangent to the bifurcation surface at the origin singles out the infinite-wavelength bifurcation mode (shown in the inset and appearing as a rigid translation).
%
\begin{figure}[htb!]
    \centering
    \begin{subfigure}{0.49\textwidth}
        \centering
        \caption{\label{fig:ellipticity_domains_10_10_pi2}$\alpha=\pi/2$}
        \includegraphics[width=0.95\linewidth]{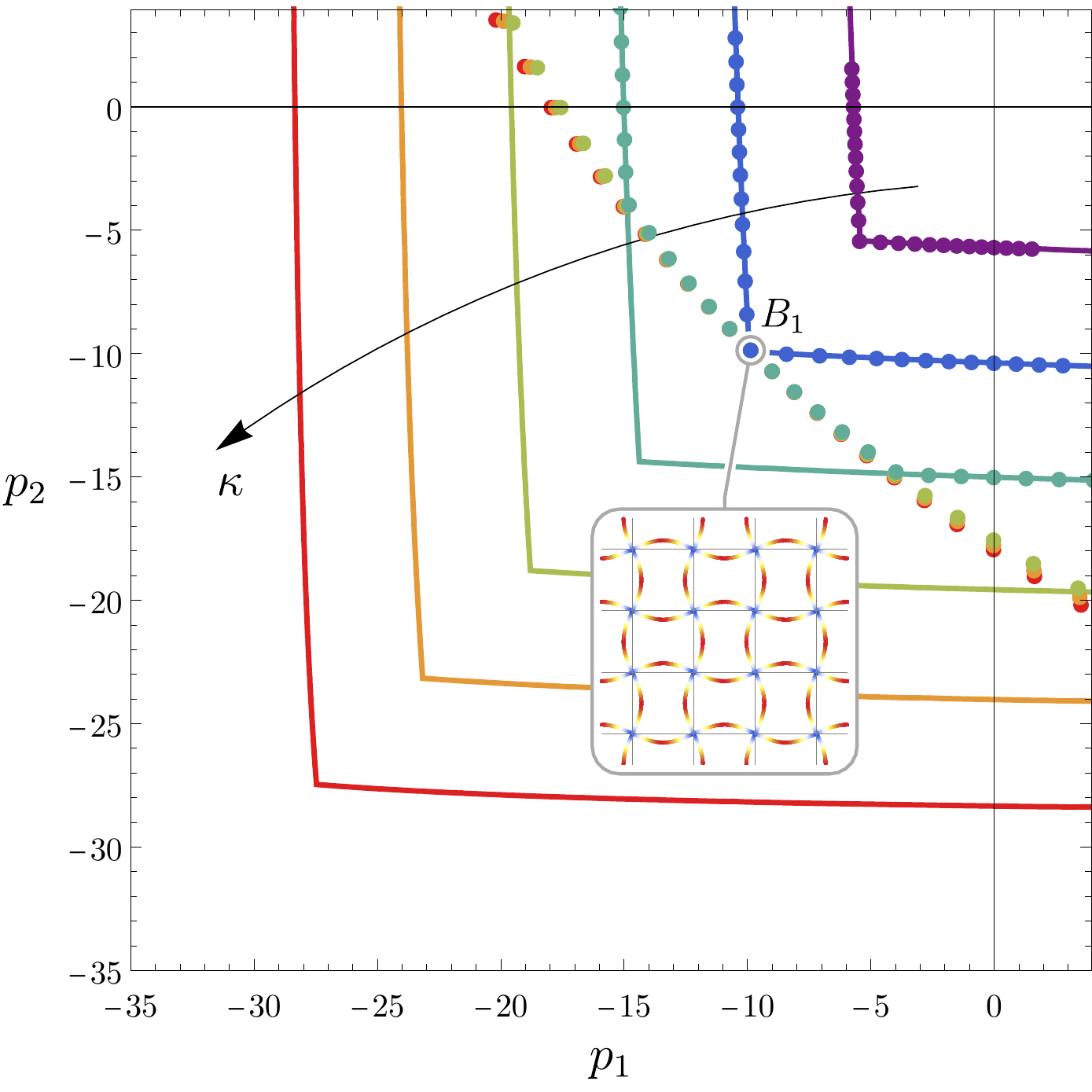}
    \end{subfigure}
    \begin{subfigure}{0.49\textwidth}
        \centering
        \caption{\label{fig:ellipticity_domains_10_10_pi3}$\alpha=\pi/3$}
        \includegraphics[width=0.95\linewidth]{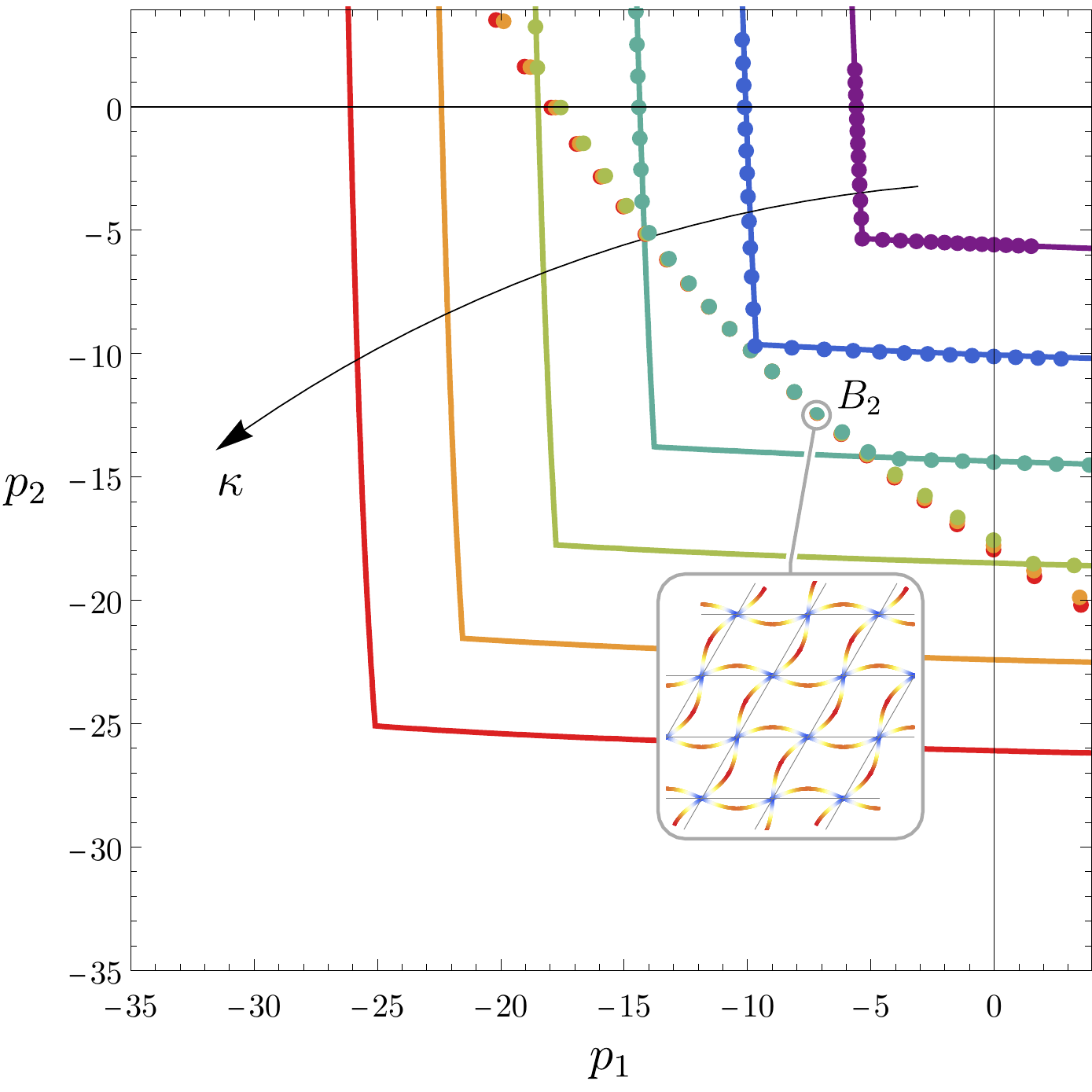}
    \end{subfigure}\\
    \begin{subfigure}{0.49\textwidth}
        \centering
        \caption{\label{fig:ellipticity_domains_10_10_pi4}$\alpha=\pi/4$}
        \includegraphics[width=0.95\linewidth]{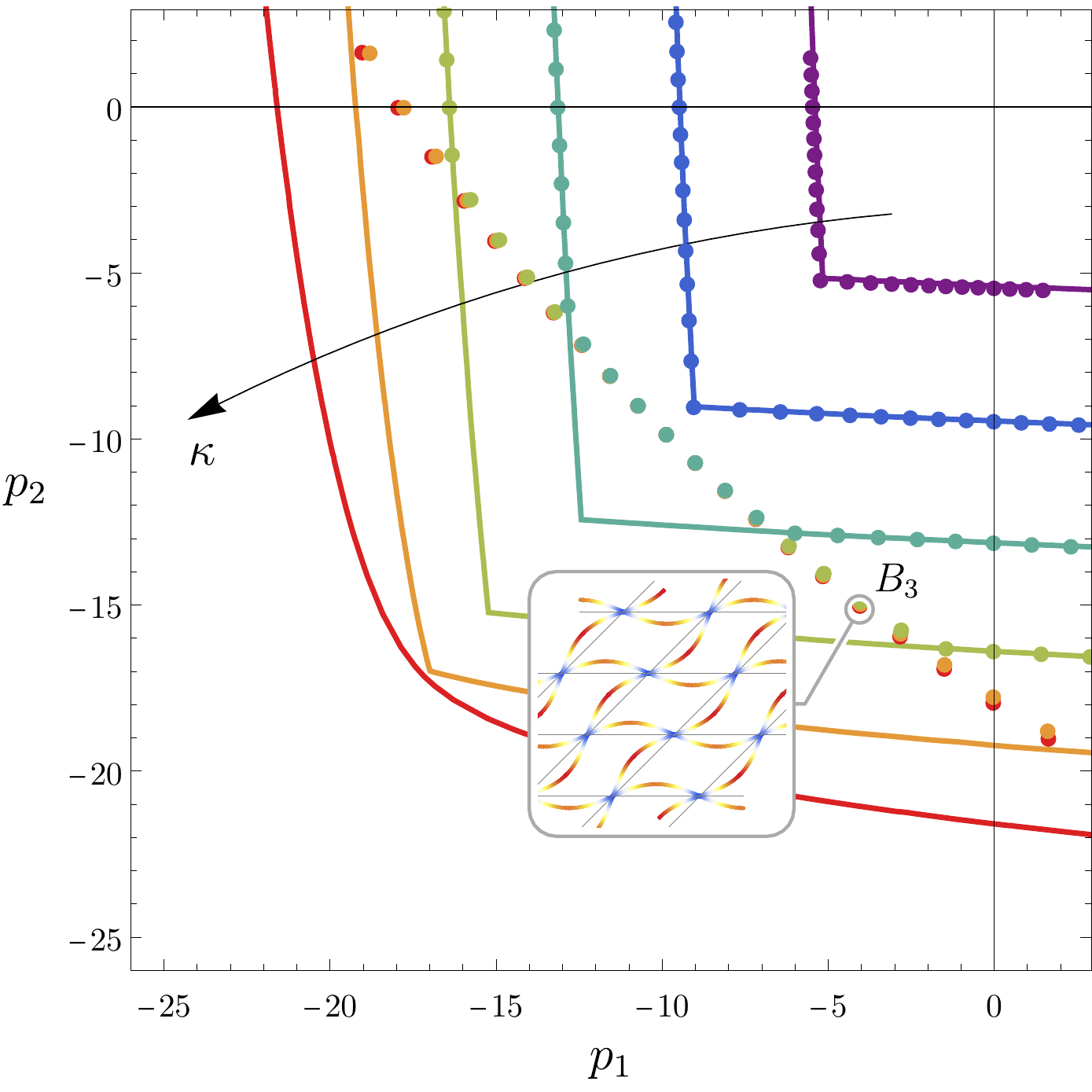}
    \end{subfigure}
    \begin{subfigure}{0.49\textwidth}
        \centering
        \caption{\label{fig:ellipticity_domains_10_10_pi6}$\alpha=\pi/6$}
        \includegraphics[width=0.95\linewidth]{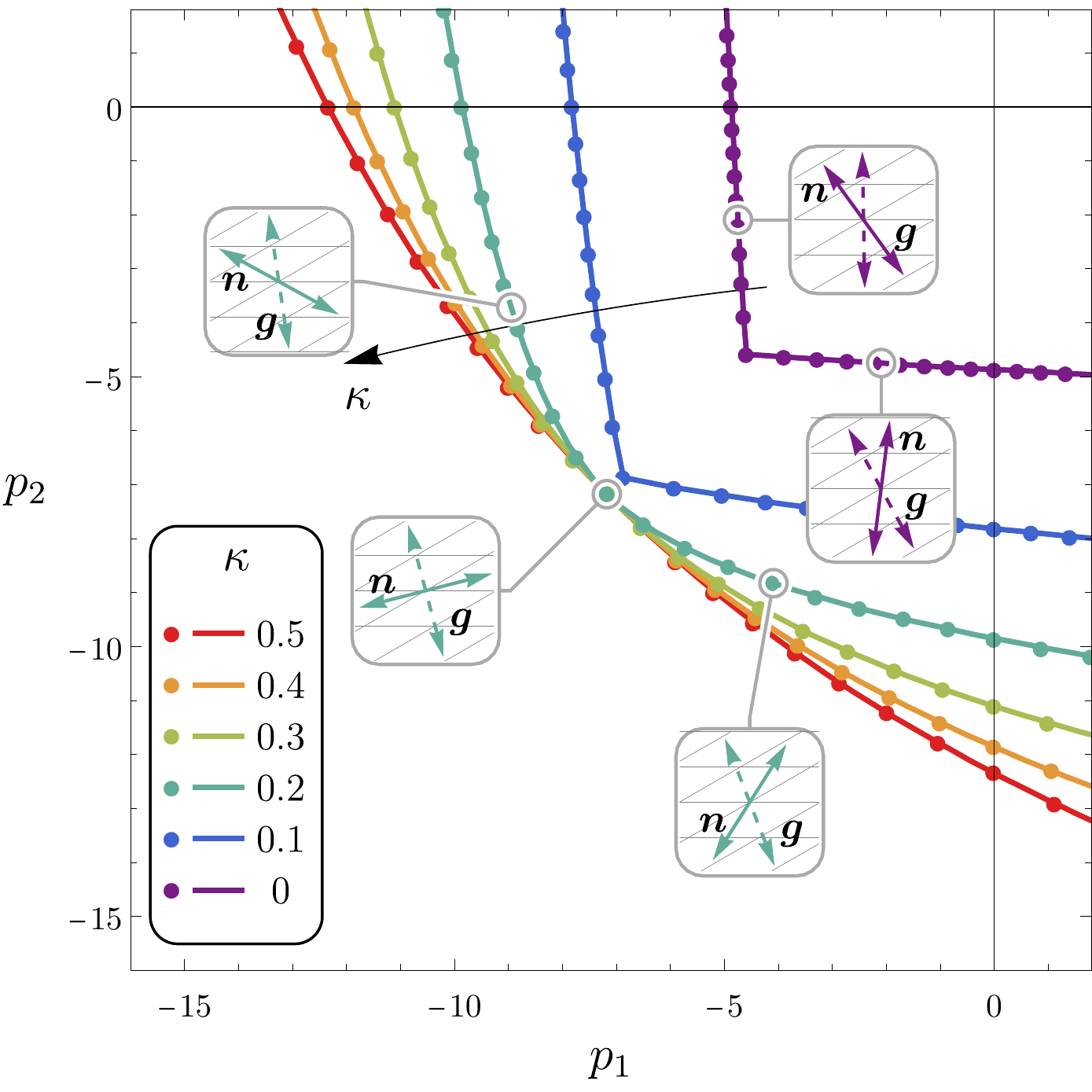}
    \end{subfigure}%
    \caption{\label{fig:ellipticity_domains_10_10}
    Strong ellipticity domains (corresponding to macro-bifurcations, continuous lines) and uniqueness domains for micro-bifurcation of the lattice (circular markers) for an orthotropic lattice of prestressed elastic rods with $\Lambda_1=\Lambda_2=10$, at different rod angles $\alpha$ and stiffness $\kappa$ of the diagonal springs.
    Points $B_1,B_2,B_3$ on the stability boundaries have been selected for the computation of the associated critical bifurcation mode (shown in the insets).
    Table~\ref{tab:buckling_modes_points} collects the critical loads and the critical wave vectors  for each bifurcation mode.
    At small grid angles, for instance the reported value of $\alpha=\pi/6$, failure of ellipticity coincides with micro-bifurcation in the lattice, so that the bifurcation mode is always characterized by an infinite wavelength. 
    For these grid configurations, the direction of ellipticity loss exhibits a \textit{`super-sensitivity'} with respect to the load directionality, shown in the insets of part~(\subref{fig:ellipticity_domains_10_10_pi6}), reporting the critical dyads $\bn \otimes \bg$ for failure of ellipticity.
    }
\end{figure}

\begin{figure}[htb!]
    \centering
    \begin{subfigure}{0.49\textwidth}
        \centering
        \caption{\label{fig:ellipticity_domains_7_15_pi2}$\alpha=\pi/2$}
        \includegraphics[width=0.95\linewidth]{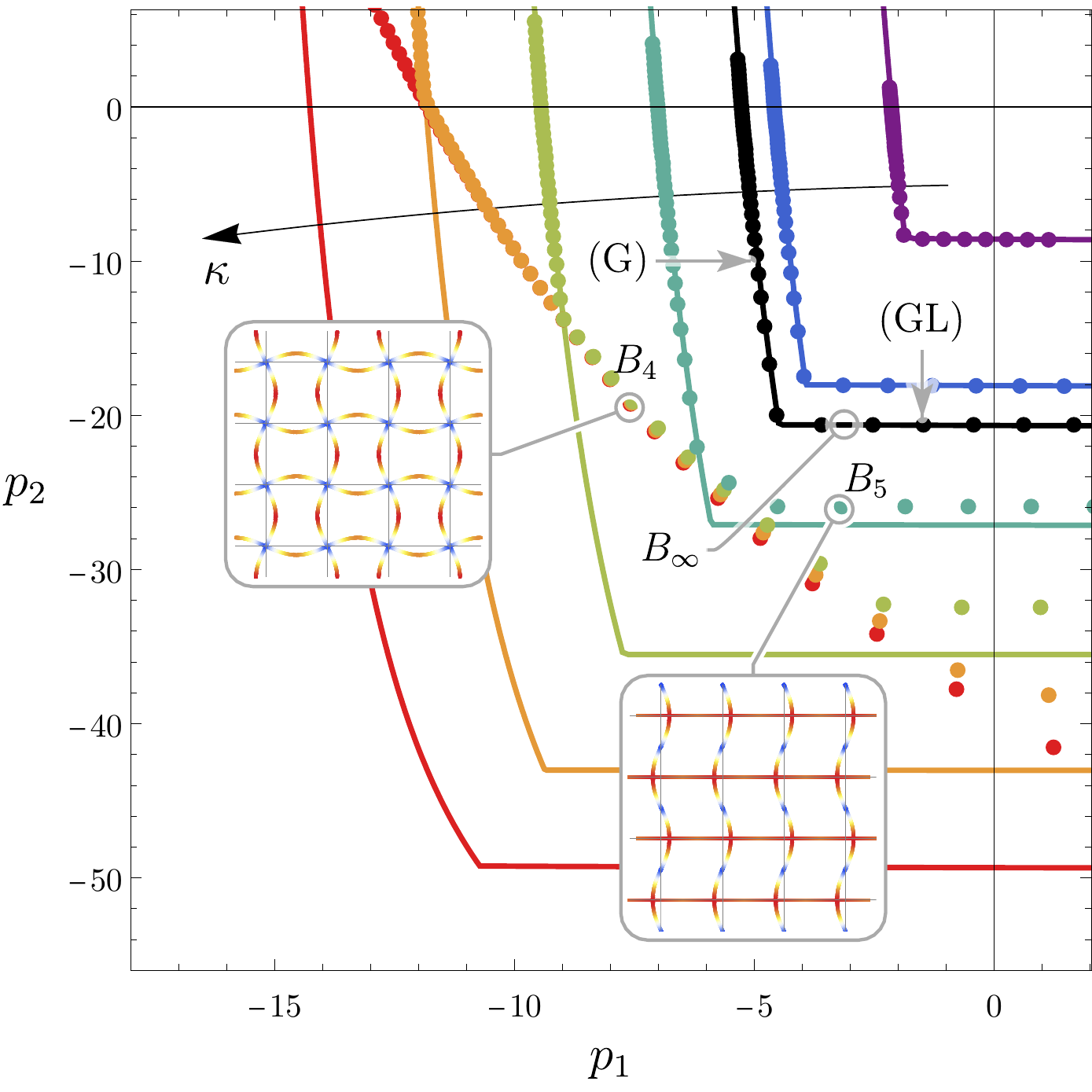}
    \end{subfigure}
    \begin{subfigure}{0.49\textwidth}
        \centering
        \caption{\label{fig:ellipticity_domains_7_15_pi3}$\alpha=\pi/3$}
        \includegraphics[width=0.95\linewidth]{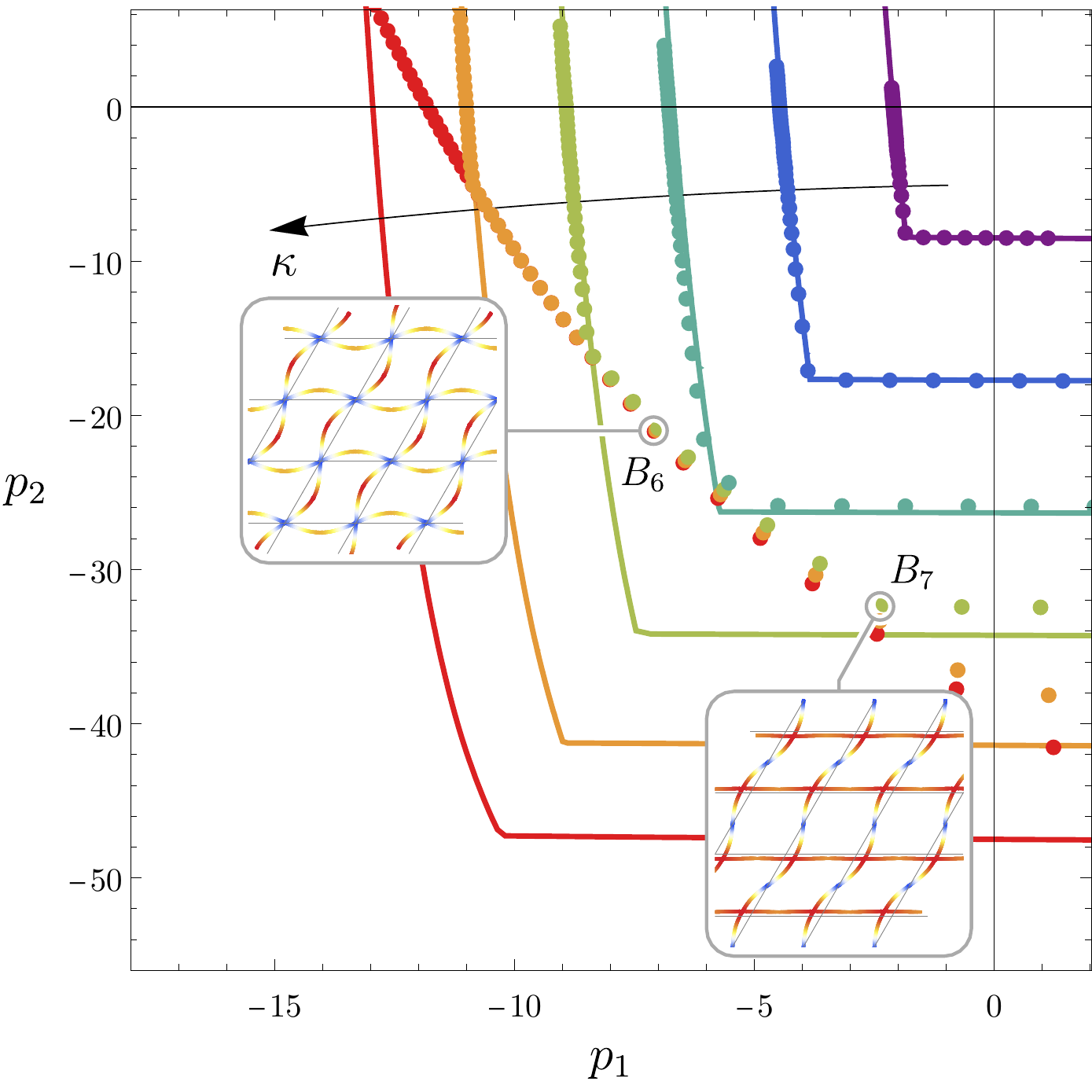}
    \end{subfigure}\\
    \begin{subfigure}{0.49\textwidth}
        \centering
        \caption{\label{fig:ellipticity_domains_7_15_pi4}$\alpha=\pi/4$}
        \includegraphics[width=0.95\linewidth]{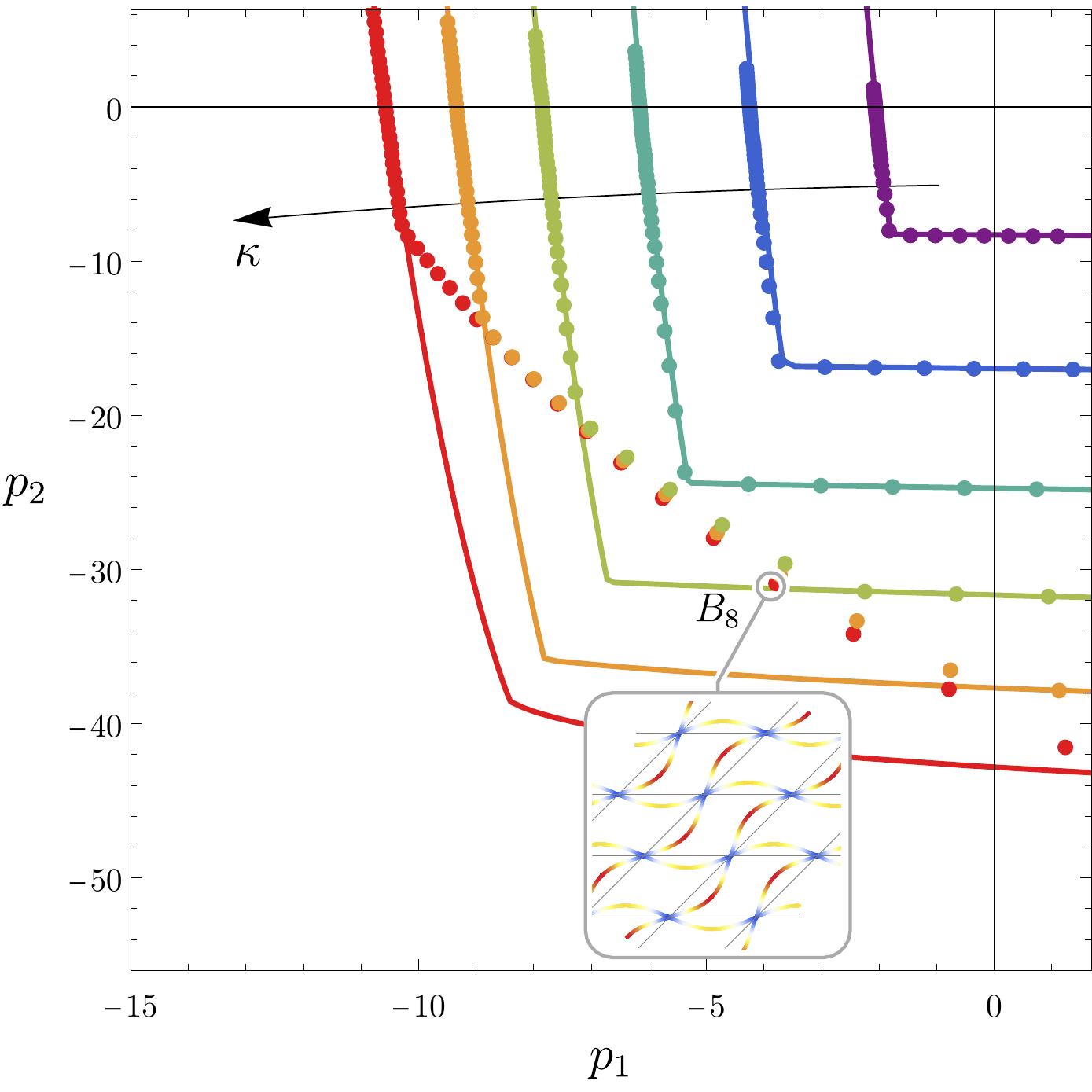}
    \end{subfigure}
    \begin{subfigure}{0.49\textwidth}
        \centering
        \caption{\label{fig:ellipticity_domains_7_15_pi6}$\alpha=\pi/6$}
        \includegraphics[width=0.95\linewidth]{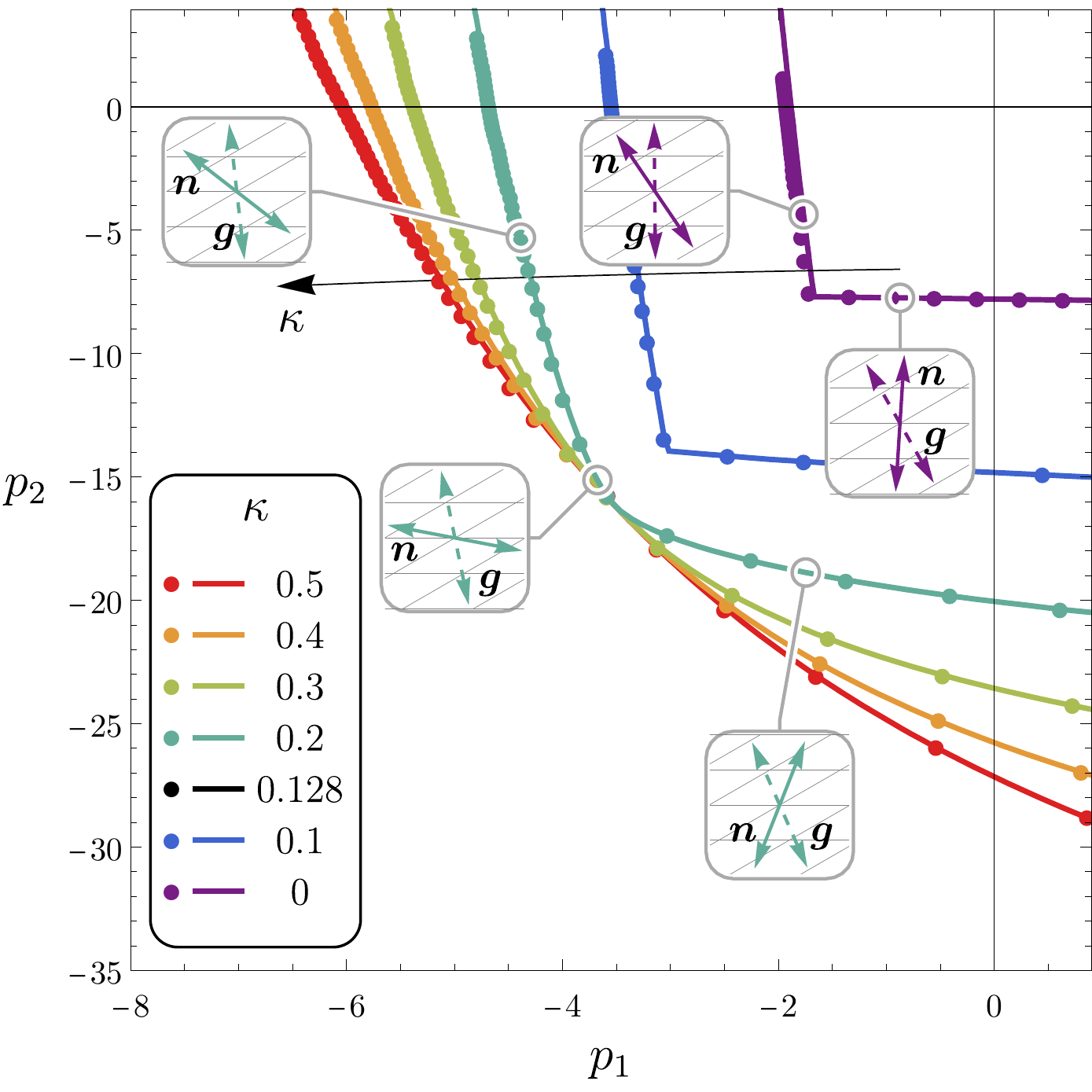}
    \end{subfigure}%
    \caption{\label{fig:ellipticity_domains_7_15}
		As for Fig. \ref{fig:ellipticity_domains_10_10}, except that $\Lambda_1=7,\,\Lambda_2=15$ and that the points on the stability boundary for which the critical bifurcation modes have been computed are labeled $B_4,B_5,B_6,B_7,B_8$. 
		Note also that, the typical microscopic bifurcation modes of the anisotropic grid exhibit widely different deformations dictated by the prestress direction (see insets in parts~\subref{fig:ellipticity_domains_7_15_pi2}, \subref{fig:ellipticity_domains_7_15_pi3}, \subref{fig:ellipticity_domains_7_15_pi4} corresponding to points labeled $B_4,B_5,B_6,B_7,B_8$ and compare for instance mode $B_4$ to $B_5$ or $B_6$ to $B_7$).
    }
\end{figure}

Further results on uniqueness domains for the orthotropic and the anisotropic grid are reported in Figs.~\ref{fig:ellipticity_domains_10_10} and~\ref{fig:ellipticity_domains_7_15}, respectively. 
The strong ellipticity boundary (corresponding to macro-bifurcation) in the equivalent solid is denoted with a continuous line, while the circular markers identify the boundary of the uniqueness/stability region for micro bifurcations, which remain undetected in the equivalent solid.
Moreover, critical bifurcation modes have been reported in insets of Figs.~\ref{fig:ellipticity_domains_10_10} and~\ref{fig:ellipticity_domains_7_15}, which refer to some specific points on the stability boundary (labelled as $B_1,B_2,B_3$ in the former figure and $B_4,B_5,B_6,B_7,B_8$ in the latter).
The critical loads and the critical wave vectors for each bifurcation mode are reported in Table~\ref{tab:buckling_modes_points}.
%
%

From Figs.~\ref{fig:ellipticity_domains_10_10} and~\ref{fig:ellipticity_domains_7_15} the following features can be highlighted. 
\begin{itemize}
    \item For the orthotropic grid the strong ellipticity boundary is symmetric with respect to the bisector defined by the condition $p_1=p_2$, which is the principal direction of orthotropy for the grid when $\Lambda_1=\Lambda_2$ (a symmetry which is broken for the anisotropic grid);
    \item For every value of the grid angle $\alpha$, the effect of the diagonal springs essentially consists in an enlargement of the strong ellipticity region (see the arrow in Fig. \ref{fig:ellipticity_domains_7_15} denoting increasing values of stiffness $\kappa$);
    \item The stiffening induced by increasing the spring stiffness $\kappa$ is much more effective for nearly orthogonal grids ($\alpha\approx\pi/2$) than for small values of  inclination $\alpha$ (compare Fig.~\ref{fig:ellipticity_domains_10_10_pi2} to Fig.~\ref{fig:ellipticity_domains_10_10_pi6} and Fig.~\ref{fig:ellipticity_domains_7_15_pi2} to Fig.~\ref{fig:ellipticity_domains_7_15_pi6});
    \item For every value of the spring stiffness $\kappa$, the deviation from orthogonality of the grid always reduces the size of the strong ellipticity region, so that the largest strong ellipticity region is attained for $\alpha=\pi/2$. 
\end{itemize}

\begin{table}[htb!]
    \centering
    \begin{tabular}{c|ccccccc}
        \toprule
        Label & $\Lambda_1$ & $\Lambda_2$ & $\alpha$ & $\kappa$ & $p_1$     & $p_2$     & $\bk_{\text{cr}}$ \\ 
        \midrule
        $B_1$ & 10          & 10          & $\pi/2$  & $0.2$    & $-\pi^2$  & $-\pi^2$  & $\pi\bb_1+\pi\bb_2$ \\
        $B_2$ & 10          & 10          & $\pi/3$  & $0.3$    & $-7.16$   & $-12.40$  & $\pi\bb_1+\pi\bb_2$ \\
        $B_3$ & 10          & 10          & $\pi/4$  & $0.7$    & $-4.05$   & $-15.13$  & $\pi\bb_1+\pi\bb_2$ \\
        $B_4$ & 7           & 15          & $\pi/2$  & $0.4$    & $-7.72$   & $-18.64$  & $\pi\bb_1+\pi\bb_2$ \\
        $B_5$ & 7           & 15          & $\pi/2$  & $0.2$    & $-3.41$   & $-25.91$  & $\pi\bb_2$ \\
        $B_6$ & 7           & 15          & $\pi/3$  & $0.3$    & $-6.98$   & $-20.93$  & $\pi\bb_1+\pi\bb_2$ \\
        $B_7$ & 7           & 15          & $\pi/3$  & $0.3$    & $-2.12$   & $-32.40$  & $\pi\bb_2$ \\
        $B_8$ & 7           & 15          & $\pi/4$  & $0.5$    & $-4.00$   & $-30.37$  & $\pi\bb_1+\pi\bb_2$ \\
        $B_\infty$ & 7      & 15          & $\pi/2$  & $0.128$  & $-3.44$   & $-20.62$  & $\eta_2\bb_2\,\forall\eta_2$ \\
        \bottomrule
    \end{tabular}
    \caption{
    Critical bifurcation modes $\bk_{\text{cr}}$ for several configurations of the orthotropic ($\Lambda_1=\Lambda_2=10$) and anisotropic ($\Lambda_1=7,\,\Lambda_2=15$) lattice.
    Plots of the corresponding deformation fields are reported as insets in Figs.~\ref{fig:ellipticity_domains_10_10} and~\ref{fig:ellipticity_domains_7_15}.
    }
    \label{tab:buckling_modes_points}
\end{table}

The stability boundaries (circular markers in Fig.~\ref{fig:ellipticity_domains_10_10} and \ref{fig:ellipticity_domains_7_15}), evidence the following characteristics.
\begin{itemize}
    \item At small values of spring stiffness $\kappa$, the first bifurcation is always global, so that \textit{the strong ellipticity and the stability boundaries coincide independently of the prestress direction}; a feature visible for $\kappa=0,0.1$ (purple and blue) in Fig.~\ref{fig:ellipticity_domains_10_10_pi2} and Fig.~\ref{fig:ellipticity_domains_7_15_pi2});
    \item An increase in the spring stiffness $\kappa$ leads to a first bifurcation of local type (the critical mode is characterized by a finite wavelength), \textit{so that the stability region lies inside the elliptic boundary};
    \item Fig.~\ref{fig:ellipticity_domains_10_10_pi6} and Fig.~\ref{fig:ellipticity_domains_7_15_pi6} show that, at sufficiently small values of grid angle (for instance at $\alpha=\pi/6$), \textit{failure of strong ellipticity dictates the first bifurcation independently of the stiffness of the diagonal springs} (see circular markers of the stability boundary overlapping with the elliptic boundary);
    \item The typical \textit{microscopic bifurcation modes of the orthotropic grid are characterized by a pure rotational deformation of the junctions of the grid} (see insets in Fig.~\ref{fig:ellipticity_domains_10_10_pi2}, \subref{fig:ellipticity_domains_10_10_pi3}, \subref{fig:ellipticity_domains_10_10_pi4} corresponding to the points labeled as $B_1,B_2,B_3$);
    \item Typical \textit{microscopic bifurcation modes of the anisotropic grid exhibit widely different deformations dictated by the prestress direction} (see insets in Fig.~\ref{fig:ellipticity_domains_7_15_pi2}, \subref{fig:ellipticity_domains_7_15_pi3}, \subref{fig:ellipticity_domains_7_15_pi4} corresponding to points labeled $B_4,B_5,B_6,B_7,B_8$ and compare for instance mode $B_4$ to $B_5$ or $B_6$ to $B_7$).
    \item A bifurcation always occurs for every lattice geometry at an equibiaxial load $\{p_1,p_2\}=\{-\pi^2,-\pi^2\}$ (point $B_1$ in Fig.~\ref{fig:ellipticity_domains_10_10_pi2}) regardless of the values of $\Lambda_1$, $\Lambda_2$, $\kappa$, and $\alpha$. 
    This bifurcation can be explained by the fact that the normalized load $P l^2/B=-\pi^2$ corresponds to the buckling load of a simply supported Euler-Bernoulli beam, and thus, when all the rods of an arbitrary grid are prestressed at this level, a purely flexural buckling mode becomes available (shown in the inset of Fig.~\ref{fig:ellipticity_domains_10_10_pi2}).
\end{itemize}

Despite the complex influence of the geometrical and mechanical parameters on the stability of the prestressed lattice, two important `transitions' characterize the nature of the first bifurcation, namely: 
\begin{enumerate}[label=(\roman*)]
    \item a \textit{macro-to-micro} transition of the critical bifurcation mode occurs at increasing stiffness of the diagonal springs $\kappa$;
    \item a \textit{micro-to-macro} transition of the critical bifurcation mode occurs at decreasing the rod's inclination $\alpha$.
\end{enumerate}
The above transitions will be exploited in Section~\ref{sec:static_response} to investigate the static response induced by a concentrated force applied to a lattice preloaded closely to a bifurcation (both global and local bifurcations will be considered).

\subsection{A single localization band with a highly tunable inclination}
\label{sec:tunable_localization}
A remarkable characteristic is associated to the micro-to-macro bifurcation transition obtained at decreasing angle $\alpha$, namely, a \textit{super-sensitivity of the localization band normal, represented by the unit vector $\bn$, with respect to the state of pre-load, while the localization mode $\bg$ results only weakly affected}.

For instance, at $\alpha=\pi/6$ and sufficiently high spring stiffness $\kappa$, the insets in Figs.~\ref{fig:ellipticity_domains_10_10_pi6} and \ref{fig:ellipticity_domains_7_15_pi6} show that the relative inclinations between the localization band normal $\bn$ and the localization mode $\bg$ strongly vary as a function of the load state in the lattice. 

When the spring stiffness vanishes, $\kappa=0$, the localization band is essentially set by the grid inclination as it is almost perfectly aligned parallel to the inclinations $0$ and $\pi/6$, so that failure of ellipticity occurs in a direction $\bn$ that is almost orthogonal to the rods.
On the contrary, at $\kappa=0.2$ a single localization band occurs, whose inclination strongly depends on the load directionality and is essentially unrelated to the underlying grid pattern (shown in the insets corresponding to $\kappa=0.2$ in Figs.~\ref{fig:ellipticity_domains_10_10_pi6} and~\ref{fig:ellipticity_domains_7_15_pi6}).
\textit{The super-sensitivity of the localization direction provides an enhanced tunability of the macroscopic localization pattern by means of a simple modification of the load applied to the lattice}.

It is worth noting that the localization direction can also be designed by constructing a lattice with a suitable value of rods' angle $\alpha$, but this approach would not be easily reconfigurable,  as the structure geometry has to be defined in advance.
 
\begin{figure}[htb!]
    \centering
    \centering
    \begin{subfigure}{0.45\textwidth}
        \centering
            \caption{\label{fig:contour_3d_infinite_buckling_modes}}
            \includegraphics[width=0.95\linewidth]{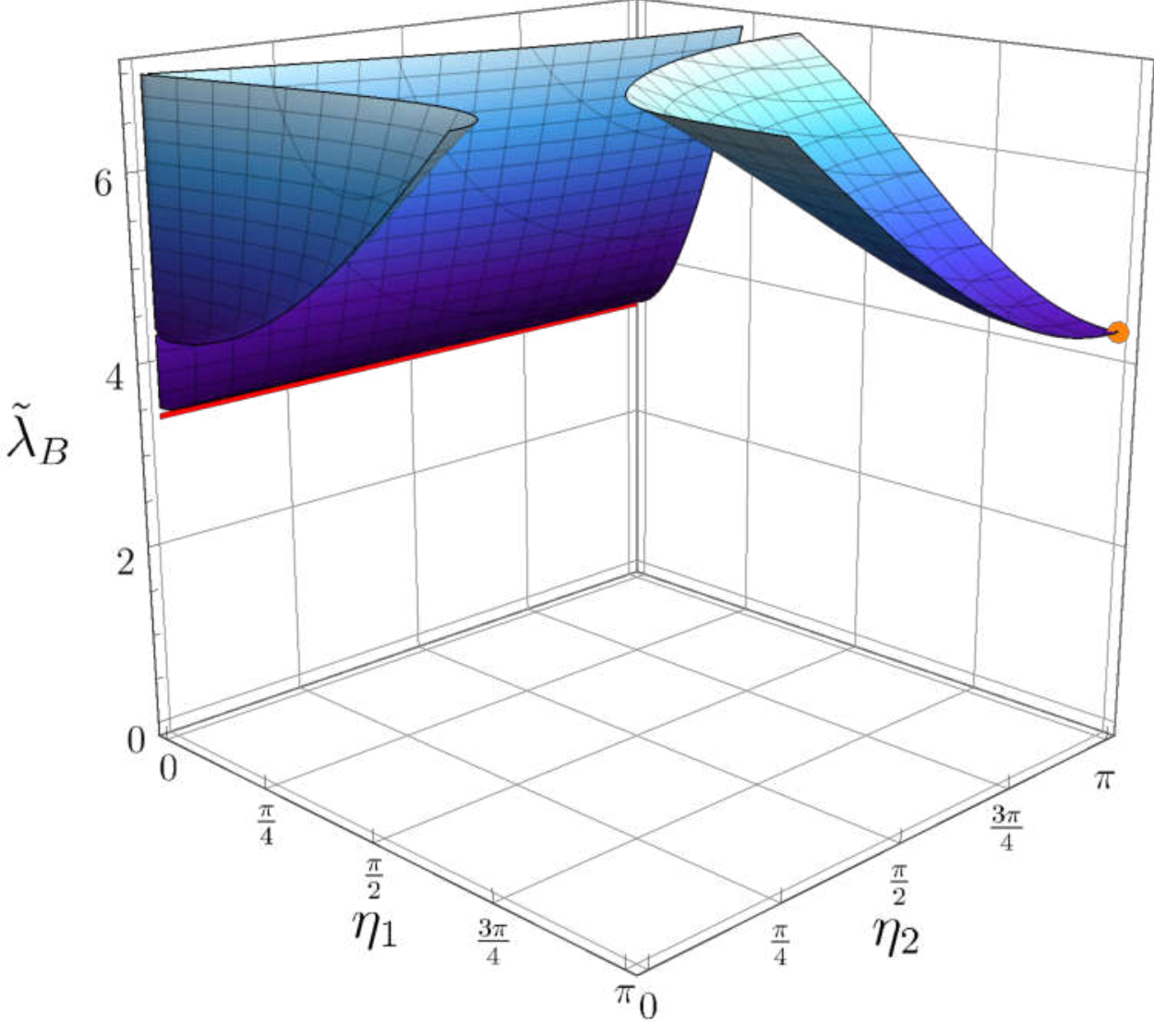}
    \end{subfigure}
    \begin{subfigure}{0.45\textwidth}
        \centering
            \caption{\label{fig:contour_2d_infinite_buckling_modes}}
            \includegraphics[width=\linewidth]{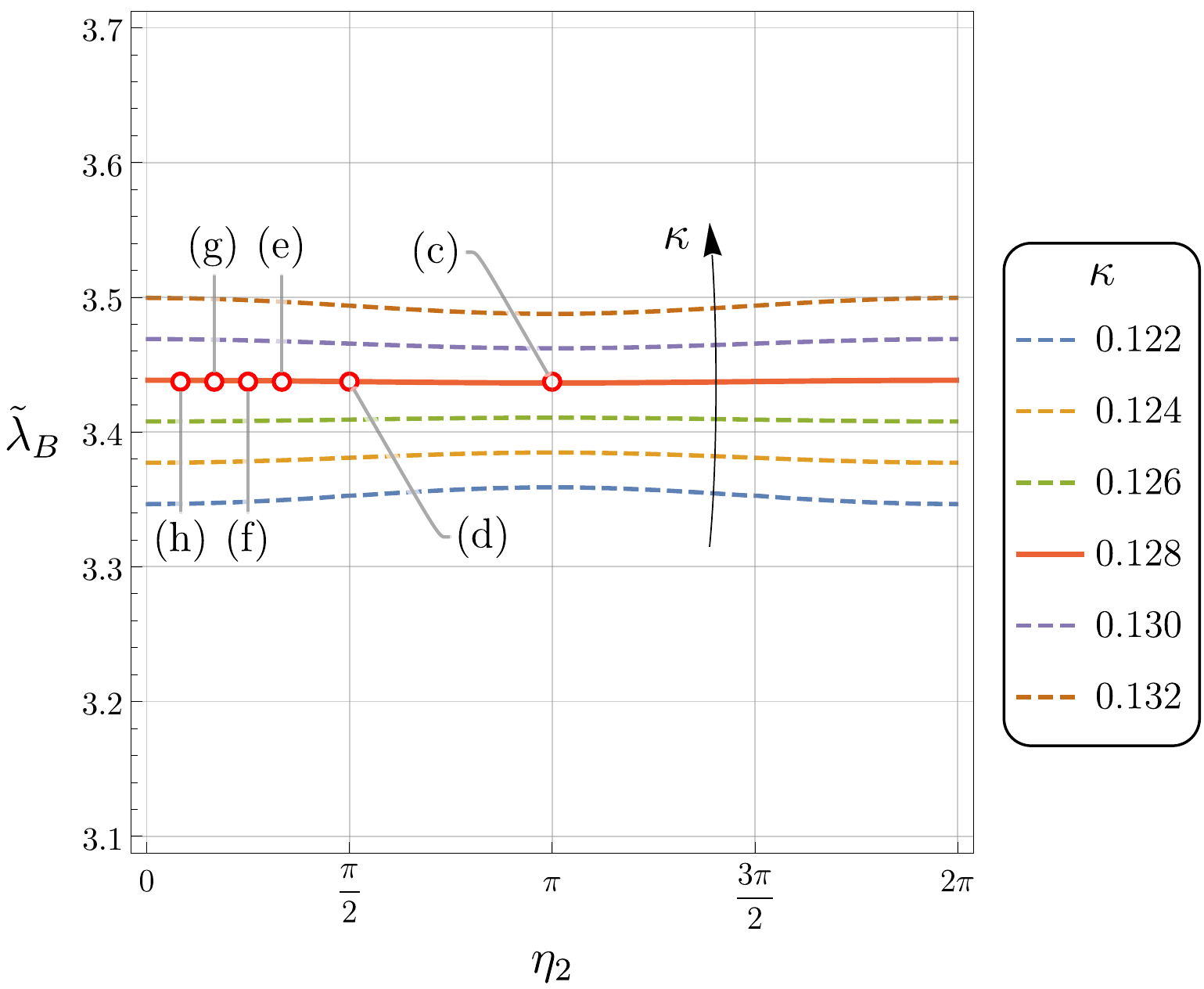}
    \end{subfigure}\\ \vspace{4mm}
    \begin{subfigure}{0.32\textwidth}
        \centering
        \caption{\label{fig:buckling_mode_7_15_k2_pi1}$\bk_{\text{cr}}=\pi\,\bb_2$}
        \includegraphics[width=0.98\linewidth]{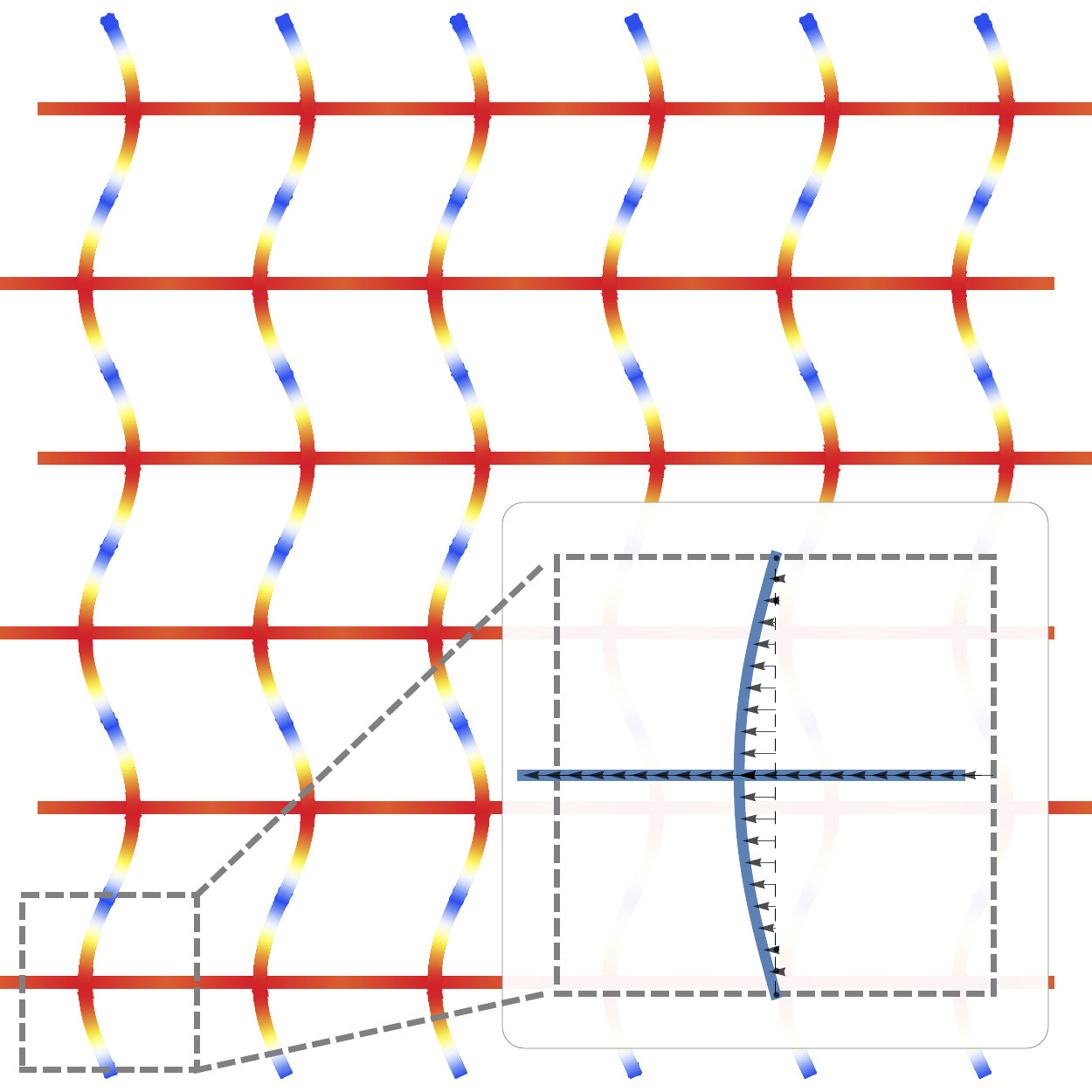}
    \end{subfigure}
    \begin{subfigure}{0.32\textwidth}
        \centering
        \caption{\label{fig:buckling_mode_7_15_k2_pi2}$\bk_{\text{cr}}=\pi/2\,\bb_2$}
        \includegraphics[width=0.98\linewidth]{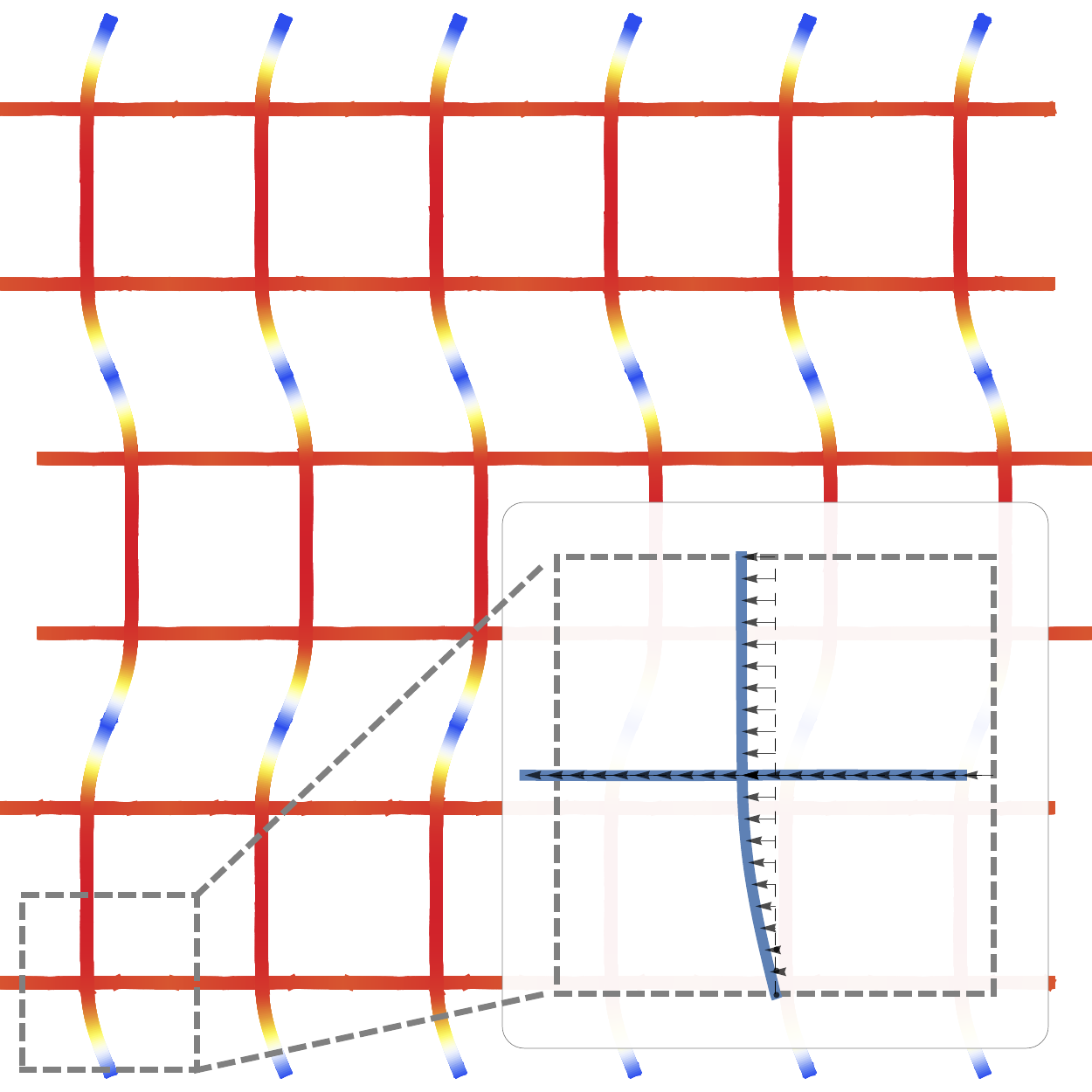}
    \end{subfigure}
    \begin{subfigure}{0.32\textwidth}
        \centering
        \caption{\label{fig:buckling_mode_7_15_k2_pi3}$\bk_{\text{cr}}=\pi/3\,\bb_2$}
        \includegraphics[width=0.98\linewidth]{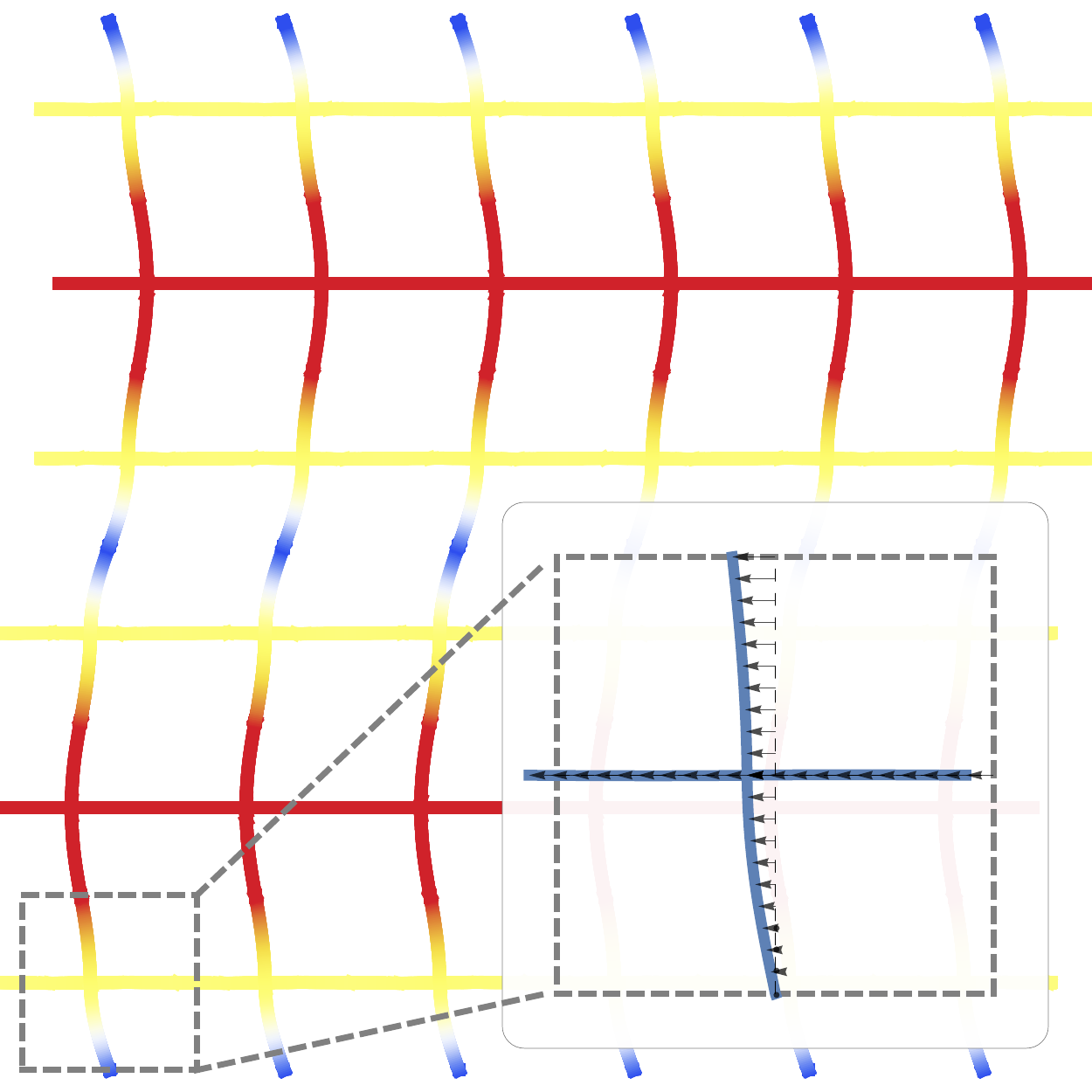}
    \end{subfigure}\\ \vspace{4mm}
    \begin{subfigure}{0.32\textwidth}
        \centering
        \caption{\label{fig:buckling_mode_7_15_k2_pi4}$\bk_{\text{cr}}=\pi/4\,\bb_2$}
        \includegraphics[width=0.98\linewidth]{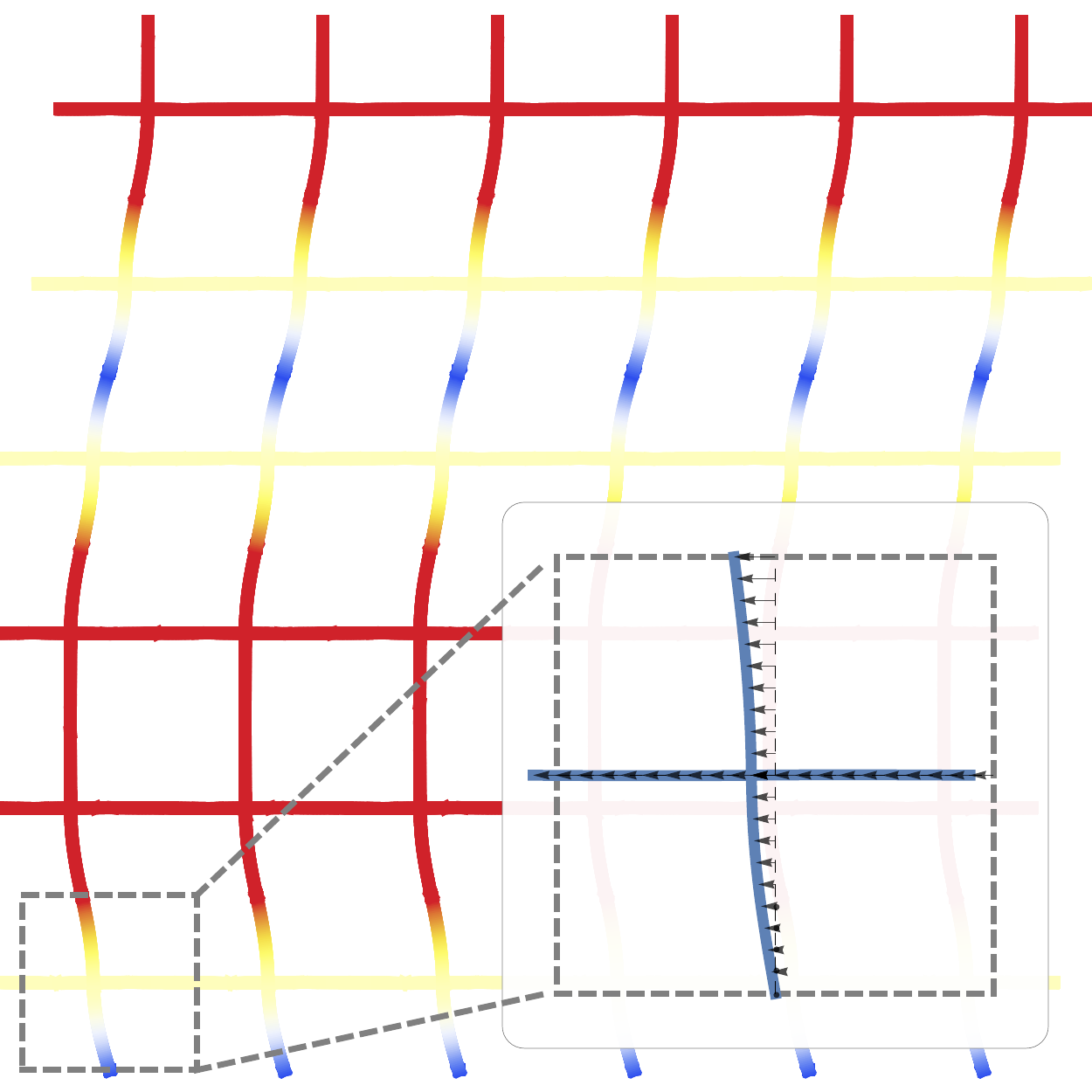}
    \end{subfigure}
    \begin{subfigure}{0.32\textwidth}
        \centering
        \caption{\label{fig:buckling_mode_7_15_k2_pi6}$\bk_{\text{cr}}=\pi/6\,\bb_2$}
        \includegraphics[width=0.98\linewidth]{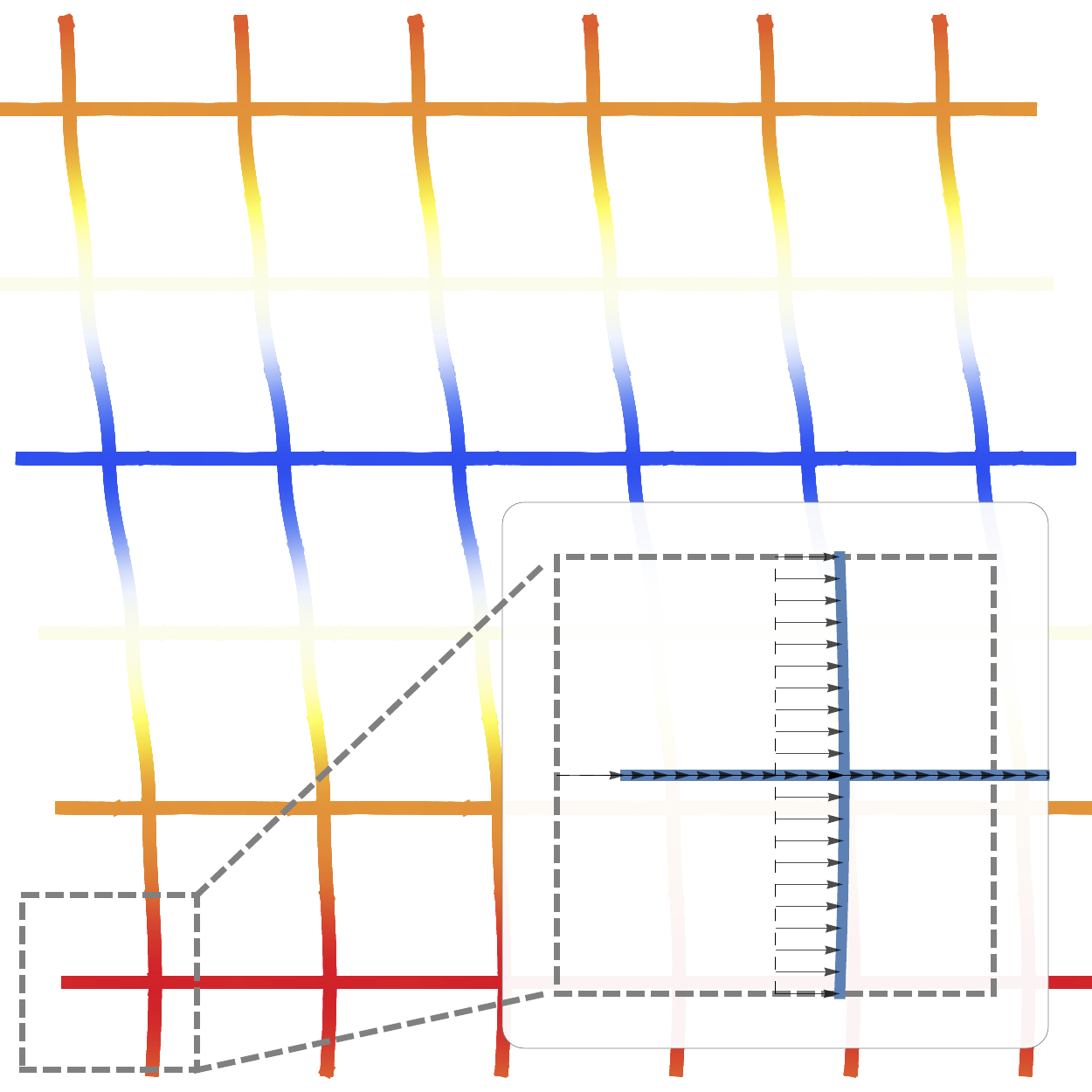}
    \end{subfigure}
    \begin{subfigure}{0.32\textwidth}
        \centering
        \caption{\label{fig:buckling_mode_7_15_k2_pi12}$\bk_{\text{cr}}=\pi/12\,\bb_2$}
        \includegraphics[width=0.97\linewidth]{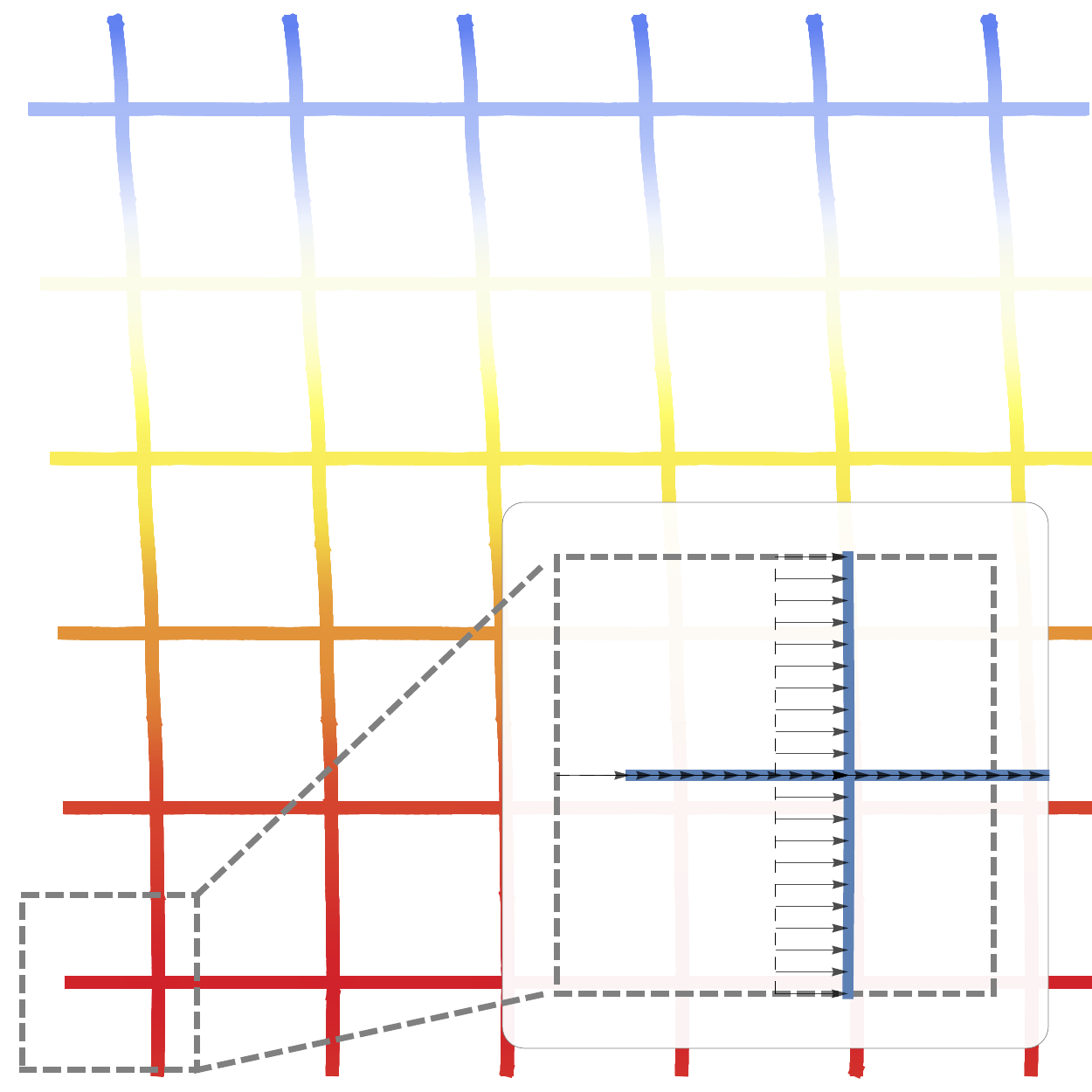}
    \end{subfigure}
    \caption{\label{fig:infinite_buckling_modes}
        Conditions showing a perfect equivalence between the lattice and the corresponding  continuum, so that when the latter looses ellipticity, the former exhibits bifurcation occurring with infinite modes covering all wavelengths, a situation which is revealed by the flat line (highlighted in red) in the bifurcation diagram (part~\subref{fig:contour_3d_infinite_buckling_modes}). 
        The perfect equivalence is obtained through accurate tuning of the stiffness of the diagonal springs ($\kappa \approx 0.128$ for a square grid with $\Lambda_1=7$ and $\Lambda_2=15$ and a loading $\{p_1,p_2\} \approx 3.44\{-1,-6\}$). 
        Part~(\subref{fig:contour_2d_infinite_buckling_modes}) represents a section of the bifurcation surface at $\eta_1=0$ detailing the flat minimum of the curve occurring at $\kappa \approx 0.128$.
        Parts (\subref{fig:buckling_mode_7_15_k2_pi1})--(\subref{fig:buckling_mode_7_15_k2_pi12}) present selected bifurcation modes documenting a transition at increasing wavelength of the bifurcation modes from a local bifurcation~(\subref{fig:buckling_mode_7_15_k2_pi1}) to a shear-band-type instability~(\subref{fig:buckling_mode_7_15_k2_pi12}).
    }
\end{figure}

\subsection{Infinite set of bifurcation wavelengths in a lattice: perfect equivalence with the continuum} 
\label{sec:infinite_buckling_modes}
Loss of ellipticity in a solid occurs at modes of all (namely, infinite,) wavelengths, while the corresponding condition in the lattice \textit{usually} is that bifurcation occurs only in a mode involving an infinite wavelength.  
In this sense the equivalent continuous body has a response differing from the lattice, a circumstance which may be expected as a consequence of the homogenization procedure, which is applied to a discrete lattice. 

Surprisingly, it is shown in the following that special conditions can be found in which the lattice bifurcates similarly to the equivalent continuum, by displaying infinite modes, covering every wavelength.
In this case a \textit{perfect equivalence between the bifurcation in the lattice structure and failure of ellipticity in the effective continuum occurs}.

For a square grid (with $\alpha=\pi/2$, $\Lambda_1=7$, and $\Lambda_2=15$), the perfect equivalence  was obtained at a fixed value of load by varying the stiffness of the diagonal springs $\kappa$, thus obtaining $\kappa\approx0.128$. This value was calculated by numerically solving equation~\eqref{eq:local_buckling_multiplier_simpler} between $\kappa=0.1$ and $\kappa=0.2$, because these two values pinpoint the threshold of separation between macro and micro bifurcation. 
For these values of the lattice parameters and loads the bifurcation mode is unique and involves only the infinite wavelength (macro bifurcation) along the curved boundary denoted as $\text{(G)}$ in Fig.~\ref{fig:ellipticity_domains_7_15_pi2}, while on the boundary denoted as $\text{(GL)}$ in the same figure \textit{an infinite number of bifurcation modes of arbitrary wavelength is present for every critical loading state}, as detailed for the point $B_\infty$ in Fig.~\ref{fig:infinite_buckling_modes}.

Fig.~\ref{fig:contour_3d_infinite_buckling_modes} reports the three-dimensional plot in the space $\{\eta_1,\eta_2,\gamma\}$ satisfying the bifurcation condition [of vanishing of the determinant in Eq.~\eqref{eq:local_buckling_multiplier_simpler}] where $\gamma$ is the loading multiplier for $\{p_1,p_2\}=\gamma\{-1,-6\}$, so that the critical value $\gamma_{\text{B}}$ leading to bifurcation is highlighted as a red line marking the minimum of the bifurcation surface.
A section of this surface at $\eta_1=0$ is reported in Fig.~\ref{fig:contour_2d_infinite_buckling_modes} to show the dependence of the critical multiplier on the stiffness $\kappa$, so that for $\kappa<0.128$ the critical wave vector is $\bk_{\text{cr}}=0$ (macro instability), while for $\kappa>0.128$ the critical wave vector is $\bk_{\text{cr}}=\pi\,\bb_2$ (micro instability), and for $\kappa=0.128$ every wave vector of the form $\bk_{\text{cr}}=\eta_2\,\bb_2$ (with arbitrary $\eta_2$) identifies a different bifurcation mode occurring at the same load multiplier $\gamma_{\text{B}}\approx3.44$.
Within this infinite set of bifurcation modes, a few bifurcation modes (see the labelled points on the red contour of Fig.~\ref{fig:contour_2d_infinite_buckling_modes}) are reported in order to show the transition of the bifurcation mode from a local bifurcation (Fig.~\ref{fig:buckling_mode_7_15_k2_pi1}) to a global shear-band type instability (Fig.~\ref{fig:buckling_mode_7_15_k2_pi12}).

\section{Macroscopic and microscopic bifurcation localizations via perturbative approach}\label{sec:static_response}
The correlation between the incremental response of the lattice and of the equivalent solid is now investigated close to the conditions of instability using the `perturbative approach' introduced in~\cite{bigoni_2002a}.
Following this approach, the response of the lattice to an applied static concentrated load (in terms of a force or a force dipole) is numerically evaluated via finite elements (using the commercial code COMSOL Multiphysics\textsuperscript{\circledR}) and compared to the response of the equivalent solid by computing the Green's function associated to the operator $\diver\fC[\grad(\bullet)]$.

The two-dimensional Green's tensor $\mG$ needed to perturb the equivalent material and corresponding to a Dirac delta function centred at $\bx=\bzero$ is \cite{bigoni_2012}
\begin{equation}
\label{eq:Green_function}
    \mG(\hat{\bx}) = -\frac{1}{4\pi^2} \oint_{\abs{\bn}=1} \left(\bA^{(\fC)}(\bn)\right)^{-1} \log\abs{\hat{\bx}\scalp\bn} \,,
\end{equation}
where the position vector $\bx$ has been made dimensionless through division by the rod's length $l$, so that $\hat{\bx}=\bx/l$.
Note that $\mG=\trans{\mG}$ due to the symmetry of the acoustic tensor.

Numerical simulations are performed to analyze the lattice by considering a finite square computational domain of width $350l$, where $l$ is discretized in 10 finite elements with cubic shape functions. 
The selected mesh has been defined by performing a number of simulations with three different mesh refinements, namely 5, 10, and 20 elements for $l$, and then adopting 10 elements, as 20 provided no significant improvement, but substantial computational burden.
As the numerical simulations are meant to be compared to the infinite-body Green's function, the size of the domain has been calibrated in order to minimize boundary disturbances with clamped conditions at the four edges of the square domain. 
The governing equation for the prestressed Euler-Bernoulli rod, Eq.~\eqref{eq:governing_beam_EB_v}, used in the finite element scheme has been implemented by modifying the bending moment contribution with an additional geometric term representing the load multiplied by the transverse displacement of the rod.

The investigation presented below will reveal that:
\begin{enumerate}[label=(\roman*)]
    \item The localization of deformation connected to macro bifurcation in the lattice and to failure of ellipticity in the equivalent solid are strictly similar; 
    \item The lattice response close to a micro bifurcation evidences a \textit{`microscopic' type of localization}, which remains completely undetected in the homogenized material.
\end{enumerate}
These two different mechanical behaviours are analyzed by exploiting the macro-to-micro transition of the first bifurcation mode, which is controlled by the increase in the stiffness of the diagonal springs of the lattice considered in Section~\ref{sec:grid}. The grid is subject now to an equibiaxial compression loading. 
Hence, in Section~\ref{sec:macroscopic_localization} the lattice is considered in the absence of diagonal springs ($\kappa=0$), while in Section~\ref{sec:microscopic_localization} the lattice is reinforced with a springs' stiffness $\kappa=0.4$.

\subsection{Macroscopic bifurcations on the verge of ellipticity loss}
\label{sec:macroscopic_localization}
The lattice configurations selected for the following analysis are reported in Table~\ref{tab:cases_analyzed_ellipticity}, together with the values of the preload $\bp_{\text{E}}$ corresponding to loss of ellipticity in the equivalent continuum (obtained by numerically solving equation~\eqref{eq:failure_E_multiplier} assuming a radial path $\bp=\{p_1,p_1\}$) or, in other words, to a macro bifurcation in the lattice.
As explained in the previous section, the stiffness of the diagonal springs is set to zero in order to ensure that a macroscopic bifurcation is critical.
The table reports also the inclination $\theta_{\text{cr}}$ of the normal $\bn$ to the localization band, defined as $\bn_{\text{E}}=\be_1\cos\theta_{\text{cr}}+\be_2\sin\theta_{\text{cr}}$. 
%
\begin{table}[htb!]
    \centering
    \begin{tabular}{lllcl}
        \toprule
        \textit{Geometry}       & \textit{Rods slenderness}          & \textit{Symmetry}   & $\bp_{\text{E}}$        & $\theta_{\text{cr}}$\\ \midrule
        Square $\alpha=\pi/2$   & $\Lambda_1=\Lambda_2=10$      & Cubic               & $-5.434\,\{1,1\}$     & $0^\circ,90^\circ$\\
                                & $\Lambda_1=7,\,\Lambda_2=15$  & Orthotropic         & $-2.071\,\{1,1\}$     & $0^\circ$\\
        Rhombus $\alpha=\pi/3$  & $\Lambda_1=\Lambda_2=10$      & Orthotropic         & $-5.345\,\{1,1\}$     & $88.2^\circ,151.8^\circ$\\
                                & $\Lambda_1=7,\,\Lambda_2=15$  & Anisotropic         & $-2.043\,\{1,1\}$     & $151.4^\circ$\\
        \bottomrule
    \end{tabular}
    \caption{\label{tab:cases_analyzed_ellipticity}
    Equibiaxial compression loads $\bp_{\text{E}}$ and inclinations $\theta_{\text{cr}}$ of $\bn_{\text{E}}$ corresponding to failure of ellipticity in the equivalent material, corresponding to a macro-bifurcation in the lattice, for different grid configurations (see Fig.~\ref{fig:geometry_grid_and_unit_cell}), in the absence of diagonal springs ($\kappa=0$).
    }
\end{table}
%
\begin{figure}[htb!]
\centering
\begin{subfigure}{0.24\textwidth}
\centering
\phantomcaption{\label{fig:square_10_10_dipole_0}}
\includegraphics[width=0.98\linewidth]{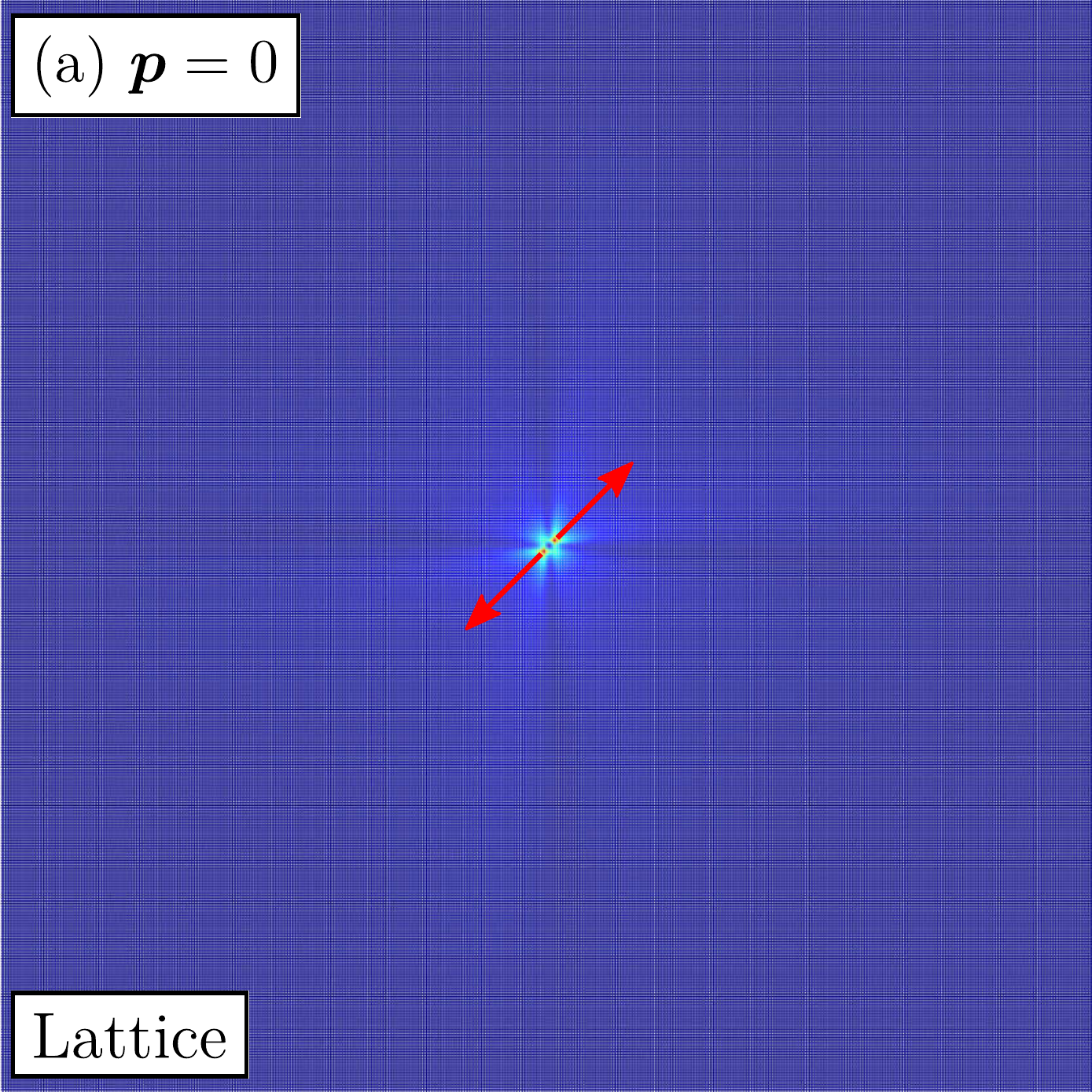}
\end{subfigure}
\begin{subfigure}{0.24\textwidth}
\centering
\phantomcaption{\label{fig:square_10_10_dipole_80}}
\includegraphics[width=0.98\linewidth]{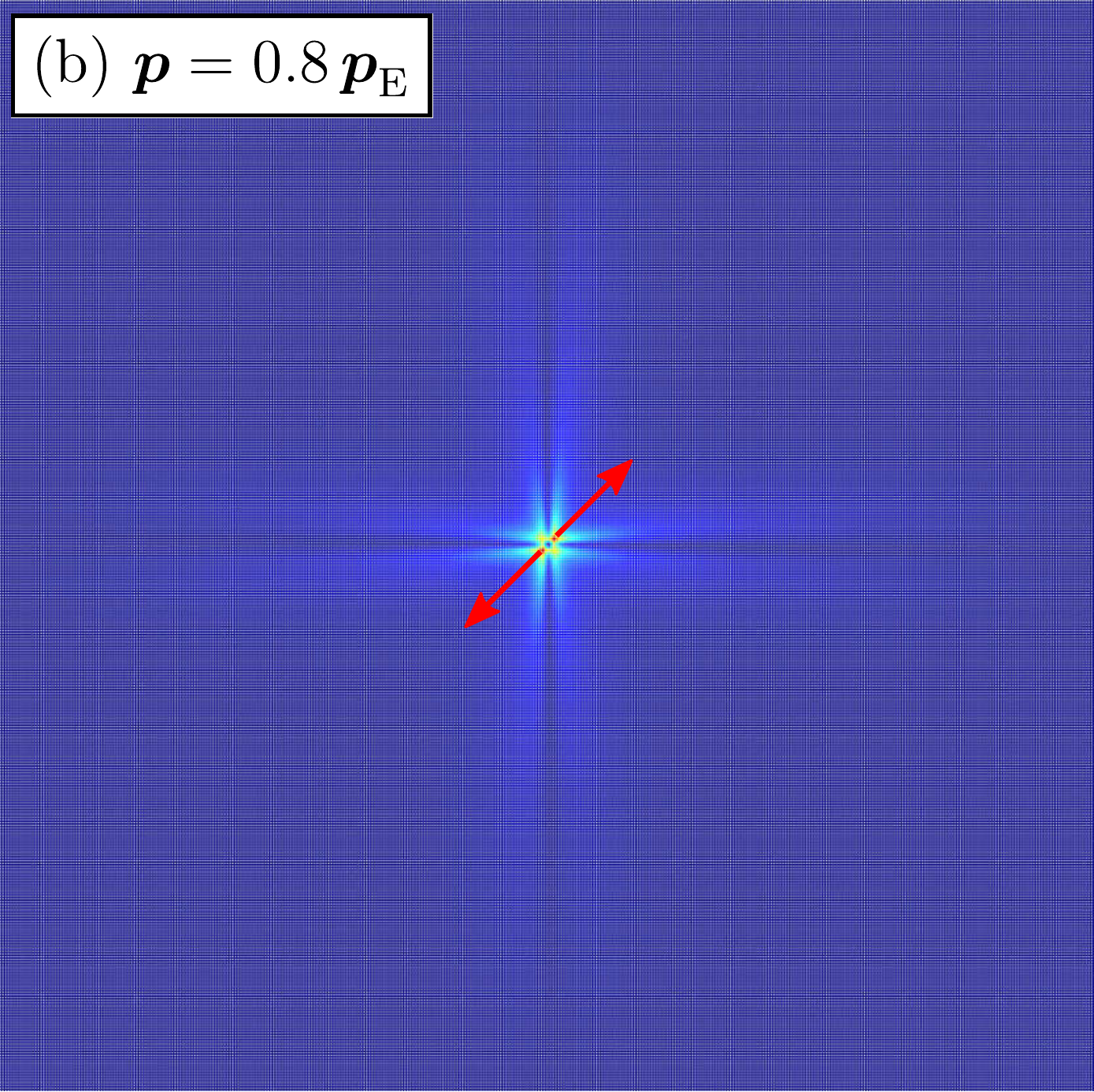}
\end{subfigure}
\begin{subfigure}{0.24\textwidth}
\centering
\phantomcaption{\label{fig:square_10_10_dipole_90}}
\includegraphics[width=0.98\linewidth]{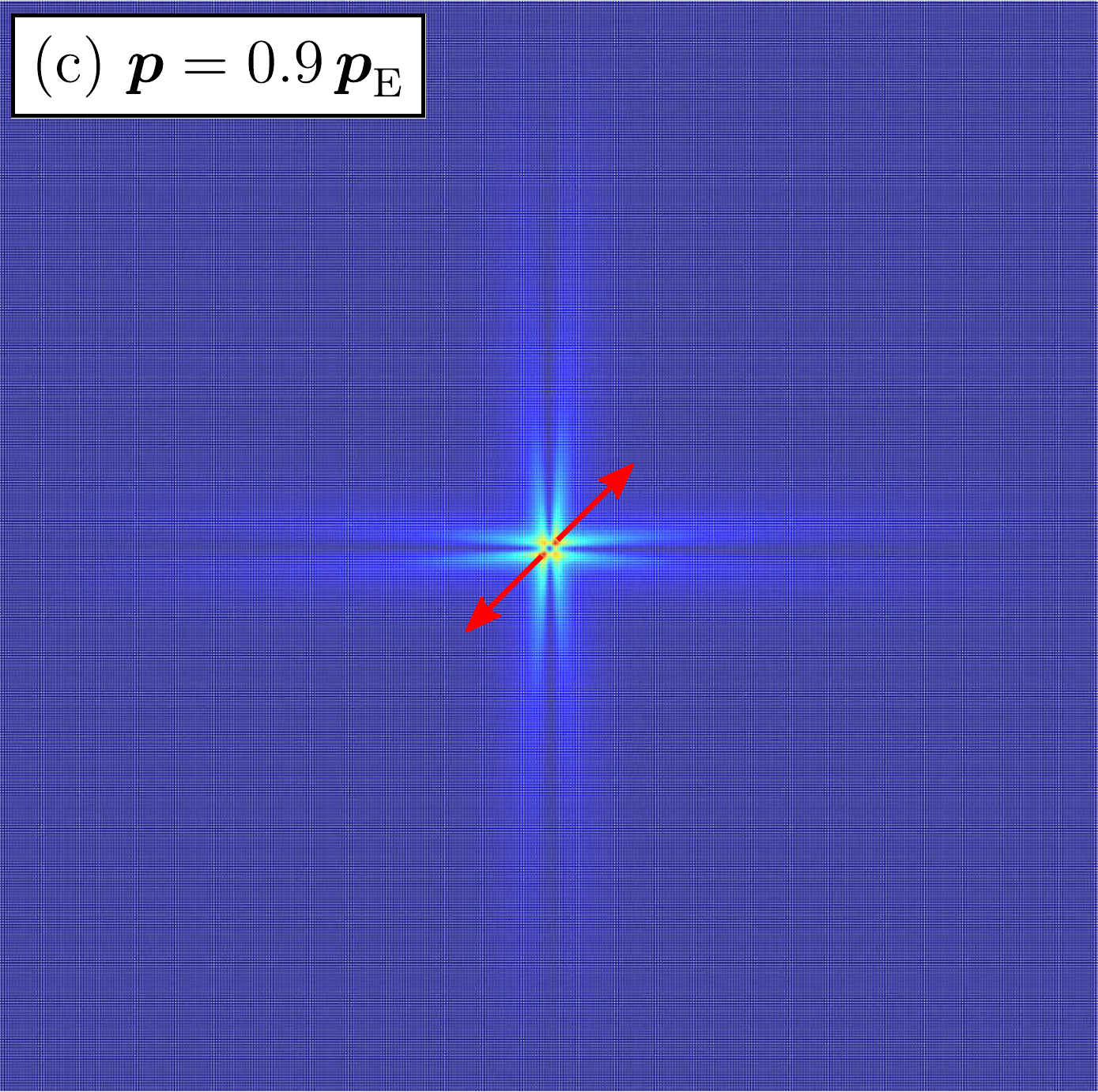}
\end{subfigure}
\begin{subfigure}{0.24\textwidth}
\centering
\phantomcaption{\label{fig:square_10_10_dipole_99}}
\includegraphics[width=0.98\linewidth]{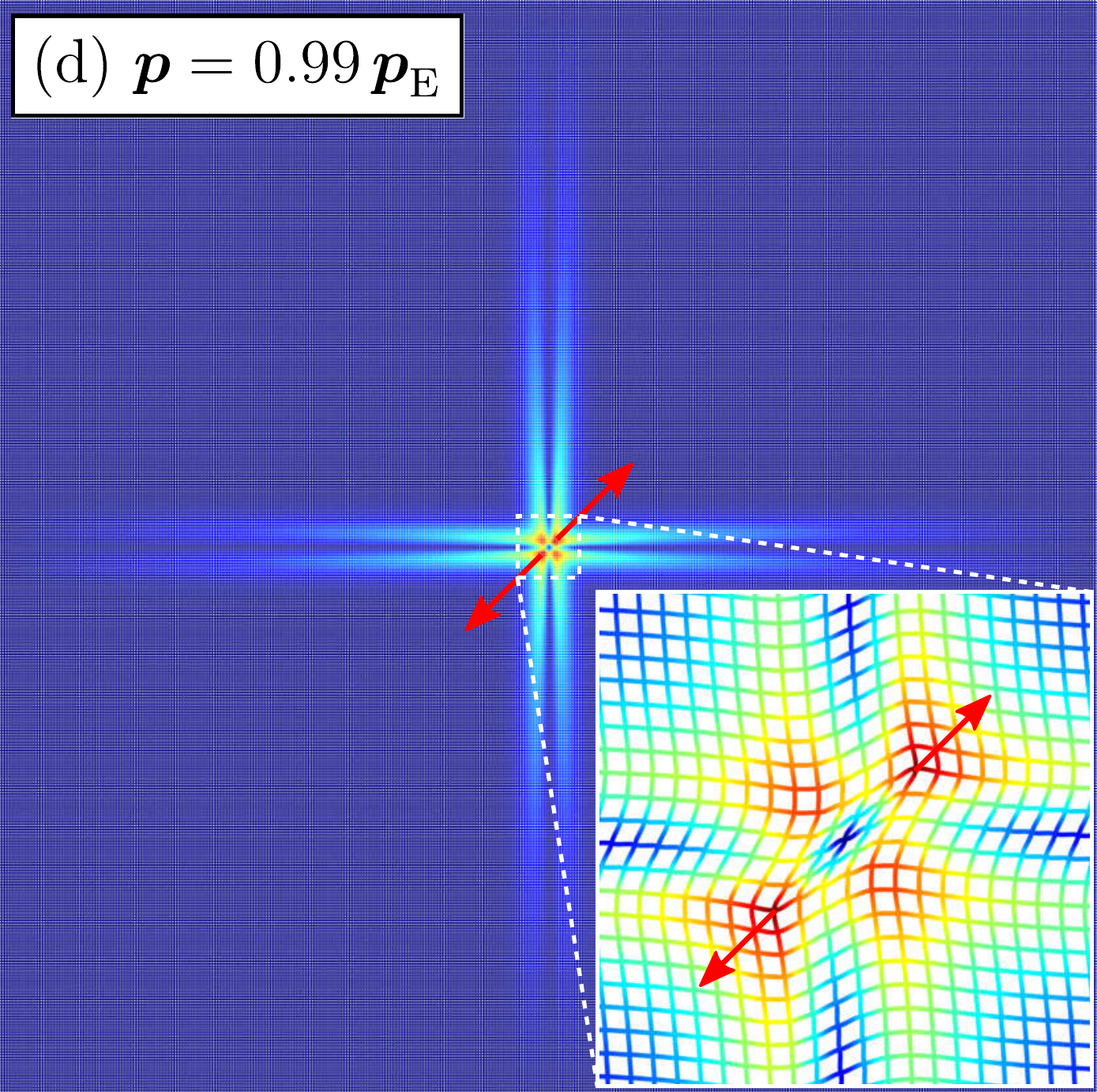}
\end{subfigure}\\
\vspace{0.01\linewidth}
\begin{subfigure}{0.24\textwidth}
\centering
\phantomcaption{\label{fig:square_10_10_dipole_0_gf}}
\includegraphics[width=0.98\linewidth]{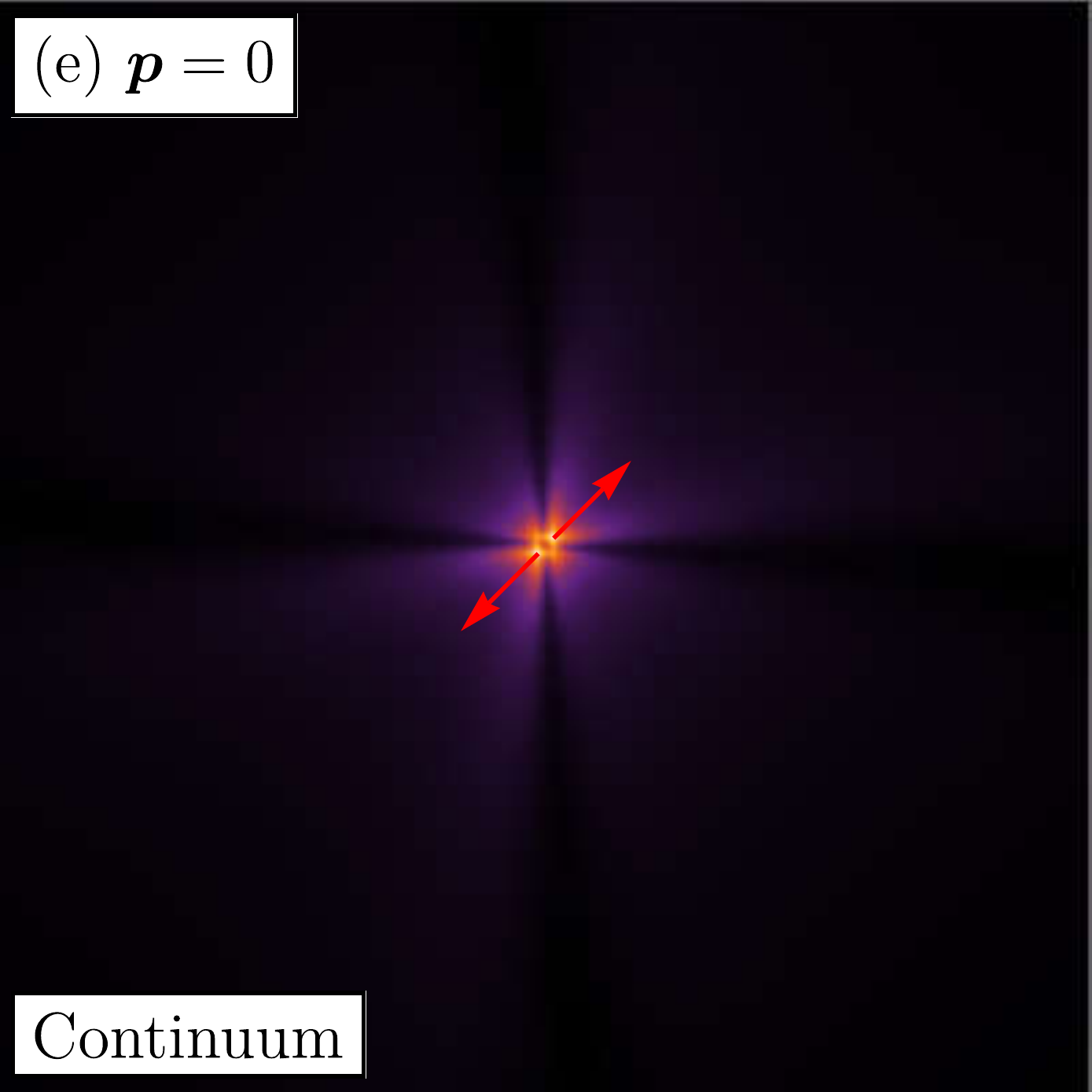}
\end{subfigure}
\begin{subfigure}{0.24\textwidth}
\centering
\phantomcaption{\label{fig:square_10_10_dipole_80_gf}}
\includegraphics[width=0.98\linewidth]{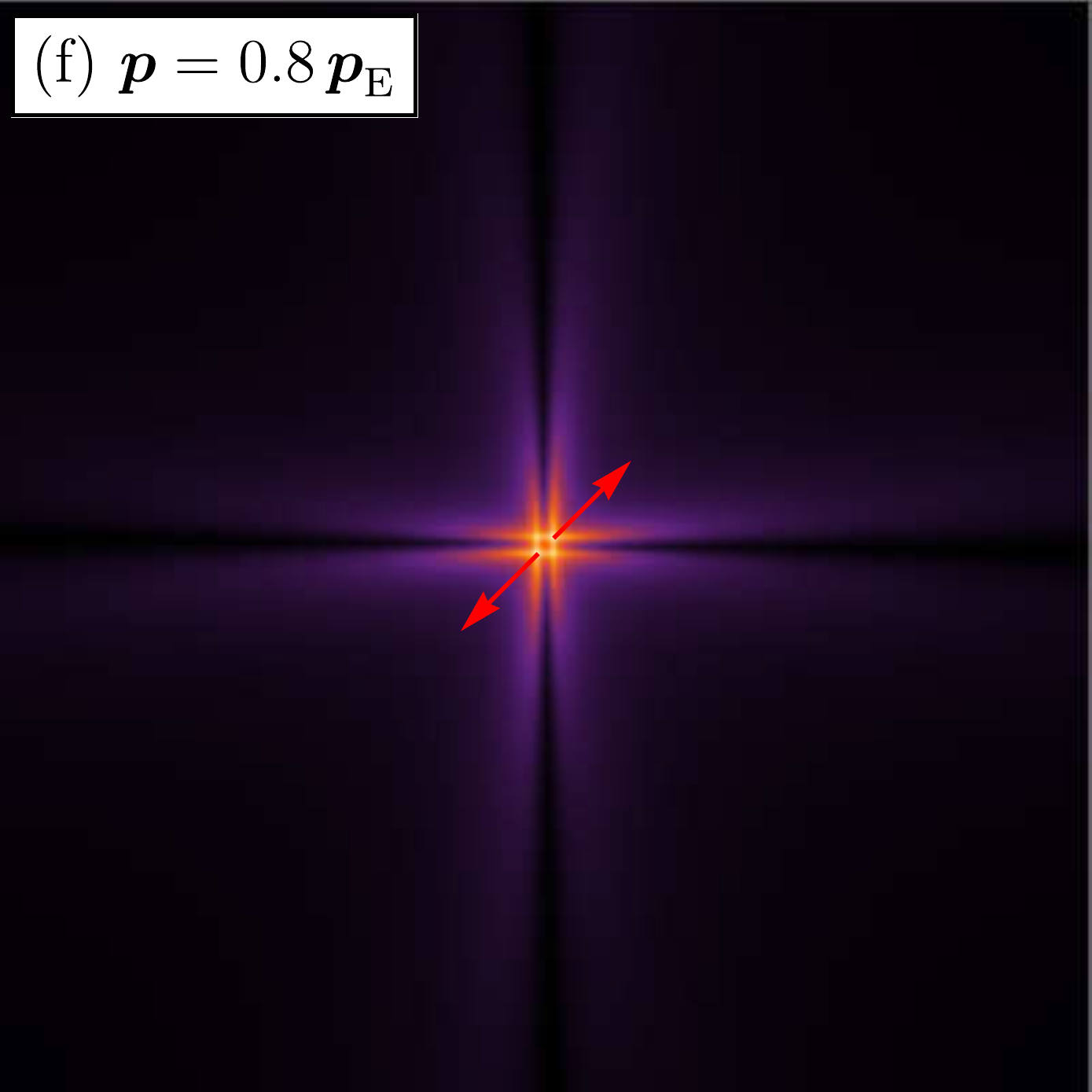}
\end{subfigure}
\begin{subfigure}{0.24\textwidth}
\centering
\phantomcaption{\label{fig:square_10_10_dipole_90_gf}}
\includegraphics[width=0.98\linewidth]{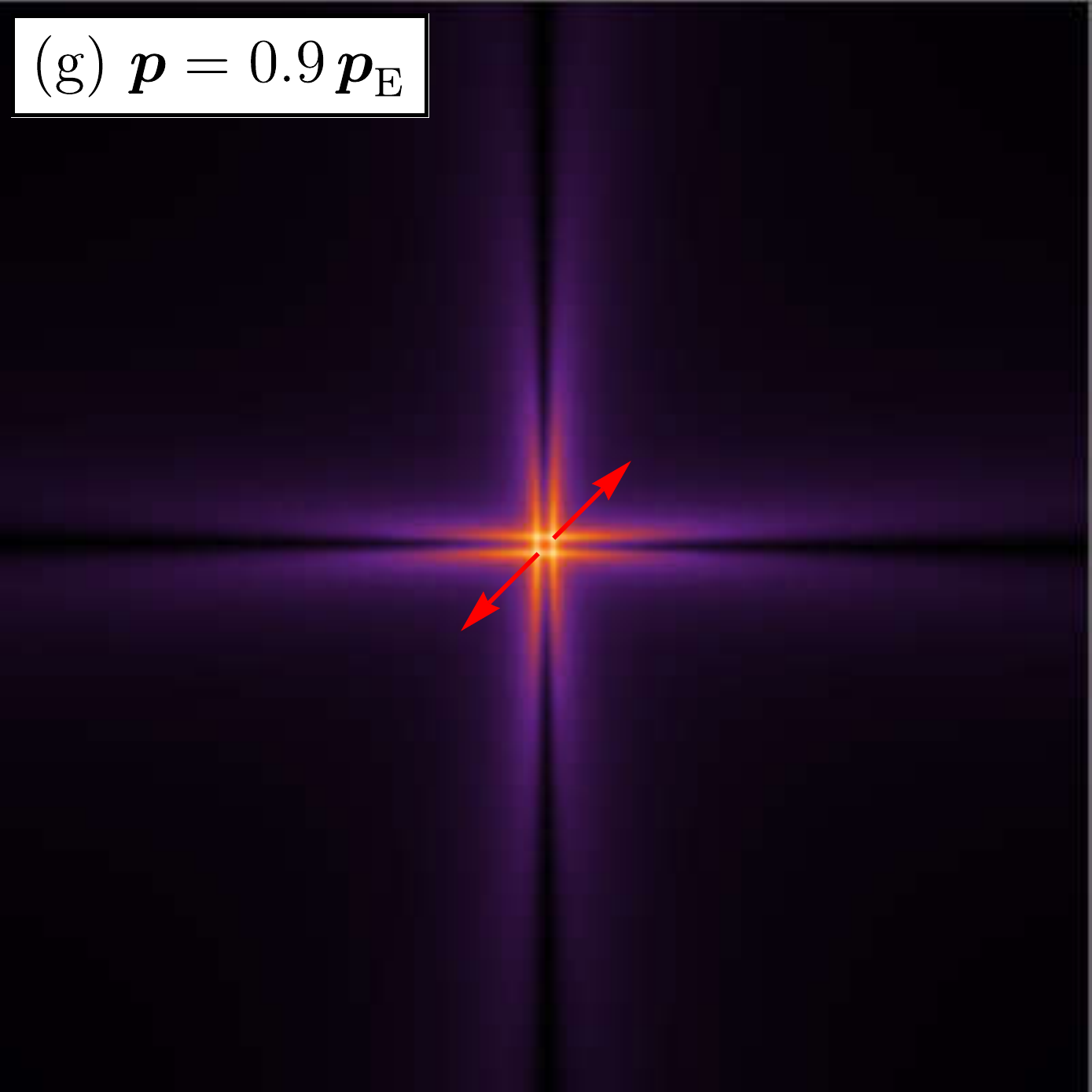}
\end{subfigure}
\begin{subfigure}{0.24\textwidth}
\centering
\phantomcaption{\label{fig:square_10_10_dipole_99_gf}}
\includegraphics[width=0.98\linewidth]{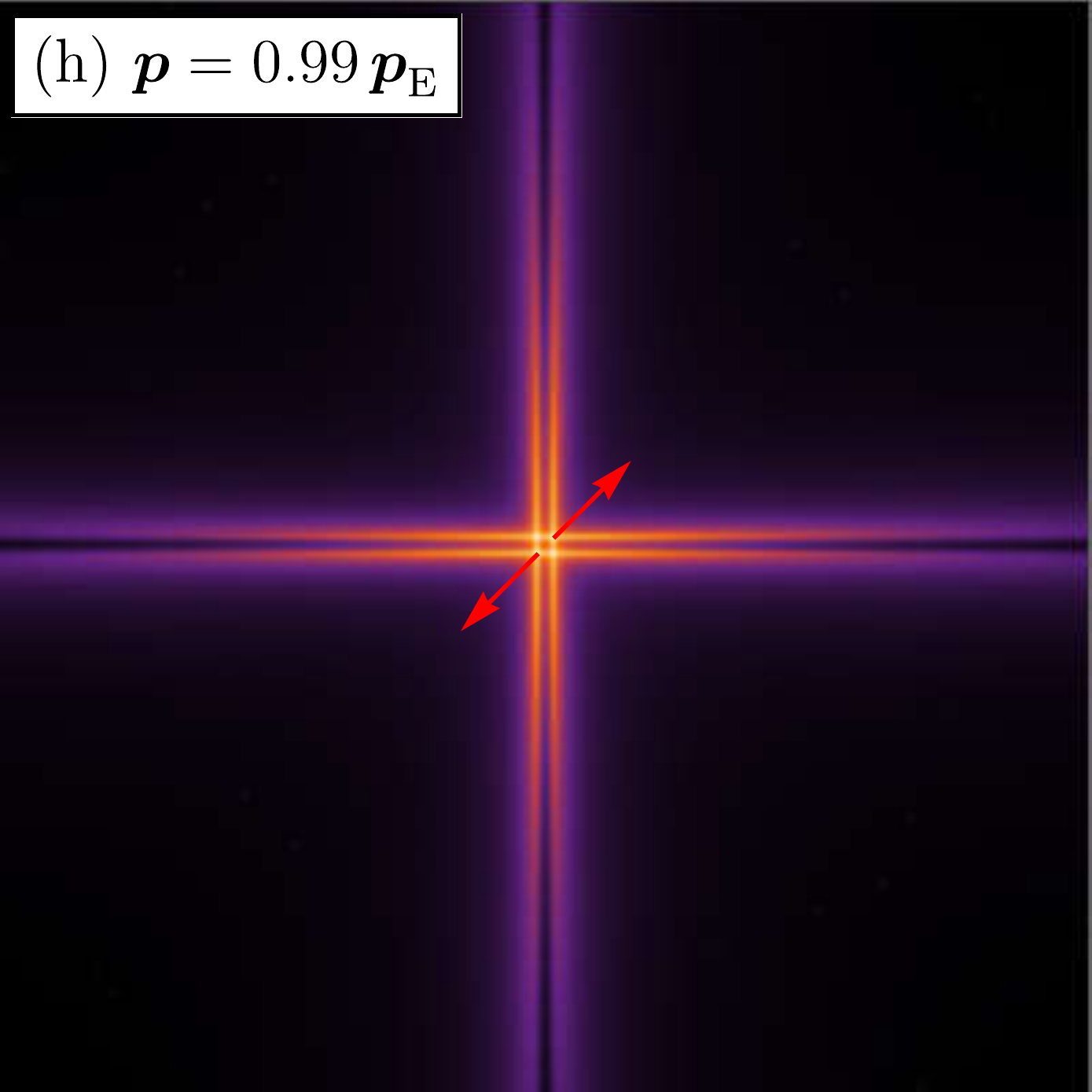}
\end{subfigure}
\caption{\label{fig:square_10_10_dipole}
	Progressive emergence at increasing load of two orthogonal shear bands visible in the displacement field generated by a diagonal force dipole applied to a square lattice (with cubic symmetry, $\Lambda_1=\Lambda_2=10$, upper part,  \subref{fig:square_10_10_dipole_0}--\subref{fig:square_10_10_dipole_99}, simulated via f.e.m.) compared to the response of the homogenized continuum (lower part  \subref{fig:square_10_10_dipole_0_gf}--\subref{fig:square_10_10_dipole_99_gf}). 
	From left to right the load increases towards failure of strong ellipticity $\bp_{\text{E}}$. 
	Shear bands are aligned parallel to the directions predicted at failure of ellipticity ($\theta_{\text{cr}}=0^\circ,90^\circ$).
}
\end{figure}

\begin{figure}[htb!]
\centering
\begin{subfigure}{0.24\textwidth}
\centering
\phantomcaption{\label{fig:square_7_15_dipole_0}}
\includegraphics[width=0.98\linewidth]{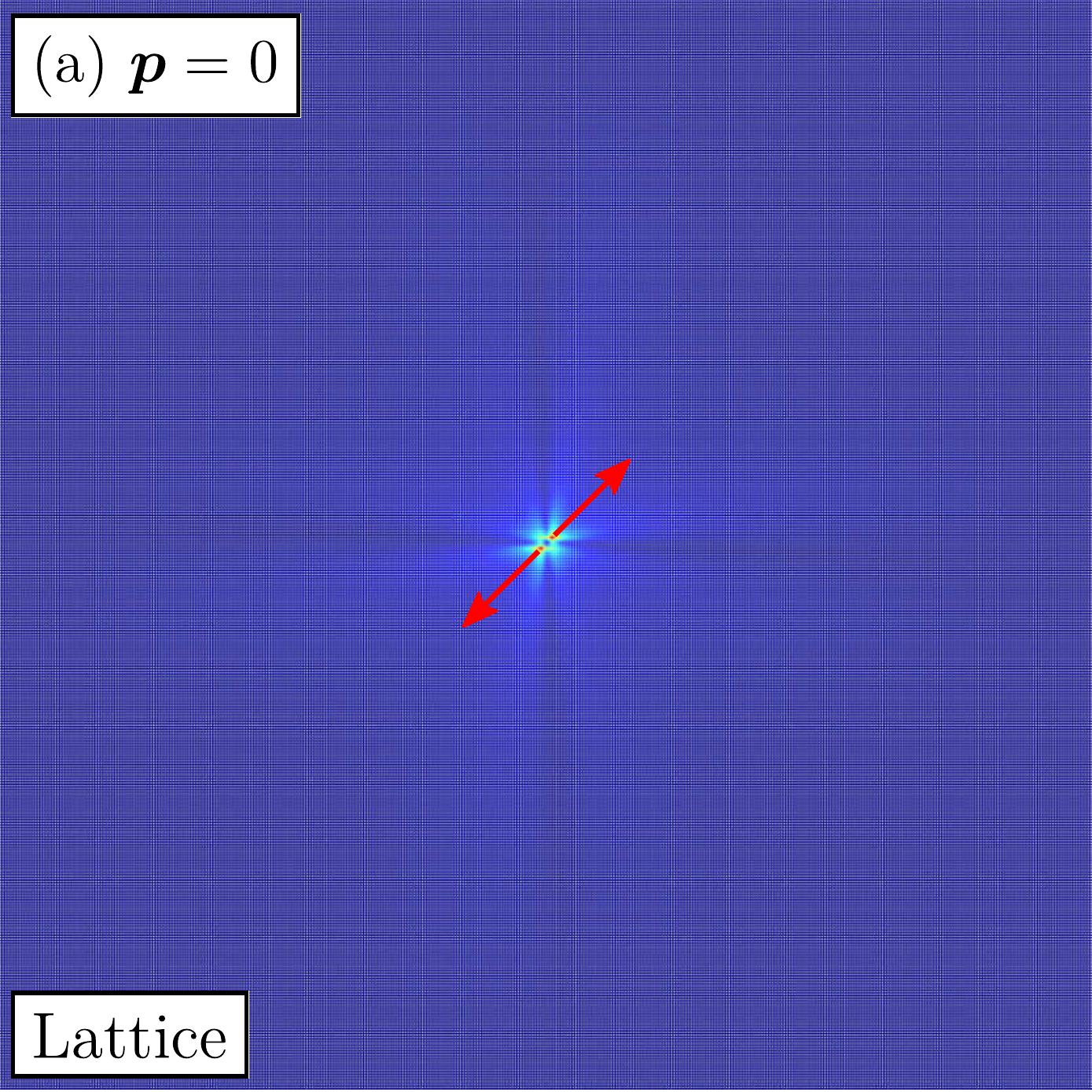}
\end{subfigure}
\begin{subfigure}{0.24\textwidth}
\centering
\phantomcaption{\label{fig:square_7_15_dipole_80}}
\includegraphics[width=0.98\linewidth]{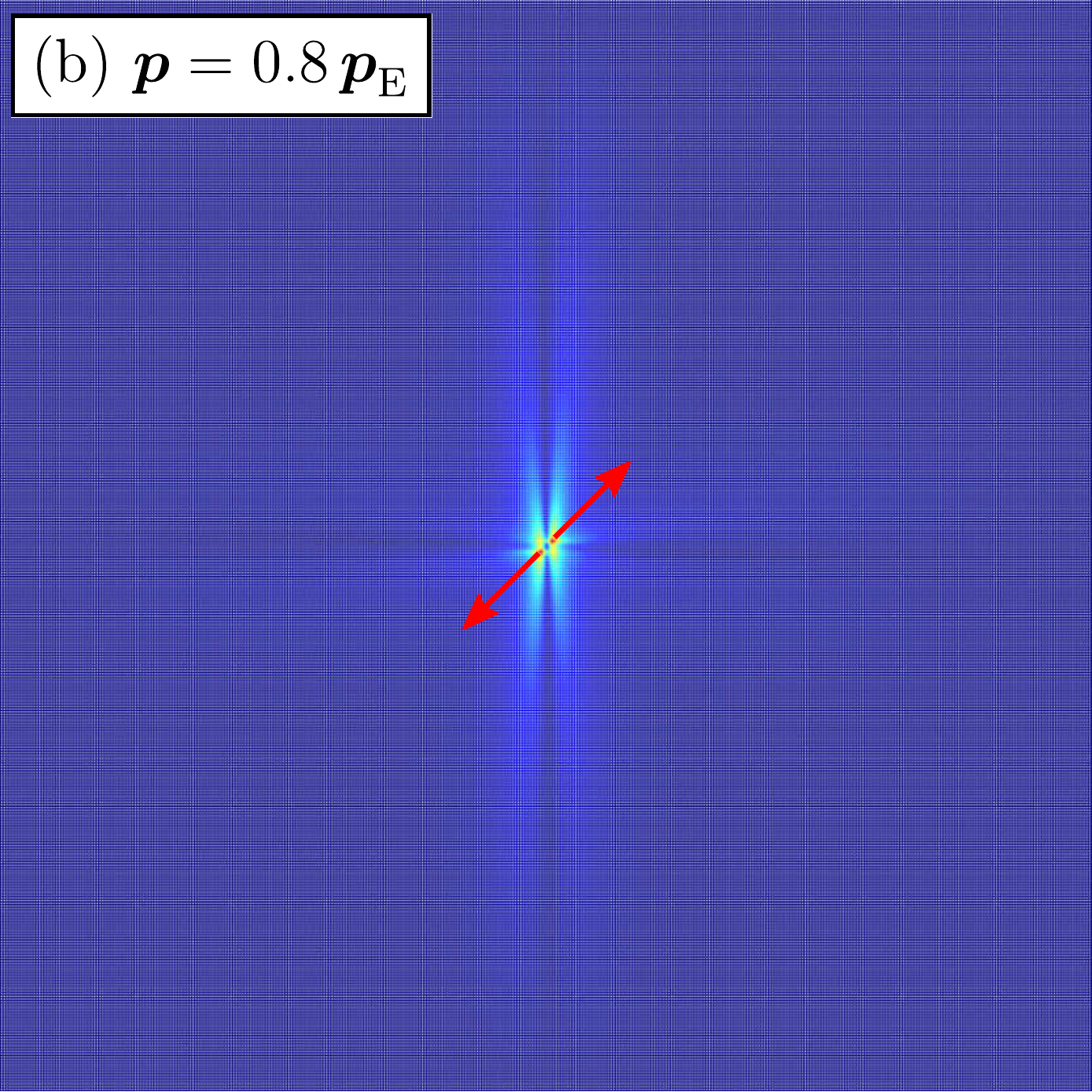}
\end{subfigure}
\begin{subfigure}{0.24\textwidth}
\centering
\phantomcaption{\label{fig:square_7_15_dipole_90}}
\includegraphics[width=0.98\linewidth]{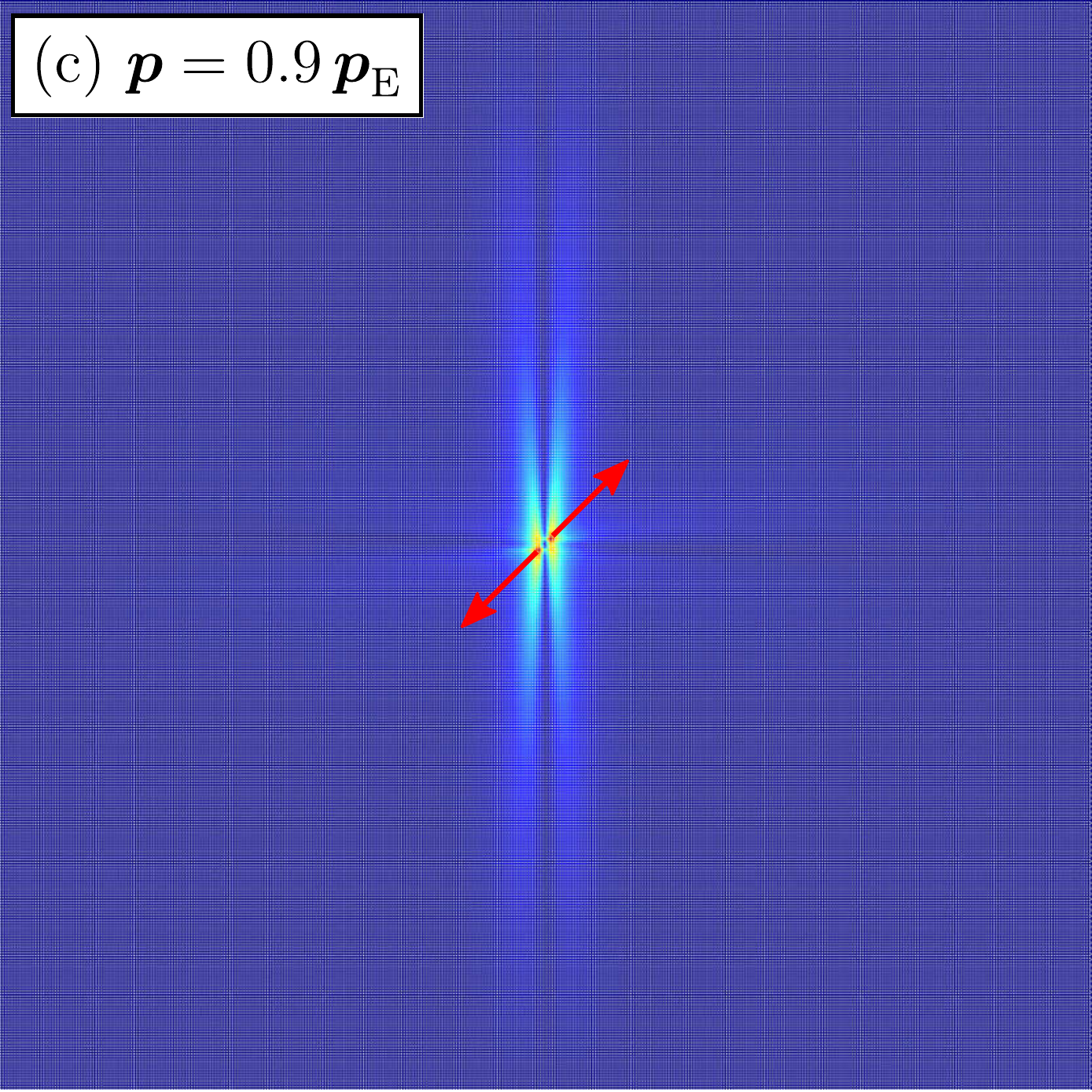}
\end{subfigure}
\begin{subfigure}{0.24\textwidth}
\centering
\phantomcaption{\label{fig:square_7_15_dipole_99}}
\includegraphics[width=0.98\linewidth]{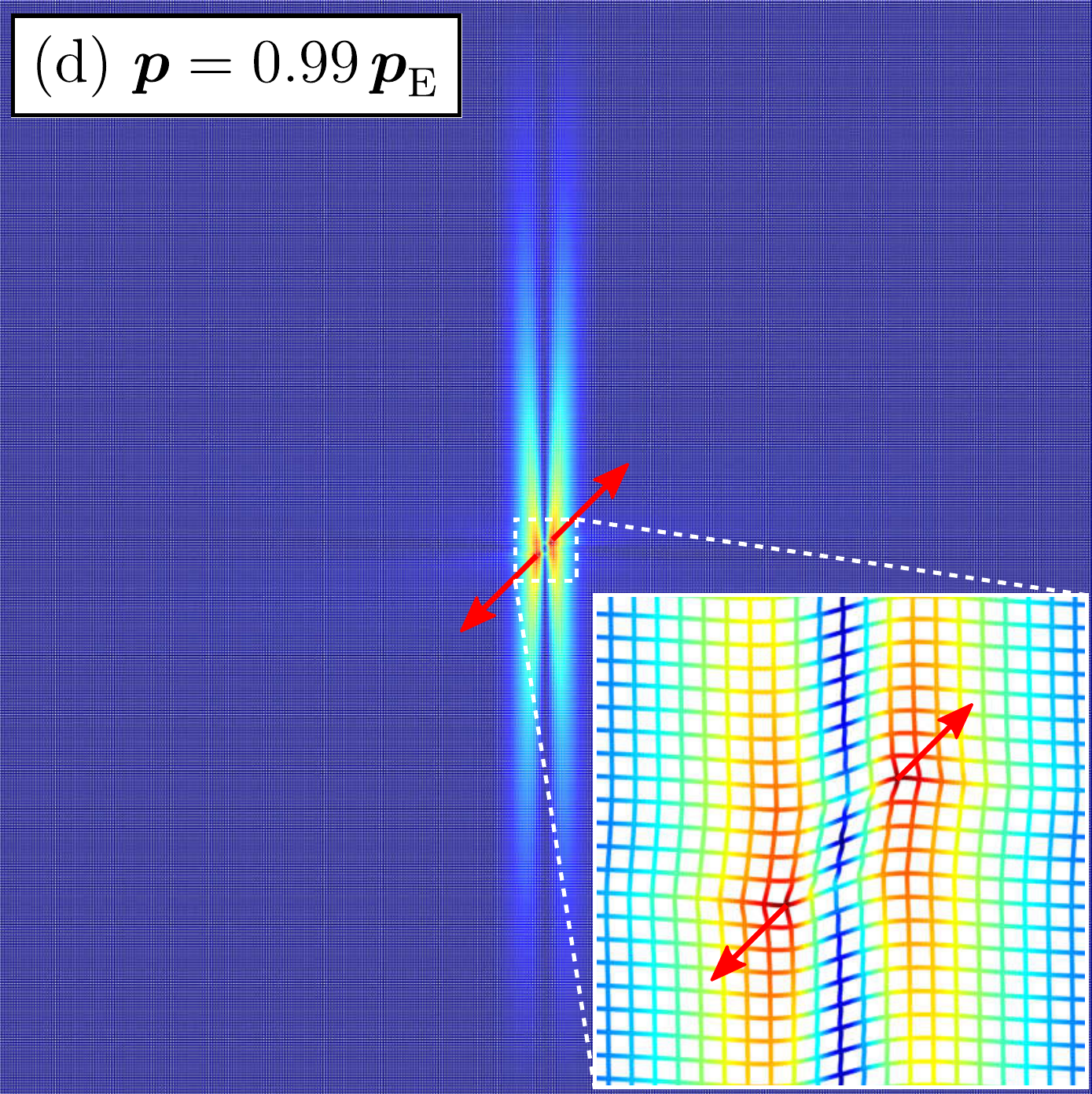}
\end{subfigure}\\
\vspace{0.01\linewidth}
\begin{subfigure}{0.24\textwidth}
\centering
\phantomcaption{\label{fig:square_7_15_dipole_0_gf}}
\includegraphics[width=0.98\linewidth]{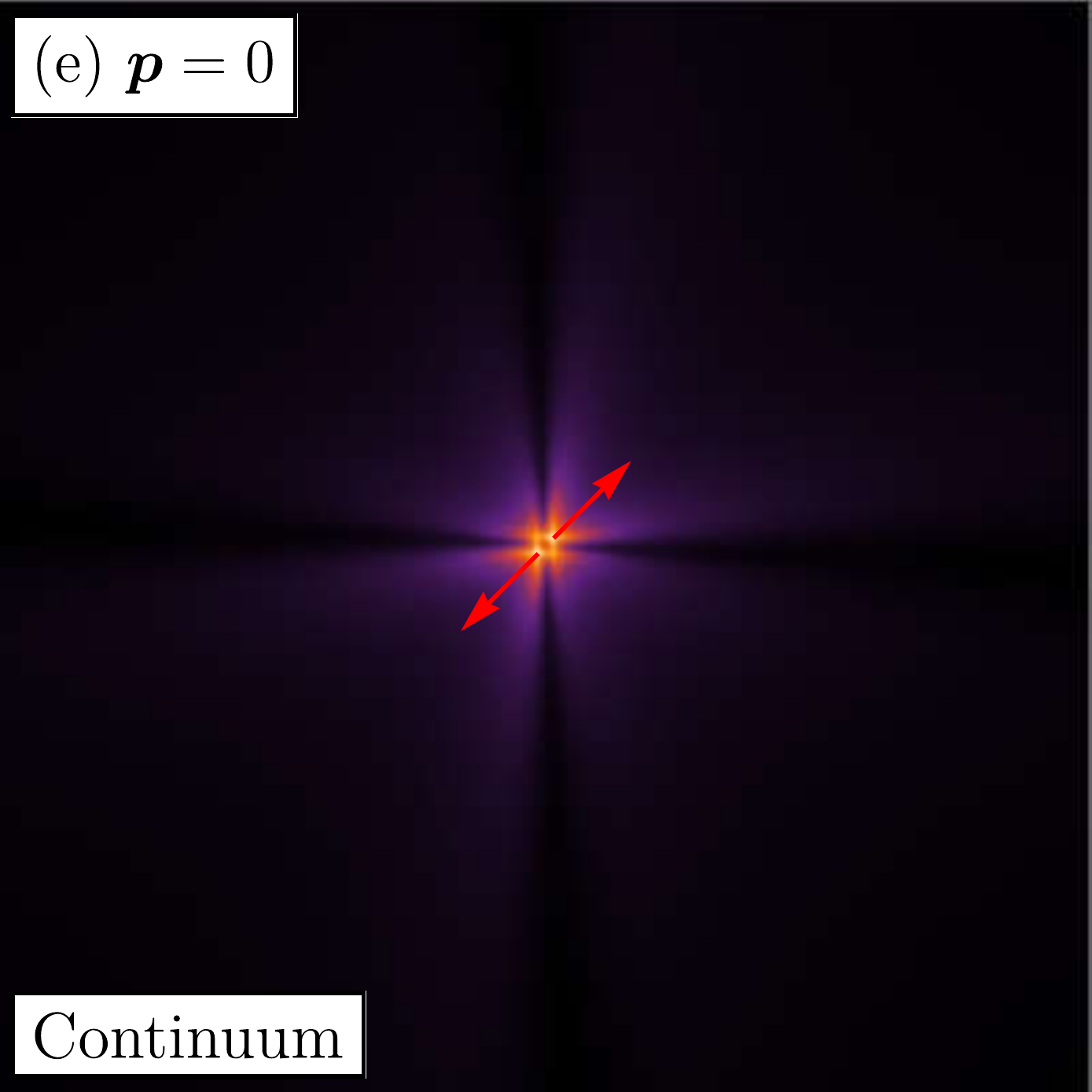}
\end{subfigure}
\begin{subfigure}{0.24\textwidth}
\centering
\phantomcaption{\label{fig:square_7_15_dipole_80_gf}}
\includegraphics[width=0.98\linewidth]{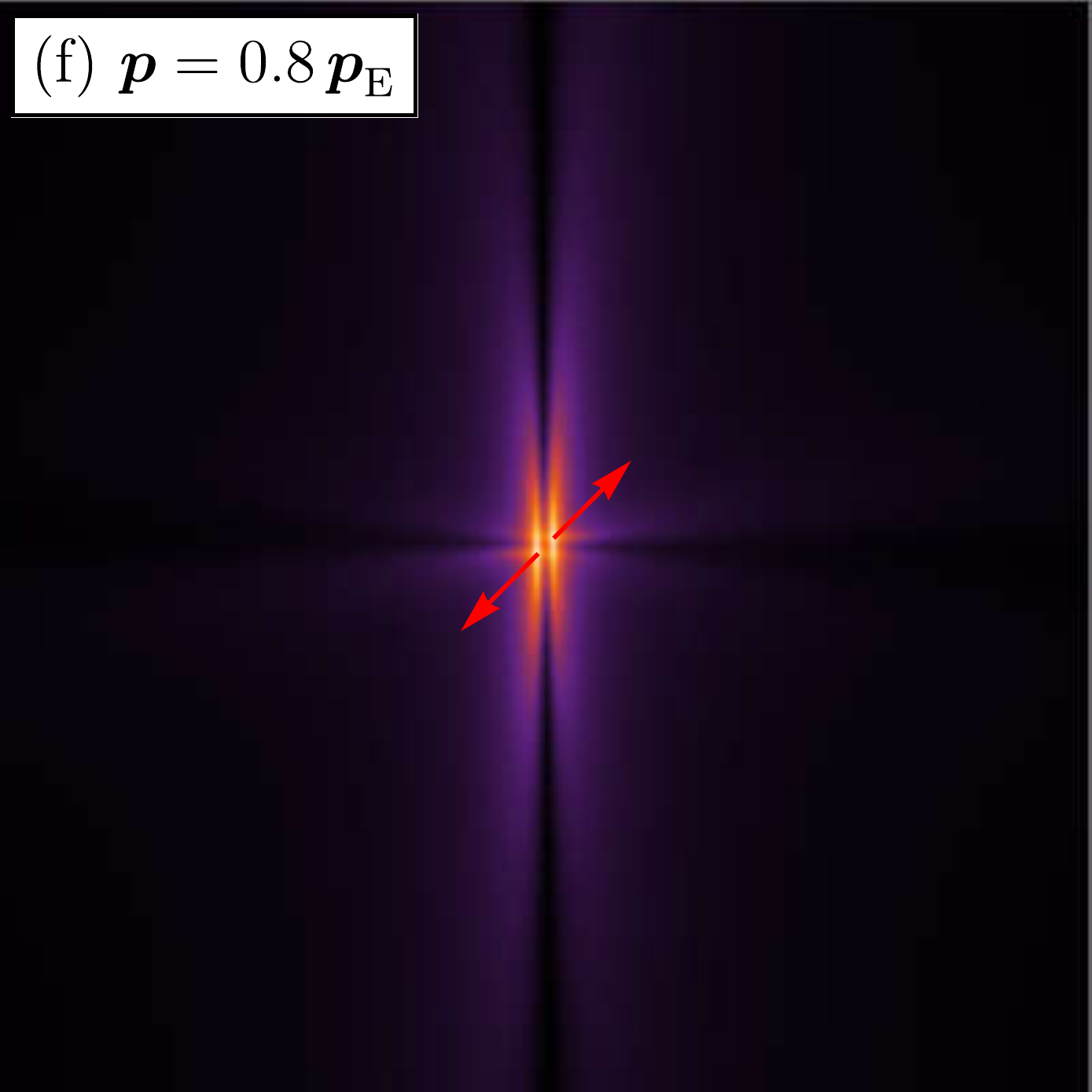}
\end{subfigure}
\begin{subfigure}{0.24\textwidth}
\centering
\phantomcaption{\label{fig:square_7_15_dipole_90_gf}}
\includegraphics[width=0.98\linewidth]{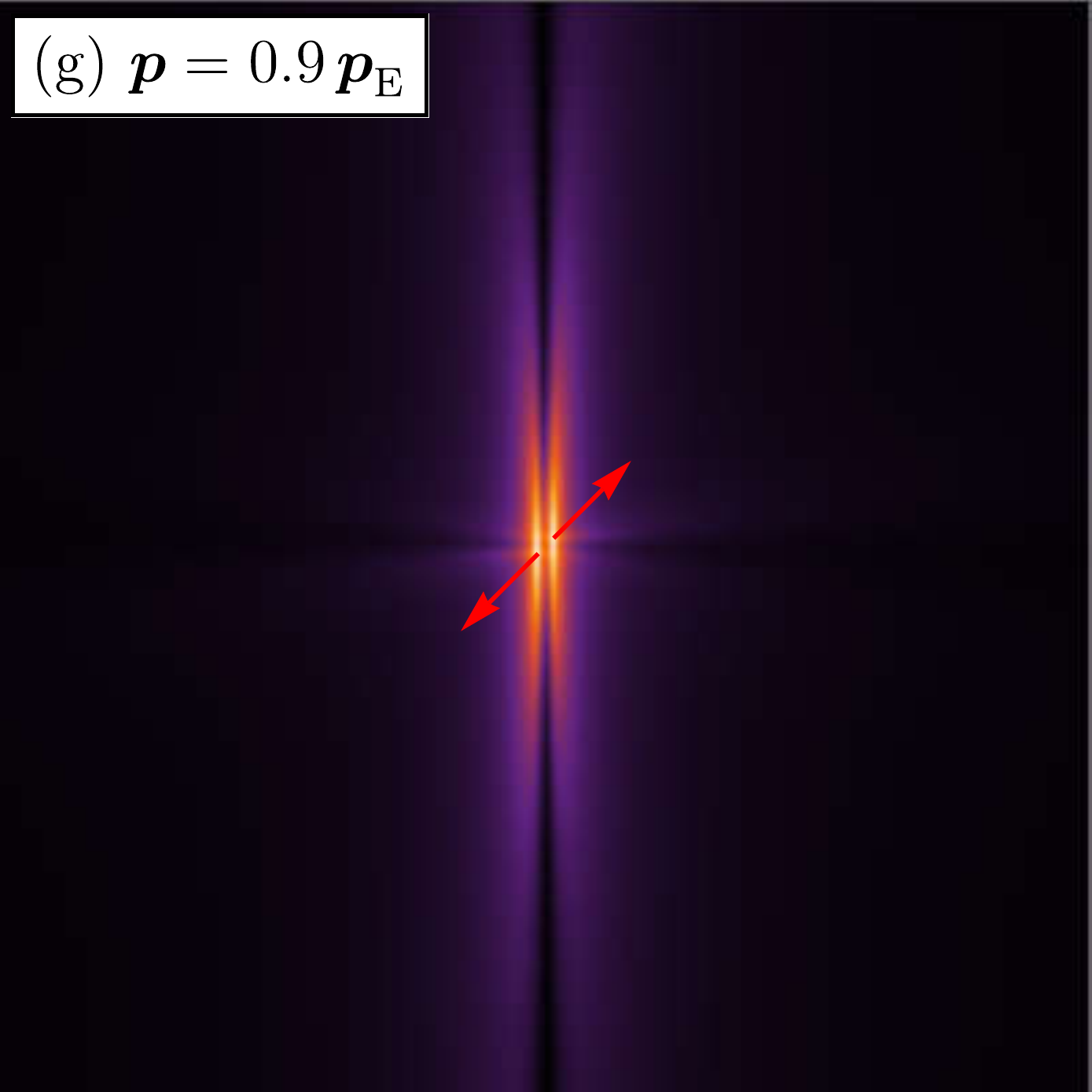}
\end{subfigure}
\begin{subfigure}{0.24\textwidth}
\centering
\phantomcaption{\label{fig:square_7_15_dipole_99_gf}}
\includegraphics[width=0.98\linewidth]{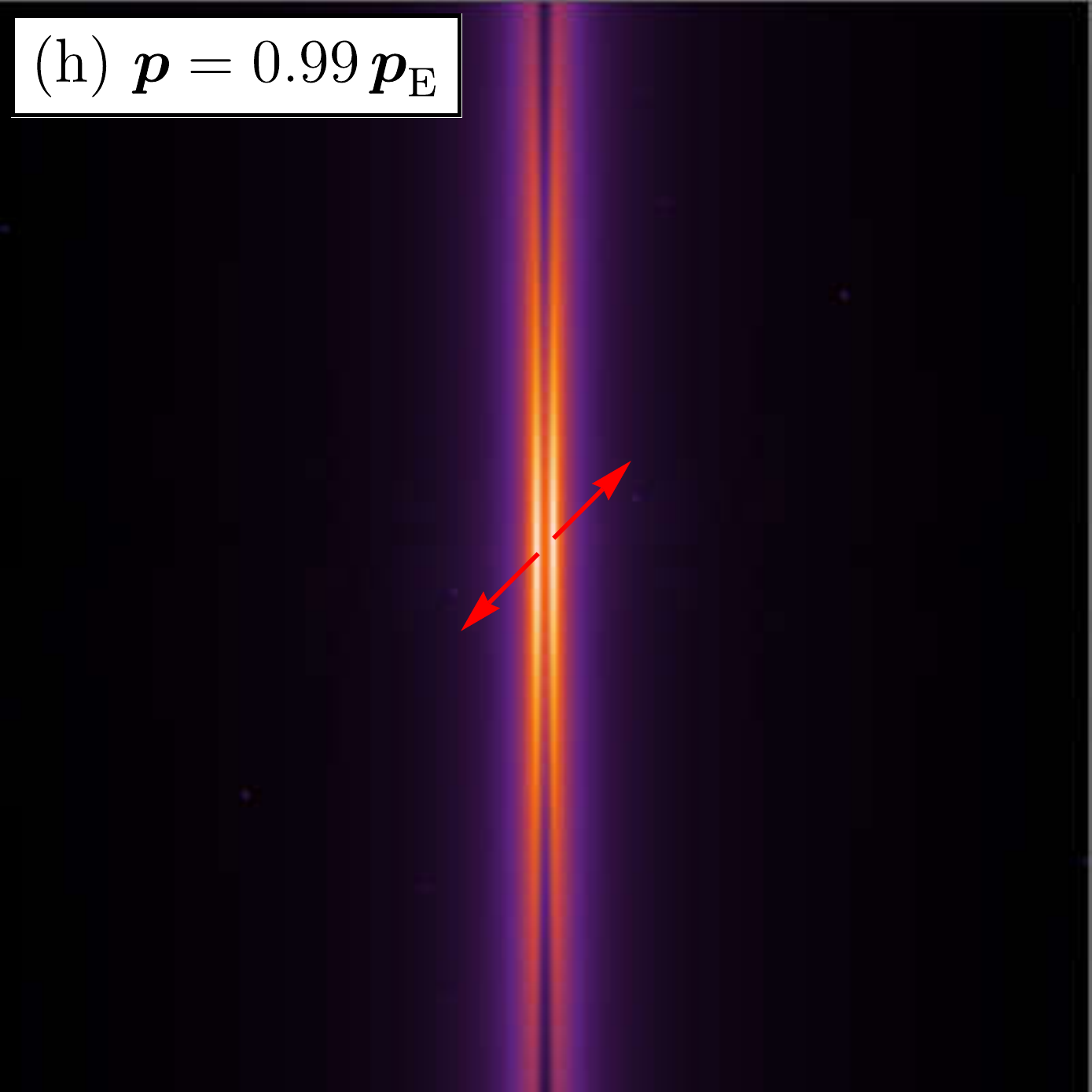}
\end{subfigure}
\caption{\label{fig:square_7_15_dipole}
	As for Fig. \ref{fig:square_10_10_dipole}, but for an orthotropic square lattice ($\Lambda_1=7,\,\Lambda_2=15$), where a single and vertical, $\theta_{\text{cr}}=90^\circ$, shear band forms.
}
\end{figure}

\begin{figure}[htb!]
\centering
\begin{subfigure}{0.24\textwidth}
\centering
\phantomcaption{\label{fig:rhombus_10_10_dipole_0}}
\includegraphics[width=0.98\linewidth]{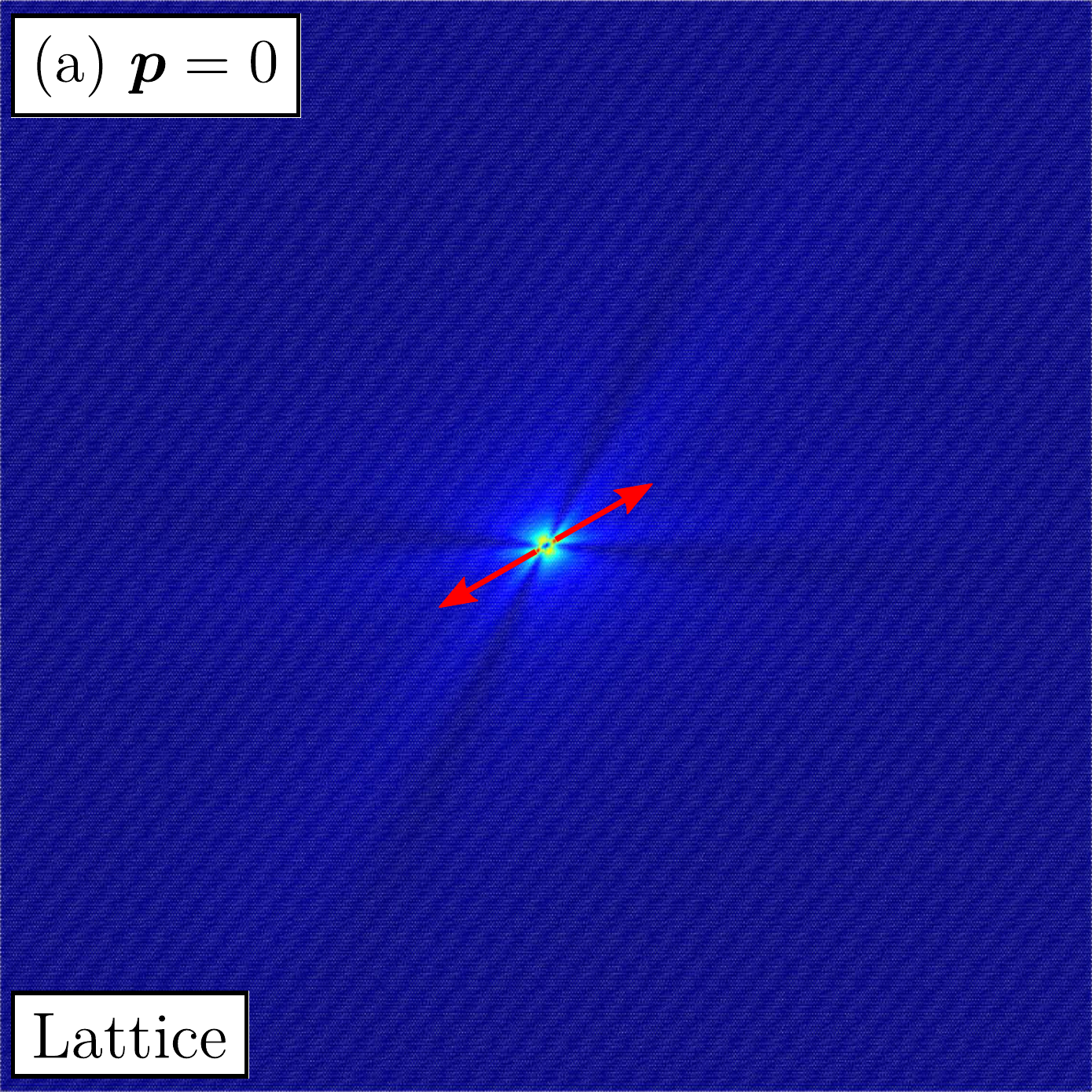}
\end{subfigure}
\begin{subfigure}{0.24\textwidth}
\centering
\phantomcaption{\label{fig:rhombus_10_10_dipole_80}}
\includegraphics[width=0.98\linewidth]{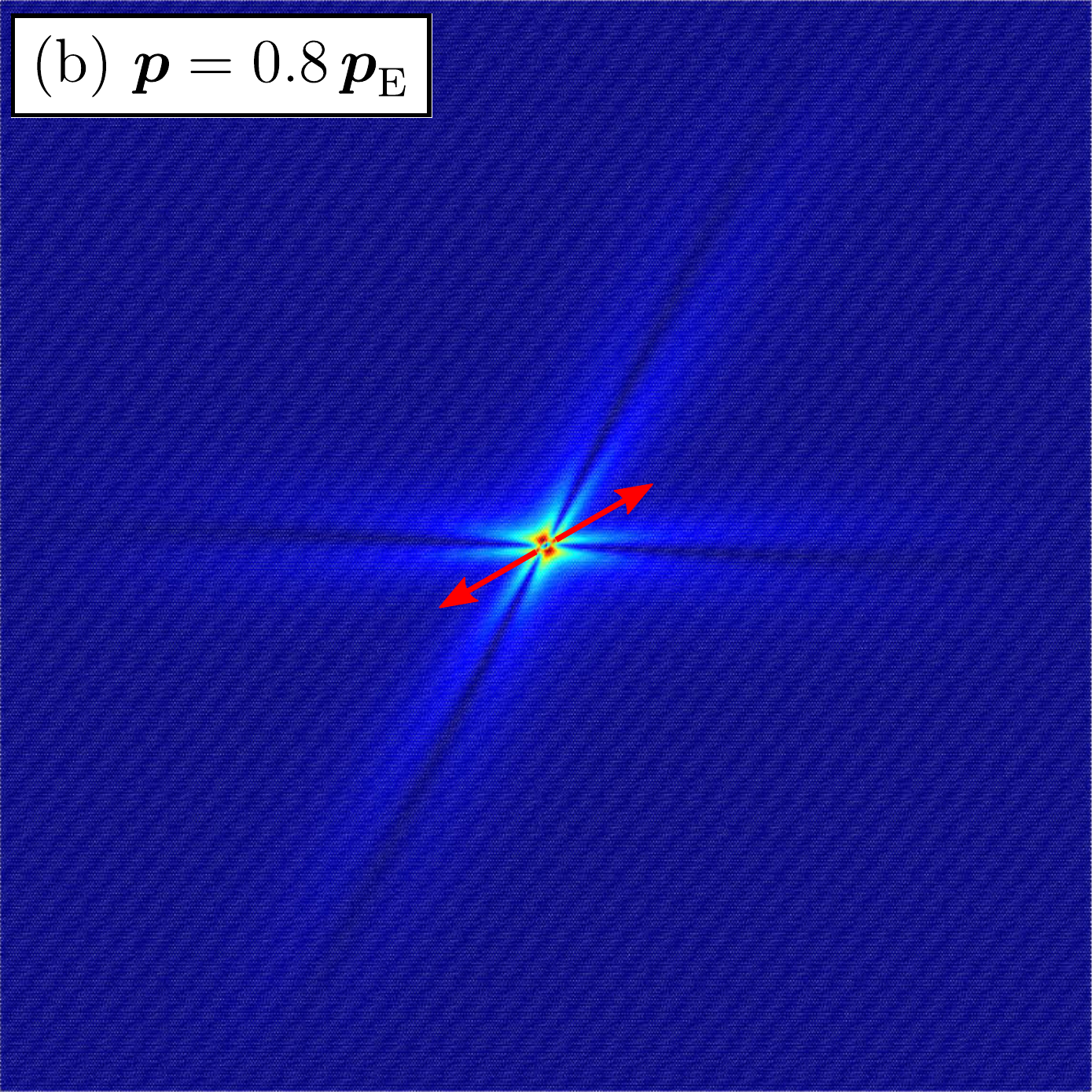}
\end{subfigure}
\begin{subfigure}{0.24\textwidth}
\centering
\phantomcaption{\label{fig:rhombus_10_10_dipole_90}}
\includegraphics[width=0.98\linewidth]{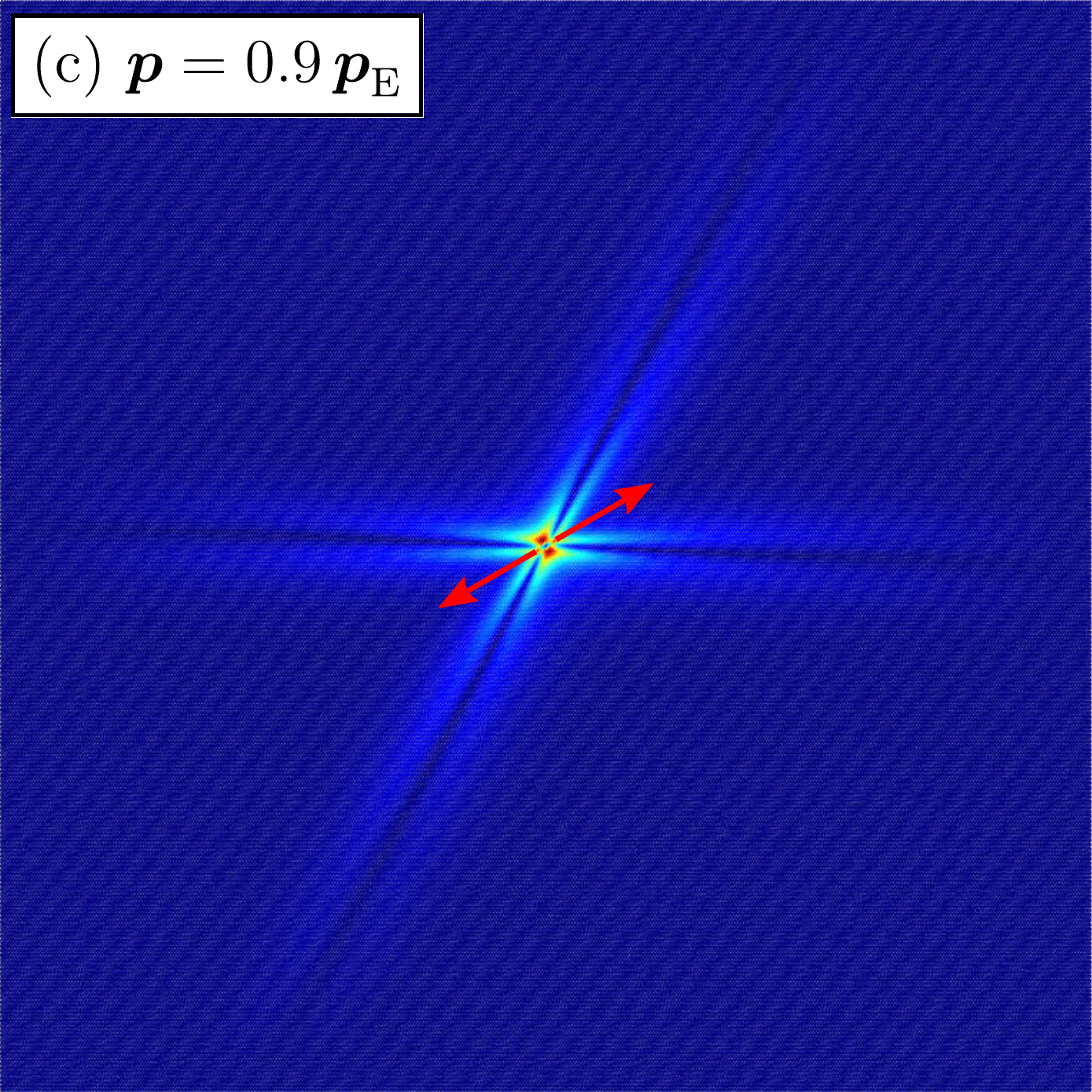}
\end{subfigure}
\begin{subfigure}{0.24\textwidth}
\centering
\phantomcaption{\label{fig:rhombus_10_10_dipole_99}}
\includegraphics[width=0.98\linewidth]{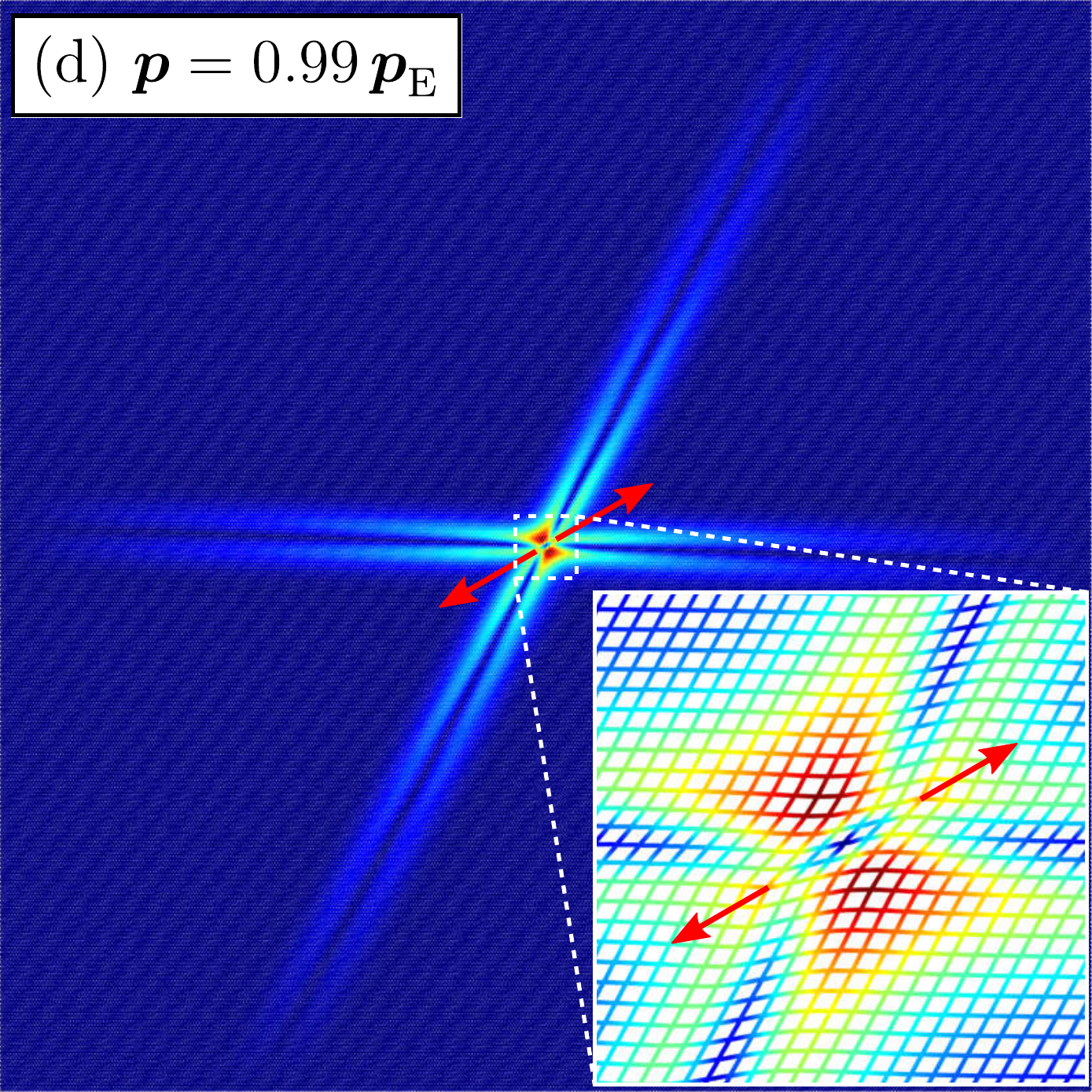}
\end{subfigure}\\
\vspace{0.01\linewidth}
\begin{subfigure}{0.24\textwidth}
\centering
\phantomcaption{\label{fig:rhombus_10_10_dipole_0_gf}}
\includegraphics[width=0.98\linewidth]{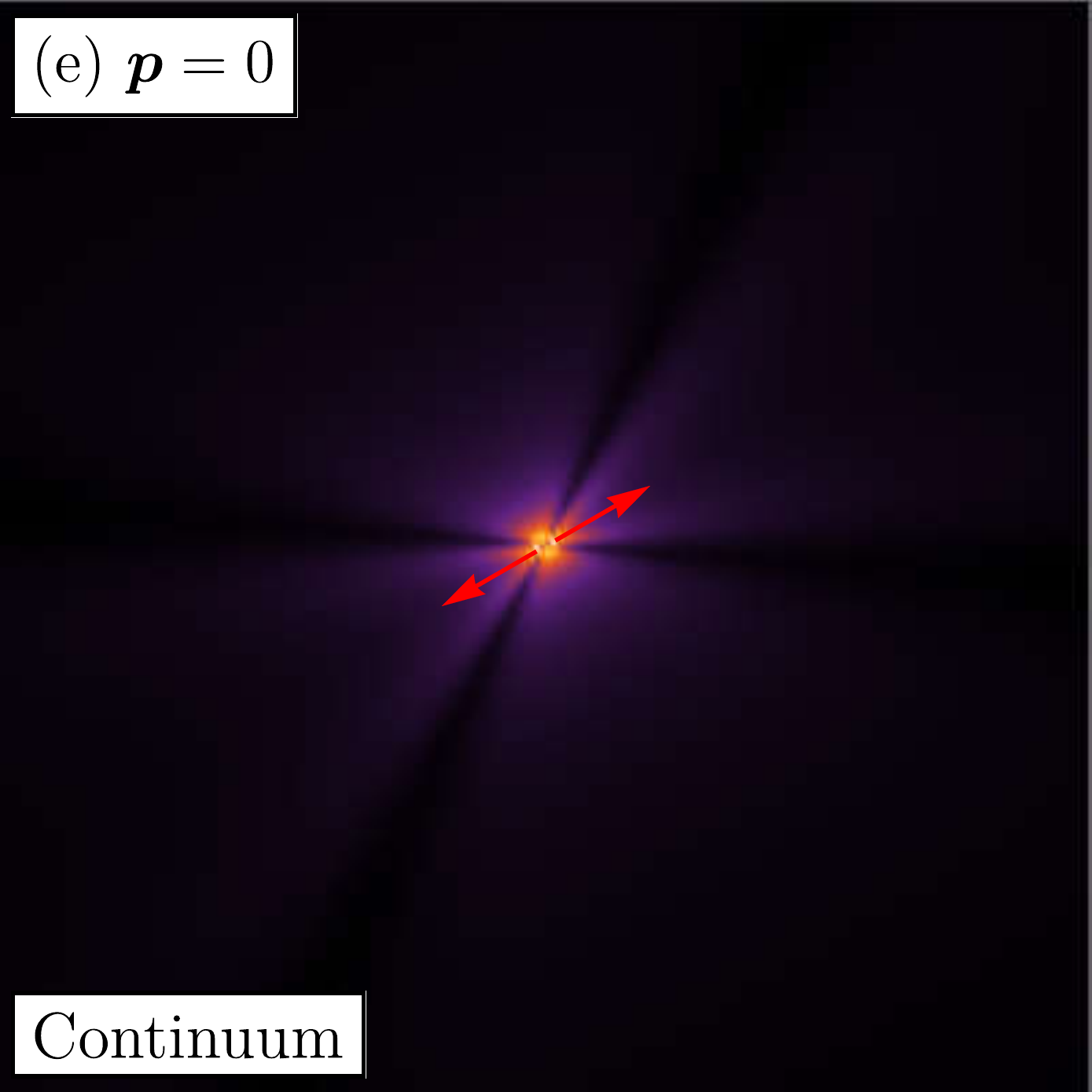}
\end{subfigure}
\begin{subfigure}{0.24\textwidth}
\centering
\phantomcaption{\label{fig:rhombus_10_10_dipole_80_gf}}
\includegraphics[width=0.98\linewidth]{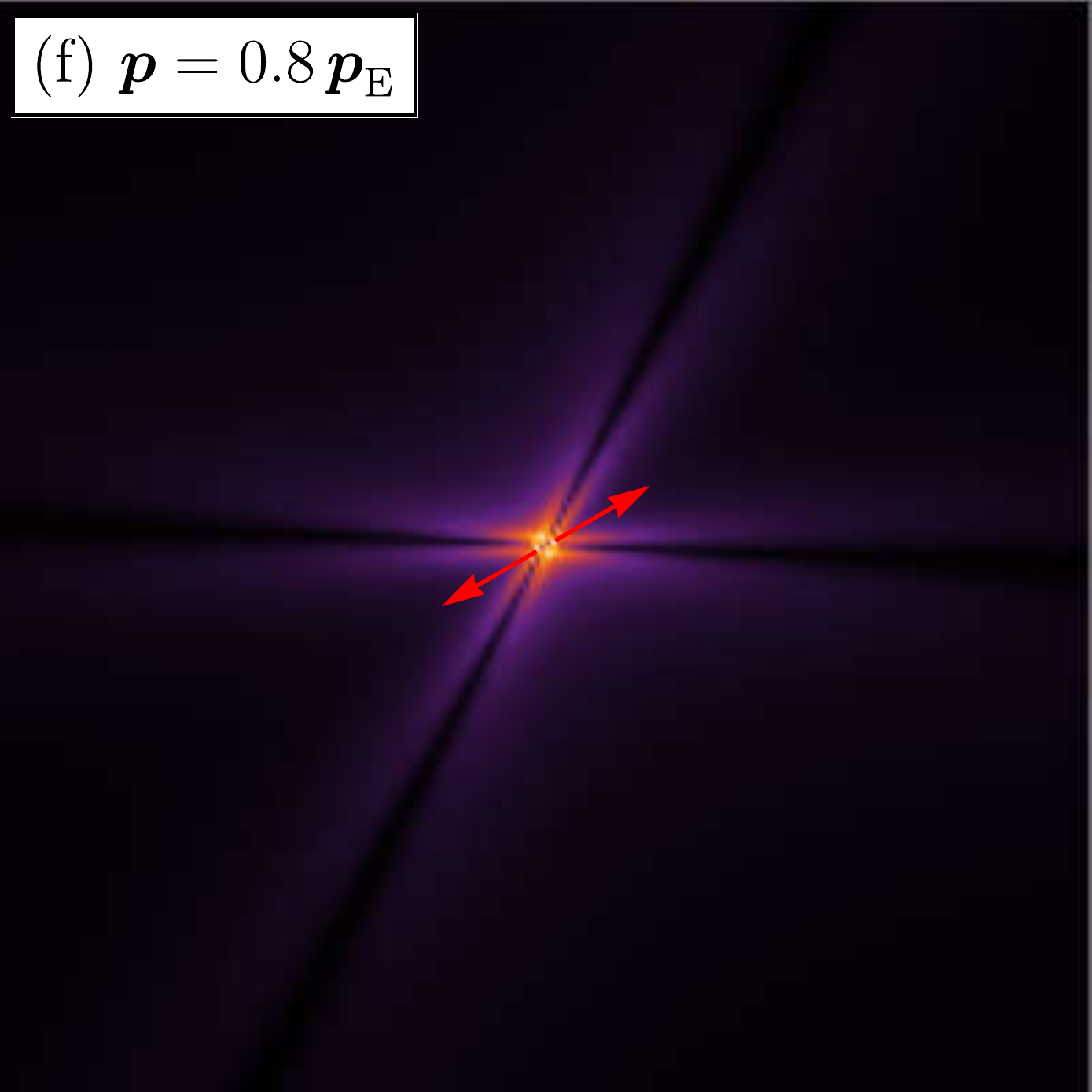}
\end{subfigure}
\begin{subfigure}{0.24\textwidth}
\centering
\phantomcaption{\label{fig:rhombus_10_10_dipole_90_gf}}
\includegraphics[width=0.98\linewidth]{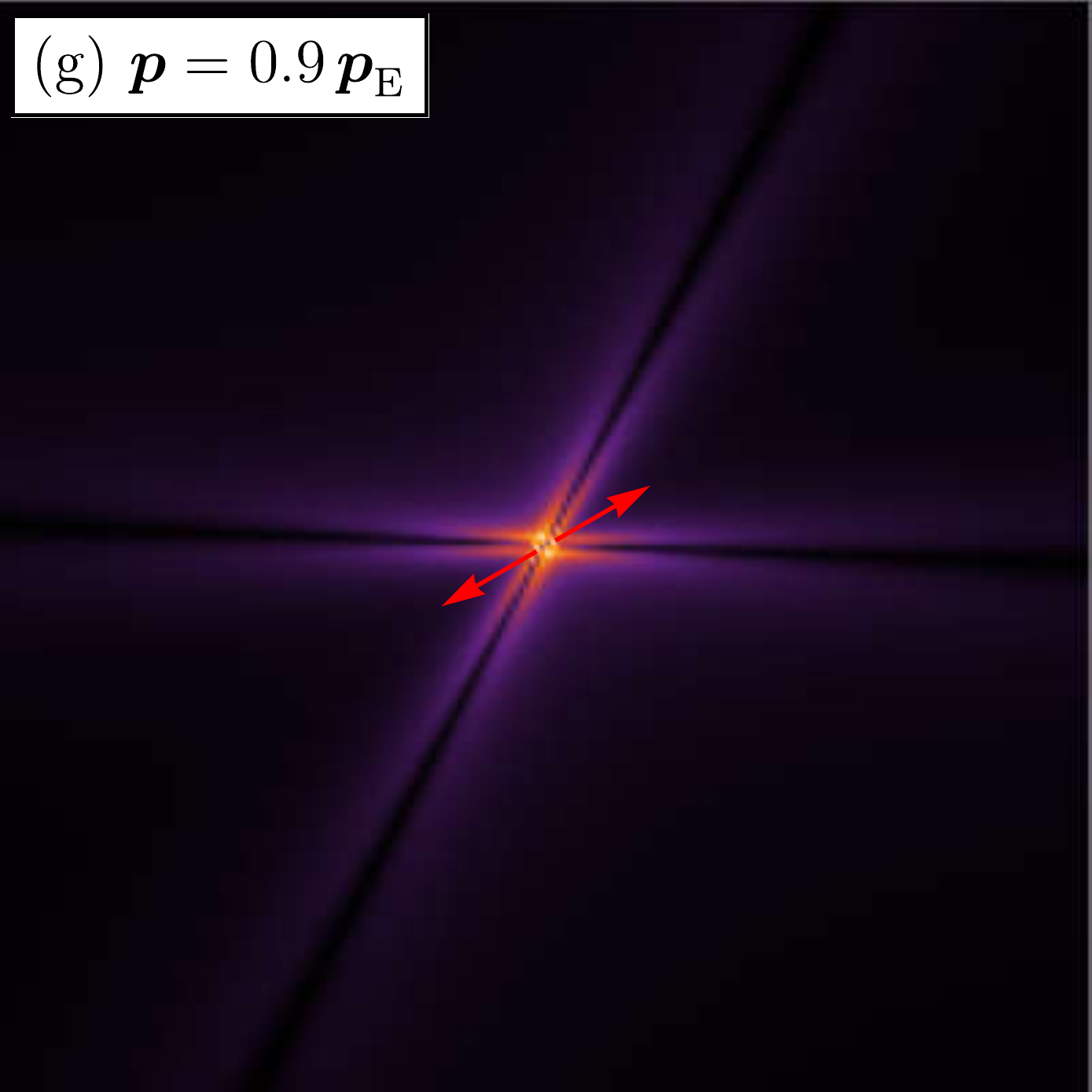}
\end{subfigure}
\begin{subfigure}{0.24\textwidth}
\centering
\phantomcaption{\label{fig:rhombus_10_10_dipole_99_gf}}
\includegraphics[width=0.98\linewidth]{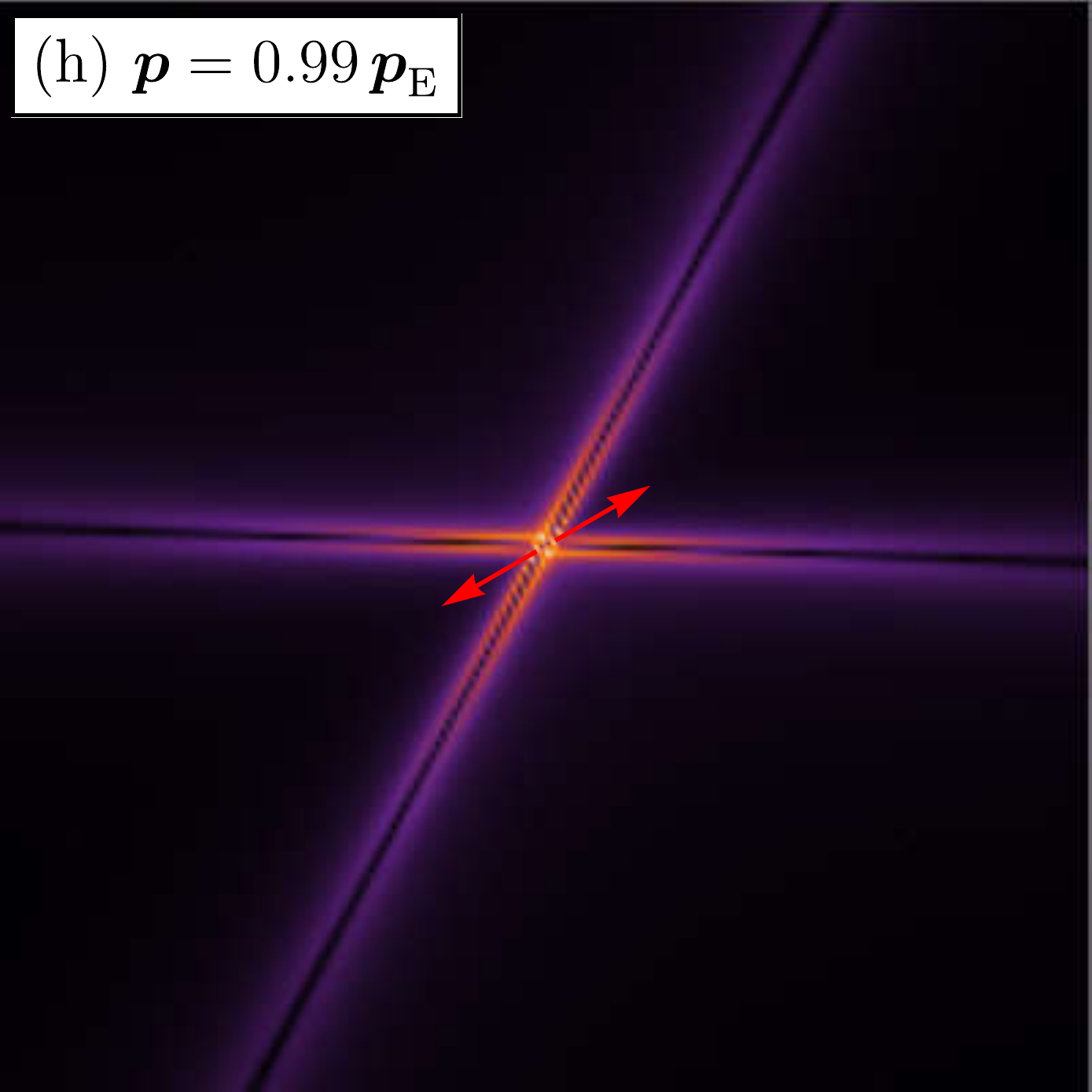}
\end{subfigure}
\caption{\label{fig:rhombus_10_10_dipole}
		As for Fig. \ref{fig:square_10_10_dipole}, but for an orthotropic rhombic lattice ($\Lambda_1=\Lambda_2=10$), where shear bands are inclined at angles $\theta_{\text{cr}}=88.2^\circ, 151.8^\circ$.
}
\end{figure}

\begin{figure}[htb!]
\centering
\begin{subfigure}{0.24\textwidth}
\centering
\phantomcaption{\label{fig:rhombus_7_15_dipole_0}}
\includegraphics[width=0.98\linewidth]{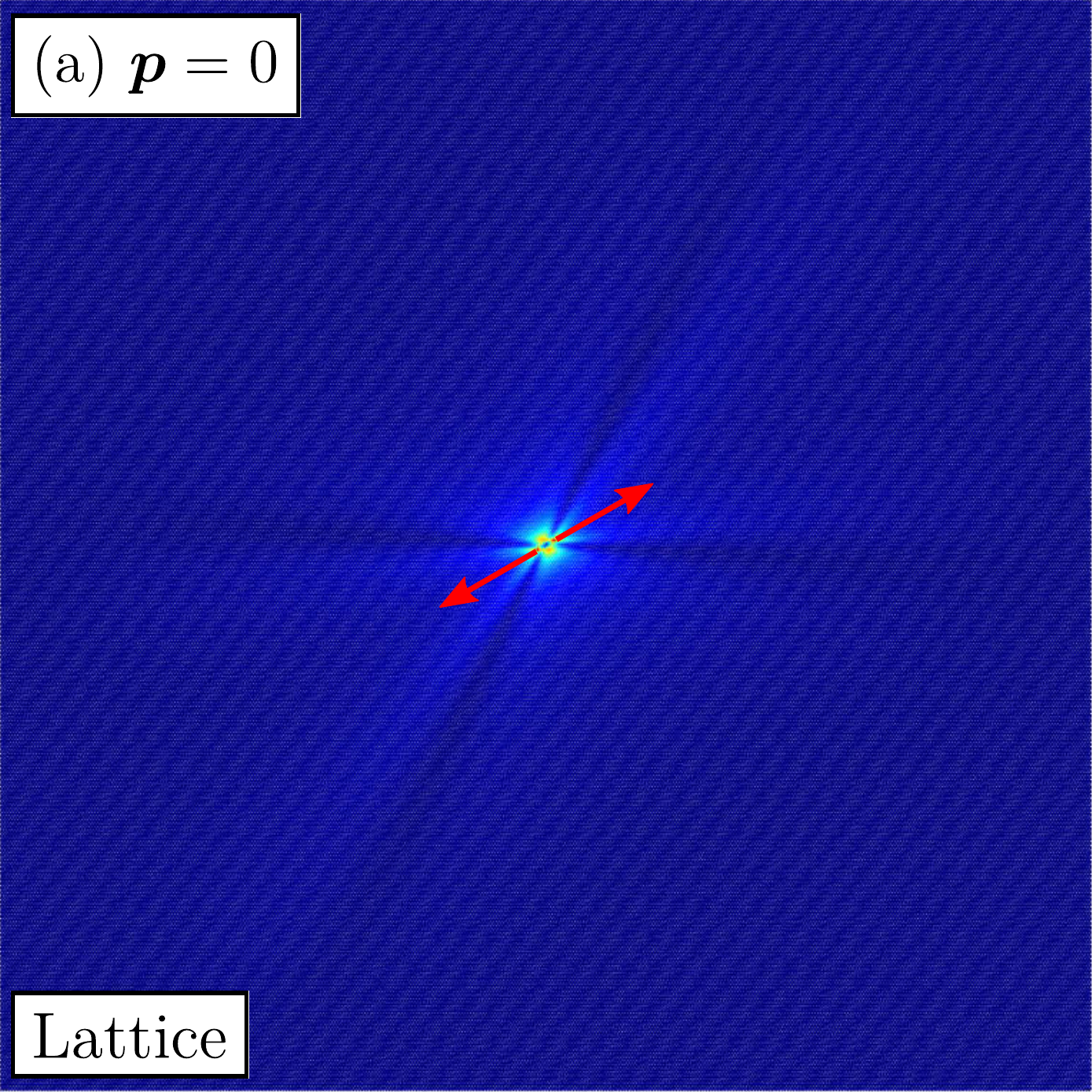}
\end{subfigure}
\begin{subfigure}{0.24\textwidth}
\centering
\phantomcaption{\label{fig:rhombus_7_15_dipole_80}}
\includegraphics[width=0.98\linewidth]{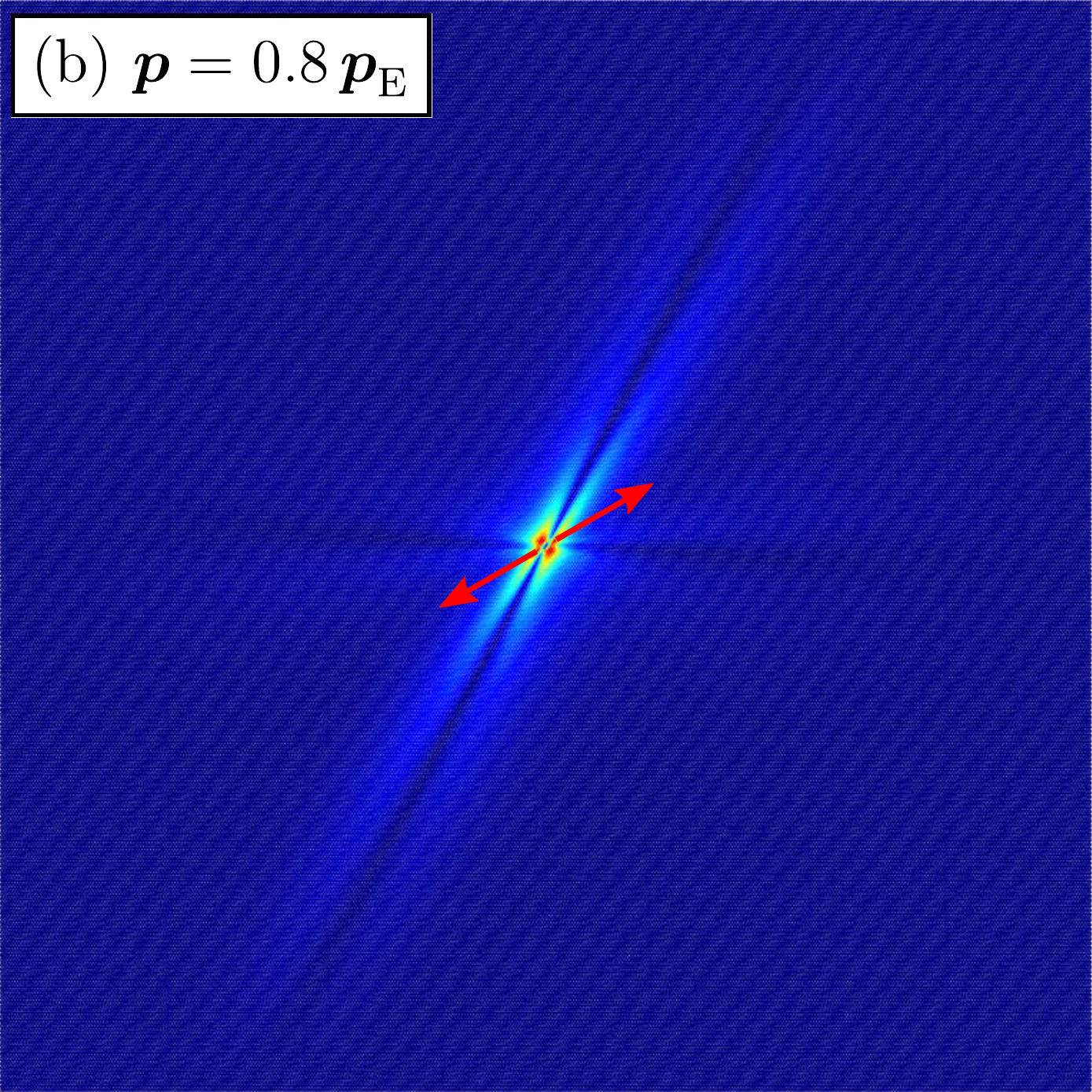}
\end{subfigure}
\begin{subfigure}{0.24\textwidth}
\centering
\phantomcaption{\label{fig:rhombus_7_15_dipole_90}}
\includegraphics[width=0.98\linewidth]{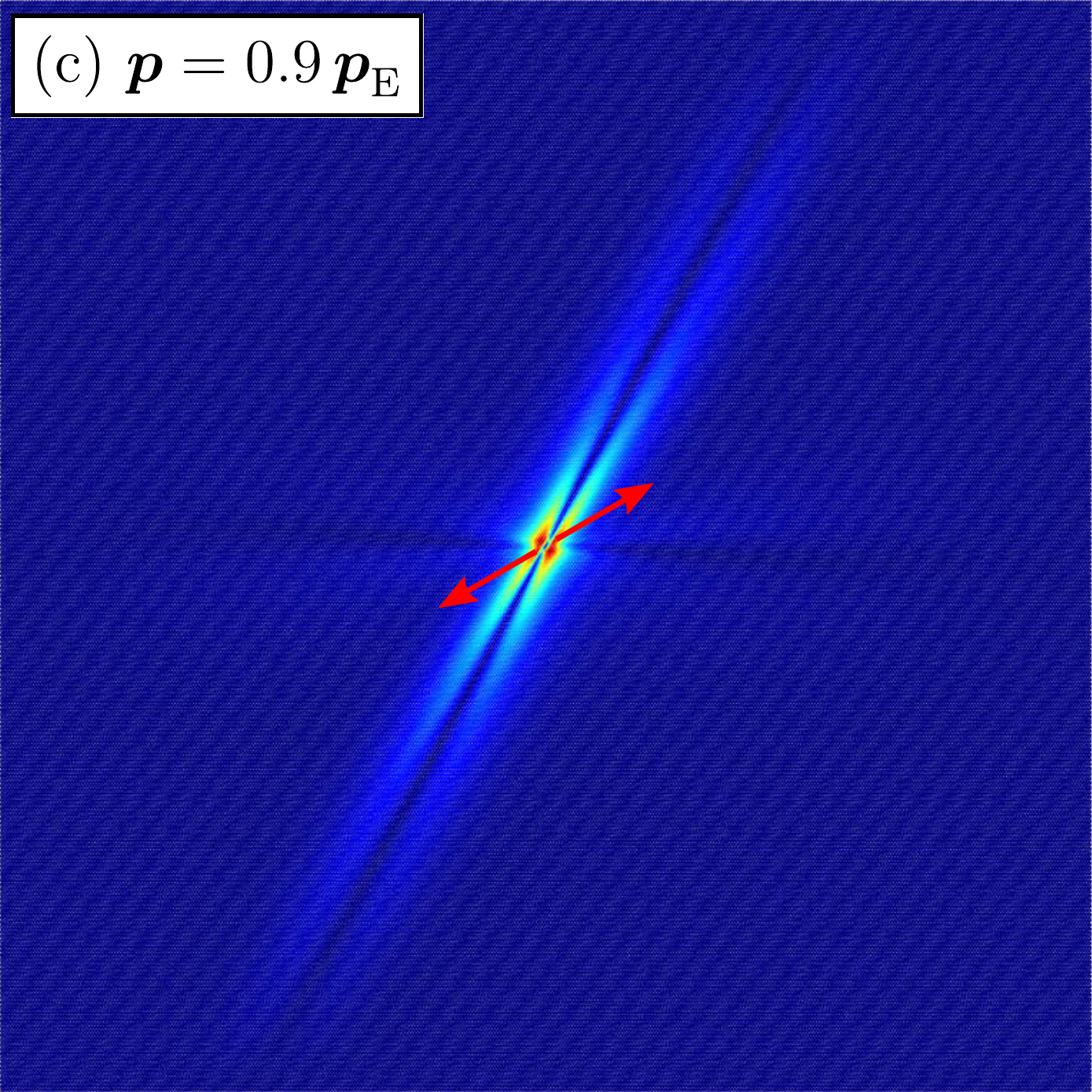}
\end{subfigure}
\begin{subfigure}{0.24\textwidth}
\centering
\phantomcaption{\label{fig:rhombus_7_15_dipole_99}}
\includegraphics[width=0.98\linewidth]{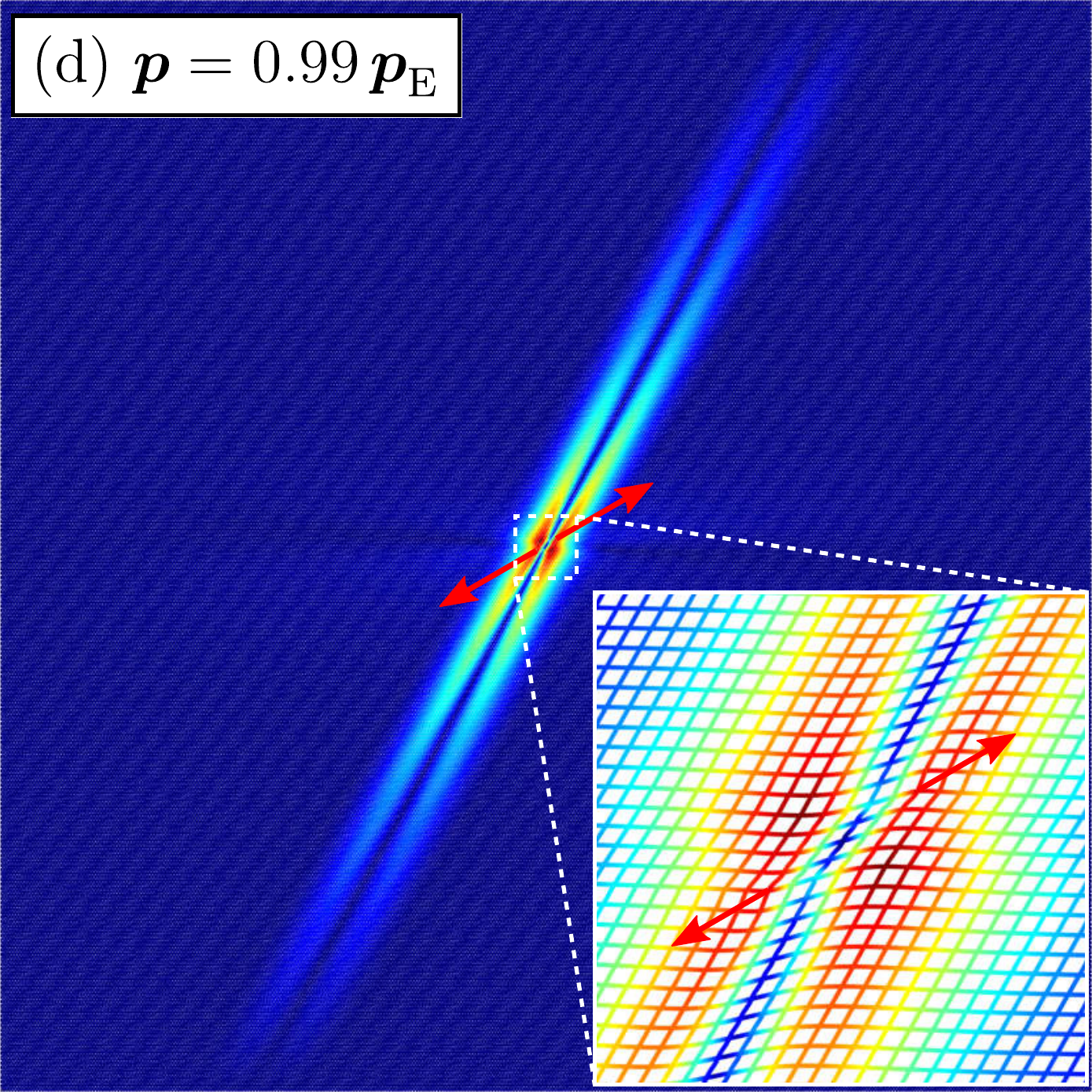}
\end{subfigure}\\
\vspace{0.01\linewidth}
\begin{subfigure}{0.24\textwidth}
\centering
\phantomcaption{\label{fig:rhombus_7_15_dipole_0_gf}}
\includegraphics[width=0.98\linewidth]{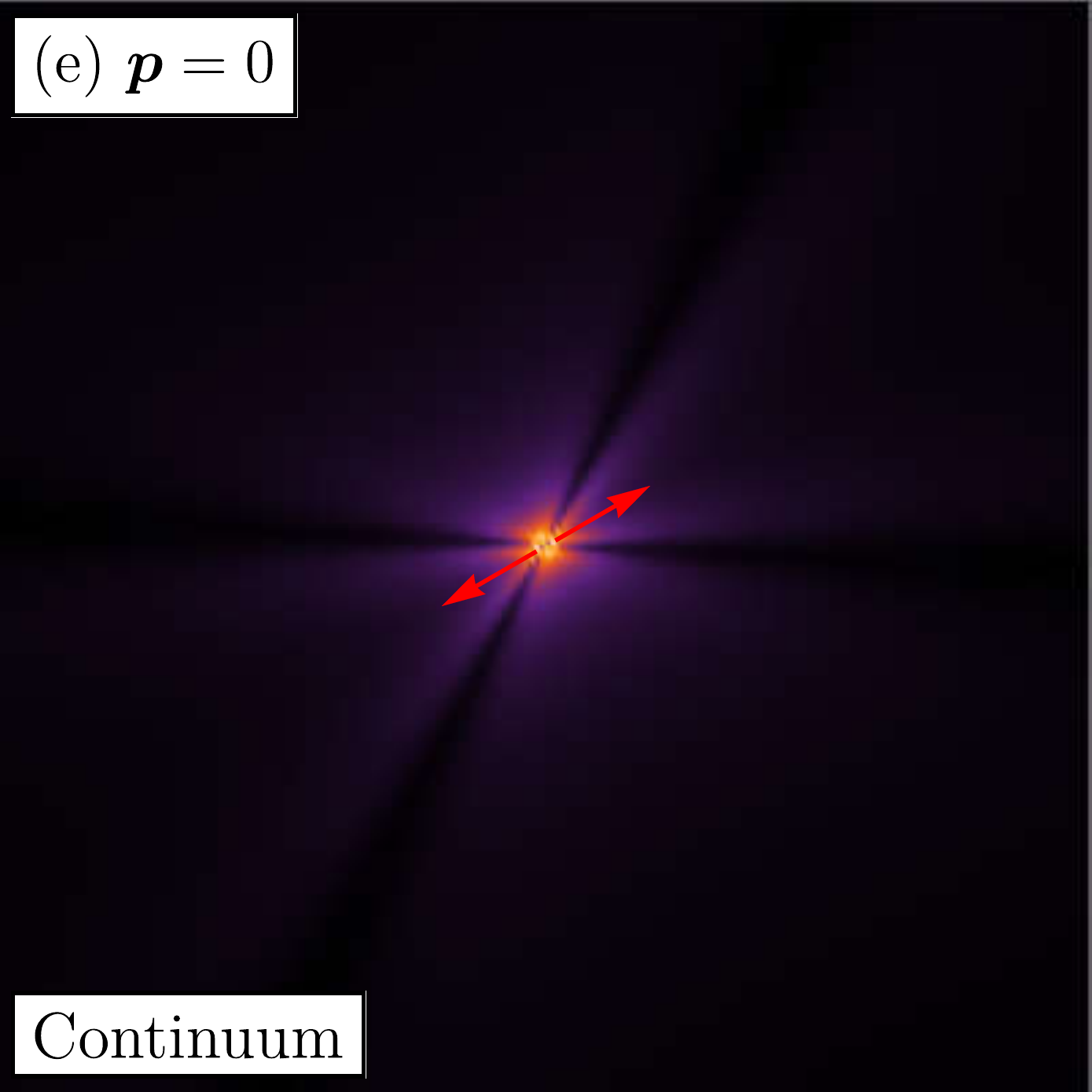}
\end{subfigure}
\begin{subfigure}{0.24\textwidth}
\centering
\phantomcaption{\label{fig:rhombus_7_15_dipole_80_gf}}
\includegraphics[width=0.98\linewidth]{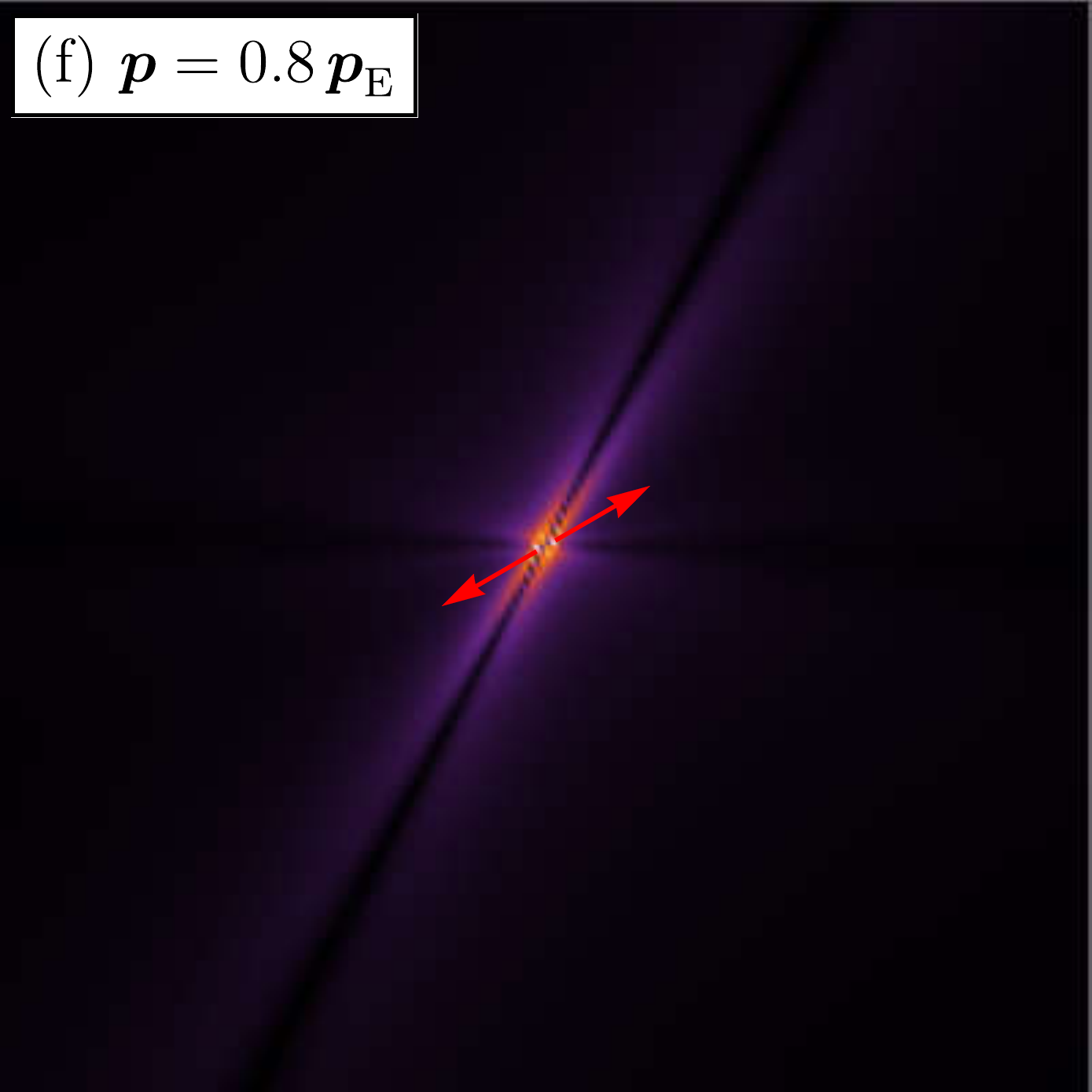}
\end{subfigure}
\begin{subfigure}{0.24\textwidth}
\centering
\phantomcaption{\label{fig:rhombus_7_15_dipole_90_gf}}
\includegraphics[width=0.98\linewidth]{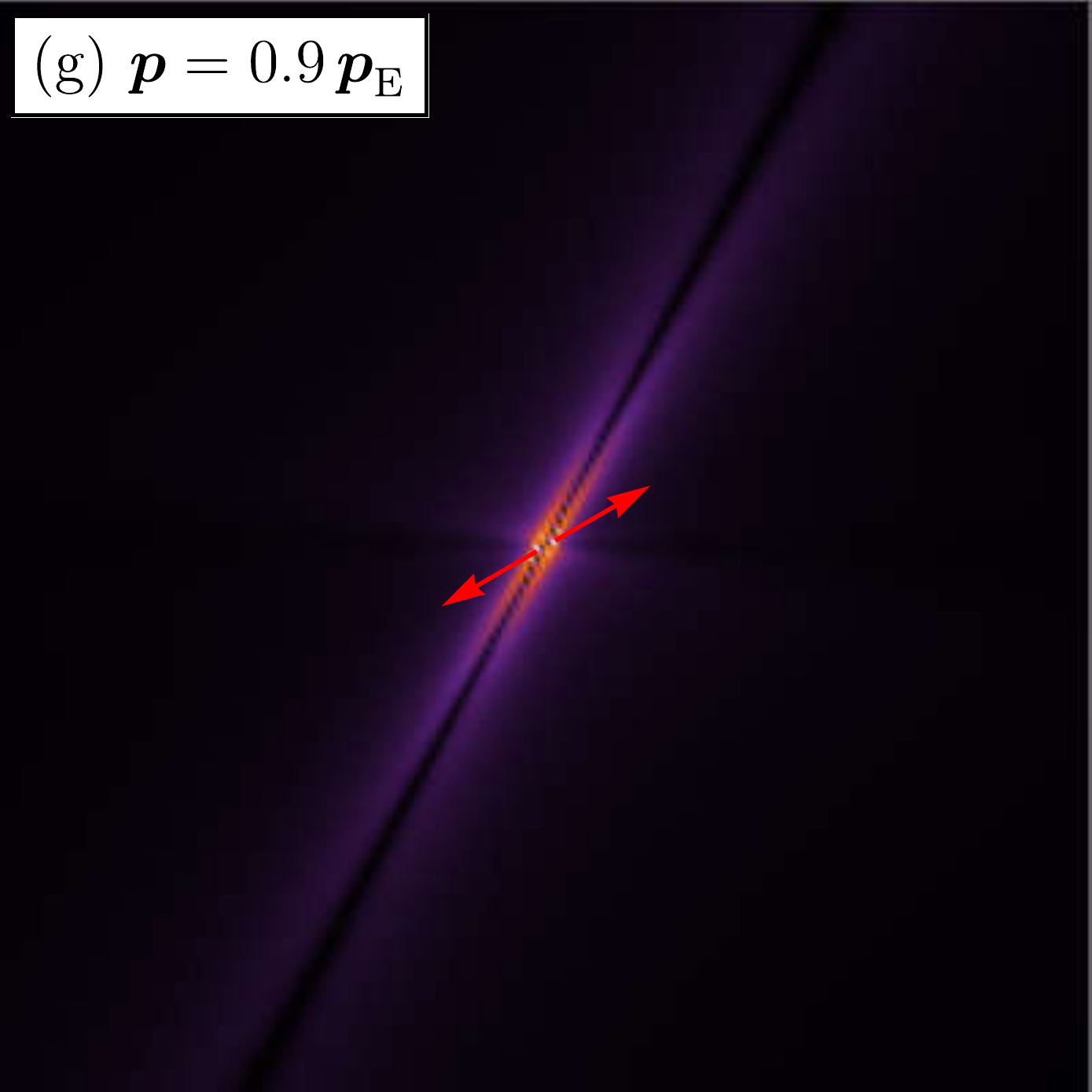}
\end{subfigure}
\begin{subfigure}{0.24\textwidth}
\centering
\phantomcaption{\label{fig:rhombus_7_15_dipole_99_gf}}
\includegraphics[width=0.98\linewidth]{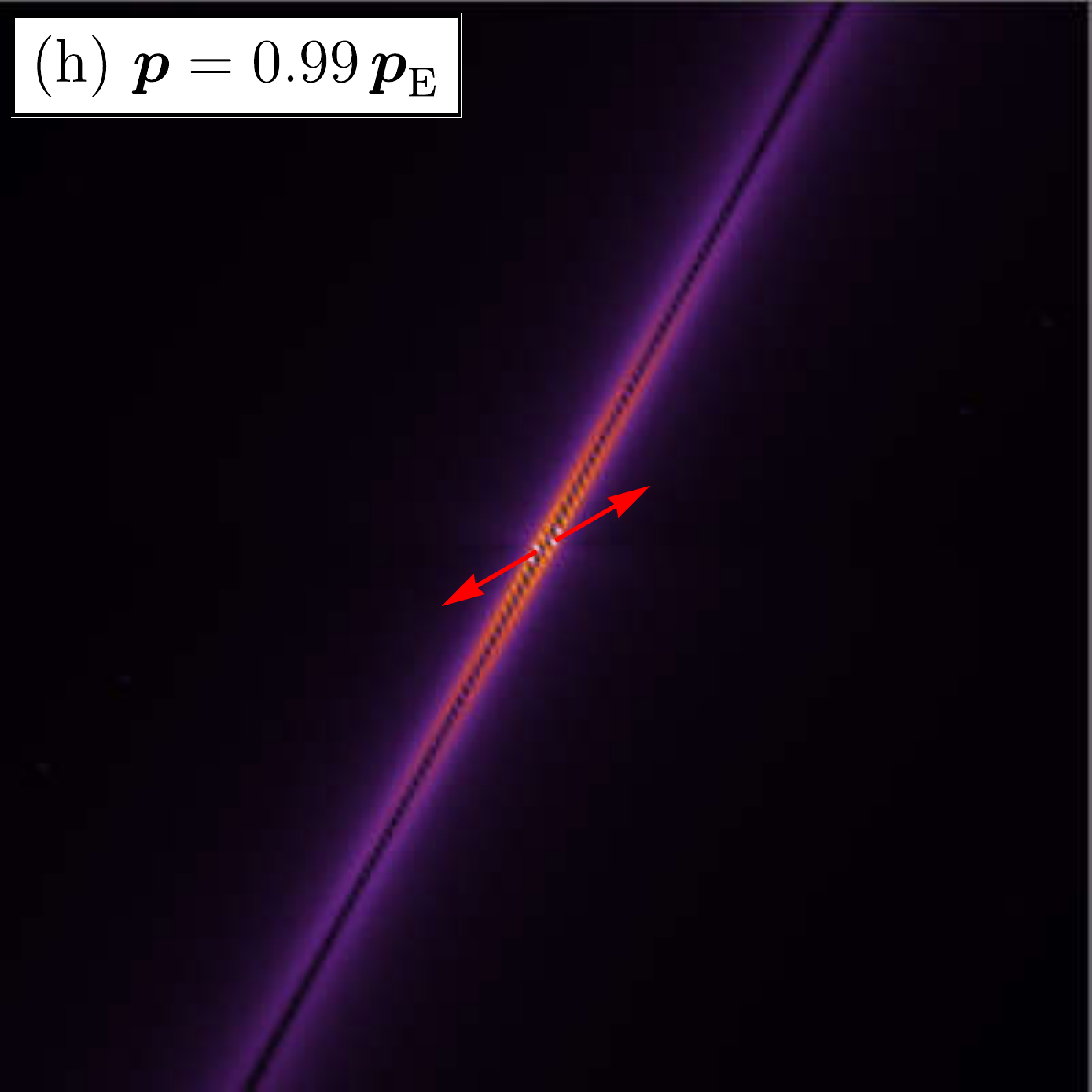}
\end{subfigure}
\caption{\label{fig:rhombus_7_15_dipole}
	As for Fig. \ref{fig:square_10_10_dipole}, but for an anisotropic rhombic lattice ($\Lambda_1=7, ~\Lambda_2=15$), where a single shear band forms inclined at an angle $\theta_{\text{cr}}=151.4^\circ$.
}
\end{figure}

A comparison is presented between the response of the lattice loaded with a concentrated force dipole and a dipole Green's function of the effective solid, in terms of maps of incremental displacements.
The results are presented as contour plots in Figs. \ref{fig:square_10_10_dipole}--\ref{fig:rhombus_7_15_dipole}, where the color scale has been conveniently normalized according to the norm of the computed displacement field.
In the upper part of the figures, results pertaining to the discrete lattice structure are presented, while, in the lower part, results are relative to the equivalent continuum, obtained via homogenization. 
The figures from left to right correspond to the application of increasing preloads, which approach the strong ellipticity boundary in the equivalent solid in situations where failure of ellipticity corresponds also to the occurrence of a macro bifurcation of infinite wavelength. 
The part (d) of each figure ($\bp=0.99\bp_{\text{E}}$) also illustrates a magnification of the lattice response in the neighborhood of the loading zone, thus disclosing the microscopic deformation pattern associated to the extreme mechanical response of the material when close to elliptic boundary.
%

%
%
%
%
In the conditions analyzed in Figs. \ref{fig:square_10_10_dipole}--\ref{fig:rhombus_7_15_dipole}, the equivalent solid is found to be fully representative of the lattice structure, so that approaching failure of ellipticity the perturbative approach reveals, both in the continuum and in the real lattice, the formation of localized incremental deformation in the form of single or double localization bands.
These can be horizontal, vertical or inclined. 
The correspondence between the behaviour of lattice and of the equivalent continuum is found to be excellent so that the maps reported in the upper part of the figures are practically identical to the corresponding maps in the lower part of the figures.

\clearpage
\subsection{Micro bifurcations in the lattice and effects on the equivalent solid}
\label{sec:microscopic_localization}
Micro-bifurcations occurring when the equivalent solid is still in the strong ellipticity range are investigated in this section, with reference to an equibiaxially compressed square lattice  with cubic symmetry $\Lambda_1=\Lambda_2=10$ and diagonal springs of stiffness $\kappa=0.4$.
With the assumed spring stiffness, a microscopic bifurcation is critical, namely, it occurs when the equivalent solid is still in the strong elliptic domain. 
%
\begin{figure}[htb!]
\centering
\begin{subfigure}{0.32\textwidth}
\centering
\phantomcaption{\label{fig:square_10_10_k_04_quadrupole}}
\includegraphics[width=0.98\linewidth]{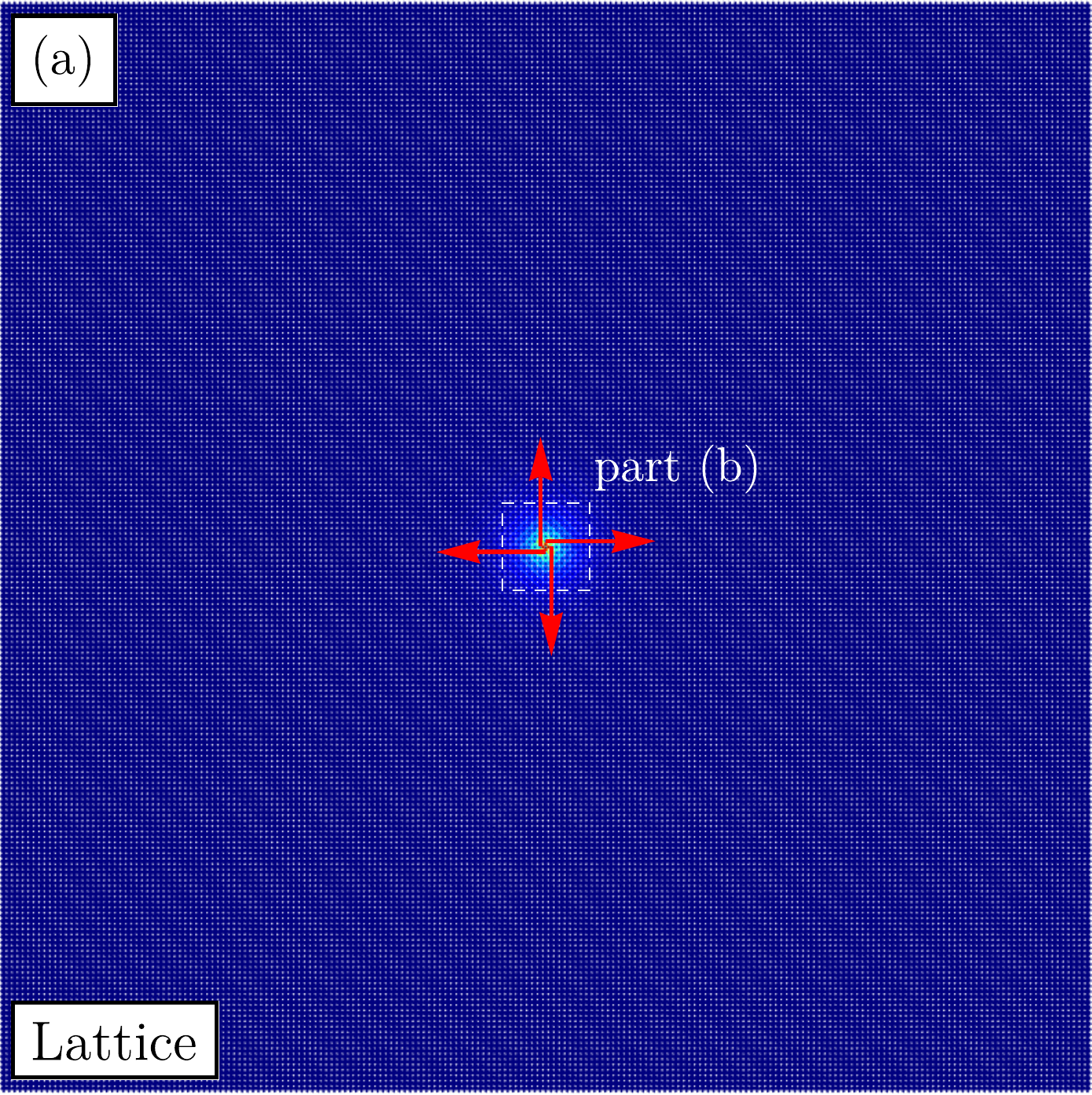}
\end{subfigure}
\begin{subfigure}{0.32\textwidth}
\centering
\phantomcaption{\label{fig:square_10_10_k_04_quadrupole_zoom}}
\includegraphics[width=0.98\linewidth]{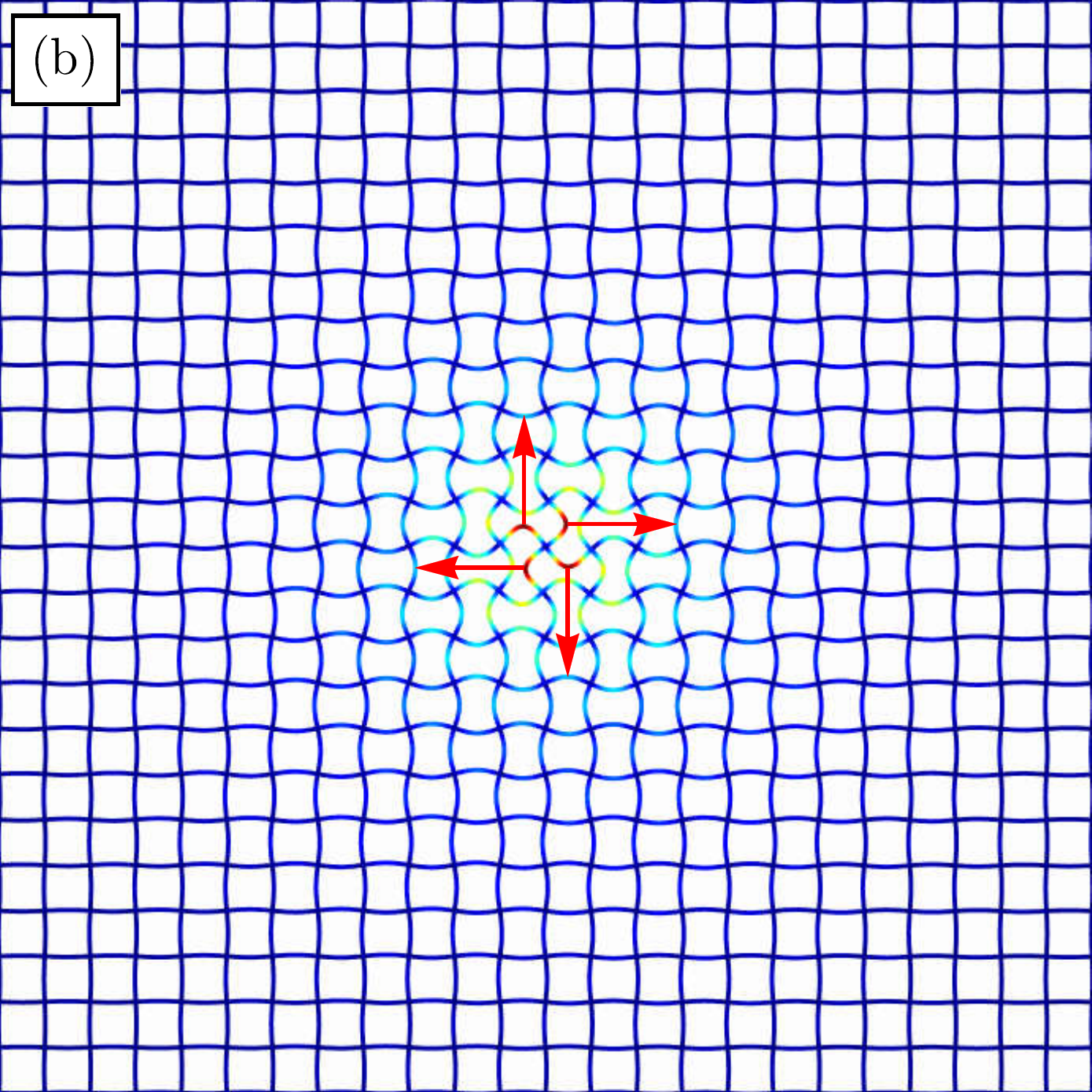}
\end{subfigure}\\
\vspace{0.01\linewidth}
\begin{subfigure}{0.32\textwidth}
\centering
\phantomcaption{\label{fig:square_10_10_k_04_quadrupole_gf}}
\includegraphics[width=0.98\linewidth]{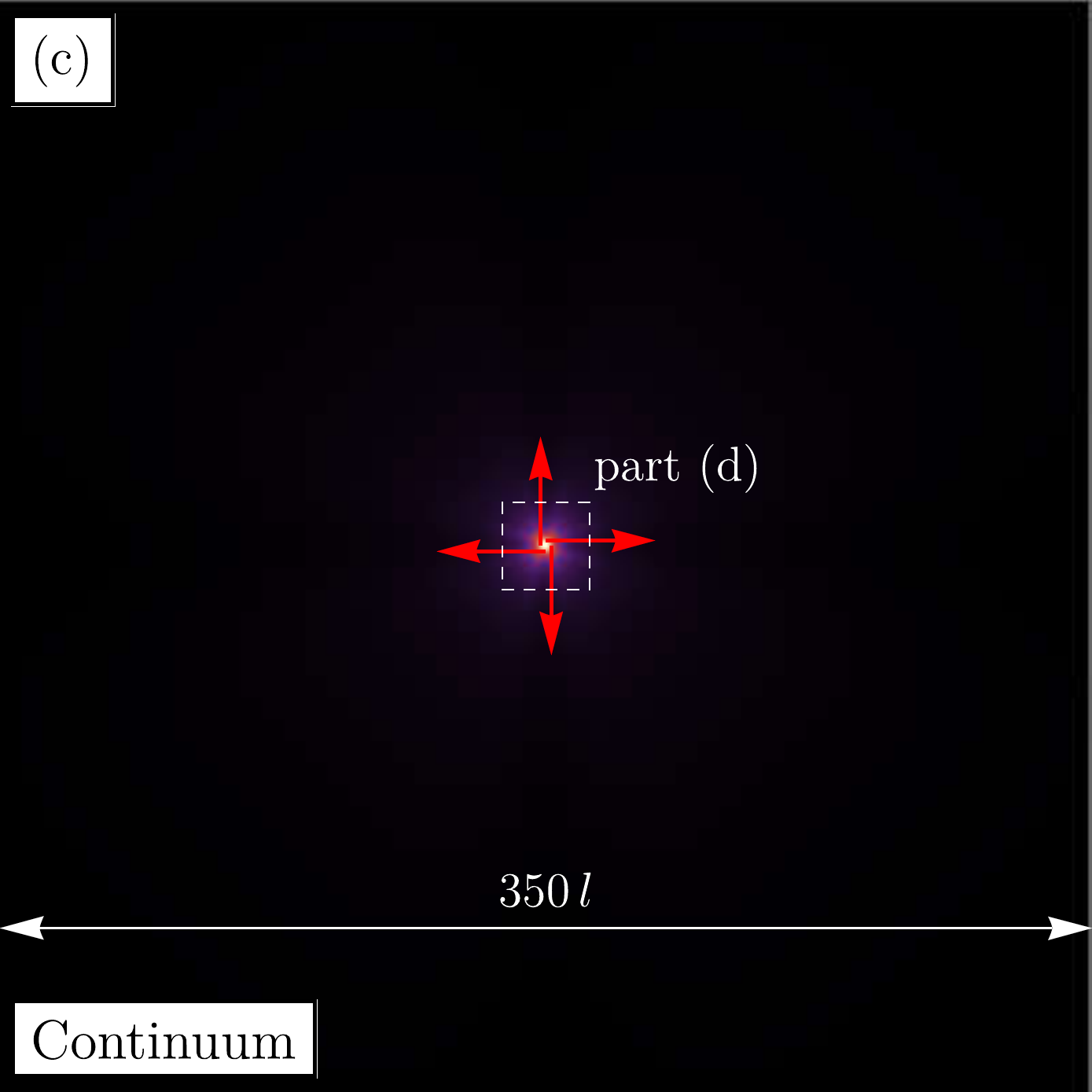}
\end{subfigure}
\begin{subfigure}{0.32\textwidth}
\centering
\phantomcaption{\label{fig:square_10_10_k_04_quadrupole_zoom_gf}}
\includegraphics[width=0.98\linewidth]{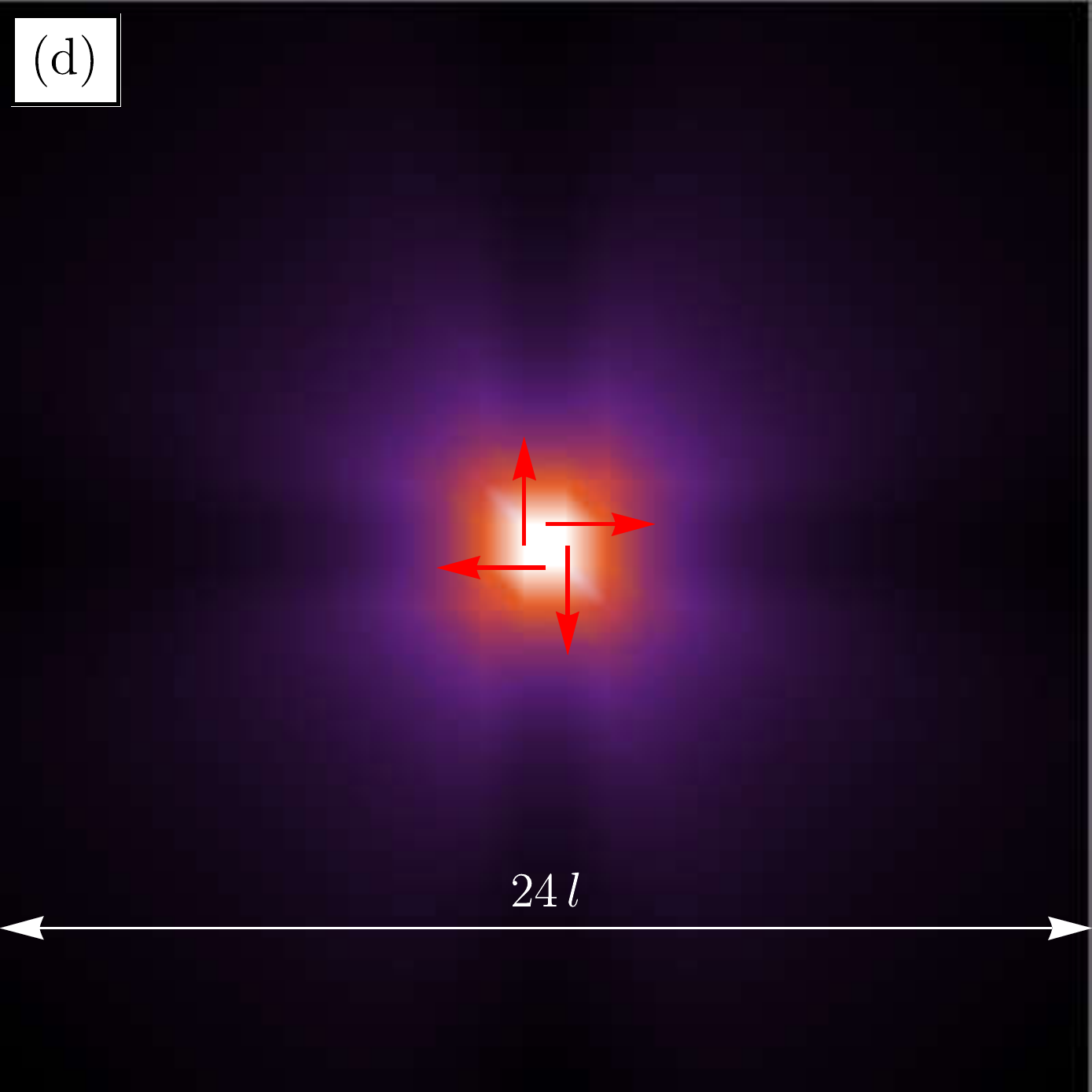}
\end{subfigure}
\caption{\label{fig:square_10_10_k_04_local_buckling}
	Microscopic localization of the bifurcation mode evidenced in the incremental displacement map relative to a square lattice (cubic symmetry, $\Lambda_1=\Lambda_2=10$, upper part) and in the equivalent continuum (lower part) at an equibiaxial compression load corresponding to bifurcation, $\bp_{\text{B}}=\{-\pi^2,-\pi^2\}$, under the action of a `quadrupole' of forces applied at the midpoints of the rods.
    The quadrupole activates a highly localized `rotational' bifurcation mode (labeled as in $B_1$ in Fig.~\ref{fig:ellipticity_domains_10_10_pi2} and Table~\ref{tab:buckling_modes_points}), which leaves the lattice and the equivalent solid `macroscopically' almost undeformed, while the \textit{inter-node deformation is predominant} at the scale of the unit cell. 
    The latter feature cannot be detected by the equivalent solid.
}
\end{figure}

The incremental displacement maps in the lattice (\textit{at} the critical load for micro-bifurcation) and in the equivalent continuum (still in the strong elliptic range) generated by the application of a force quadrupole are shown in Fig.~\ref{fig:square_10_10_k_04_local_buckling}, where the upper parts (lower parts) refer to the lattice (to the continuum) and the parts on the right are a magnification of the zone around the quadrupole shown on the left. 

The figure shows that the incremental response of the prestressed lattice is highly localized, so that only a strong magnification reveals buckling of the elastic rods. 
Even if the equivalent continuum is not at bifurcation, but still within the uniqueness/stability domain, the distribution of displacements in it somehow resembles that in the lattice, so that the homogenization is still representative of the response of the discrete structure, even though the \textit{inter-node deformation} cannot be captured.
%
\begin{figure}[htb!]
\centering
\begin{subfigure}{50.5mm}
\centering
\phantomcaption{\label{fig:square_10_10_k_04_p_100_dipole}}
\includegraphics[width=\linewidth]{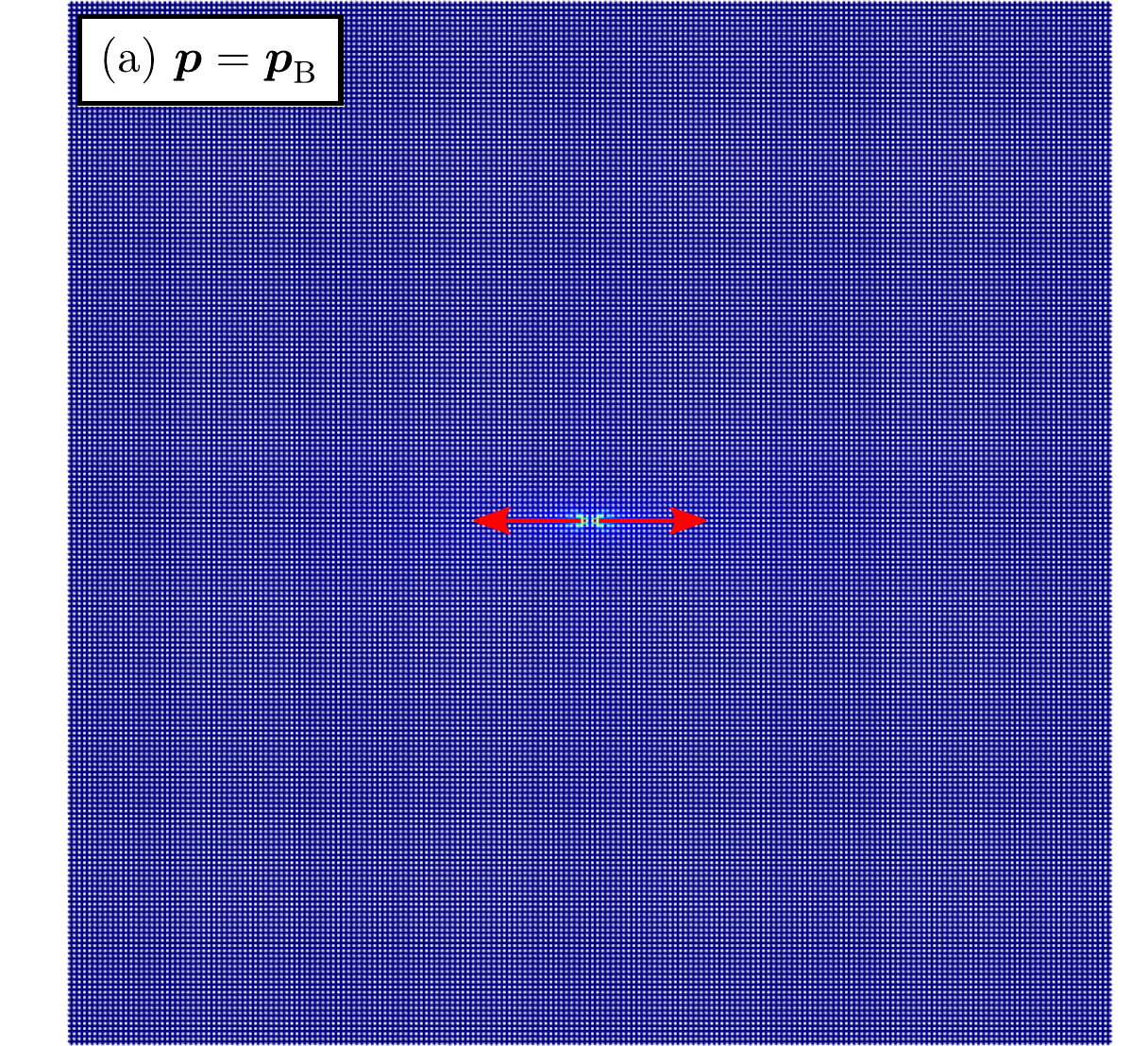}
\end{subfigure}
\begin{subfigure}{48.3mm}
\centering
\phantomcaption{\label{fig:square_10_10_k_04_p_105_dipole}}
\includegraphics[width=\linewidth]{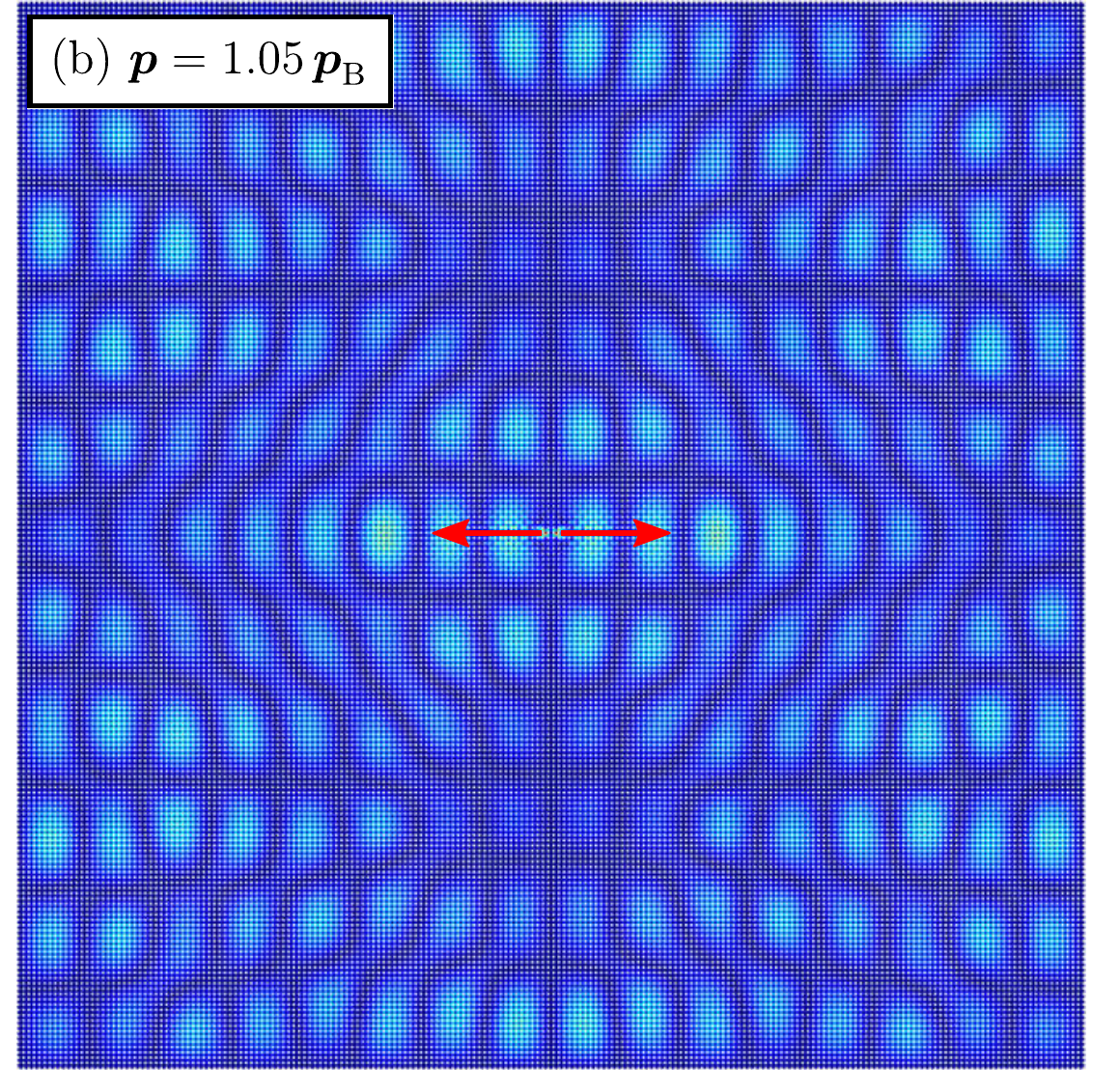}
\end{subfigure}
\begin{subfigure}{48.3mm}
\centering
\phantomcaption{\label{fig:square_10_10_k_04_p_110_dipole}}
\includegraphics[width=\linewidth]{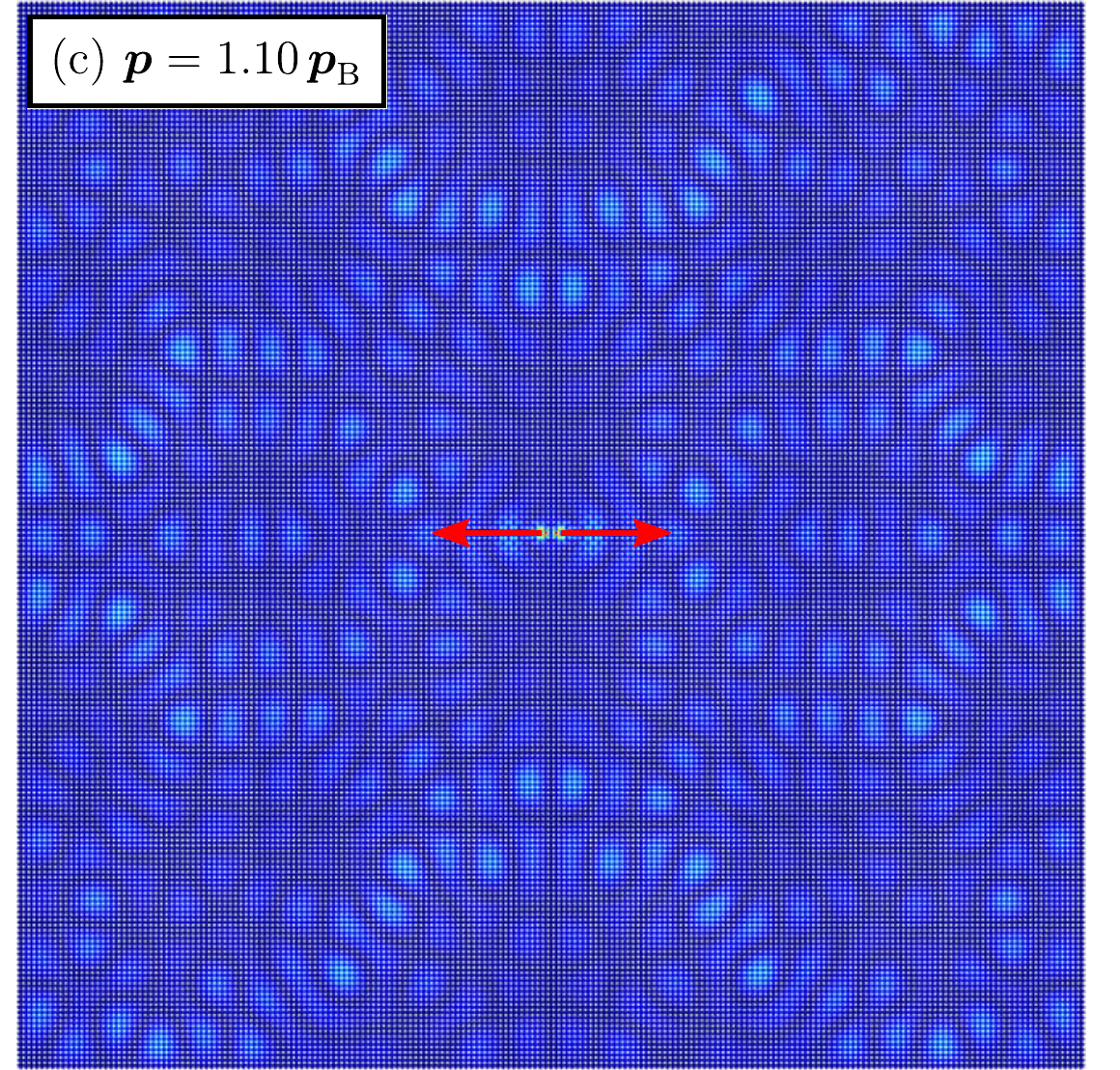}
\end{subfigure}\\
\vspace{1.5mm}
\centering
\begin{subfigure}{50.5mm}
\centering
\phantomcaption{\label{fig:square_10_10_k_04_p_100_dipole_FFT}}
\includegraphics[width=\linewidth]{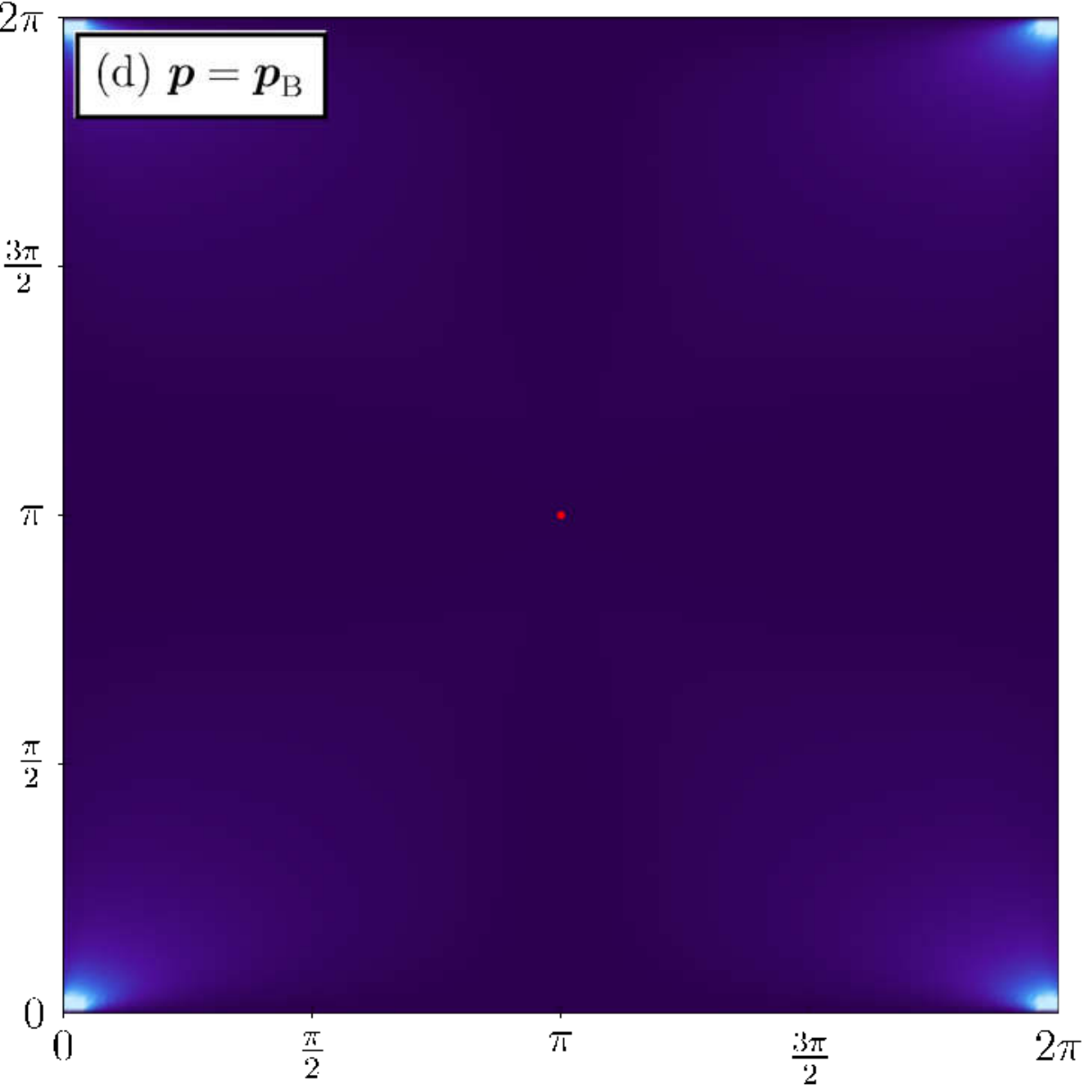}
\end{subfigure}
\begin{subfigure}{48.3mm}
\centering
\phantomcaption{\label{fig:square_10_10_k_04_p_105_dipole_FFT}}
\includegraphics[width=\linewidth]{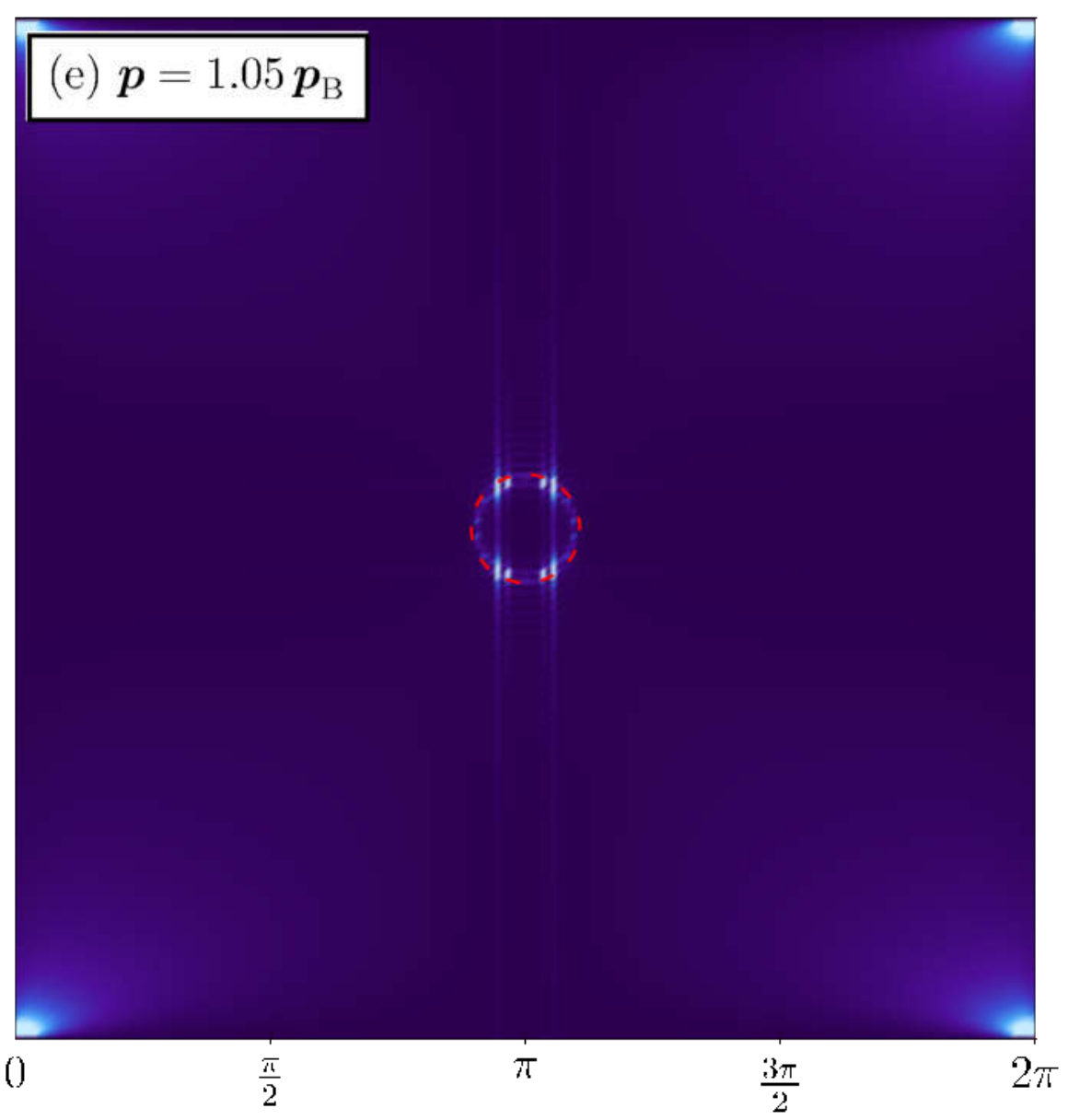}
\end{subfigure}
\begin{subfigure}{48.3mm}
\centering
\phantomcaption{\label{fig:square_10_10_k_04_p_110_dipole_FFT}}
\includegraphics[width=\linewidth]{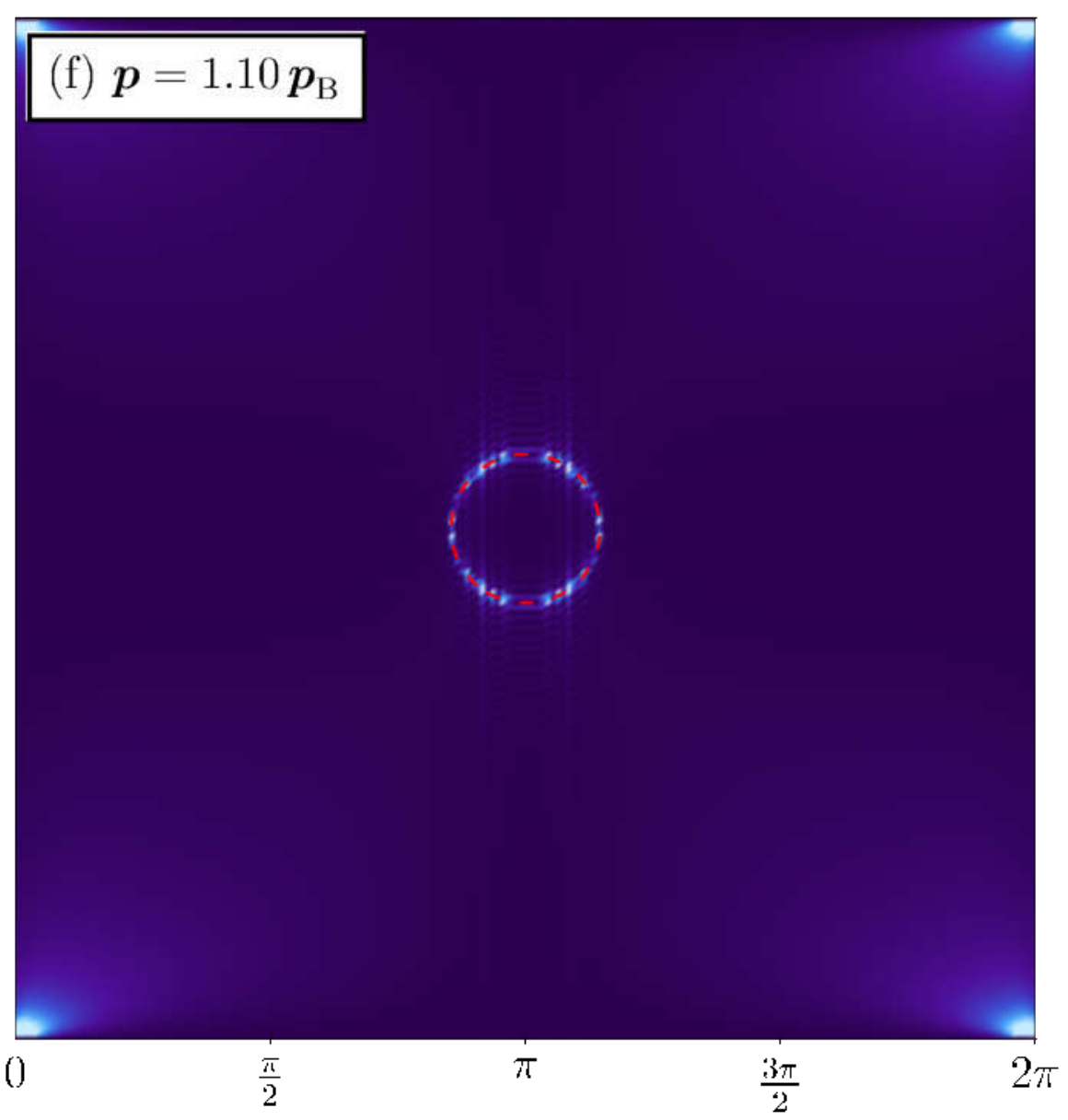}
\end{subfigure}
\caption{\label{fig:square_10_10_k_04_dipole_local_buckling}
    Displacement map (\subref{fig:square_10_10_k_04_p_100_dipole})--(\subref{fig:square_10_10_k_04_p_110_dipole}) and corresponding Fourier transform (\subref{fig:square_10_10_k_04_p_100_dipole_FFT})--(\subref{fig:square_10_10_k_04_p_110_dipole_FFT}) showing the response of the lattice at a load corresponding to microscopic instability, $\bp=\bp_{\text{B}}$ and beyond, $\bp=1.05\bp_{\text{B}}, ~ 1.10 \bp_{\text{B}}$. 
    The slowness contour at null frequency, evaluated through the bifurcation condition is superimposed in red. While at the critical load the perturbation is so localized that results almost invisible, at higher loads an `explosive' instability involving the whole lattice and extending up to the boundary of the domain is clearly observed.
}
\end{figure}

The situation depicted in Fig.~\ref{fig:square_10_10_k_04_local_buckling} completely changes when the lattice is loaded with forces beyond the critical value for micro bifurcation in the lattice, as shown in Fig.~\ref{fig:square_10_10_k_04_dipole_local_buckling}, only referred to the lattice loaded with a horizontal force dipole at different biaxial compression loadings (\textit{at} the critical load $\bp=\bp_{\text{B}}$ for micro-buckling and beyond, namely, at $\bp=1.05\,\bp_{\text{B}}$ and at $\bp=1.10\,\bp_{\text{B}}$). 

This figure shows displacement maps (upper part) and the corresponding Fourier transform (obtained via FFT of nodal displacements, lower part), with superimposed slowness contours corresponding to null frequency. 
The slowness contour (highlighted in red in the figure) was obtained from the bifurcation condition, Eq.~\eqref{eq:det_local_buckling}. The fact that the slowness contour is superimposed to the peaks of the transform (reported white in the figure), is a validation of the good correspondence between calculations performed via Floquet-Bloch and finite element simulations. 

It can be concluded from Fig.~\ref{fig:square_10_10_k_04_dipole_local_buckling} that, while at bifurcation load the incremental perturbation induced by the force dipole is practically so small and highly localized that results almost invisible, an `explosive instability' is found in the lattice, which does not decay and extends to the whole domain occupied by the structure considered in the analysis. 
This is a special behaviour which remains unobserved in the equivalent continuum (still in the strong elliptic range) and cannot therefore be revealed through homogenization.

\section{Conclusions}
Homogenization of the incremental response of a lattice of elastic rods, axially pre-loaded to an arbitrary amount, has been shown to provide a superb tool for the design of cellular elastic materials of tunable properties and capable of extreme localized deformations. 
In particular, the perturbative approach to material instability reveals that strain localization in the composite is almost coincident with that occurring in the equivalent solid, which remains unaffected by micro bifurcations, possibly occurring in the lattice. However, the developed homogenization approach allows to play with geometries and stiffnesses of the composite in a way to inhibit or promote strain localization with respect to other micro instabilities. 
The vibrational properties of the lattice and the ability of the homogenization scheme to correctly capture them is a final crucial aspect in the design of cellular materials, that will be addressed in Part II of this study.

\section*{Acknowledgements}
Financial support is acknowledged from: 
the ERC Advanced Grant `Instabilities and nonlocal multiscale modelling of materials' ERC-2013-ADG-340561-INSTABILITIES (G.B. and L.C.),
PRIN 2015 2015LYYXA8-006 and ARS01-01384-PROSCAN (D.B. and A.P.).
The authors also acknowledge support from the Italian Ministry of Education, University and Research (MIUR) in the frame of the `Departments of Excellence' grant L. 232/2016.

\printbibliography

\appendix
\numberwithin{equation}{section}            
\numberwithin{figure}{section}              

\section{Linearized equilibrium of an axially pre-stretched elastica}
\label{sec:linearized_elastica}
The linearized equilibrium of an axially stretchable Euler-Bernoulli elastic beam can be obtained through a linearization (around a stretched equilibrium configuration) of the equations governing the equilibrium of large deflections and flexure of an elastic rod. 

Denoting the stress-free, straight configuration of the elastic rod with a local axial coordinate $x_0$, the potential energy is defined in the reference configuration as 
\begin{equation}
\label{eq:potential_energy_elastica}
    \mV = \int_0^{l_0} \left( \psi_\lambda(\lambda) + \psi_\chi(\chi) - P\,u'(x_0) \right) dx_0 \,,
\end{equation}
where $l_0$ is the initial length, while $\psi_\lambda$ and $\psi_\chi$ are strain-energy functions for, respectively, axial  and flexural deformations.
The axial stretch $\lambda$ and the curvature $\chi$ are defined by the kinematics of an extensible unshearable elastica as
\begin{subequations}
\label{eq:strain_measures_elastica}
\begin{align}
    \label{eq:axial_stretch}
    \lambda &= (1+u'(x_0))\cos{\theta(x_0)} + v'(x_0)\sin{\theta(x_0)} \,, \\
    \label{eq:curvature}
    \chi &= \theta'(x_0) = \deriv{}{x_0} \left(\arctan\left(\frac{v'(x_0)}{1+u'(x_0)}\right)\right) \,,
\end{align}
\end{subequations}
where in~\eqref{eq:curvature} the unshearability constraint $\theta = \arctan\left(\frac{v'}{1+u'}\right)$ has been explicitly substituted.

The linearized response around \textit{straight, but axially stretched, configurations} can be obtained through the second-order expansion of the functional~\eqref{eq:potential_energy_elastica} with respect to the independent displacement fields $\{u, v\}$ around the deformed configuration $\{u_0,v_0\} =\{(\lambda_0 - 1) x_0, 0\}$.
Hence, by substituting~\eqref{eq:strain_measures_elastica} into~\eqref{eq:potential_energy_elastica} and neglecting an arbitrary constant term, the following expansion is obtained
\begin{equation}
\label{eq:potential_energy_elastica_second_order}
\begin{aligned}
    \mV(u_0+\delta u, v_0 + \delta v) &\sim
    \int_0^{l_0} \left(\psi_\lambda'(\lambda_0)-P\right)\delta u'(x_0) dx_0 +\\
    & +\frac{1}{2} \int_0^{l_0} \psi_\lambda''(\lambda_0)\delta u'(x_0)^2 dx_0
    +\frac{1}{2} \int_0^{l_0}\left( \frac{\psi_\lambda'(\lambda_0)}{\lambda_0}\delta v'(x_0)^2 +
    \frac{\psi_\chi''(0)}{\lambda_0^2}\delta v''(x_0)^2 \right) dx_0 \,,
\end{aligned}
\end{equation}
where it has been assumed that the residual bending moment is absent in the unloaded configuration $\psi_\chi'(0) = 0$.

Since the first-order term of~\eqref{eq:potential_energy_elastica_second_order} has to vanish when the configuration $\{u_0,v_0\} =\{(\lambda_0 - 1) x_0, 0\}$ satisfies equilibrium, the pre-stretch $\lambda_0$ is the solution of the condition $\psi_\lambda'(\lambda_0)-P=0$, indicating that the applied load P is indeed equal to the axial prestress.
Moreover, it is important to note that the second-order term of~\eqref{eq:potential_energy_elastica_second_order} involves the strain energy functions only in terms of second derivatives, $\psi_\lambda''(\lambda_0)$ and $\psi_\chi''(0)$, evaluated on the straight stretched configuration.

It is now instrumental to update the reference configuration from the stress-free configuration to the stretched configuration, so that the second-order functional~\eqref{eq:potential_energy_elastica_second_order} can be adopted to govern the incremental response of the rod.
This can be performed by changing the variable of integration from $x_0$ to the current stretched coordinate $s = \lambda_0 x_0$ and expressing the fields $\{u, v\}$ as functions of $s$.
Thus, the second-order terms in eqs~\eqref{eq:potential_energy_elastica_second_order} become
\begin{equation}
\label{eq:potential_energy_elastica_second_order_updated}
\begin{aligned}
    \mV(u_0+\delta u, v_0 + \delta v) \sim
    & \frac{1}{2} \int_0^{l} \psi_\lambda''(\lambda_0) \lambda_0 \, \delta u'(s)^2 ds +\frac{1}{2} \int_0^{l}\left( P\delta v'(s)^2 +
    \psi_\chi''(0) \lambda_0 \, \delta v''(s)^2 \right) ds \,,
\end{aligned}
\end{equation}
where $l = \lambda_0 l_0$ is the current rod's length and the symbol $'$ has to be understood as differentiation with respect to $s$\footnote{
Note that, with a little abuse of notation, the symbols for the functions $\{u,v\}$ have been maintained even though the independent variable has changed from $x_0$ to $s$.
}.
Note also that, as $\delta u'(s)$ and $\delta v''(s)$ are, respectively, the incremental axial strain and curvature, the corresponding coefficients are effectively the \textit{current value of axial and bending stiffness}, so that they can be concisely denoted as $\psi_\lambda''(\lambda_0)\lambda_0 = A(\lambda_0)$ and $\psi_\chi''(0)\lambda_0 = B(\lambda_0)$, both functions of the current axial stretch $\lambda_0$.

As this second-order functional has been derived from the large deformation beam theory, it describes the correct incremental response superimposed upon a give pre-stretched state.
Therefore, the correct form of the equilibrium equations governing the incremental displacements can be derived employing the following \textit{incremental potential energy}
\begin{equation}
\label{eq:potential_energy_elastica_incremental}
    \mV(u, v) =
    \frac{1}{2} \int_0^{l} A(\lambda_0) \, u'(s)^2 ds
    +\frac{1}{2} \int_0^{l}\left(P \, v'(s)^2 + B(\lambda_0) \, v''(s)^2 \right) ds \,,
\end{equation}
where now the fields $\{u(s),v(s)\}$ are the current incremental fields and the dependence of the current stiffnesses $A(\lambda_0)$ and $B(\lambda_0)$ on the current axial stretch is highlighted.
The governing equations~\eqref{eq:governing_beam_EB} are directly obtained from~\eqref{eq:potential_energy_elastica_incremental} by imposing the first variation $\delta\mV(u, v)$ to vanish.
Note that when $B(\lambda_0)$ is assumed constant and the axial term of~\eqref{eq:potential_energy_elastica_incremental} is dropped, the usual result of an inextensible Euler-Bernoulli beam is recovered.

\subsection{Example of a rod made of incompressible hyperelastic materials}
The incremental potential~\eqref{eq:potential_energy_elastica_incremental} has been derived with reference to the elastica defined by two arbitrary strain-energy functions governing the current stiffnesses $A(\lambda_0)$ and $B(\lambda_0)$.
It is now shown that these two parameters can be evaluated explicitly for every incompressible elastic material selected to model the lattice's rods.

The incremental constitutive response of a rectangular block of incompressible material, deformed under plane strain and initially isotropic can be described (when a uniaxial stress state prevails in the current configuration) through \cite{bigoni_2012}
\begin{align*}
    \dot{S}_{11} &= (2\mu_* - T_1) \deriv{u_1}{x_1} + \dot{p} \,, &
    \dot{S}_{22} &= 2\mu_* \deriv{u_2}{x_2} + \dot{p} \,,
\end{align*}
where $\dot{S}_{ij}$ is the increment of the first Piola-Kirchhoff stress and $u_i$ are the incremental displacements, $\mu_*$ the incremental modulus (corresponding to shearing inclined at $45^\circ$ with respect to the axes), $T_1$ the current uniaxial Cauchy stress ($T_2=0$), and $\dot{p}$ the incremental Lagrange multiplier associated to the incompressibility constraint.
Assuming that plane stress prevails incrementally, $\dot{S}_{22} = 0$, and using the incompressibility constraint, $\dot{p}$ can be eliminated to yield
\begin{equation}
\label{eq:S_11}
    \dot{S}_{11} = (4\mu_* - T_1) \deriv{u_1}{x_1} \,.
\end{equation}

By considering the incremental equilibrium along the $x_1$ direction
\begin{equation*}
    \deriv{\dot{S}_{11}}{x_1} + \deriv{\dot{S}_{12}}{x_2} = 0 \,,
\end{equation*}
an integration over the current thickness $h$ of the block and a subsequent substitution of Eq.~\eqref{eq:S_11} lead to 
\begin{equation}
\label{eq:axial_incremental_equilibrium}
    \int_{-h/2}^{h/2} \deriv{\dot{S}_{11}}{x_1} \,dx_2 = (4\mu_* - T_1) \int_{-h/2}^{h/2} \nderiv{u_1}{x_1}{2}\,dx_2 = 0 \,,
\end{equation}
where the assumption of vanishing traction at $x_2=\pm h/2$ has been used.

The incremental flexural equilibrium can also be retrieved.
To this purpose, for a perturbation from the current unixial stress state, Biot~\cite{biot_1965} has shown that the incremental equilibrium requires
\begin{equation}
\label{eq:flexural_incremental_equilibrium}
    \nderiv{}{x_1}{2} \int_{-h/2}^{h/2} x_2 \dot{S}_{11}\,dx_2 + T_1 \nderiv{}{x_1}{2} \int_{-h/2}^{h/2} u_2\,dx_2 = 0,
\end{equation}
where the first integral can be recognized to be the incremental bending moment.

By adopting the incremental kinematics of an Euler-Bernoulli beam (satisfying the unshearability condition)
\begin{equation}
\label{eq:kinematics_EB}
    u_1(x_1,x_2) = u(x_1) - x_2 \deriv{v(x_1)}{x_1} \,, \qquad
    u_2(x_1,x_2) = v(x_1) \,,
\end{equation}
and using~\eqref{eq:S_11}, the axial and flexural equilibrium equations~\eqref{eq:axial_incremental_equilibrium} and~\eqref{eq:flexural_incremental_equilibrium} become
\begin{subequations}
\label{eq:incremental_equilibrium}
\begin{align}
    \label{eq:incremental_equilibrium_u}& (4\mu_* - T_1) h \nderiv{u(x_1)}{x_1}{2} = 0 \,, \\
    \label{eq:incremental_equilibrium_v}& (4\mu_* - T_1)\frac{h^3}{12} \nderiv{v(x_1)}{x_1}{4} - T_1 h \nderiv{v(x_1)}{x_1}{2} = 0 \,.
\end{align}
\end{subequations}
By noting that $T_1 h$ is the resultant axial load, so that $T_1 h=P$, the direct comparison between equations~\eqref{eq:incremental_equilibrium} and~\eqref{eq:governing_beam_EB} provides the identification of the current stiffnesses $A(\lambda_0)$ and $B(\lambda_0)$ as
\begin{equation}
\label{eq:current_stiffnesses}
    A(\lambda_0) = (4\mu_*(\lambda_0) - T_1(\lambda_0)) h(\lambda_0) \,, \qquad B(\lambda_0) = (4\mu_*(\lambda_0) - T_1(\lambda_0)) h(\lambda_0)^3/12 \,,
\end{equation}
where the explicit dependence on the current pre-stretch $\lambda_0$ has been highlighted.
For instance, for a Mooney-Rivlin material $\mu_*=\mu_0(\lambda_0^2+\lambda_0^{-2})/2$ and $T_1=\mu_0(\lambda_0^2-\lambda_0^{-2})$, and expressions~\eqref{eq:current_stiffnesses} become
\begin{equation*}
    A(\lambda_0) = \mu_0(\lambda_0 + 3\lambda_0^{-3})h_0 \,, \qquad B(\lambda_0) = \mu_0(\lambda_0^{-1} + 3\lambda_0^{-5})h_0^3/12 \,.
\end{equation*}
with $h_0=h/\lambda_0$ being the initial thickness, and $\mu_0$ the initial shear modulus of the material.

\section{Regime classification of the effective continuum}
\label{sec:regime_classification}
The mathematical classification of the PDE describing the incremental equilibrium of the equivalent solid provides valuable information on the number of localizations available on the elliptic boundary.
In fact, the partial differential equations governing the equilibrium of the effective continuum, in the absence of body forces, 
\begin{equation}
    \label{eq:pde}
    \diver{\fC[\grad \bv]} = \bzero ,
\end{equation}
can be classified according to the following general criterion. 
Referring to a two-dimensional setting, a solution of the system~\eqref{eq:pde} is selected in a wave form, 
\begin{equation}
 \label{eq:wave}
 \bv = \bg \exp[i (x_1 + \Omega x_2)] \,,
\end{equation}
where $\bg$ is the wave amplitude and $\Omega$ a complex angular frequency.
A substitution of~\eqref{eq:wave} in the governing equation~\eqref{eq:pde} yields the following linear algebraic system
\begin{equation*}
    \begin{bmatrix}
        \fC_{1212} \Omega^2 + 2 \fC_{1112} \Omega + \fC_{1111} & \fC_{1222} \Omega^2 + (\fC_{1122} + \fC_{1221}) \Omega + \fC_{1111} \\
        \fC_{1222} \Omega^2 + (\fC_{1122} + \fC_{1221}) \Omega + \fC_{1111} & \fC_{2222} \Omega^2 + 2 \fC_{2122} \Omega + \fC_{2121} \\
    \end{bmatrix}
    \begin{Bmatrix}
        g_1 \\ g_2 
    \end{Bmatrix} = 
    \begin{Bmatrix}
        0 \\ 0 
    \end{Bmatrix} .
\end{equation*}
\begin{figure}[htb!]
    \centering
    \begin{subfigure}{0.3\textwidth}
        \centering
        \caption{\label{fig:classification_domains_10_10_2}$\Lambda_1=\Lambda_2=10,~\alpha=\pi/6$}
        \includegraphics[width=0.98\linewidth]{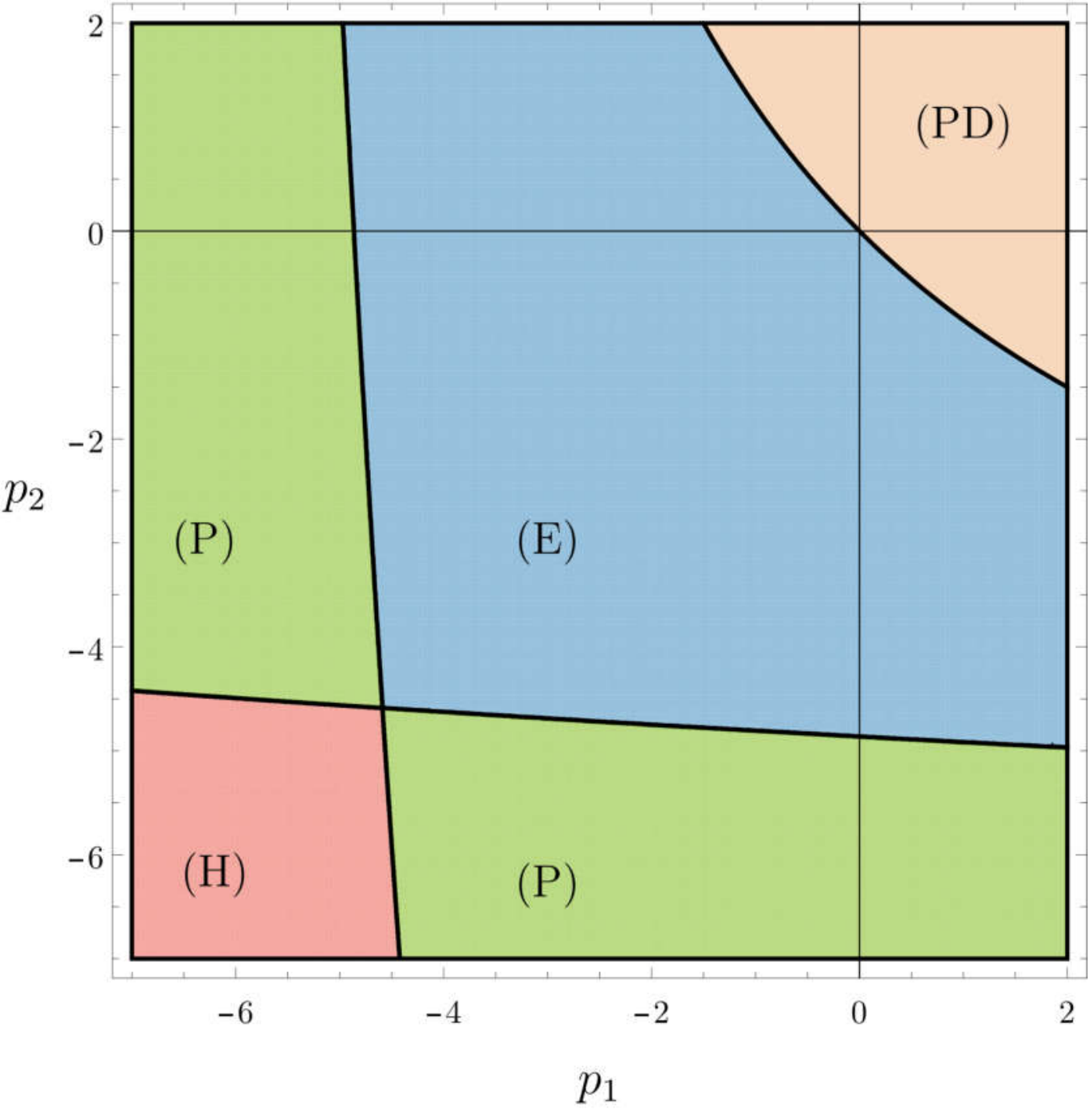}
    \end{subfigure}
    \begin{subfigure}{0.3\textwidth}
        \centering
        \caption{\label{fig:classification_domains_10_10_1}$\Lambda_1=\Lambda_2=10,~\alpha=\pi/8$}
        \includegraphics[width=0.98\linewidth]{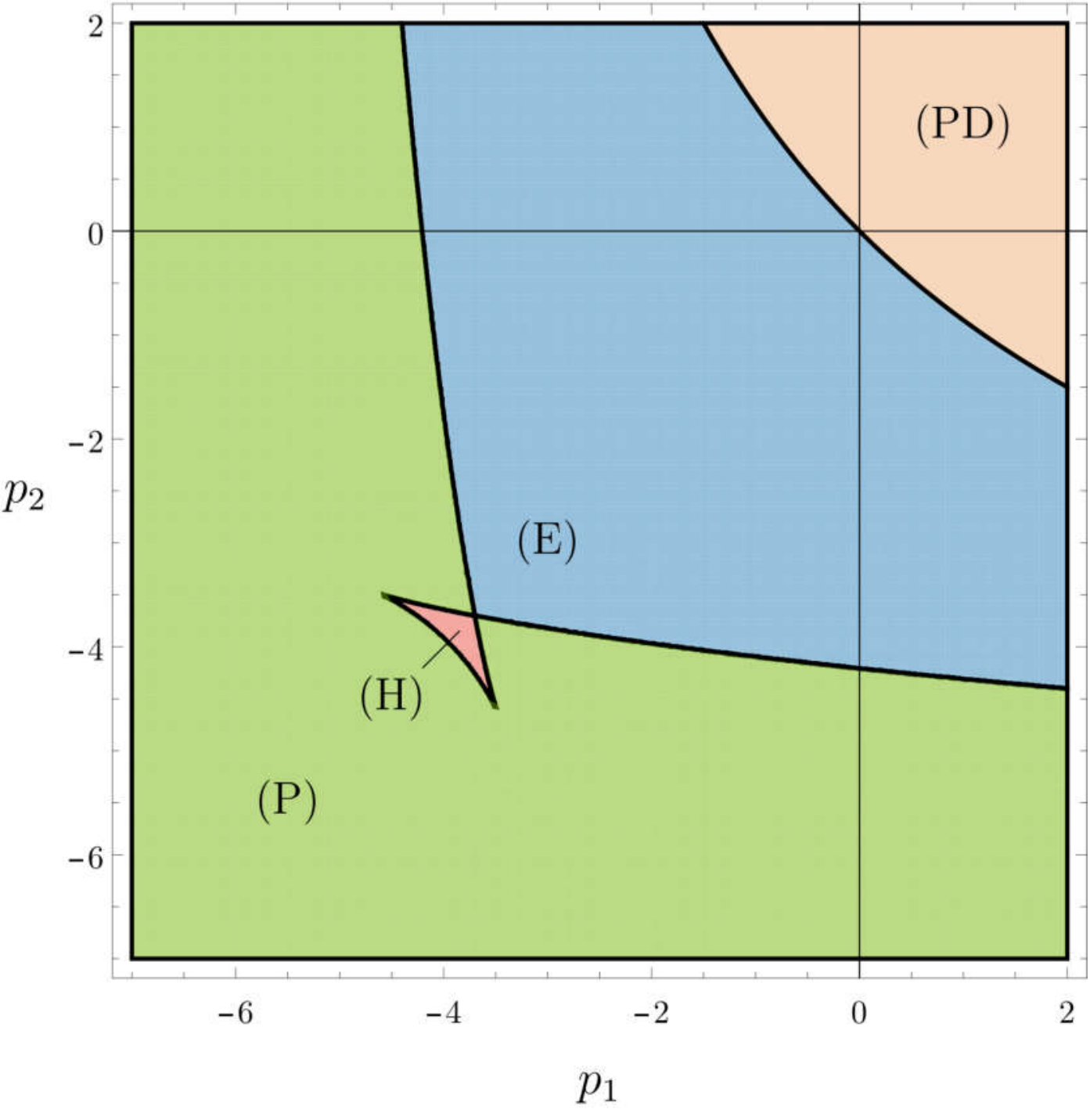}
    \end{subfigure}\\
    \begin{subfigure}{0.3\textwidth}
        \centering
        \caption{\label{fig:classification_domains_7_15_2}$\Lambda_1=7,~\Lambda_2=15,~\alpha=\pi/6$}
        \includegraphics[width=0.98\linewidth]{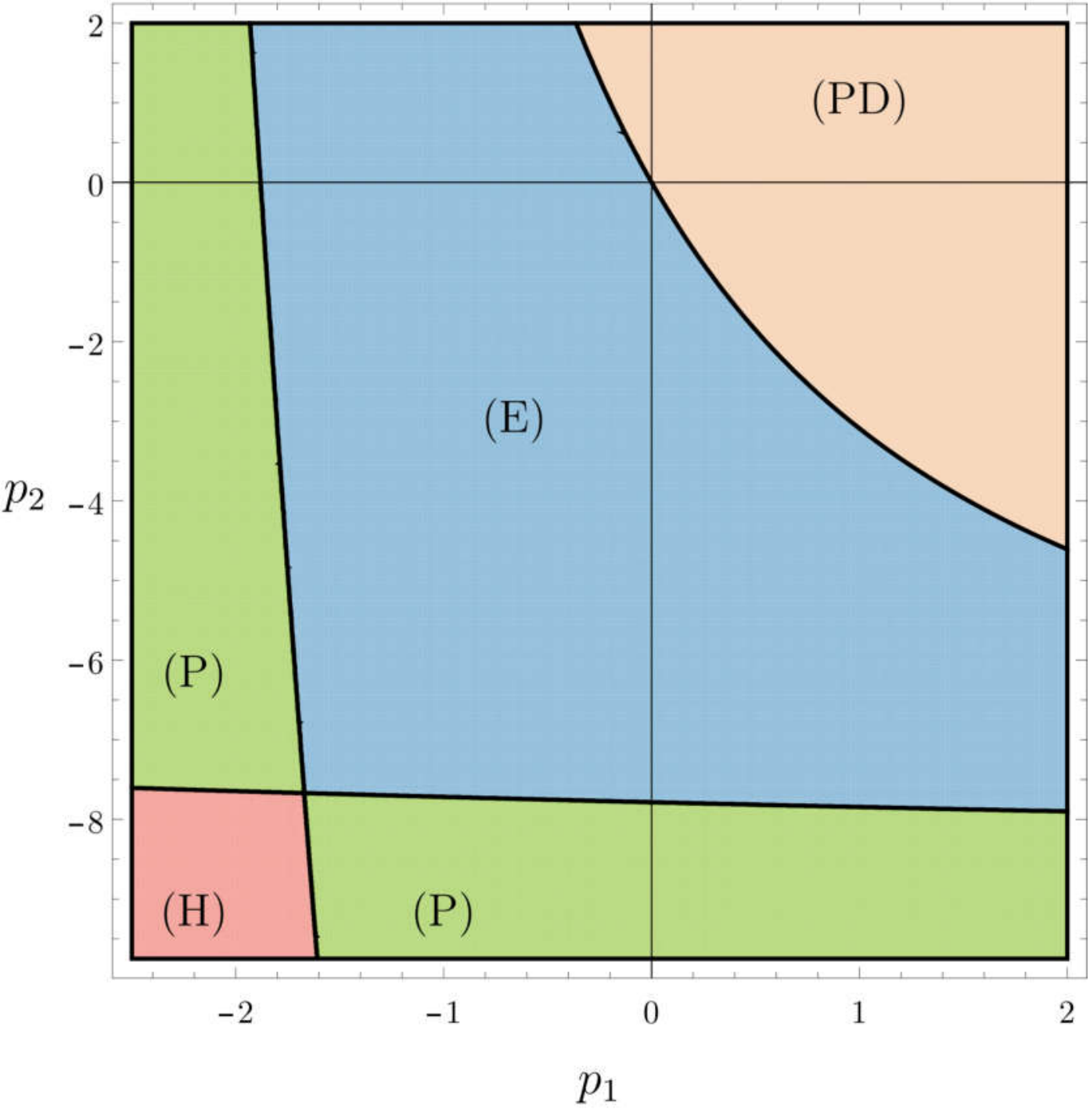}
    \end{subfigure}
    \begin{subfigure}{0.3\textwidth}
        \centering
        \caption{\label{fig:classification_domains_7_15_1}$\Lambda_1=7,~\Lambda_2=15,~\alpha=\pi/8$}
        \includegraphics[width=0.98\linewidth]{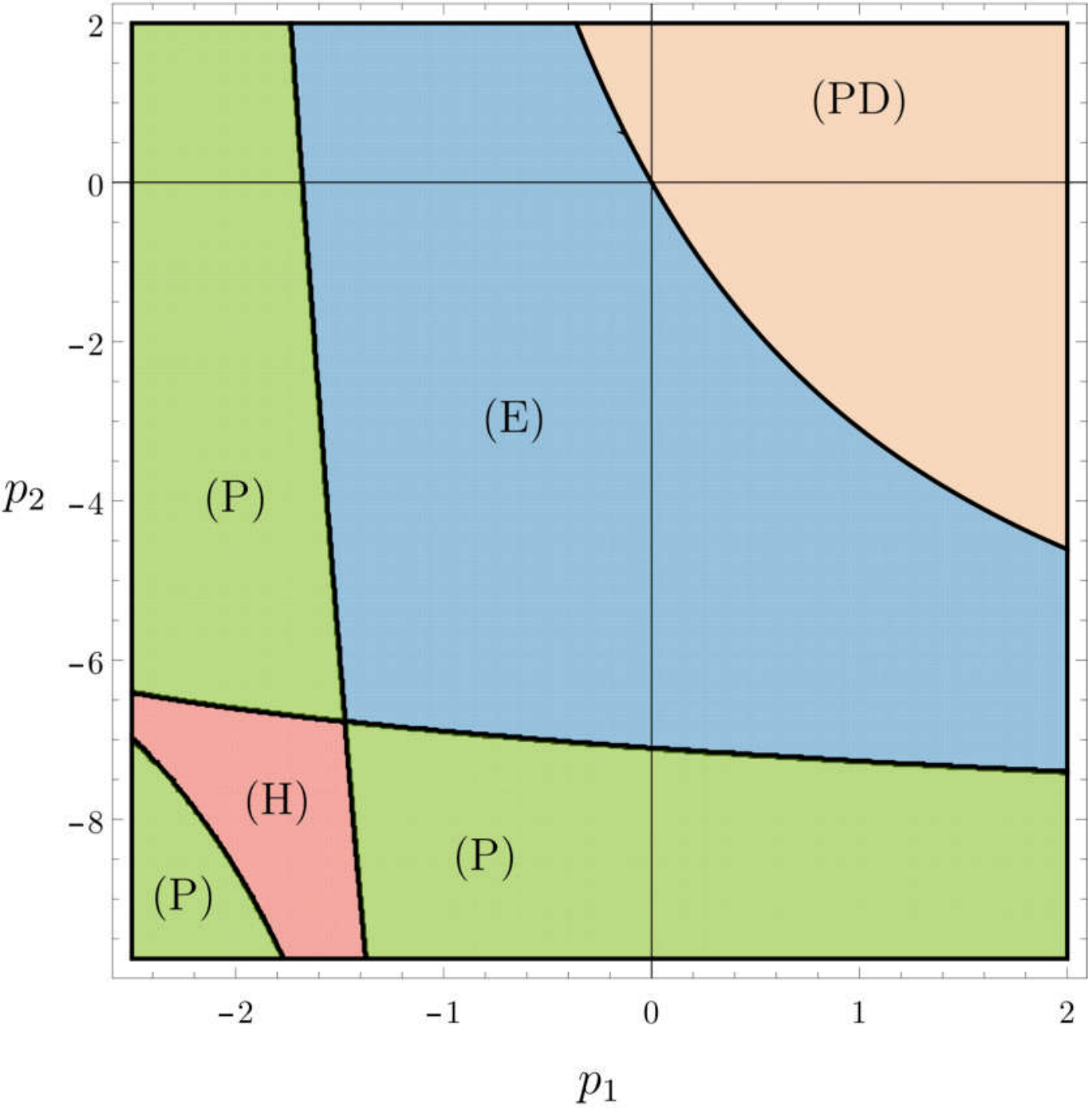}
    \end{subfigure}
    \caption{\label{fig:classification_domains}
    Regime classification of equilibrium PDE for the effective continuum equivalent of a rhombic elastic lattice as that sketched in Fig.~\ref{fig:geometry_grid_and_unit_cell} but without diagonal springs.
    The upper parts (\subref{fig:classification_domains_10_10_2}, \subref{fig:classification_domains_10_10_1}) refer to an orthotropic material ($\Lambda_1=\Lambda_2=10$) material, while the lower parts  (\subref{fig:classification_domains_7_15_2}, \subref{fig:classification_domains_7_15_1}) to a completely anisotropic material ($\Lambda_1=7, ~\Lambda_2=15$). 
    The left parts (\subref{fig:classification_domains_10_10_2}, \subref{fig:classification_domains_7_15_2}) refer to a grid with inclination $\alpha=\pi/6$ and the right $\alpha=\pi/8$ (\subref{fig:classification_domains_10_10_1}, \subref{fig:classification_domains_7_15_1}). 
    Note that the Hyperbolic and Elliptic regions `touch' at a point. 
    }
\end{figure}
This system has non-trivial solutions if and only if the determinant of the coefficient matrix is equal to zero, a condition yielding the characteristic equation in the form of a quartic
\begin{equation}
    \label{eq:quartic}
    a_4 \Omega^4 + 2 a_3 \Omega^3 + a_2 \Omega^2 + 2 a_1 \Omega + a_0 = 0 \,,
\end{equation}
where
\begin{align*}
 a_4 &= \fC_{1222}^2 - \fC_{1212} \fC_{2222} \,, \\
 a_3 &= \left(\fC_{1122} + \fC_{1221}\right) \fC_{1222} - \fC_{1212} \fC_{2122} - \fC_{1112} \fC_{2222} \,, \\
 a_2 &= \left(\fC_{1122} + \fC_{1221}\right)^2 + 2 \fC_{1121} \fC_{1222} - \fC_{1212} \fC_{2121} - 4 \fC_{1112} \fC_{2122} - \fC_{1111} \fC_{2222} \,, \\
 a_1 &= -\fC_{1121} \left(\fC_{1122} + \fC_{1221}\right) + \fC_{1112} \fC_{2121} + \fC_{1111} \fC_{2122} \,, \\
 a_0 &= \fC_{1121}^2 - \fC_{1111} \fC_{2121} \,.
\end{align*}
The nature of the roots $\Omega_j$ of the quartic~\eqref{eq:quartic} defines the regime classification according to the following nomenclature~\cite{bigoni_2012,renardy_2004}:
\begin{itemize}
 \item In the elliptical regime all the roots $\Omega_j$ are complex;
 \item In the hyperbolic regime all the roots $\Omega_j$ are real;
 \item In the parabolic regime two roots are real and two roots are complex.
\end{itemize}
According to this criterion, the regimes for the grid-like lattice of prestressed elastic rods have been classified and the results are shown in Fig.~\ref{fig:classification_domains} for both orthotropic ($\Lambda_1=\Lambda_2=10$) and anisotropic ($\Lambda_1=7,~\Lambda_2=15$) case.
For the sake of brevity, only the case $\kappa=0$ is reported.

Note that the elliptic region `touches' the hyperbolic domain only at a point.

\end{document}

%% file: definitions.tex
\DeclareMathOperator{\diver}{div}

\DeclareMathOperator{\grad}{grad}

\DeclarePairedDelimiter{\abs}{\lvert}{\rvert}
\DeclarePairedDelimiter{\norm}{\lVert}{\rVert}

\newcommand{\scalp}{\bm \cdot}
\newcommand{\trans}[1]{#1^{\intercal}}

\newcommand{\conjtrans}[1]{#1^{\mathsf{H}}}
\newcommand{\deriv}[2]{\frac{\partial #1}{\partial #2}}
\newcommand{\nderiv}[3]{\frac{\partial^{\,#3} #1}{\partial #2^{\,#3}}}

\newcommand{\Reals}{\mathbb{R}}

\newcommand{\bzero}{\bm 0}
\newcommand{\ba}{\bm a}
\newcommand{\bb}{\bm b}

\newcommand{\be}{\bm e}
\newcommand{\bef}{\bm f}
\newcommand{\bg}{\bm g}

\newcommand{\bk}{\bm k}

\newcommand{\bn}{\bm n}

\newcommand{\bp}{\bm p}
\newcommand{\bq}{\bm q}

\newcommand{\bt}{\bm t}
\newcommand{\bu}{\bm u}
\newcommand{\bv}{\bm v}

\newcommand{\bx}{\bm x}

\newcommand{\bA}{\bm A}
\newcommand{\bB}{\bm B}
\newcommand{\bC}{\bm C}
\newcommand{\bD}{\bm D}
\newcommand{\bE}{\bm E}

\newcommand{\bI}{\bm I}

\newcommand{\bK}{\bm K}
\newcommand{\bL}{\bm L}

\newcommand{\bN}{\bm N}

\newcommand{\bP}{\bm P}

\newcommand{\bS}{\bm S}
\newcommand{\bT}{\bm T}

\newcommand{\bW}{\bm W}

\newcommand{\bZ}{\bm Z}

\newcommand{\fC}{\mathbb C}

\newcommand{\fE}{\mathbb E}

\newcommand{\mB}{\mathcal B}
\newcommand{\mC}{\mathcal C}

\newcommand{\mE}{\mathcal E}

\newcommand{\mG}{\mathcal G}

\newcommand{\mV}{\mathcal V}
\newcommand{\mW}{\mathcal W}